\documentclass[%
  reprint,
  superscriptaddress,
  amsmath,
  amssymb,
  aps,
  prx,
  longbibliography,
  floatfix,
]{revtex4-2}
\usepackage{xspace, soul, color, colortbl}
\usepackage[dvipsnames]{xcolor}
\usepackage{amsmath,amssymb,wasysym,amsthm,marvosym,stackengine}
\usepackage{hyperref}
\usepackage[capitalize]{cleveref}
\usepackage{verbatim}
\usepackage{listings}
\usepackage{tikz}
\usetikzlibrary{shapes.multipart}
\usetikzlibrary{calc}
\usetikzlibrary{arrows.meta}
\usepackage{relsize}
\usepackage{subcaption}
\usepackage{mathtools}
\usepackage[noend]{algpseudocode}
\usepackage{algorithm}

\usepackage{algorithmicx}
\algrenewcommand\algorithmicrequire{\textbf{Input:}}

\usepackage{cellspace}
\newcommand{\hyperblossom}{\textsf{\textsc{HyperBlossom}}\xspace}
\newcommand{\hyperion}{\textsf{\textsc{Hyperion}}\xspace}

\newcommand{\nosection}[1]{\vspace{3pt}\noindent\textbf{#1}}

\definecolor{lightgray}{gray}{0.6}
\definecolor{lightblue}{rgb}{0.9,0.9,1}
\definecolor{aqua}{rgb}{0.0, 1.0, 1.0}

\makeatletter
\newcommand{\definitionlabel}[2]{%
  \protected@write \@auxout {}{\string \newlabel {def:#1}{{\textit{#2}}{\thepage}{#2}{def:#1}{}} }%
  \protected@write \@auxout {}{\string \newlabel {def:#1s}{{\textit{#2s}}{\thepage}{#2}{def:#1}{}} }%
  \hypertarget{def:#1}{\noindent\textit{Definition: #2.}}
}
\newcommand{\definitionlabelvalid}[2]{%
  \protected@write \@auxout {}{\string \newlabel {def:#1}{{\textit{#2}}{\thepage}{#2}{def:#1}{}} }%
  \protected@write \@auxout {}{\string \newlabel {def:valid}{{\textit{Valid}}{\thepage}{#2}{def:#1}{}} }%
  \hypertarget{def:#1}{\noindent\textit{Definition: #2.}}
}
\newcommand{\definitionlabelmatrix}[2]{%
  \protected@write \@auxout {}{\string \newlabel {def:#1-matrix}{{\textit{#2 Matrix}}{\thepage}{#2 Matrix}{def:#1-matrix}{}} }%
  \protected@write \@auxout {}{\string \newlabel {def:#1-matrices}{{\textit{#2 Matrices}}{\thepage}{#2 Matrix}{def:#1-matrix}{}} }%
  \hypertarget{def:#1-matrix}{\noindent\textit{Definition: #2 Matrix.}}
}
\makeatother
\newcommand{\refdef}[1]{\ref{def:#1}}

\makeatletter
\newcommand{\definelemma}[2]{%
  \protected@write \@auxout {}{\string \newlabel {lemma:#1}{{\textit{Lemma: #2}}{\thepage}{#2}{lemma:#1}{}} }%
  \hypertarget{lemma:#1}{\noindent\textit{Lemma: #2.}}
}
\makeatother
\newcommand{\reflemma}[1]{\ref{lemma:#1}}

\makeatletter
\newcommand{\nosectionlabel}[2]{%
  \protected@write \@auxout {}{\string \newlabel {#1}{{\textbf{#2}}{\thepage}{#2}{#1}{}} }%
  \hypertarget{#1}{\noindent\textbf{#2.}}
}
\makeatother

\makeatletter
\newcommand{\defineeqs}[2]{%
  \protected@write \@auxout {}{\string \newlabel {eqs:#1}{{#1}{\thepage}{#1}{eqs:#1}{}} }%
  \hypertarget{eqs:#1}{\vspace{2ex}\noindent\makebox[\linewidth-1mm]{{\centering \textbf{#1 (#2)}}}}
}
\makeatother
\newcommand{\refeqs}[1]{\ref{eqs:#1}}

\newcommand{\dns}{\!\!}  
\newcommand{\qns}{\!\!\!\!}  
\newcommand{\hns}{\qns\dns}  
\newcommand{\dqns}{\qns\qns}  
\newcommand{\tqns}{\qns\qns\qns}  
\newcommand{\qqns}{\qns\qns\qns\qns}  

\makeatletter
\newcommand{\definetheorem}[2]{%
  \protected@write \@auxout {}{\string \newlabel {theorem:#1}{{\textbf{Theorem: #2}}{\thepage}{#2}{theorem:#1}{}} }%
  \hypertarget{theorem:#1}{\vspace{1ex}\noindent\textbf{Theorem: #2.}}
}
\makeatother
\newcommand{\reftheorem}[1]{\ref{theorem:#1}}

\makeatletter
\newcommand{\definealgorithm}[2]{%
  \protected@write \@auxout {}{\string \newlabel {algorithm:#1}{{\textbf{Algorithm: #2}}{\thepage}{#2}{algorithm:#1}{}} }%
  \hypertarget{algorithm:#1}{\vspace{1ex}\noindent\textbf{Algorithm: #2.}}
}
\makeatother

\definecolor{primalinterfacecolor}{RGB}{29,108,171}
\definecolor{dualinterfacecolor}{RGB}{208,35,36}
\definecolor{paritydecodercolor}{RGB}{227,237,238}

\newcommand{\relaxing}{\textit{relaxing}\xspace}
\newcommand{\Relaxing}{\textit{Relaxing}\xspace}

\newcommand{\Clustering}{\textit{Clustering}\xspace}

\usepackage{graphicx}
\usepackage{dcolumn}
\usepackage{bm}
\begin{document}


\title{Minimum-Weight Parity Factor Decoder for Quantum Error Correction}

\author{Yue Wu}
\affiliation{Yale University, Department of Computer Science, New Haven, CT 06511}
\author{Binghong Li}
\affiliation{Yale University, Department of Computer Science, New Haven, CT 06511}
\author{Kathleen (Katie) Chang}
\affiliation{Yale University, Department of Applied Physics, New Haven, CT 06520}
\author{Shruti Puri}
\affiliation{Yale University, Department of Applied Physics, New Haven, CT 06520}
\author{Lin Zhong}
\affiliation{Yale University, Department of Computer Science, New Haven, CT 06511}



\thispagestyle{plain}
\pagestyle{plain}

\maketitle

\newcommand{\lemmaClusterOfDefect}[1]{
  \definelemma{#1}{Cluster of Defect Vertex}
  Each defect vertex $v \in D$ belongs to a unique \refdef{cluster} $C = \mathcal{C}(v), v \in V_C$.
}

\newcommand{\lemmaClusterofHyperblossom}[1]{
  \definelemma{#1}{Cluster of Hyperblossom}
  Each \refdef{hyperblossom} $S \in \mathcal{B}$ belongs to a unique \refdef{cluster} $C = \mathcal{C}(S), S \in \mathcal{B}_C$.
}

\newcommand{\lemmaClusterNonOverlapping}[1]{
  \definelemma{#1}{Cluster Non-overlapping}
  \refdef{clusters} have exclusive vertices, edges or \refdef{hyperblossoms}. $\forall C_1, C_2 \in \mathcal{C}, C_1 \neq C_2 \rightarrow V_{C_1} \cap V_{C_2} = \varnothing, E_{C_1} \cap E_{C_2} = \varnothing, \mathcal{B}_{C_1} \cap \mathcal{B}_{C_2} = \varnothing$.
}

\newcommand{\lemmaHairMatrixOddRowExistence}[1]{
  \definelemma{#1}{Hair Matrix Odd Row Existence}
  In a \refdef{hair-matrix}, a row of which the right most value is 1 is called an \emph{Odd} row. An Odd row always exists in a \refdef{hair-matrix}.
}

\newcommand{\lemmaSingleHairOptimalityNullityLetwo}[1]{
  \definelemma{#1}{SingleHair Optimality on Nullity$_{\le 1}$ Hypergraphs}
  For a decoding hypergraph whose incidence matrix has a nullity of 0 or 1, the \emph{SingleHair} \refdef{relaxer}-finding algorithm is optimal.
}

\newcommand{\lemmaUniqueRowIfNotARelaxer}[1] {
  \definelemma{#1}{Unique Row if Not a Relaxer}
  When $E^- = \varnothing$, i.e., there is no \refdef{relaxer} corresponding to an Odd row, then there is only one row.
}

\newcommand{\lemmaSubgraphOSubset}[1] {
  \definelemma{#1}{Subgraph $\mathcal{O}' \subseteq \mathcal{O}$}
  Given a decoding hypergraph $G = (V, E)$ and a subgraph $G' = (V', E')$ with $V' \subseteq V$, $E' \subseteq E[V']$, the set of \refdef{invalid} subgraphs $\mathcal{O}'$ of $G'$ is a subset of the set of \refdef{invalid} subgraphs $\mathcal{O}$ of $G$.
}

\newcommand{\lemmaClusterFeasibleDirection}[1] {
  \definelemma{#1}{Cluster Feasible Direction is Globally Feasible}
  Given a decoding hypergraph $G = (V, E)$ and a \refdef{cluster} $C \in \mathcal{C}$, if a \refdef{direction} $\Delta\vec{y}$ is \emph{Feasible} for the subgraph $C = (V_C, E_C)$, then it is also \emph{Feasible} for $G$.
}

\newcommand{\lemmaNullityIsHereditary}[1] {
  \definelemma{#1}{Nullity$_{\le 1}$ is a hereditary property}
  Any subgraph of a nullity$_{\le 1}$ hypergraph is also nullity$_{\le 1}$
}

\newcommand{\theoremilpequalmwpf}[1]{%
  \definetheorem{#1}{$\boldsymbol{\min\text{ILP} = \min\text{MWPF}}$}%
}

\makeatletter
\newcommand{\definemwpfcondition}[3]{%
  \protected@write \@auxout {}{\string \newlabel {condition:#1}{{\textbf{#2}}{\thepage}{#2}{condition:#1}{}} }%
  \protected@write \@auxout {}{\string \newlabel {condition:#1-short}{{\textbf{Certifiability Condition}}{\thepage}{#2}{condition:#1}{}} }%
  \hypertarget{condition:#1}{\vspace{1ex}\noindent\textbf{#3}}
}
\makeatother
\newcommand{\conditionmwpf}[1]{%
  \definemwpfcondition{#1}%
  {$\boldsymbol{\min\text{LP} = \min\text{ILP}}$}%
  {Certifiability Condition: $\boldsymbol{\min\text{LP} = \min\text{ILP}}$}%
}

\newcommand{\theoremRelaxerExistenceOrTrivialDirection}[1]{
  \definetheorem{#1}{Relaxing}
  Given a suboptimal \refeqs{DLP} $\vec{y}$, there exists a \refdef{relaxer} or a \refdef{trivial-direction}, or both.
}

\newcommand{\theoremBatchedRelaxing}[1] {
  \definetheorem{#1}{Batch Relaxing} Given \refdef{relaxers} $R'_i[T], i=1,...,n$, if there exists a \refdef{feasible-direction} $\Delta\vec{y}[T \setminus(\cup_i \mathcal{R}'_i)]$, we can compose a \refdef{feasible-direction} $\Delta'\vec{y}[T]$ such that $\sum \Delta'\vec{y} \ge \sum \Delta\vec{y}$. In the special case where $\Delta\vec{y}$ is a \refdef{relaxer},  $\Delta'\vec{y}$ is also a \refdef{relaxer} $R'[T]$ such that $\mathcal{R}' \supseteq \mathcal{R}$.
}

\newcommand{\theoremClusterOptimalityCriteria}[1]{
  \definetheorem{#1}{Optimality Criteria of Clusters}
  When all the \refdef{clusters} are \refdef{locally-optimal-clusters}, the union of all the local \refeqs{MWPF} and \refeqs{DLP} solutions are the global optimal \refeqs{MWPF} $\mathcal{E} = \cup_{C \in \mathcal{C}} \mathcal{E}_C$ and global optimal \refeqs{DLP} solution $\vec{y} = \bigcup_{C \in \mathcal{C}} \{ S: y_S | S \in \mathcal{B}_C \}$, respectively.
}

\newcommand{\theoremRelaxerExistenceOrInvalidCluster}[1]{
  \definetheorem{#1}{Relaxing with Clusters}
  Given a suboptimal \refeqs{DLP} solution $\vec{y}$, there exists a \refdef{relaxer} or an \refdef{invalid} \refdef{cluster}, or both.
}

\newcommand{\theoremHyperBlossomOptimality}[1] {
  \definetheorem{#1}{HyperBlossom Algorithm Optimality}
  There exists a \refdef{relaxer} finder so that the \hyperblossom algorithm equipped with it would find optimal \refeqs{MWPF} and \refeqs{DLP} solutions, if the decoding hypergraph and all its subgraphs satisfy \ref{condition:mwpf}.
}

\newcommand{\theoremBlossomMapping}[1] {
  \definetheorem{#1}{Blossom Dual Function}
  There exists a bijective function $f$ from a \refdef{blossom-dlp} $\vec{y^*}$ to a \refeqs{DLP} solution of the \hyperblossom algorithm $\vec{y} = f(\vec{y^*})$ while satisfying two conditions: their dual objectives are the same $\sum_{S \in \mathcal{O}} y_S = \sum_{S^* \in \mathcal{O}^*} y^*_{S^*}$, and, when a syndrome-graph edge $(u, v)$ is tight~\cite{wu2023qce} in $\vec{y^*}$, then the minimum-weight path between $u$ and $v$ in the \refeqs{DLP} solution $\vec{y} = f(\vec{y^*})$ consists of all \refdef{tight-edges}.
}

\newcommand{\theoremAlternatingTreeReconstruction}[1] {
  \definetheorem{#1}{Alternating Tree Reconstruction}
  It is possible to reconstruct alternating trees from a \refdef{blossom-dlp} $\vec{y^*}$.
}

\newcommand{\theoremSimpleGraphOptimality}[1] {
  \definetheorem{#1}{Simple Graphs \ref{condition:mwpf}}
}

\newcommand{\theoremSingledofGraphOptimality}[1] {
  \definetheorem{#1}{Nullity$_{\le 1}$ Hypergraphs \ref{condition:mwpf}}
}

\section*{abstract}

Fast and accurate quantum error correction (QEC) decoding is crucial for scalable fault-tolerant quantum computation.
Most-Likely-Error (MLE) decoding, while being near-optimal, is intractable on general quantum Low-Density Parity-Check (qLDPC) codes and typically relies on approximation and heuristics.
We propose HyperBlossom, a unified framework that formulates MLE decoding as a Minimum-Weight Parity Factor (MWPF) problem and generalizes the blossom algorithm to hypergraphs via a similar primal-dual linear programming model with certifiable proximity bounds.
HyperBlossom unifies all the existing graph-based decoders like (Hypergraph) Union-Find decoders and Minimum-Weight Perfect Matching (MWPM) decoder, thus bridging the gap between heuristic and certifying decoders.

We implement HyperBlossom in software, namely Hyperion.
Hyperion achieves a 4.8x lower logical error rate compared to the MWPM decoder on the distance-11 surface code and 1.6x lower logical error rate compared to a fine-tuned BPOSD decoder on the $[[90, 8, 10]]$ bivariate bicycle code under code-capacity noise.
It also achieves an almost-linear average runtime scaling on both the surface code and the color code, with numerical results up to sufficiently large code distances of 99 and 31 for code-capacity noise and circuit-level noise, respectively.

\section{Introduction}

Quantum error correction (QEC) is crucial for realizing scalable and fault-tolerant quantum computation.
A fundamental challenge in QEC is decoding.
Most-Likely-Error (MLE) decoding provides near-optimal accuracy by finding the error pattern with maximum probability.
While MLE decoding for surface codes can be efficiently solved as a Minimum-Weight Perfect Matching (MWPM)~\cite{dennis2002topological} problem via the blossom algorithm~\cite{edmonds1965paths,higgott2025sparse,wu2023qce}, extending this approach to general quantum Low-Density Parity-Check (qLDPC) codes remains intractable given its NP-hardness on general hypergraphs~\cite{berlekamp1978inherent}.

Existing qLDPC decoders, such as Union-Find (UF) \cite{delfosse2021almost,delfosse2022toward} and Belief Propagation with Ordered Statistics (BPOSD) \cite{roffe2020decoding,Roffe_LDPC_Python_tools_2022} decoders, offer fast heuristic solutions but lack rigorous guarantees and often require careful parameter tuning.
Meanwhile, certifying decoders that provide rigorous optimality or proximity bounds have been largely limited to special cases, such as surface codes.
This gap between heuristic efficiency and certifiable correctness has hindered the design of fast and accurate QEC decoders.

We address this challenge by introducing a unified mathematical framework, called \hyperblossom, that formulates the MLE decoding problem of qLDPC codes as a Minimum-Weight Parity Factor (MWPF) problem on the decoding hypergraph.
This framework generalizes the principles of the blossom algorithm to hypergraphs through three key innovations:

\begin{itemize}
  \item \textbf{Primal-dual linear programming (LP)} formulation that provides a certifiable proximity bound and guarantees optimality for certain hypergraphs, e.g., simple graphs and nullity$_{\le 1}$ hypergraphs.
  \item \textbf{Relaxing} and modular relaxer-finding algorithms that decompose the complex LP optimization problem into tractable relaxer finders.
  \item \textbf{Clustering} technique that exploits the locality of MLE decoding to achieve an almost-linear average runtime while preserving global optimality under the \ref{condition:mwpf-short}.
\end{itemize}

We propose \emph{SingleHair}, an efficient but suboptimal relaxer-finding algorithm for general hypergraphs.
Our software implementation, \hyperion, realizes the \hyperblossom framework and \emph{SingleHair} relaxer finder in practice.
In our evaluation, \hyperion achieves a 4.8x lower logical error rate compared to the MWPM decoder on the distance-11 surface code
and 1.6x lower logical error rate compared to a fine-tuned BPOSD decoder on the $[[90, 8, 10]]$ bivariate bicycle code under code-capacity noise.
Note that the MWPM decoder is not an optimal MLE decoder for surface codes with depolarizing noise because it cannot handle the Pauli $Y$ errors, which is why \hyperion achieves higher accuracy.
It also achieves an almost-linear average runtime scaling on both the surface code and the color code, with numerical results up to sufficiently large code distances of 99 and 31 for code-capacity noise and circuit-level noise, respectively.

More importantly, the \hyperblossom framework unifies UF, MWPM, and Hypergraph UF (HUF)~\cite{delfosse2022toward} decoders within a single mathematical formulation.
Within this framework, the MWPM decoder is realized through a \emph{Blossom} relaxer finder, while the UF and HUF decoders correspond to a \emph{UnionFind} relaxer finder.
In particular, UF can be viewed as a special case of HUF on simple graphs, and both can be interpreted as simplified MWPF decoders with a trivial relaxer finder.
Thus, the \hyperblossom framework not only generalizes these decoders but also provides a unified lens through which their designs and performance trade-offs can be understood in previously impossible ways.

\section{Background}

\subsection{Most Likely Error (MLE) Decoding}\label{ssec:MLE}
An MLE decoder aims to determine the error pattern that is most likely to happen given a noise model, while agreeing with the measured syndrome.
Using protocols like Pauli twirling~\cite{emerson2007symmetrized}, a noise model can be decomposed into a set of independent physical error sources, each of which is represented by Pauli operators acting on qubits.
If a physical error occurs, it flips the stabilizer measurements intended to monitor this error source.
Since a stabilizer may monitor multiple error sources, it may be flipped multiple times.
In QEC, a stabilizer measures a \textit{defect} if it is flipped an odd number of times.

\begin{figure}[t]
  \centering
  \begin{subfigure}{.49\linewidth}
    \centering
    \includegraphics[width=0.9\textwidth]{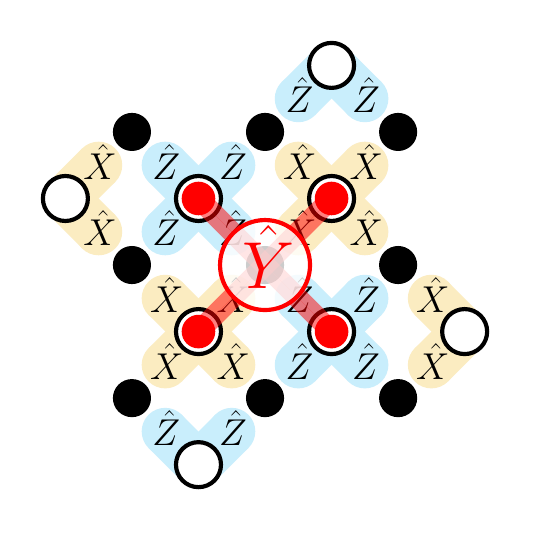}
    \caption{Rotated Surface Code.}
    \label{fig:rotated-surface-code}
  \end{subfigure}
  \begin{subfigure}{.49\linewidth}
    \centering
    \includegraphics[width=0.9\textwidth]{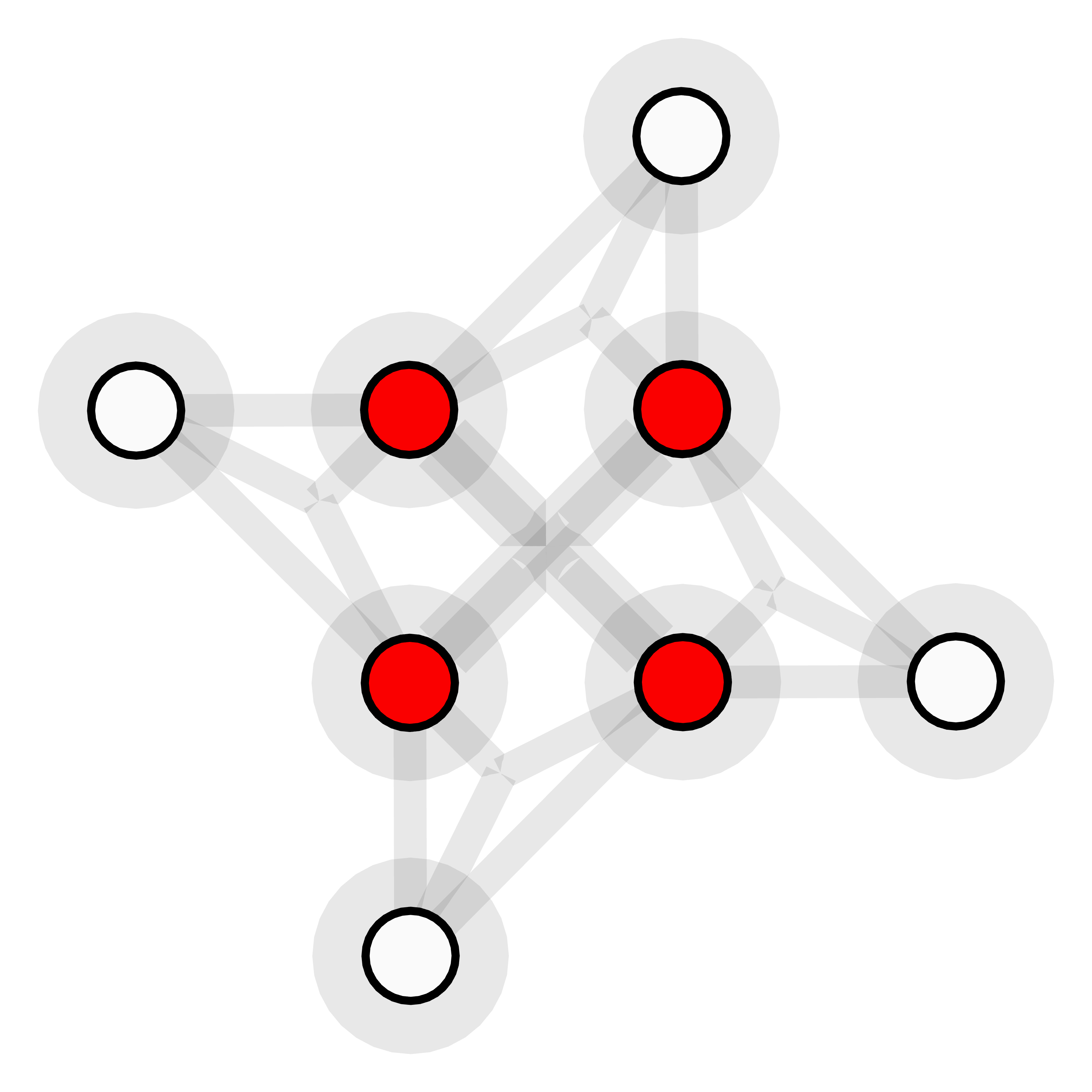}
    \caption{Decoding Hypergraph.}
    \label{fig:decoding-hypergraph}
  \end{subfigure}

  \begin{subfigure}[t]{\linewidth}
    \centering
    \includegraphics[width=1\linewidth]{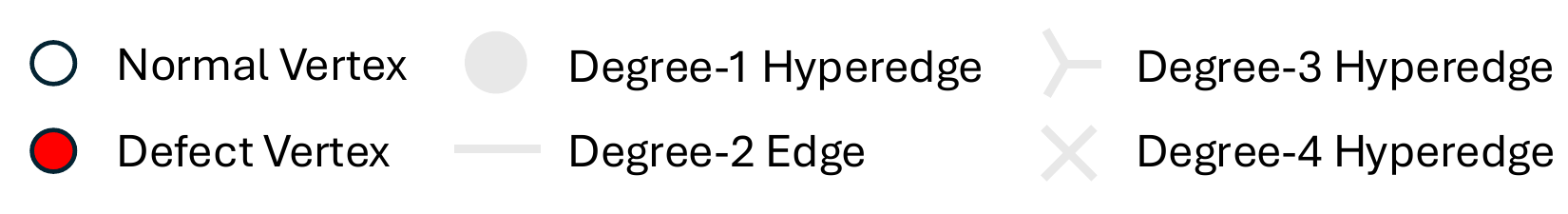}
  \end{subfigure}

  \caption{Example decoding hypergraph of rotated surface code with depolarizing noise. Black circles are data qubits. Red and white circles are ancilla qubits.}
  \label{fig:decoding-hypergraph-example}
\end{figure}

We represent the MLE decoding problem on a hypergraph $G = (V, E)$, where each stabilizer measurement corresponds to a vertex $v \in V$, and each independent physical error corresponds to a hyperedge $e \in E$ that connects the defect vertices it generates when it occurs alone.
\autoref{fig:decoding-hypergraph-example} shows an example decoding hypergraph.
In our visualization, we fill the defect vertices $D$ with solid red, and keep the non-defect vertices in white.
For a degree-1 hyperedge $e = \{ v \}$, we draw a circle centered at $v$.
For other hyperedges, we use the Tanner graph visualization by connecting every vertex $v \in e$ to a center point (hidden for visual simplicity).

MLE decoding on general decoding hypergraphs is notoriously difficult given its NP-hardness~\cite{berlekamp1978inherent}.

A possible solution for the MLE decoding problem is a subset of physical errors, called the error pattern $\mathcal{E} \subseteq E$.
It can be denoted by $\vec{x}=(x_e)_{e\in E} \in \mathbb{F}_2^{|E|}$ where $x_e = 1$ if $e$ is part of the error pattern, i.e., $e \in \mathcal{E}$. The measured syndrome denoted by the defects $D \subseteq V$ puts constraints on the solutions.
Let $\mathcal{D}(\mathcal{E})$ denote the defects of the error pattern $\mathcal{E}$, which is defined below.

\vspace{1ex}\definitionlabel{defect-error-pattern}{Defects of Error Pattern} Given any error pattern $\mathcal{E} \subseteq E$, $\mathcal{D}(\mathcal{E})$ is the set of defects it generates.
\begin{align*}
  \mathcal{D}(\mathcal{E}) = \{ v \in V |\ \mathcal{E} \cap E(v)\ \text{has an odd cardinality} \}
\end{align*}

\vspace{1ex}\definitionlabel{parity-factor}{Parity Factor}
Given syndrome $D$, a parity factor is an error pattern $\mathcal{E} \subseteq E$ such that $\mathcal{D}(\mathcal{E}) = D$.

\vspace{1ex}
This term was first introduced in~\cite{lovasz1972factorization} and then further discussed in~\cite{yu2010graph,akiyama2011factors}.
The general form, known as the parity $(g, f)$-factor~\cite{akiyama2011factors}, describes a subgraph $\mathcal{E} \subseteq E$ such that for every vertex $v \in V$, the degree satisfies $g(v) \le |\mathcal{E} \cap E(v)| \le f(v)$ and $|\mathcal{E} \cap E(v)| \equiv g(v) \mod 2$.
In the decoding context, only the parity constraint is enforced, which corresponds to setting $f(v) = +\infty$ and $g(v) = 1$ for defect vertices and $g(v) = 0$ otherwise.

\vspace{1ex}
The solution for the MLE decoding problem must be a \refdef{parity-factor}.
Additionally, MLE decoding aims to identify the parity factor of the highest probability.
Given the independence between the error sources, the probability of a parity factor $\mathcal{E}$ is:
\begin{align*}
  P(\mathcal{E}) &= \left(\prod_{e \in \mathcal{E}} p_e\right) \left(\prod_{e \in E \setminus \mathcal{E}} (1 - p_e)\right)
  \propto \prod_{e \in \mathcal{E}} \frac{p_e}{1 - p_e}
\end{align*}

We define the weight of an edge so that the sum of the edge weights relates to the probability of the error pattern.
Maximizing probability $P(\mathcal{E})$ is equivalent to minimizing the sum of edge weights in $\mathcal{E}$, i.e., $\min\sum_{e \in \mathcal{E}} w_e$.

\vspace{1ex}\definitionlabel{edge-weight}{Weight of Edge} For an error source of edge $e \in E$ with probability $p_e$, we define its edge weight as:
\begin{align*}
  w_e = \log\frac{1 - p_e}{p_e}
\end{align*}

\vspace{1ex}\definitionlabel{error-pattern-weight}{Weight of Error Pattern} For an error pattern $\mathcal{E} \subseteq E$, we define its weight as $W(\mathcal{E})$.
\begin{align*}
  W(\mathcal{E}) = \sum_{e \in \mathcal{E}} w_e = \log \prod_{e \in \mathcal{E}} \frac{1 - p_e}{p_e} = -\log P(\mathcal{E}) + C
\end{align*}

Thus, the goal of MLE decoding problem is to find a parity factor $\mathcal{E} \subseteq E, \mathcal{D}(\mathcal{E}) = D$ with minimum $W(\mathcal{E})$.
That is, to solve the MLE decoding problem is to find a \emph{Minimum Weight Parity Factor} (MWPF) in the decoding hypergraph.

One caveat is that the edge weights $w_e$ can be negative when its error probability $p_e > 50\%$.
In this case, we can convert the problem to an equivalent one with $w_e \ge 0, \forall e \in E$ in polynomial time~\cite{higgott2025sparse}.
Specifically, if $p_e > 50\%$, we can treat $e$ as an ``always-occurring'' error and redefine the probability to $p'_e = 1 - p_e \le 50\%$.
We can solve the MLE $\mathcal{E}'$ on a modified graph of $w'_e = -w_e \ge 0$ and a modified syndrome $D' = D \oplus e$.
The MLE of the original problem is then $\mathcal{E} = \mathcal{E'} \oplus \{ e \}$.
In the rest of the paper, we assume all edge weights are non-negative.

\subsection{MWPF Problem Formulation}
\label{ssec:mwpf}
In the above, we showed that the MLE decoding problem is equivalent to the \textbf{Minimum Weight Parity Factor} (MWPF) problem on the decoding hypergraph. That is,
maximizing the probability of a parity factor $\mathcal{E} \subseteq E, \mathcal{D}(\mathcal{E}) = D$ is equivalent to minimizing its weight $W(\mathcal{E})$.
Inspired by the blossom algorithm, we formulate the MWPF problem by introducing a set of variables $x_e,\ \forall e \in E$, with which an error pattern $\mathcal{E}$ is represented: $x_e = 1,\ \forall e \in \mathcal{E}$ and $x_e = 0,\ \forall e \in E\setminus \mathcal{E}$.
Given the set of defect vertices $D \subseteq V$, the non-defect vertices are $\overline{D} = V \setminus D$.
Unlike the blossom algorithm, which solves the MWPM problem on the syndrome graph, we formulate the MWPF problem and solve it on the decoding hypergraph.

\defineeqs{MWPF}{Minimum-Weight Parity Factor}
\begin{align}
  \min\;\;\quad \sum_{e \in E} w_e & x_e & \\
  \text{subject to}\qquad\quad\ x_e &\in \{ 0, 1 \} &\forall e \in E \nonumber \tag{\theequation a}\label{eq:mwpf-constraint-1}\\
  \sum_{e \in E(v)} x_e &= 1 \mod 2 &\forall v \in D \nonumber \tag{\theequation b}\label{eq:mwpf-constraint-2}\\
  \sum_{e \in E(v)} x_e &= 0 \mod 2 &\forall v \in \overline{D} \nonumber \tag{\theequation c}\label{eq:mwpf-constraint-3}
\end{align}

The formulation above involves mixing $\mathbb{F}_2$ constraints (\cref{eq:mwpf-constraint-2} and \cref{eq:mwpf-constraint-3}) with an objective function defined in the real field $\mathbb{R}$.
Note that this formulation is widely considered in prior work~\cite{fawzi2021linear,takada2023highly,hillmann2024localized,berent2024decoding,ott2025decision}, albeit not in this particular form.

\subsection{Minimum-Weight Perfect Matching (MWPM)}\label{ssec:mwpm}

When the decoding hypergraph is a simple graph, i.e., $\deg(e) = 2, \forall e \in E$, the MLE decoding problem reduces to the minimum-weight perfect matching (MWPM) problem on a simple graph called syndrome graph~\cite{dennis2002topological,wu2022interpretation}.
The syndrome graph $G^* = (D, E^*)$ is a complete graph over the defect vertices $D$, where each edge $(u, v) \in E^* = \{ (u, v) | u, v \in D, u \neq v \}$ is weighted by the weight of the minimum-weight path between $u$ and $v$ in the decoding graph.
We use a superscript $*$ to distinguish notations specific to the syndrome graph from that of the decoding graph.
The MWPM decoder~\cite{fowler2012towards} solves this special case using the blossom algorithm with polynomial complexity~\cite{edmonds1973matching,kolmogorov2009blossom}.

However, it is commonly believed that the same reduction does not work for hypergraphs, for two reasons.
First, the reduced ``syndrome hypergraph'' would have exponentially many hyperedges because the trick of path compression is only applicable to simple graphs.
Second, hypergraph perfect matching is a well-known NP-hard problem~\cite{stockmeyer1982np}.

Parity Blossom~\cite{wu2023qce} is one of the fastest MWPM decoder implementations, using the famous blossom algorithm~\cite{kolmogorov2009blossom}.
It structures the decoder into two phases: the Primal phase and the Dual phase, working on the primal solution $\vec{x^*}$ and dual solution $\vec{y^*}$, respectively.
By implementing the Dual phase on the decoding graph instead of the syndrome graph, it achieves an almost-linear average time complexity.

\section{HyperBlossom Framework}\label{sec:math}

We introduce a mathematical framework for solving the MWPF problem for QEC decoding, called \hyperblossom.
The framework exploits the sparsity of the edges and defect vertices in quantum error correction to accelerate the \textbf{common} case.
It draws inspiration from the blossom algorithm family~\cite{edmonds1965paths,kolmogorov2009blossom}, especially Parity Blossom~\cite{wu2023qce}.
We formally define the linear programming problems in \S\ref{ssec:problem-definitions}.
We then discuss the \emph{clustering} technique (\S\ref{ssec:algo-cluster}) and \emph{relaxing} technique (\S\ref{ssec:cascaded-relaxing}), both preserving optimality and simplifying the problem.
The mathematical framework provides the following benefits that distinguish from other qLDPC decoders:
\begin{itemize}
  \item Unlike heuristic decoders, it rigorously proves a proximity bound along with each solution (\S\ref{ssec:problem-definitions}).
  \item It interoperates with MWPM decoders (\S\ref{ssec:interoperability-mwpm}) for decoding heterogeneous QEC architectures~\cite{stein2025hetec}.
  \item With \emph{clustering} technique (\S\ref{ssec:algo-cluster}), it decomposes a large problem into small clusters when $p \ll 1$, with the potential to achieve an almost-linear average decoding time to $|D| \propto p |V|$.
  \item It reduces the complex linear programming problem into simpler relaxer finding algorithms (\S\ref{ssec:cascaded-relaxing}). Using the \emph{relaxing} technique, relaxer finding algorithms can compose and complement each other. We discuss some relaxer finding algorithms in \S\ref{sec:subroutine} and we expect more to be found in the future.
\end{itemize}

\begin{figure*}[th]
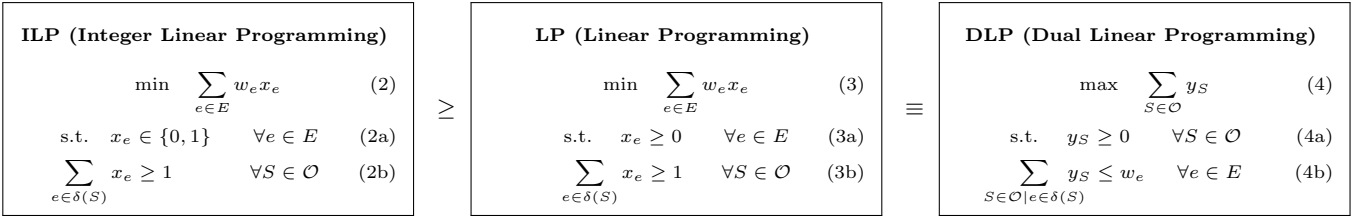

  \fbox{
    \begin{minipage}{0.28\textwidth}
      \scriptsize
      \defineeqs{ILP}{Integer Linear Programming}
      \begin{gather}
        \min\quad\sum_{e \in E} w_ex_e  \\
        \text{s.t.}\quad x_e\in \{ 0, 1 \}\qquad\forall e \in E \tag{\theequation a}\label{eq:ilp-constraint-1}\nonumber\\
        \!\!\sum_{e \in \delta(S)} x_e\ge 1 \ \ \quad\qquad\forall S\in\mathcal{O} \tag{\theequation b}\label{eq:ilp-constraint-2}\nonumber
      \end{gather}
    \end{minipage}
  }
  \hfill$\geq$\hfill
  \fbox{
    \begin{minipage}{0.28\textwidth}
      \scriptsize
      \defineeqs{LP}{Linear Programming}
      \begin{gather}
        \min\quad\sum_{e \in E} w_ex_e\\
        \text{s.t.}\quad\ x_e\ge 0 \qquad\forall e \in E \tag{\theequation a}\label{eq:lp-constraint-1}\nonumber\\
        \quad\quad\!\sum_{e \in \delta(S)}x_e\ge 1 \qquad\!\forall S \in \mathcal{O} \tag{\theequation b}\label{eq:lp-constraint-2}\nonumber
      \end{gather}
    \end{minipage}
  }
  \hfill$\equiv$\hfill
  \fbox{
    \begin{minipage}{0.28\textwidth}
      \scriptsize
      \defineeqs{DLP}{Dual Linear Programming}
      \begin{gather}
        \max\quad \sum_{S \in \mathcal{O}} y_S\\
        \text{s.t.}\quad\ y_S\ge 0 \qquad \forall S \in \mathcal{O} \tag{\theequation a}\label{eq:dual-constraint-1} \nonumber\\
        \qns\qns\,\sum_{S \in \mathcal{O} | e \in \delta(S)}\hns y_S\le w_e\quad\ \,\forall e \in E \tag{\theequation b}\label{eq:dual-constraint-2}\nonumber
      \end{gather}
    \end{minipage}
  }
  \caption{The \hyperblossom framework is based on relaxing the \refeqs{ILP} formulation (Left) of \refeqs{MWPF} (\S\ref{ssec:mwpf}) into an \refeqs{LP} problem (Middle) and solving the latter's dual formulation (\refeqs{DLP}) (Right) along with the \refeqs{MWPF} problem. This approach is inspired by the blossom algorithm but formulates the problem on the decoding hypergraph, instead of the syndrome graph.}
  \label{fig:formulations}
\end{figure*}

\subsection{Important Concepts}
\label{ssec:concepts}

We first introduce important concepts and notations for the problem definition.
Following convention, we use $E(V_S) = \{ e \in E | e \cap V_S \neq \varnothing \}$ and $E[V_S] = \{ e \in E | e \subseteq V_S \}$ to denote the set of edges incident to $V_S\subseteq V$ and the set of edges within it, respectively. We note that the above also uses $e$ to stand for the subset of vertices incident to the edge $e$ as is convention.

\vspace{1ex}
\definitionlabelvalid{invalid}{Invalid} For a decoding hypergraph $G(V, E)$ and syndrome $D \subseteq V$, we say it is \emph{Invalid} if and only if there exists no parity factor within it.
\begin{align*}
  \forall \mathcal{E} \subseteq E, \mathcal{D}(\mathcal{E}) \neq D
\end{align*}

\vspace{1ex}
For a subgraph $S = (V_S, E_S)$, $V_S \subseteq V, E_S \subseteq E[V_S]$ of a decoding hypergraph $G(V, E)$, we say it is an \emph{Invalid Subgraph} if $(V_S, E_S)$ is \refdef{invalid} for the partial syndrome $D \cap V_S$.
$\mathcal{O}$ denotes the set of all \refdef{invalid} subgraphs of $G$.
The decoding hypergraph $G(V,E)$ is \refdef{valid} by definition, thus $G \notin \mathcal{O}$.

\vspace{1ex}
\definitionlabel{hair}{Hair} The \textit{Hair} of an \refdef{invalid} subgraph $S$ is the set of edges not in $E_S$ but incident to at least one vertex in $V_S$.
\begin{align*}
  \delta(S) = E(V_S) \setminus E_S
\end{align*}

Note that here $E_S$ could be any subset of $E[V_S]$.
This is different from the blossom algorithm (\S\ref{ssec:mwpm}), where $E_S$ is always $E[V_S]$ so that $\delta(S) = E(V_S) \setminus E[V_S] = \{ (u, v) \in E | u \in V_S \land v \notin V_S \}$ as defined in the original blossom algorithm~\cite{kolmogorov2009blossom}.
We introduce this difference for two reasons.
First, \refeqs{MWPF} on hypergraphs requires a more complicated polytope to describe the parity constraints, and we show that removing flexibility in $E_S$ for certain hypergraphs leads to suboptimal polytope (\S\ref{ssec:why-need-ES}).
Second, determining the existence of a parity factor on a subgraph $(V_S, E_S)$ (using parity matrix in \S\ref{ssec:parity-matrix}) is much easier than finding the minimum-weighted one, while determining the existence of a perfect matching is almost as hard as finding the minimum-weighted one, even if the subgraph $(V_S, E_S)$ is a simple graph.

\subsection{Problem Definitions}\label{ssec:problem-definitions}

We next introduce an Integer Linear Programming (ILP) problem that is equivalent to the \refeqs{MWPF} problem described in \S\ref{ssec:mwpf}. As shown in \autoref{fig:formulations} (Left),
the ILP problem is reminiscent of the LP formulation used by the blossom algorithm. However, our ILP problem is formulated on the decoding hypergraph while the LP formulation in the blossom algorithm is on the syndrome graph.


The constraints \cref{eq:ilp-constraint-2} simply say that for each \refdef{invalid} subgraph $S \in \mathcal{O}$, at least one edge of its hair $\delta(S)$ must appear in a parity factor.
This is quite intuitive: an \refdef{invalid} subgraph $S$ must resort to at least one external edge $\delta(S)$ to satisfy the parity constraints posed by $V_S$.

We further prove that \refeqs{MWPF} to \refeqs{ILP} is an equivalent transformation with the following theorem in \S\ref{ssec:minilp-equal-mwpf}.

\theoremilpequalmwpf{ilp-equal-mpwf} The optimal objective value of the two problems are equal.

\vspace{1ex}
Note that not every feasible \refeqs{ILP} solution is a feasible parity factor.
For example, $x_e = 1, \forall e \in E$ is feasible in \refeqs{ILP} but not necessarily a feasible parity factor.

Next, we relax the integer constraints from $x_e \in \{ 0, 1 \}$ to $x_e \in \mathbb{R}_+$ to derive a Linear Programming (LP) problem, as shown in \autoref{fig:formulations} (Middle).


As a relaxation, we have $\min \text{\refeqs{LP}} \le \min \text{\refeqs{ILP}}$.
For some specific classes of hypergraphs, like simple graphs (\S\ref{ssec:interoperability-mwpm}) and nullity$_{\le 1}$ hypergraphs (defined in \S\ref{ssec:biased-single-dof}), we prove that $\min \text{\refeqs{LP}} = \min \text{\refeqs{ILP}}$.
For more general hypergraphs, we do not have a proof of equality, nor have we identified a counterexample despite extensive efforts to construct one analytically and numerically.
Thus, we will not assume the equality throughout the paper.

Again drawing inspiration from the blossom algorithm (\S\ref{ssec:mwpm}), we then consider the dual LP problem~\cite{dantzig1951maximization,matouvsek2007understanding}, as shown in \autoref{fig:formulations} (Right).

\begin{figure}
  \centering
  \begin{subfigure}[t]{.32\linewidth}
    \centering
    \includegraphics[width=1\textwidth]{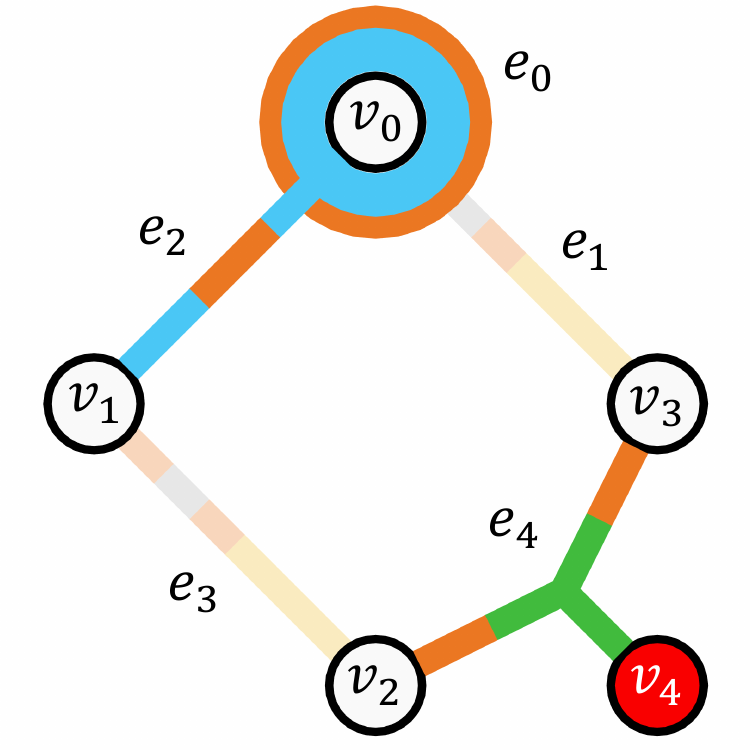}
    \caption{Visual of $\vec{y}$.}
    \label{fig:dual-visual-split-0}
  \end{subfigure}
  \begin{subfigure}[t]{.64\linewidth}
    \centering
    \includegraphics[width=1\textwidth]{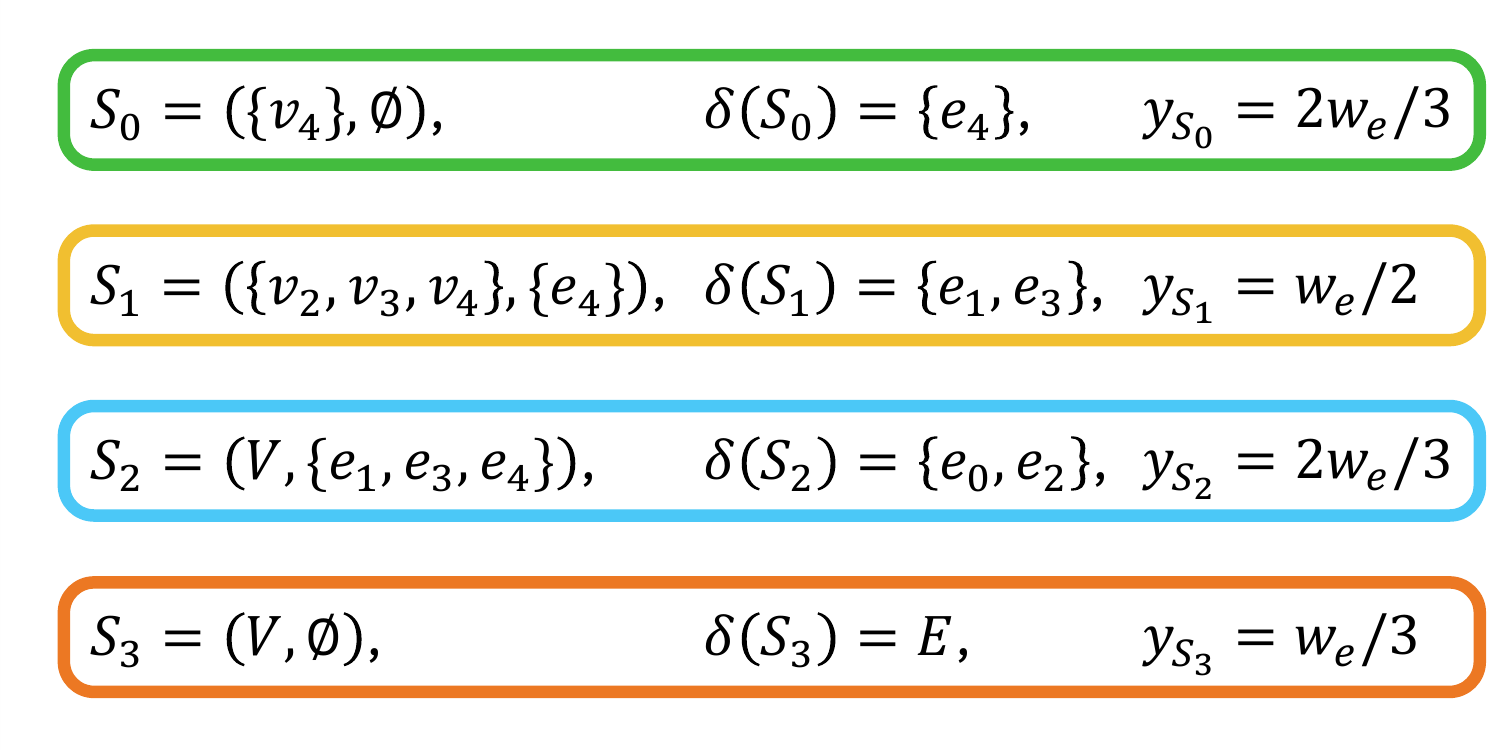}
    \caption{\refdef{hyperblossoms} $\mathcal{B}$ and $y_S, \forall S \in \mathcal{B}$.}
    \label{fig:dual-visual-split-1}
  \end{subfigure}
  \caption{
    Visualization of \refeqs{DLP} solution $\vec{y}$ on the decoding hypergraph $G = (V, E)$ with a uniform edge weight of $w_e$.
    For each \refdef{invalid} subgraph $S \in \mathcal{O}$, we visualize its dual variable $y_S$ as colored segments occupying a $y_S/w_e$ portion of each edge $e \in \delta(S)$.
    By definition, \refdef{tight-edges} are those fully occupied, e.g., $T = \{e_0, e_2, e_4\}$ highlighted in solid color instead of the greyed colors.
  }
  \label{fig:dual-visual}
\end{figure}

Similar to $\vec{x}$, we use $\vec{y}$ to denote the vector of dual variables $y_S, \forall S \in \mathcal{O}$.
We can visualize the dual variables on the decoding hypergraph as shown in \autoref{fig:dual-visual}.

According to the Duality Theorem~\cite{winston2004operations}, $\max \text{\refeqs{DLP}} = \min\text{\refeqs{LP}}$.
Thus, the optimal \refeqs{DLP} solution has an objective value of $\sum_{S \in \mathcal{O}} y_S = \min\text{\refeqs{LP}}$. Taken together, we have this chain of inequality:

\begin{gather}
  \begin{gathered}
    \text{\refeqs{MWPF}}\\
    \sum_{e \in E} w_e x_e \\\vspace{-1.4ex}
  \end{gathered}
  \ge \min\text{\refeqs{MWPF}}=\min\text{\refeqs{ILP}}\ge \min\text{\refeqs{LP}}\ge
  \begin{gathered}
    \text{\refeqs{DLP}}\\
    \sum_{S \in \mathcal{O}} y_S \\\vspace{-1.4ex}
  \end{gathered}
  \label{eq:mwpf-chain}
\end{gather}

Equality holds for the left and right $\ge$ when the \refeqs{MWPF} and \refeqs{DLP} solutions are optimal, respectively.



\subsection{More Important Concepts}

Astute readers might notice that the cardinality $|\mathcal{O}|$ grows exponentially with $|V|$ and $|E|$.
Fortunately, although the number of dual variables $|\mathcal{O}|$ is exponential, there exists a trivial feasible \refeqs{DLP} solution $y_S = 0, \forall S\in \mathcal{O}$.
Thus, we only need to track those $y_S > 0$, as the blossom algorithm does.

\vspace{1ex}
\definitionlabel{hyperblossom}{Hyperblossom}
A \textit{Hyperblossom} is an \refdef{invalid} subgraph $S$ whose corresponding dual variable $y_S$ is positive. We denote the set of all \refdef{hyperblossoms} with $\mathcal{B} = \{S \in \mathcal{O} | y_S > 0 \}$.

\vspace{1ex}
The notion of \refdef{hyperblossom} is analogous to that of \emph{blossom} in the blossom algorithm family, both representing positive dual variables (\ref{eq:dual-constraint-1}).
We note that \refdef{hyperblossoms} are defined on the decoding hypergraph, while \emph{blossoms} are defined on the syndrome graph.

\vspace{1ex}
\definitionlabel{tight-edge}{Tight Edge} A hyperedge $e \in E$ is \emph{Tight} when its dual constraint (\cref{eq:dual-constraint-2}) takes the equal sign: $\sum_{S\in\mathcal{O} | e\in\delta(S)} y_S = w_e$. We use $T \subseteq E$ to represent the set of \refdef{tight-edges}.

\vspace{1ex}
\definitionlabel{direction}{Direction} A \emph{Direction} $\Delta\vec{y}$ updates the \refeqs{DLP} solution $\vec{y}$ by $\vec{y}' \coloneqq \vec{y} + l \Delta\vec{y}$ given a length $l>0$.

\vspace{1ex}
\definitionlabel{feasible-direction}{Feasible Direction} A \refdef{direction} $\Delta\vec{y}$ is called \emph{Feasible} when it can grow a small positive length without violating any \refeqs{DLP} constraint. We denote a \emph{Feasible} \refdef{direction} with $\Delta\vec{y}[G, \mathcal{B}, T]$.
When the context is clear, we elide $G$, $\mathcal{B}$, $T$ and write $\Delta\vec{y}$ or $\Delta\vec{y}[T]$ for $\Delta\vec{y}[G, \mathcal{B}, T]$.
\stepcounter{equation}
\begin{gather}
  \forall S \in \mathcal{O} \setminus \mathcal{B}, \quad\; \Delta y_S \ge 0 \tag{\theequation a}\label{eq:feasible-a} \\
  \forall e \in T, \sum_{S\in \mathcal{O}|e \in \delta(S)}\dqns \Delta y_S \le 0 \tag{\theequation b}\label{eq:feasible-b}
\end{gather}

\cref{eq:feasible-a} says that a \refdef{feasible-direction} will not decrease dual variables that are already $0$. \cref{eq:feasible-b} says that it will not over-grow a \refdef{tight-edge}.

\vspace{1ex}
\definitionlabel{useful-direction}{Useful Direction} A \refdef{feasible-direction} $\Delta\vec{y}$ is \emph{Useful} if $\sum \Delta\vec{y} = \sum_{S \in \mathcal{O}} \Delta y_S > 0$.

\vspace{1ex}
Applying a \refdef{useful-direction} for a positive length will improve a suboptimal \refeqs{DLP} solution.

\vspace{1ex}
\definitionlabel{trivial-direction}{Trivial Direction} A \refdef{feasible-direction} $\Delta\vec{y}$ is \emph{Trivial} if $\Delta\vec{y} = \{ \Delta y_S: +1 \}$ where $S \in \mathcal{O}$.

\vspace{1ex}
Applying a \refdef{trivial-direction} will only grow one \refdef{invalid} subgraph $S$ while leaving others unchanged.
We have $\delta(S) \cap T = \varnothing$ because $\Delta\vec{y}$ is \emph{Feasible} and must satisfy \cref{eq:feasible-b}.
By definition, a \refdef{trivial-direction} is also a \refdef{useful-direction}.

\subsection{HyperBlossom Algorithm Overview}
\label{ssec:algo-overview}

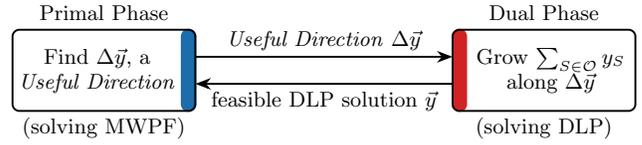
\begin{figure}
  \centering
  \scalebox{0.9}{
    \begin{tikzpicture}%
\path[primalinterfacecolor,thick,draw,fill=primalinterfacecolor,rounded corners=1mm] (2.4999999999999996,0.0) rectangle (2.6999999999999997,1.2);%
\path[thick,draw,rounded corners=1mm] (0.0,0.0) rectangle (2.6999999999999997,1.2);%
\node[color=black,align=center,font=\small] at (1.2499999999999998,0.6) {Find $\Delta\vec{y}$, a\\\refdef{useful-direction}};%
\path[dualinterfacecolor,thick,draw,fill=dualinterfacecolor,rounded corners=1mm] (6.5,0.0) rectangle (6.7,1.2);%
\path[thick,draw,rounded corners=1mm] (6.5,0.0) rectangle (9.2,1.2);%
\node[color=black,align=center,font=\small] at (7.949999999999999,0.6) {Grow $\sum_{S \in \mathcal{O}} y_S$\\along $\Delta\vec{y}$};%
\node[color=black,align=center,font=\small] at (1.3499999999999999,1.45) {Primal Phase};%
\node[color=black,align=center,font=\small] at (7.85,1.45) {Dual Phase};%
\node[color=black,align=center,font=\small] at (1.3499999999999999,-0.25) {(solving \refeqs{MWPF})};%
\node[color=black,align=center,font=\small] at (7.85,-0.25) {(solving \refeqs{DLP})};%
\path[thick,-{Stealth},draw] (2.6999999999999997,0.8) -- (6.5,0.8);%
\node[color=black,align=center,font=\small] at (4.6,1.05) {\refdef{useful-direction} $\Delta\vec{y}$};%
\path[thick,-{Stealth},draw] (6.5,0.39999999999999997) -- (2.6999999999999997,0.39999999999999997);%
\node[color=black,align=center,font=\small] at (4.6,0.14999999999999997) {feasible \refeqs{DLP} solution $\vec{y}$};%
\end{tikzpicture}
  }
  \caption{Overview of the \hyperblossom algorithm. The Primal phase solves the \refeqs{MWPF} problem, while the Dual phase solves the \refeqs{DLP} problem. They exchange information through a narrow interface. This interaction is inspired by the blossom algorithm.}
  \label{fig:primal-dual-interface}
\end{figure}

Built on top of the concepts and problem definitions introduced so far, the \hyperblossom algorithm features three levels of innovations. At the highest level, drawing inspiration from the blossom algorithm, \hyperblossom consists of two phases:
the Primal phase works on the \refeqs{MWPF} problem as formulated in \S\ref{ssec:mwpf}; the Dual phase works on the \refeqs{DLP} problem formulated in \S\ref{ssec:problem-definitions}. The two phases exchange information using a narrow interface, as shown in \autoref{fig:primal-dual-interface}.

\begin{table*}[t]
  \normalsize
  \centering
  \caption{Relationship between the Blossom Algorithm (\S\ref{ssec:mwpm}) and the \hyperblossom Algorithm}\label{tab:mwpm-mwpf}
  \begin{tabular}{@{}Sc@{\hskip 5ex}Sc@{\hskip 5ex}Sc@{}}
    \hline
    & Blossom~\cite{kolmogorov2009blossom,wu2023qce,higgott2025sparse} & \hyperblossom \\
    \hline
    Problem Solved & MWPM on syndrome graph & MWPF on decoding hypergraph\\
    \hline
    Primal Variables & $x^*_e$ per syndrome graph edge $e \in E^*$ & $x_e$ per decoding hypergraph edge $e \in E$ \\
    \hline
    \begin{tabular}{@{}c@{}} Primal Constraints \\
      (Dual Variables)
    \end{tabular} &
    \begin{tabular}{@{}c@{}} $\mathcal{O}^* = \Big\{ S^* \subseteq D\ \Big|\ |S^*| = 1\mod 2 \Big\}$
    \end{tabular} &
    \begin{tabular}{@{}c@{}} $\mathcal{O} = \{ S = (V_S, E_S)\ |\ V_S \subseteq V, $ \\ $E_S \subseteq E[V_S], \forall \mathcal{E} \subseteq E_S, \mathcal{D}(\mathcal{E}) \neq D \cap V_S \}$
    \end{tabular} \\
    \hline
    (Hyper)Blossoms & \emph{Blossoms} $\mathcal{B}^* = \{ S^* \in \mathcal{O}^*\ |\ y^*_{S^*} > 0 \}$ & \refdef{hyperblossoms} $\mathcal{B} = \{ S \in \mathcal{O}\ |\ y_S > 0 \}$ \\
    \hline
    Dual Updates & along \refdef{direction} $\Delta \vec{y^*} \in \{ 0, +1, -1 \}^{|\mathcal{O}^*|}$ &
    \begin{tabular}{@{}l@{}} along \refdef{direction} $\Delta \vec{y} \in \mathbb{R}^{|\mathcal{O}|}$
    \end{tabular} \\
    \hline
    Find more (Hyper)Blossoms & alternating tree~\cite{kolmogorov2009blossom} & \emph{relaxing} (\S\ref{ssec:cascaded-relaxing}) (\S\ref{sec:subroutine}) \\
    \hline
  \end{tabular}
\end{table*}

The Primal phase maintains parity factors consisting of \refdef{tight-edges} and informs the Dual phase new \refdef{useful-directions} $\Delta \vec{y}$ to find better \refeqs{DLP} solutions.
The Dual phase tracks the \refdef{hyperblossoms}
and grows them along the \refdef{useful-directions} $\Delta \vec{y}$ from the Primal phase, and informs the Primal phase the improved \refeqs{DLP} solution $\vec{y}$.

According to \cref{eq:mwpf-chain}, given a feasible \refeqs{DLP} solution $\vec{y}$ and a feasible \refeqs{MWPF} solution (parity factor) $\vec{x}$, their objective values are the lower and upper bounds, respectively, of that of the optimal \refeqs{MWPF} solution.
We name $\sum_{e \in E} w_e x_e - \sum_{S \in \mathcal{O}} y_S$ the \textit{primal-dual gap}.
The \hyperblossom algorithm seeks to reduce this gap and get closer to optimality.
Especially, when the primal-dual gap is zero, the optimality is certified by the following theorem.

\definetheorem{provable-optimum}{Certifying Optimum}
A feasible \refeqs{DLP} $\vec{y}$ with the same objective value as the weight of a parity factor $\mathcal{E}$ certifies the optimality of $\mathcal{E}$ as an \refeqs{MWPF}.
\begin{gather*}
  W(\mathcal{E}) = \sum_{S \in \mathcal{O}} y_S \quad\Longrightarrow\quad \text{$\mathcal{E}$ is an MWPF}
\end{gather*}

Note that in the blossom algorithm, the Primal and Dual phases also work on two related problems: ILP formulation of MWPM and the dual of a relaxed LP problem of the ILP problem (DLP), respectively~\cite{wu2023qce}.
However, because the ILP and DLP problems have the same objective value, the blossom algorithm always finds the optimal solution for the MWPM problem.
In contrast, the \hyperblossom algorithm is subject to the inequality chain of \cref{eq:mwpf-chain}: there might exist cases where $\min\text{\refeqs{LP}} < \min\text{\refeqs{ILP}}$.
Only when the \ref{condition:mwpf-short} below holds, the \hyperblossom algorithm guarantees to certify the \refeqs{MWPF} solution.

\conditionmwpf{mwpf}

\vspace{1ex}
We prove that two classes of decoding hypergraphs satisfy the above condition: simple graphs and nullity$_{\le 1}$ hypergraphs (whose incidence matrix has a nullity of 0 or 1).
We prove the theorems below in \S\ref{ssec:mwpf-condition-simple-graph} and \S\ref{ssec:biased-single-dof}, respectively.

\vspace{1ex}
\theoremSimpleGraphOptimality{simple-graph-optimality}

\vspace{1ex}
\theoremSingledofGraphOptimality{single-dof-graph-optimality}

\vspace{1ex}
When \ref{condition:mwpf}, according to the Complementary Slackness Theorem~\cite{winston2004operations}, for any optimal \refeqs{MWPF} $\vec{x}$ and \refeqs{DLP} $\vec{y}$:
\begin{align}
  \sum_{S\in\mathcal{O}|e\in \delta(S)}\dqns y_S < w_e \  &\Longrightarrow \ x_e = 0, \qquad\forall e \in E \tag{C1}\label{eq:cs-must-tight}
\end{align}
That is, an optimal \refeqs{MWPF} solution includes only \refdef{tight-edges} of an optimal \refeqs{DLP} solution.
As a heuristic, the Primal Phase can consider only \refdef{tight-edges} in searching for an MWPF, even though \ref{condition:mwpf} is not always true. We call this the \refdef{tight-edge} heuristic.

\vspace{2ex}
At the second level, \hyperblossom features two important optimizations.
\Relaxing (\S\ref{ssec:cascaded-relaxing}) turns the search for an optimal \refeqs{DLP} solution into a sequence of calls to a relaxer finding algorithm.
\Clustering (\S\ref{ssec:algo-cluster}) exploits spatial locality in defects to divide the decoding hypergraph into clusters and conquer them independently.
The Primal phase implements \relaxing because it is responsible for computing the \refdef{direction}.
Both phases are aware of the clusters, using cluster-oriented data structures.

At the third and lowest level, \hyperblossom features several \refdef{relaxer}-finding algorithms (\S\ref{sec:subroutine}).
We show that the blossom algorithm (\S\ref{ssec:mwpm}) can be derived from one of those \refdef{relaxer}-finding algorithms (\S\ref{ssec:blossom-subroutine}), demonstrating the power of \relaxing.
Importantly, we propose a new, simple \refdef{relaxer}-finding algorithm called the \textit{SingleHair} (\S\ref{ssec:single-hair-subroutine}) for general hypergraphs.
We analytically prove the optimality of the \emph{SingleHair} algorithm in certain conditions (\S\ref{ssec:optimal-1-dof}) and also show the limitations of it (\S\ref{ssec:failure-hypergraph} and \S\ref{ssec:failure-simple-graph}).
Empirically, we show that the \emph{SingleHair} algorithm is an accurate and fast \refdef{relaxer}-finding algorithm for a variety of qLDPC codes (\S\ref{sec:evaluation}).

We compare the blossom algorithm and the \hyperblossom algorithm in \cref{tab:mwpm-mwpf}.

\subsection{Relaxing}\label{ssec:cascaded-relaxing}

The Primal phase helps the Dual phase by finding a \refdef{useful-direction} so that the latter can improve its current solution toward optimality, as shown in \autoref{fig:primal-dual-interface}.
The challenge is that the dimensionality of $\vec{y}$ or $\Delta\vec{y}$ is $|\mathcal{O}|$: there are too many dual variables.
\emph{Relaxing} converts the problem of finding a \refdef{useful-direction} into finding \refdef{relaxers}, as defined below.

\vspace{1ex}
\definitionlabel{relaxer}{Relaxer}
Given a decoding hypergraph $G$, its \refdef{tight-edges} $T$ and \refdef{hyperblossoms} $\mathcal{B}$, a \emph{Relaxer} $R[G,\mathcal{B},T]$ is a \refdef{feasible-direction} $\Delta\vec{y}$ that satisfies the following conditions.
(\ref{eq:relaxer-a}) Applying it relaxes a non-empty set of \refdef{tight-edges} $\mathcal{R}(R) \subseteq T$. That is, these edges will no longer be tight after growing along $\Delta \vec{y}$. (\ref{eq:relaxer-b}) It does not reduce the \refeqs{DLP} objective.
When the context is clear, we elide $G, \mathcal{B}, T$ and write $R$ or $R[T]$ for $R[G,\mathcal{B},T]$.
\stepcounter{equation}
\begin{gather}
  \mathcal{R}(R)\neq \varnothing,\ \ \forall e \in \mathcal{R}(R) \subseteq T, \sum_{S\in \mathcal{O}|e \in \delta(S)}\dqns \Delta y_S < 0 \tag{\theequation a}\label{eq:relaxer-a} \\
  \sum_{S \in \mathcal{O}} \Delta y_S \ge 0 \tag{\theequation b}\label{eq:relaxer-b}
\end{gather}

\vspace{1ex}
Applying a \refdef{relaxer} reduces the number of \refdef{tight-edges}.
As a result, it reduces the search space of the \refeqs{MWPF} problem. Importantly, it also helps find a \refdef{useful-direction} with the following theorem proved in \S\ref{ssec:optimality-of-relaxing}:

\vspace{1ex}
\theoremRelaxerExistenceOrTrivialDirection{relaxing}

\vspace{1ex}
A simple algorithm to compute the \refdef{trivial-direction} can iterate over $S\in\mathcal{O}$ and find one with $\delta(S) \cap T=\varnothing$.
As noted earlier, $\Delta\vec{y}=\{\Delta y_S: +1\}$ is \emph{Feasible} (and therefore \emph{Trivial}).
Later, we will present a much more efficient algorithm in \cref{algo:hyperblossom-primal-phase}, line \ref{line:trivial-direction-hyperblossom}.

Applying the theorem, the Primal Phase will find a sequence of \refdef{relaxers} $R_i[T_i]$, $i=1,2,...,n$, before it could no longer find any \refdef{relaxer}.
$T_i = T \setminus (\cup_{j=1}^{i-1} \mathcal{R}_j)$ is the set of \refdef{tight-edges} after applying the  first $(i-1)$ \refdef{relaxers}.
At this point, according to the theorem, the Primal phase can find a \refdef{trivial-direction} $\Delta\vec{y}[T_{n+1}]$ for the Dual phase.
Since a \refdef{trivial-direction} is \emph{Useful} by definition, the theorem guarantees that the Dual phase makes progress via \emph{relaxing}.
Because each \refdef{relaxer} reduces the number of \refdef{tight-edges} by at least 1, we have $|T_{n+1}|<|T_n|<...<|T_1|=|T|$.
That is, $n$ is upper bounded by $|T|$.
In the end, the Primal phase will find a \refdef{trivial-direction} for the Dual phase to make progress after at most $|T|$ \refdef{relaxers}.

\subsubsection{Dual Phase Optimization}

\reftheorem{relaxing}, however, does not guarantee the Dual phase terminates.
To tackle this, we introduce an optimization in the Dual phase.
Instead of growing dual variables strictly along $\Delta \vec{y}$, the Dual phase infers from $\Delta \vec{y}$ what new \refdef{hyperblossoms} will be formed if this \refdef{direction} is applied.
It then calls an LP solver to solve the \refeqs{DLP} problem only using dual variables corresponding to \refdef{hyperblossoms} ever observed as far, called \refdef{history}.

\vspace{1ex}
\definitionlabel{history}{History}
The \emph{History} $\mathcal{B}^H$ includes all \refdef{hyperblossoms} that have formed so far during the Dual phase operation.

\vspace{1ex}
When using an LP solver to solve the \refeqs{DLP} problem, the Dual phase fixes the dual variables corresponding to $\mathcal{O} \setminus \mathcal{B}^H$ to $0$.
As the LP solver always finds a \refeqs{DLP} solution that maximizes $\sum_{S \in \mathcal{B}^H} y_S$, further progress by the Dual phase according to \reftheorem{relaxing} must introduce new \refdef{hyperblossoms}.
That is, each time the Dual phase receives a \refdef{useful-direction}, $|\mathcal{B}^H|$ increases while  $|\mathcal{O} \setminus \mathcal{B}^H|$ decreases.
Therefore, the Dual phase must terminate with at most $|\mathcal{O}|$ \refdef{useful-directions} from the Primal phase.

\subsubsection{Batch Relaxing}

\begin{figure}[t]

  \begin{algorithm}[H]
    \caption{Compose Relaxers}\label{algo:compose}
    \begin{algorithmic}[1]
    \Require{$R'_i[T], i=1,2,...,n$ and  $\Delta\vec{y}[T \setminus (\cup_i \mathcal{R}'_i)$)]}
    \Ensure{$\Delta'\vec{y}[T]$ in \reftheorem{cascaded-relaxing}}
    \Procedure{Compose}{$\{ R'_i[T] \}, \Delta\vec{y}$}\label{line:compose}
    \State $\Delta'\vec{y} \gets \Delta\vec{y}$
    \label{line:initial-direction}
    \For{$e \in \cup_i \mathcal{R}'_i \cap T$}\label{line:for-loop-fix}
    \State $\alpha \gets \sum_{S \in \mathcal{O} | e \in \delta(S)} \Delta'y_S$ \label{line:growing-alpha}
    \If{$\alpha > 0$} \Comment{if $\Delta\vec{y}$ violates (\ref{eq:dual-constraint-2}) of $e$}\label{line:violates-tight-edge}
    \State $\Delta^e\vec{y} \gets R'_k\ \text{where } e \in \mathcal{R}'_k$, $1\leq k\leq n$
    \label{line:find-relaxer}
    \State $\Delta'\vec{y} \gets \Delta'\vec{y} - \frac{\alpha}{\qns\sum\limits_{S \in \mathcal{O}| e \in \delta(S)}\dqns \Delta^e y_S} \Delta^e\vec{y}$ \Comment{fix violation}\label{line:fix-violation}
    \EndIf
    \EndFor
    \State \Return {$\Delta'\vec{y}$}
    \EndProcedure
  \end{algorithmic}
\end{algorithm}

\begin{algorithm}[H]
  \caption{Batched Relaxing}\label{algo:relaxing}
  \begin{algorithmic}[1]
    \Require{$G = (V, E)$ (decoding hypergraph), $\vec{y}$ (\refeqs{DLP} solution)}
    \Ensure{$\{ R'[T] \}$ (a set of at most $|T|$ \refdef{relaxers} that maximally relaxes \refdef{tight-edges} $T$)}
    \Procedure{BatchedRelaxing}{$G, \vec{y}$}
    \State $Rs' \gets \varnothing$ \Comment{set of \refdef{relaxers}}\label{line:relaxer-pool-init}
    \State $T' \gets T$ \Comment{remaining \refdef{tight-edges} $T_i$}
    \While{$(R[T'] \gets \Call{FindRelaxer}{G, \mathcal{B}, T'}) \neq \textsc{Nil}$} \label{line:find-relaxers}
    \State $R'[T] \gets \Call{Compose}{Rs', R[T']}$ \label{line:call-compose}
    \State $Rs' \gets Rs' \cup \{R'\}$ \label{line:relaxer-pool}
    \State $T' \gets T' \setminus \mathcal{R}'$ \label{line:reduced-tight-edges}
    \EndWhile
    \State \Return {$Rs'$}
    \EndProcedure
  \end{algorithmic}
\end{algorithm}

\end{figure}

\begin{figure*}[t]
  \centering
  \begin{subfigure}[t]{0.23\linewidth}
    \centering
    \includegraphics[width=\textwidth,page=1]{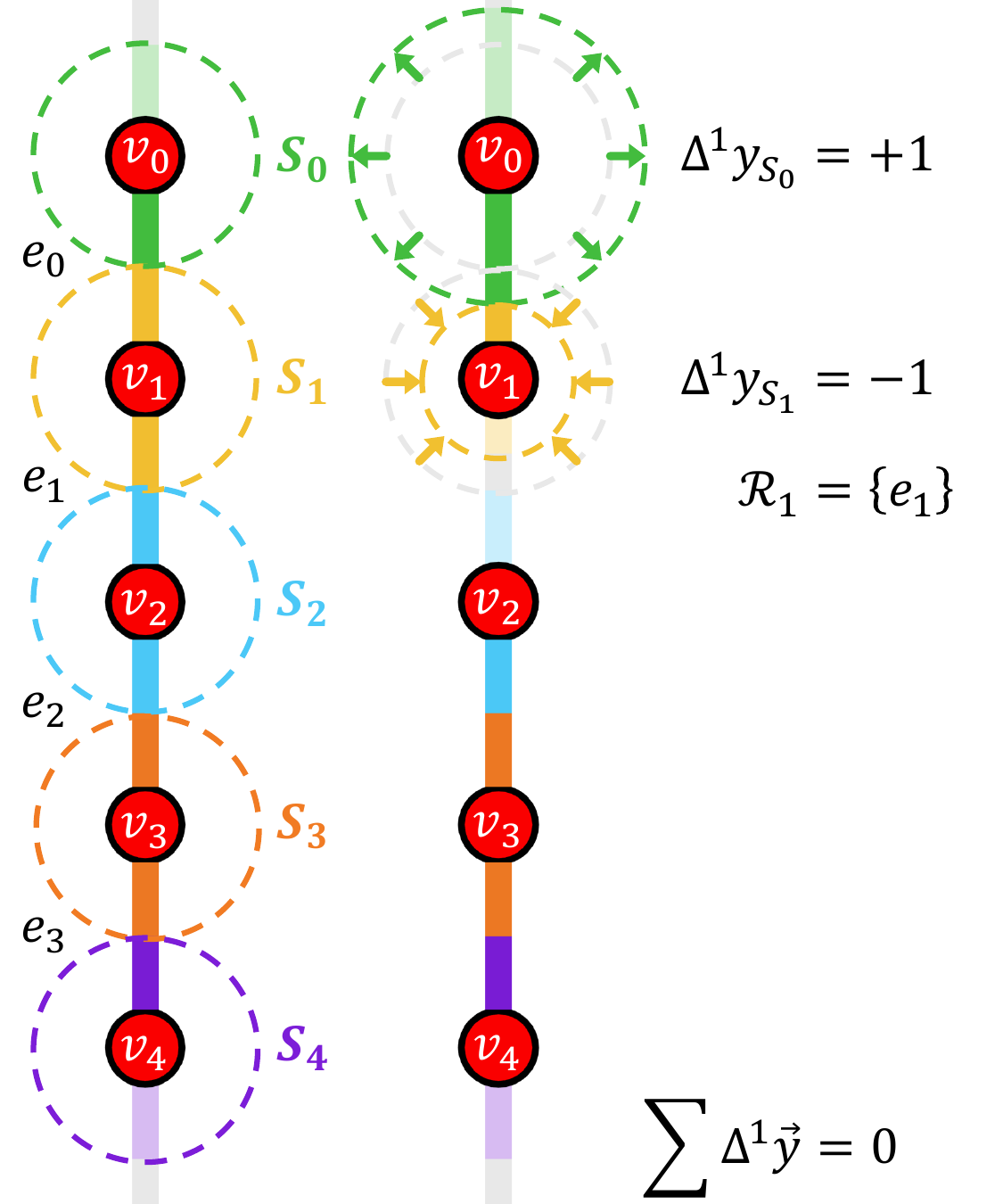}
    \caption{\refdef{relaxer} $R_1[T_1]=\Delta^1\vec{y}$}
    \label{fig:cascaded-relaxing-0}
  \end{subfigure}
  \hspace{1ex}
  \begin{subfigure}[t]{0.23\linewidth}
    \centering
    \includegraphics[width=\textwidth,page=2]{figures/cascaded-relaxing/cascaded-relaxing.pdf}
    \caption{\refdef{relaxer} $R_2[T_2]=\Delta^2\vec{y}$}
    \label{fig:cascaded-relaxing-1}
  \end{subfigure}
  \hspace{1ex}
  \begin{subfigure}[t]{0.23\linewidth}
    \centering
    \includegraphics[width=\textwidth,page=3]{figures/cascaded-relaxing/cascaded-relaxing.pdf}
    \caption{\refdef{trivial-direction} $\Delta\vec{y}[T_3]$}
    \label{fig:cascaded-relaxing-2}
  \end{subfigure}
  \hspace{1ex}
  \begin{subfigure}[t]{0.23\linewidth}
    \centering
    \includegraphics[width=\textwidth,page=4]{figures/cascaded-relaxing/cascaded-relaxing.pdf}
    \caption{\refdef{useful-direction} $\Delta'\vec{y}[T]$}
    \label{fig:cascaded-relaxing-3}
  \end{subfigure}
  \caption{An example of \emph{batch relaxing}. The radius of the circle centered at a vertex $v_i$ represents the corresponding dual variable $y_S, S = (\{v\}, \varnothing)$. (a) The initial \refeqs{DLP} solution has \refdef{tight-edges} $T_1 = T = \{ e_0, e_1, e_2, e_3 \}$. We find a \refdef{relaxer} $R_1[T_1]$ that increases $y_{S_0}$, decreases $y_{S_1}$, and leaves others unchanged. We have $\mathcal{R}_1 = \{ e_1 \}$. (b) Given the remaining \refdef{tight-edges} $T_2 = T_1 \setminus \mathcal{R}_1 =\{ e_0, e_2, e_3 \}$, we find a \refdef{relaxer} $R_2[T_2]$ with $\mathcal{R}_2= \{ e_3 \}$. (c) For the remaining \refdef{tight-edges} $T_3 = T_2 \setminus \mathcal{R}_2 = \{ e_0, e_2 \}$, we find a \refdef{trivial-direction} $\Delta\vec{y}[T_3]$ to increase the dual sum: growing $y_{S_4}$. (d) We compose these \refdef{directions} $R_1[T_1]$, $R_2[T_2]$ and $\Delta\vec{y}[T_3]$ into a single \refdef{useful-direction} $\Delta'\vec{y}[T]$ using \reftheorem{cascaded-relaxing} (\cref{algo:compose}).}
  \label{fig:cascaded-relaxing}
\end{figure*}

Applying a \refdef{relaxer} $R=\Delta\vec{y}$ is expensive because (\textit{i}) it will convert $S\in \mathcal{O}$ into a \refdef{hyperblossom} if $\Delta y_S>0$ and therefore increase $|\mathcal{B}^H|$;
and (\textit{ii}) the time complexity of the LP solver is cubic in the number of the \refdef{history} $|\mathcal{B}^H|$. On the other hand,
\reftheorem{relaxing} requires multiple \refdef{relaxers} to be applied before a \refdef{trivial-direction} can be found.
Fortunately, with \emph{batch relaxing}, the Primal phase no longer needs to send each of $R_i[T_i]$, $i=1,2,...,n$ and the \refdef{trivial-direction} $\Delta\vec{y}[T_{n+1}]$ to the Dual phase to apply. Rather, it can compute a \refdef{useful-direction} $\Delta'\vec{y}[T]$ from $R_i[T_i]$ and $\Delta\vec{y}[T_{n+1}]$ so that the Dual phase only needs to grow once, instead $n+1$ times, and will grow the \refdef{history} $|\mathcal{B}^H|$ only if necessary.

\theoremBatchedRelaxing{cascaded-relaxing}

\cref{algo:compose} describes a method to compose the \refdef{feasible-direction} $\Delta'\vec{y}[T]$ and therefore, serves as a constructive proof of the theorem.

\vspace{1ex}
We next describe how the Primal phase computes a \refdef{useful-direction} $\Delta'\vec{y}$ from  $R_i[T_i], i=1,2,...,n$, and $\Delta\vec{y}$.
First, it computes
$R'_i[T], i=1,2,..n$ from $R_i[T_i]$ such that $R'_i[T]$ relaxes all the edges that $R_i[T_i]$ relaxes.
Because $T_1 = T$, the first \refdef{relaxer} $R'_1[T]$ is simply $R_1[T_1]$.
Assuming we have all the \refdef{relaxers} $R'_j[T]$, $1\le j<i$,  we can compose $R'_i[T]$ that relaxes $\mathcal{R}'_i \supseteq \mathcal{R}_i$, according to \reftheorem{cascaded-relaxing}.
Overall, we have $(\cup_i \mathcal{R}_i)\subseteq (\cup_i \mathcal{R}'_i)$.
\cref{algo:relaxing} is a pseudo code of the above process, by repeatedly finding \refdef{relaxers} $R_i[T_i]$ (line \ref{line:find-relaxers}) and computing $R'_i[T]$ (line \ref{line:call-compose}).

Second, the Primal phase computes a \refdef{useful-direction} $\Delta'\vec{y}[T]$ from \refdef{relaxers} $R'_i[T], i=1,2,...,n$ and the \refdef{trivial-direction} $\Delta\vec{y}[T \setminus (\cup_i \mathcal{R}_i)]$.
Because $T \setminus (\cup_i \mathcal{R}'_i)\subseteq T \setminus (\cup_i \mathcal{R}_i)$, $\Delta\vec{y}$ is also \emph{Feasible} on $T \setminus (\cup_i \mathcal{R}'_i)$.
With \refdef{relaxers} $R'_i[T], i=1,2,...,n$, the Primal phase can compose a \refdef{feasible-direction} $\Delta'\vec{y}[T]$ such that $\sum\Delta'\vec{y} \ge \sum\Delta\vec{y} > 0$, according to \reftheorem{cascaded-relaxing} (\cref{algo:compose}).
That is, $\Delta'\vec{y}$ is a \refdef{useful-direction}.

In order to keep the \refdef{history} $\mathcal{B}^H$ small, \cref{algo:compose} takes special care so that applying the \refdef{useful-direction} $\Delta'\vec{y}[T]$ will not create unnecessary \refdef{hyperblossoms} in the Dual phase.
Initially (line \ref{line:initial-direction}), the \refdef{trivial-direction} $\Delta'\vec{y}$ would introduce a single \refdef{hyperblossom} $S$ because $\Delta\vec{y}=\{\Delta y_S: +1\}$.
However, $\Delta\vec{y}$ is \emph{Feasible} on $[T\setminus (\cup_i\mathcal{R}_i)]$, not on $T$. Applying it on $T$ may violate \cref{eq:dual-constraint-2} for some edges in $\cup_i\mathcal{R}_i \cap T$.
The algorithm fixes this violation using a for-loop (line \ref{line:for-loop-fix}) over all edges in $\cup_i\mathcal{R}_i \cap T$ and include a \refdef{relaxer} to relax each (lines \ref{line:find-relaxer} and \ref{line:fix-violation}).
Instead of relaxing all edges $\cup_i\mathcal{R}_i \cap T$, \cref{eq:composed-direction} only relax $e$ if $e$ will violate \cref{eq:dual-constraint-2} (line \ref{line:violates-tight-edge}), and includes the \refdef{hyperblossoms} of this \refdef{relaxer} in $\Delta'\vec{y}$.
As a result, the output $\Delta'\vec{y}[T]$ usually uses a small number of \refdef{relaxers} from $\{R'_i\}$ and will only create new \refdef{hyperblossoms} from them.
We note that our implementation actually employs an even more sophisticated algorithm whose output creates even fewer \refdef{hyperblossoms} and therefore achieves higher performance.

We show an example of \emph{batch relaxing} in \autoref{fig:cascaded-relaxing}.

\subsection{Clustering}\label{ssec:algo-cluster}

When the physical error rate is sufficiently low $p \ll 1$, a defect vertex is likely to be caused by a single incident hyperedge.
Many have exploited this property of locality to speed up decoding~\cite{fowler2012towards,fowler2013minimum,delfosse2021almost,higgott2025sparse,wu2023qce,delfosse2022toward,wu2025asplos}.
They group defect vertices into clusters and seek to find parity factors within the same cluster first.
For example, in the blossom algorithm~\cite{kolmogorov2009blossom}, alternating trees and matched pairs can be considered as clusters, while the (Hypergraph) Union-Find decoders~\cite{delfosse2021almost,delfosse2022toward} explicitly define clusters.
We formally define the concept of clusters and prove that the \emph{clustering} technique preserves global optimality under certain conditions.

The \hyperblossom framework organizes vertices, \refdef{tight-edges} and \refdef{hyperblossoms} into \refdef{clusters}.
Both the Primal and Dual phases operate on the \refdef{clusters} and the interface in \cref{ssec:algo-overview} becomes per-\refdef{cluster}.

\vspace{1ex}
\definitionlabel{cluster}{Cluster}
A \emph{Cluster} $C = (V_C, E_C)$ is a subgraph of the decoding hypergraph defined as:
\begin{itemize}
  \item A defect vertex $v \in D$ not connected by any \refdef{tight-edge} is a \emph{Cluster} $C = (\{v\}, \varnothing)$.
  \item A maximally connected subgraph of \refdef{tight-edges} $E_C \subseteq T$ is a \emph{Cluster} $C = (\cup_{e \in E_C} e, E_C)$.
  \item A \refdef{hyperblossom} $S \in \mathcal{B}$ spanning \refdef{clusters} $\mathcal{C}_S = \{ C \in \mathcal{C} | V_S \cap V_C \neq \varnothing \}$ merges them into a single \refdef{cluster} $C' = (\cup_{C \in \mathcal{C}_S} V_C \cup V_S, \cup_{C \in \mathcal{C}_S} E_C)$.
\end{itemize}

\vspace{1ex}
The edges of a \refdef{cluster} must be \refdef{tight-edges}, i.e., $E_C\subseteq T$.
By defining clusters this way, \hyperblossom implicitly applies the \refdef{tight-edge} heuristic (\cref{eq:cs-must-tight}).

The \refdef{hyperblossoms} of a \refdef{cluster} $C \in \mathcal{C}$ is the set of \refdef{hyperblossoms} whose vertices overlap with that of the \refdef{cluster}, represented by $\mathcal{B}_C = \{ S \in \mathcal{B} | V_S \cap V_C \neq \varnothing \}$.
According to the definition of \refdef{clusters}, the \refdef{hyperblossoms} of different \refdef{clusters} do not overlap, i.e., $\forall C_1, C_2 \in \mathcal{C}, C_1 \neq C_2 \implies \mathcal{B}_{C_1} \cap \mathcal{B}_{C_2} = \varnothing$.

\begin{figure}[t]
  \centering
  \includegraphics[width=\linewidth]{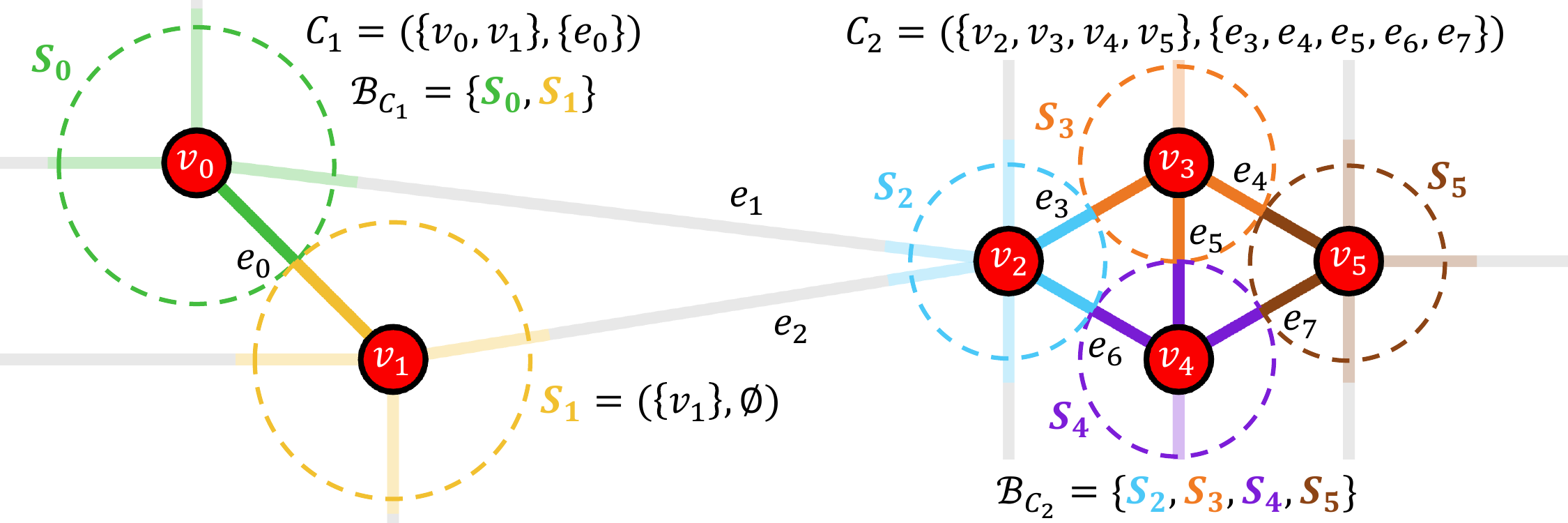}
  \caption{An example of \emph{clustering}. Two \refdef{clusters} $C_1$ and $C_2$ are not connected by any \refdef{tight-edge} (solid color), so they can be solved independently. The two \refdef{clusters} merge into one if either $e_1$ or $e_2$ becomes tight.}
  \label{fig:clustering}
\end{figure}

By definition, no two \refdef{clusters} share any vertex, \refdef{tight-edge} or \refdef{hyperblossom}.
A \refdef{cluster} only includes \refdef{tight-edges}, i.e., $E_C\subseteq T$.
\autoref{fig:clustering} illustrates a few examples of \refdef{clusters}.

\vspace{1ex}
\definitionlabel{locally-optimal-cluster}{Locally Optimal Cluster}
A \refdef{cluster} $C$ is called \emph{locally optimal} if there exists a parity factor $\mathcal{E}_C \subseteq E_C$ such that $\mathcal{D}(\mathcal{E}_C) = D \cap V_C$ and $W(\mathcal{E}_C) = \sum_{S \in \mathcal{B}_C} y_S$.

\begin{figure}[t]
  \begin{algorithm}[H]
    \caption{Check Cluster Local Optimality}\label{algo:clustering}
    \begin{algorithmic}[1]
      \Require{$C$ (\refdef{cluster})}
      \Ensure{whether $C$ is locally optimal}
      \Procedure{IsLocallyOptimal}{$C$}
      \If{$C \in \mathcal{O}$} \Comment{\refdef{invalid} \refdef{cluster}} \label{line:cluster-no-parity-factor}
      \State \Return {$\textsc{False}$}
      \EndIf
      \State $\mathcal{E}_C \gets \text{\refeqs{MWPF} of subgraph } C = (V_C, E_C)$ \label{line:clustering-finding-local-mwpf}
      \If{$W(\mathcal{E}_C) = \sum_{S \in \mathcal{B}_C} y_S$} \label{line:clustering-local-optimality}
      \State \Return {$\textsc{True}$}
      \EndIf
      \State \Return {$\textsc{False}$}
      \EndProcedure
    \end{algorithmic}
  \end{algorithm}

  \begin{algorithm}[H]
    \caption{Merge Clusters}\label{algo:merge-cluster}
    \begin{algorithmic}[1]
      \Require{$\mathcal{C}$ (\refdef{clusters}), $\mathcal{B}$ (\refdef{hyperblossoms}), $T$ (\refdef{tight-edges})}
      \Ensure{$\mathcal{C}$ (merged \refdef{clusters})}
      \Procedure{Merge}{$\mathcal{C}, \mathcal{B}, T$}
      \While{$\exists C \in \mathcal{C}, \exists e \in T, e \cap V_C \neq \varnothing, e \nsubseteq V_C$}
      \State $V_C \gets V_C \cup e$ \Comment{merge vertices into \refdef{clusters}}
      \EndWhile
      \State \Comment{no need to merge $\mathcal{B}$ because a per-\refdef{cluster} \refdef{relaxer} finder only finds \refdef{hyperblossom} $S = (V_S \subseteq V_C, E_S \subseteq E_C)$}
      \While{$\exists C_1, C_2 \in \mathcal{C}, V_{C_1} \cap V_{C_2} \neq \varnothing$}
      \State $\mathcal{C} \gets \mathcal{C} \cup \{ (V_{C_1} \cup V_{C_2}, E_{C_1} \cup E_{C_2}) \} \setminus \{C_1, C_2\}$\Comment{merge two \refdef{clusters}}
      \EndWhile
      \For{$C \in \mathcal{C}$}
      \State $E_C \gets E[V_C] \cap T$
      \State $\mathcal{B}_C \gets \{ S \in \mathcal{B} | V_S \cap V_C \neq \varnothing \}$
      \EndFor
      \State \Return {$\mathcal{C}$}
      \EndProcedure
    \end{algorithmic}
  \end{algorithm}

  \begin{algorithm}[H]
    \caption{Optimal Solutions from Clusters}\label{algo:cluster-optimal-solution}
    \begin{algorithmic}[1]
      \Require{$\mathcal{C}$ (\refdef{clusters})}
      \Ensure{$\mathcal{E}$, $\vec{y}$ (optimal \refeqs{MWPF} and \refeqs{DLP} solutions when all the \refdef{clusters} are locally optimal checked by \cref{algo:clustering})}
      \Procedure{OptimalSolutions}{$\mathcal{C}$}
      \State $\mathcal{E} \gets \varnothing$
      \State $\vec{y} \gets \varnothing$ \Comment{default $y_S = 0$ for all $S \in \mathcal{O}$}
      \For{$C \in \mathcal{C}$}
      \State $\mathcal{E}_C \gets \text{\refeqs{MWPF} of subgraph } C = (V_C, E_C)$ \label{line:optimal-solution-mwpf}
      \State $\mathcal{E} \gets \mathcal{E} \cup \mathcal{E}_C$
      \State $\vec{y} \gets \vec{y} \cup \{ S : y_S | S \in \mathcal{B}_C \}$
      \EndFor
      \State \Return {$\mathcal{E}, \vec{y}$}
      \EndProcedure
    \end{algorithmic}
  \end{algorithm}
\end{figure}

\vspace{1ex}
\theoremClusterOptimalityCriteria{cluster-optimality-criteria}

\vspace{1ex}
With the above criteria proved in \S\ref{ssec:cluster-optimality-proof}, we can use \cref{algo:clustering} to determine whether every \refdef{cluster} is a \refdef{locally-optimal-cluster}.
If not, we can optimize each suboptimal \refdef{cluster} individually by improving the \refdef{clusters} and then using \cref{algo:merge-cluster} to merge the \refdef{clusters}.
Once all the \refdef{clusters} are \refdef{locally-optimal-clusters}, we then use \cref{algo:cluster-optimal-solution} to find the global optimal solution.
In the special case where the \ref{condition:mwpf-short} is satisfied, \hyperblossom framework can find optimal \refeqs{MWPF} and \refeqs{DLP} solutions by operating on each \refdef{cluster} individually (\S\ref{ssec:hyperblossom-algorithm}).

A \hyperblossom algorithm starts with \refdef{clusters} of a single defect vertex; as it makes progress, it merges them as \refdef{tight-edges} (and \refdef{hyperblossoms}) are created.
In theory, when \refdef{tight-edges} become untight, due to relaxing or DLP progress, some \refdef{clusters} will split.
However, in implementation, as is in \cref{algo:cluster-optimal-solution}, we can forgo splitting and only merge (and grow) \refdef{clusters} in order to simplify data structure maintenance.
Because they do not overlap, the above theorem still holds.

With the \emph{clustering} technique, we can design optimal decoders that adaptively consider \refdef{clusters} by their size, starting with small ones.
This has the potential to achieve an average runtime almost-linear to the number of defects $|D| \propto p|V|$, like those fast MWPM decoders~\cite{fowler2012towards,higgott2025sparse,wu2023qce,wu2025asplos}.
As illustrated in \autoref{fig:clustering}, defect vertices remain isolated until a \refdef{tight-edge} connects them, at which point they are merged into a single \refdef{cluster}.
The \emph{clustering} technique does not improve the worst-case time complexity because a \refdef{cluster} may span the entire decoding hypergraph

\subsection{HyperBlossom Algorithm}\label{ssec:hyperblossom-algorithm}

The \hyperblossom algorithm combines \emph{relaxing} (\S\ref{ssec:cascaded-relaxing}) and \emph{clustering} (\S\ref{ssec:algo-cluster}) as described by \cref{algo:hyperblossom}.
Specifically, it applies \emph{batch relaxing} (\cref{algo:relaxing}) to each \refdef{cluster}. Using a stronger version of \reftheorem{relaxing}, it is able to find a \refdef{trivial-direction} much more efficiently, as proved in \S\ref{ssec:hyperblossom-optimality}.

\vspace{1ex}
\theoremRelaxerExistenceOrInvalidCluster{relaxer-existence-invalid-cluster}

\vspace{1ex}
Using the above theorem, the Primal phase, described by \cref{algo:hyperblossom-primal-phase},  can easily find a \refdef{trivial-direction} by checking whether the \refdef{cluster} is still \refdef{valid} after relaxing a maximal set of edges in line \ref{line:trivial-direction-hyperblossom}.
In the special case where the decoding hypergraph and all its subgraphs satisfy the \ref{condition:mwpf-short}, \cref{algo:hyperblossom} finds the optimal \refeqs{MWPF} and \refeqs{DLP} solutions, according to the following theorem proved in \S\ref{ssec:hyperblossom-optimality}.

\theoremHyperBlossomOptimality{hyperblossom-algorithm-optimality}

\vspace{1ex}
We note that when a property is true for a graph and all its subgraphs, the property is called \emph{hereditary}~\cite{farrugia2005factorizations}.
For example, being ``simple'' is a hereditary property because any subgraph of a simple graph is also a simple graph.
Given \reftheorem{simple-graph-optimality}, \ref{condition:mwpf} is a hereditary property of any simple graph.

Nullity$_{\le 1}$ is also a hereditary property of hypergraphs, as proved in \S\ref{ssec:optimal-1-dof}.
We define a \emph{nullity$_{\le 1}$ hypergraph} as a hypergraph whose incidence matrix $M_I$ has a nullity of at most 1, i.e., its null space ($M_I \vec{x} = 0$) has a dimensionality of at most 1.
Thus, the \hyperblossom algorithm is optimal for both simple graphs and nullity$_{\le 1}$ hypergraphs.

\begin{figure}[t]
  \begin{algorithm}[H]
    \caption{\hyperblossom Algorithm}\label{algo:hyperblossom}
    \begin{algorithmic}[1]
      \Require{$G$ (decoding hypergraph), $D$ (defects)}
      \Ensure{$\mathcal{E}, \vec{y}$ (a pair of optimal \refeqs{MWPF} and \refeqs{DLP} solutions if \ref{condition:mwpf}; otherwise, feasible solutions)}
      \Procedure{HyperBlossom}{$G, D$}
      \State $\mathcal{C} \gets \Call{InitializeDualPhase}{G, D}$
      \While{$(C, \Delta\vec{y} \gets \Call{PrimalPhase}{\mathcal{C}}) \neq \textsc{Nil}$} \label{line:dual-loop-hyperblossom}
      \State $\mathcal{C} \gets \Call{DualPhase}{\mathcal{C}, C, \Delta\vec{y}}$
      \EndWhile
      \State $\mathcal{E}, \vec{y} \gets \Call{OptimalSolutions}{\mathcal{C}}$ \Comment{\cref{algo:cluster-optimal-solution}}
      \State \Return {$\mathcal{E}, \vec{y}$}
      \EndProcedure
    \end{algorithmic}
  \end{algorithm}
  \begin{algorithm}[H]
    \caption{\hyperblossom Dual Phase}\label{algo:hyperblossom-dual-phase}
    \begin{algorithmic}[1]
      \Require{$G$ (decoding hypergraph), $D$ (defects)}
      \Ensure{$\mathcal{C}$ (\refdef{clusters})}
      \Procedure{InitializeDualPhase}{$G, D$}
      \State $\vec{y} \gets \textbf{0}$ \Comment{initial \refeqs{DLP} solution}
      \State $\mathcal{C} \gets \{ (\{v\}, \varnothing) | v \in D \}$
      \State $\forall C \in \mathcal{C}, \mathcal{B}_C^H \gets \{ (V_C, \varnothing) \}$ \Comment{\refdef{history}}
      \State $T \gets \{ e \in E | w_e = 0 \}$ \Comment{initial \refdef{tight-edges}}
      \State $\mathcal{C} \gets \Call{Merge}{\mathcal{C}, \varnothing, T}$ \Comment{initial \refdef{clusters}}
      \State \Return {$\mathcal{C}$}
      \EndProcedure
      \Procedure{DualPhase}{$\mathcal{C}, C, \Delta\vec{y}$} \Comment{update a \refdef{cluster}}
      \State $\mathcal{B}_C^H \gets \mathcal{B}_C^H \cup \{ S \in \mathcal{O} | \Delta y_S > 0 \}$ \label{line:update-history-hyperblossom}
      \State $\text{optimize partial \refeqs{DLP} of } \{ y_S | S \in \mathcal{B}_C^H \}$ \label{line:partial-dlp-hyperblossom}
      \State $\mathcal{B} \gets \{ S \in \mathcal{O} | y_S > 0 \}$ \Comment{\refdef{hyperblossoms}}
      \State $T \gets \{ e \in E | \sum_{S \in \mathcal{O} | e \in \delta(S)} y_S = w_e \}$ \Comment{\refdef{tight-edges}}
      \State $\mathcal{C} \gets \Call{Merge}{\mathcal{C}, \mathcal{B}, T}$
      \Comment{\cref{algo:merge-cluster}}
      \State \Return {$\mathcal{C}$}
      \label{line:hyperblossom-dual-return}
      \EndProcedure
    \end{algorithmic}
  \end{algorithm}
  \begin{algorithm}[H]
    \caption{\hyperblossom Primal Phase}\label{algo:hyperblossom-primal-phase}
    \begin{algorithmic}[1]
      \Require{$\mathcal{C}$ (\refdef{clusters} $C \in \mathcal{C}$ with their \refdef{hyperblossoms} $\mathcal{B}_C$ and \refdef{tight-edges} $E_C$)}
      \Ensure{A locally suboptimal \refdef{cluster} $C$ and a \refdef{useful-direction} $\Delta\vec{y}$; if not found, return \textsc{Nil}}
      \Procedure{PrimalPhase}{$\mathcal{C}$}
      \For{$C \in \mathcal{C}$}
      \If{$\Call{IsLocallyOptimal}{C}$} \Comment{(\cref{algo:clustering})}
      \State \textbf{continue} \Comment{skip locally optimal \refdef{clusters}}
      \EndIf
      \State $Rs \gets \Call{BatchedRelaxing}{C, \mathcal{B}_C, E_C}$\Comment{(Algo. \ref{algo:relaxing})}\label{line:call-relaxing-hyperblossom}
      \State $E'_C \gets E_C \setminus \cup_{R_i \in Rs} \mathcal{R}_i$ \Comment{remaining \refdef{tight-edges}} \label{line:remaining-tight-edges-hyperblossom}
      \If{$(V_C, E'_C) \in \mathcal{O}$}  \label{line:trivial-direction-hyperblossom}
      \State $S \gets \text{any \refdef{invalid} subgraph of}\ (V_C, E'_C)$
      \State $\Delta\vec{y} \gets \{ \Delta y_S: +1 \}$ \Comment{\refdef{trivial-direction}}\label{line:trivial-direction-hyperblossom-value}
      \State \Return {$C, \Call{Compose}{Rs, \Delta\vec{y}}$}\Comment{(\cref{algo:compose})}\label{line:compose-trivial-direction-hyperblossom}
      \EndIf
      \EndFor
      \State \Return {$\textsc{Nil}$}
      \EndProcedure
    \end{algorithmic}
  \end{algorithm}
\end{figure}

When the conditions to \reftheorem{hyperblossom-algorithm-optimality} are not satisfied, the \hyperblossom algorithm does not always output optimal solution.
Nonetheless, it still provides certifying information, which is the final primal-dual gap, according to the inequality chain \cref{eq:mwpf-chain}.

The worst-case time complexity of the algorithm depends on the \refdef{relaxer}-finding algorithm and is exponential with regard to the size of the decoding graph.
There are at most $|\mathcal{O}|$ calls of \textsc{PrimalPhase} (line \ref{line:dual-loop-hyperblossom}), each incrementing the number of \refdef{history} $|\cup_{C \in \mathcal{C}} \mathcal{B}_C^H|$ by at least one.
Within each call, we need to find at most $|T| \le |E|$ \refdef{relaxers} (line \ref{line:call-relaxing-hyperblossom}).
Suppose the \refdef{relaxer}-finding algorithm runs in $O(F(|V|,|E|))$ time, then the time complexity of finding a \refdef{useful-direction} is $O(|E| F(|V|,|E|))$.
We then need to run the LP solver for each iteration, consuming $O(|\mathcal{O}|^{1.5} |V|^2)$ time~\cite{karmarkar1984new} given $O(|\mathcal{O}|)$ variables and $O(|V|)$ constraints.
Overall, the time complexity of \cref{algo:hyperblossom} is $O(|\mathcal{O}|^{2.5} |V|^2 + |\mathcal{O}| |E| F(|V|,|E|))$.

\section{Relaxer-Finding Algorithms}\label{sec:subroutine}

\begin{figure}[t]
  \centering
  \begin{subfigure}[t]{0.4\linewidth}
    \centering
    \includegraphics[width=0.7\textwidth,page=1]{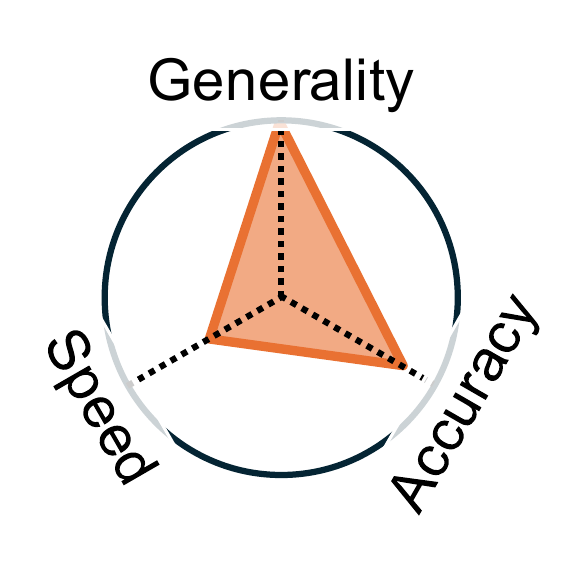}
    \caption{\emph{SingleHair} (\S\ref{ssec:single-hair-subroutine})}
    \label{fig:tradeoffs-single-hair}
  \end{subfigure}
  \hspace{4mm}
  \begin{subfigure}[t]{0.4\linewidth}
    \centering
    \includegraphics[width=0.7\textwidth,page=4]{figures/slides/tradeoffs.pdf}
    \caption{\emph{UnionFind} (\S\ref{ssec:union-find-subroutine})}
    \label{fig:tradeoffs-union-find}
  \end{subfigure}\\
  \begin{subfigure}[t]{0.4\linewidth}
    \centering
    \includegraphics[width=0.7\textwidth,page=2]{figures/slides/tradeoffs.pdf}
    \caption{\emph{Blossom} (\S\ref{ssec:blossom-subroutine})}
    \label{fig:tradeoffs-blossom}
  \end{subfigure}
  \hspace{4mm}
  \begin{subfigure}[t]{0.4\linewidth}
    \centering
    \includegraphics[width=0.7\textwidth,page=3]{figures/slides/tradeoffs.pdf}
    \caption{\emph{Nullity$_{\le 1}$} (\S\ref{ssec:biased-single-dof})}
    \label{fig:tradeoffs-single-dof}
  \end{subfigure}
  \caption{Tradeoffs of different \refdef{relaxer} finders. Generality, speed and accuracy cannot be achieved simultaneously given the NP-hardness of \refeqs{MWPF}~\cite{berlekamp1978inherent}.
  }
  \label{fig:tradeoffs}
\end{figure}

With the \emph{relaxing} (\S\ref{ssec:cascaded-relaxing}) and \emph{clustering} (\S\ref{ssec:algo-cluster}) techniques, we reduce the MWPF decoding problem to \refdef{relaxer}-finding algorithms that find \refdef{relaxers} for each \refdef{cluster} $C \in \mathcal{C}$.
We present four \refdef{relaxer}-finding algorithms or \refdef{relaxer} finders that make different tradeoffs in generality, speed, and accuracy, as illustrated by \autoref{fig:tradeoffs}.
Here ``generality'' refers to an algorithm's ability for handling different types of decoding hypergraphs, e.g., algorithms that only works on simple graphs are less general.

\begin{itemize}
  \item \emph{SingleHair}  (\S\ref{ssec:single-hair-subroutine}) finds \refdef{relaxers} on general hypergraphs, which is an efficient algorithm that runs in $O(\text{poly}(|V| + |T| + |\mathcal{B}|))$.
  \item \emph{UnionFind} (\S\ref{ssec:union-find-subroutine}) does not find any \refdef{relaxer}, which is efficient but less accurate.
  \item \emph{Blossom}  (\S\ref{ssec:blossom-subroutine}) is efficient and optimal but only works for simple graphs.
  \item \emph{Nullity$_{\le 1}$} (\S\ref{ssec:biased-single-dof}) is efficient and optimal on a class of decoding hypergraphs whose incidence matrix has a nullity of 0 or 1.
\end{itemize}

We say a \refdef{relaxer}-finding algorithm is \emph{optimal} if the \hyperblossom algorithm equipped with it finds an optimal \refeqs{DLP} solution.
To prove a \refdef{relaxer} finder is optimal, it suffices to prove that the \refdef{relaxer} finder always finds a \refdef{relaxer} for a suboptimal \refeqs{DLP} solution or returns \textsc{Nil} when there exists an \refdef{invalid} \refdef{cluster}, according to \reftheorem{relaxer-existence-invalid-cluster}.
For the special classes of hypergraphs that they are intended to work, \emph{Blossom} and \emph{Nullity{$_{\le 1}$}} are optimal.
\emph{SingleHair} and \emph{UnionFind} work for general hypergraphs but are not optimal.
In \S\ref{sec:analytic}, we compare \textit{SingleHair} with the less general but optimal ones, i.e., \textit{Blossom} and \textit{Nullity$_{\le 1}$},
to reveal its limitations and potential improvements, which may inspire future research into better \refdef{relaxer}-finding algorithms.

Note that we describe the \refdef{relaxer} finders using the language of the decoding hypergraph $(G, \mathcal{B}, T)$ for simplicity (See \cref{algo:relaxing} line \ref{line:find-relaxers}).
With the \emph{clustering} technique, these \refdef{relaxer} finders naturally apply to each \refdef{cluster} $C \in \mathcal{C}$ individually, by passing in $(C, \mathcal{B}_C, E_C)$ as the input (See \cref{algo:hyperblossom-primal-phase} line \ref{line:call-relaxing-hyperblossom}).

Moreover, \emph{relaxing} allows each time one or more \refdef{relaxer} finders to be used, i.e., at line \ref{line:find-relaxers} of \cref{algo:relaxing}.
Therefore, the \hyperblossom framework can easily combine different \refdef{relaxer} finders to achieve desirable tradeoffs.
For example, using the faster \emph{Nullity$_{\le 1}$} first and then the more general \emph{SingleHair} will speed up decoding biased noise models.

\subsection{\emph{SingleHair} Relaxer Finder}\label{ssec:single-hair-subroutine}

We present a simple \refdef{relaxer} finding algorithm for general hypergraphs, also described by \cref{algo:single-hair}.
When $(V, T)$ is an \refdef{invalid} subgraph, it returns \textsc{Nil}.
Otherwise, it attempts to find \refdef{relaxers} of the form $\Delta\vec{y} = \{ \Delta y_{S}: -1, \Delta y_{S^+}: +1 \}$: one \refdef{hyperblossom} $S \in \mathcal{B}$ shrinks and another \refdef{invalid} subgraph $S^+ \in \mathcal{O}$ grows.
The algorithm examines each \refdef{hyperblossom} $S \in \mathcal{B}$, constructs as many $S^+$ using its parity matrix (\S\ref{ssec:parity-matrix}), and checks if $\Delta\vec{y} = \{ \Delta y_{S}: -1, \Delta y_{S^+}: +1 \}$ is a \refdef{relaxer}.

\subsubsection{Matrix Representation}\label{ssec:parity-matrix}

The \emph{SingleHair} \refdef{relaxer} finder uses a matrix representation of the parity constraints called \textit{parity matrix}.
A \textit{parity matrix} $M_G$ is an augmented matrix in linear algebra with $|V|$ rows and $|E| + 1$ columns and defined on the modulo two field $\mathbb{F}_2$.
The first $|E|$ columns correspond to the incidence matrix of the decoding hypergraph, while the last column is the syndrome $D$.
The parity constraints set by $D$ can be represented by $M_G (\vec{x}, 1)^T = 0$.

\begin{figure}
  \centering
  \begin{subfigure}[t]{.98\linewidth}
    \centering
    \includegraphics[width=1\textwidth]{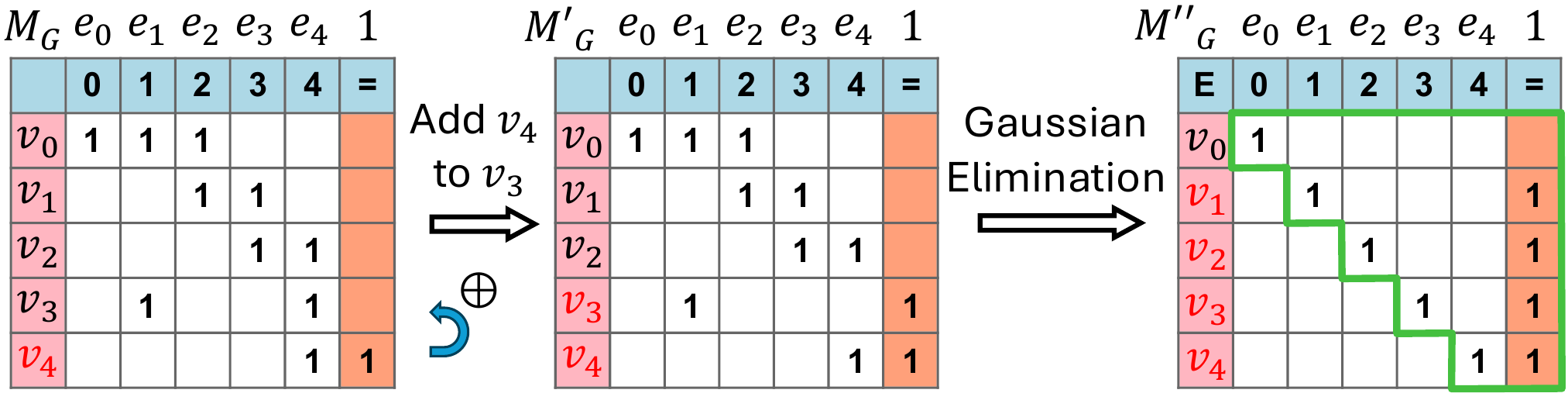}
    \caption{Equivalent Parity Matrices $M_G \cong M'_G \cong M''_G$.}
    \label{fig:example-matrix}
  \end{subfigure}

  \renewcommand{\thesubfigure}{$G$}
  \begin{subfigure}[t]{.31\linewidth}
    \centering
    \includegraphics[width=1\textwidth]{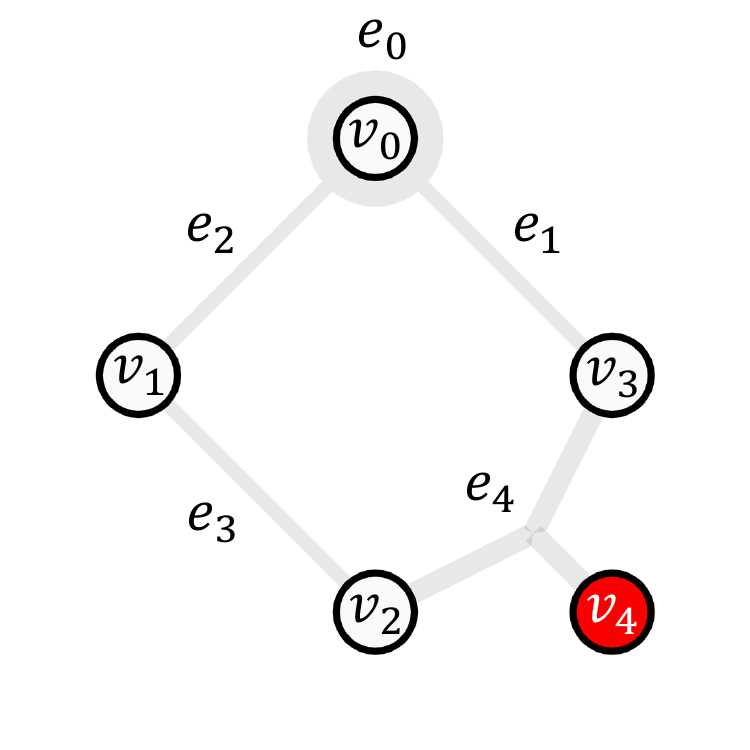}
    \caption{$\equiv M_G$.}
    \label{fig:example}
  \end{subfigure}
  \renewcommand{\thesubfigure}{$G'$}
  \begin{subfigure}[t]{.31\linewidth}
    \centering
    \includegraphics[width=1\textwidth]{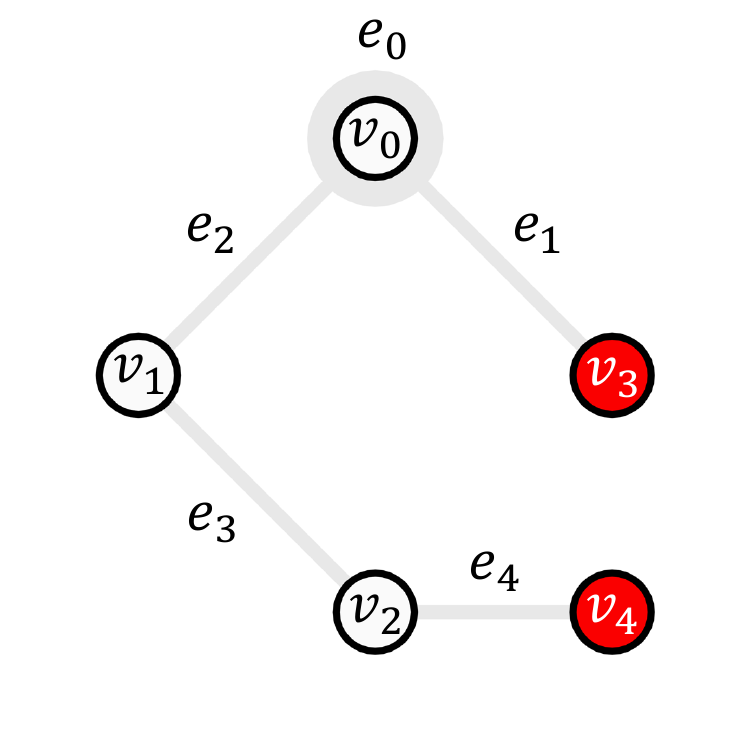}
    \caption{$\equiv M'_G$.}
    \label{fig:example2}
  \end{subfigure}
  \renewcommand{\thesubfigure}{$G''$}
  \begin{subfigure}[t]{.31\linewidth}
    \centering
    \includegraphics[width=1\textwidth]{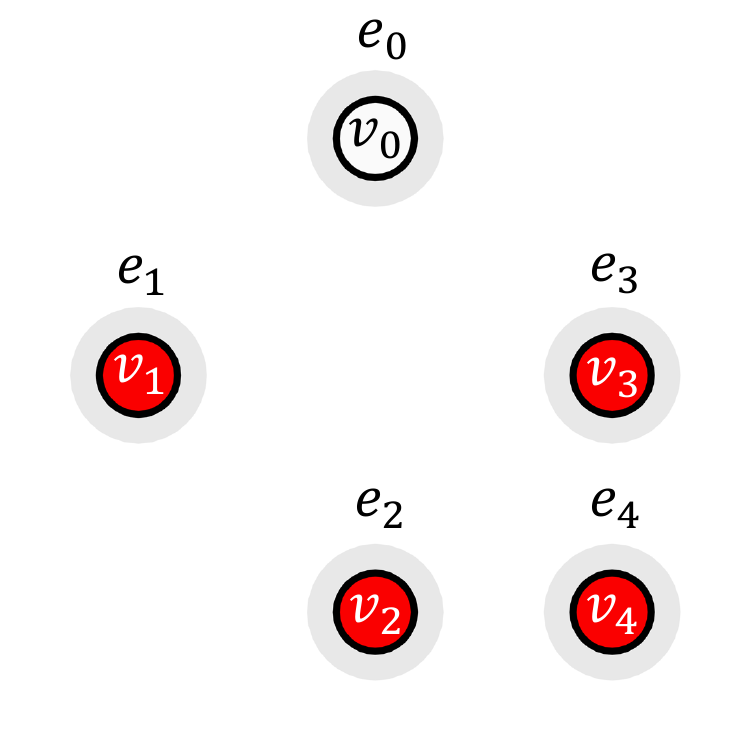}
    \caption{$\equiv M''_G$.}
    \label{fig:example3}
  \end{subfigure}

  \caption{
    Examples of equivalent parity matrices $M_G \cong M'_G \cong M''_G$ and their corresponding decoding hypergraphs $G \cong G' \cong G''$.
    We construct $M'_G$ by adding row $v_4$ to row $v_3$ of $M_G$.
    The same transformation can be understood as adding the vertex $v_4$ to $v_3$ in the decoding hypergraph $G$ while updating the adjacent hyperedges of $v_3$.
    Applying Gauss-Jordan elimination on $M'_G$ results in an \emph{Echelon Matrix} $M''_G$ and a simplified decoding hypergraph $G''$.
  }
  \label{fig:parity-matrix-example}
\end{figure}

We show an example of a parity matrix in \autoref{fig:example-matrix}.

\begin{figure*}[t]
  \centering

  \begin{subfigure}[t]{0.144\linewidth}
    \centering
    \includegraphics[width=\textwidth]{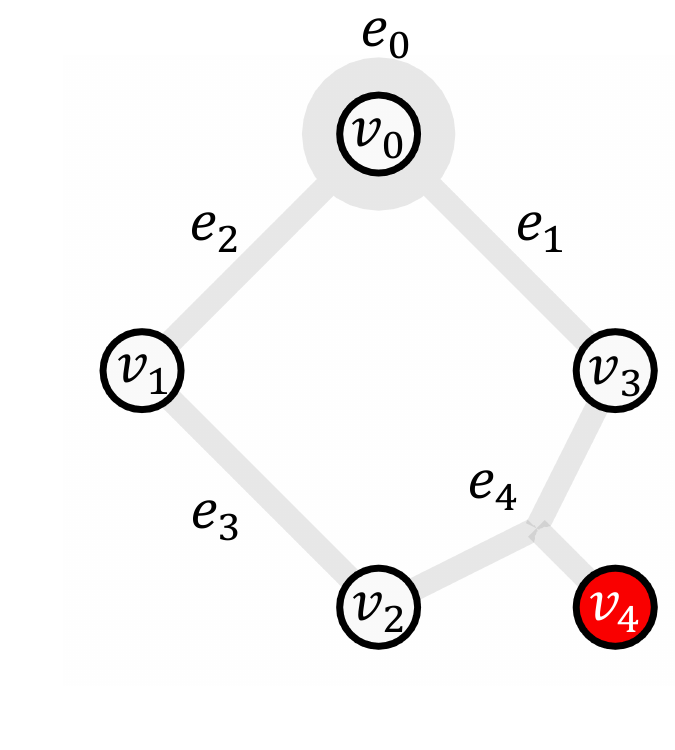}
    \caption{Initial $\vec{y} = 0$}
    \label{fig:union-find-split-0}
  \end{subfigure}
  \begin{subfigure}[t]{0.245\linewidth}
    \centering
    \includegraphics[width=\textwidth]{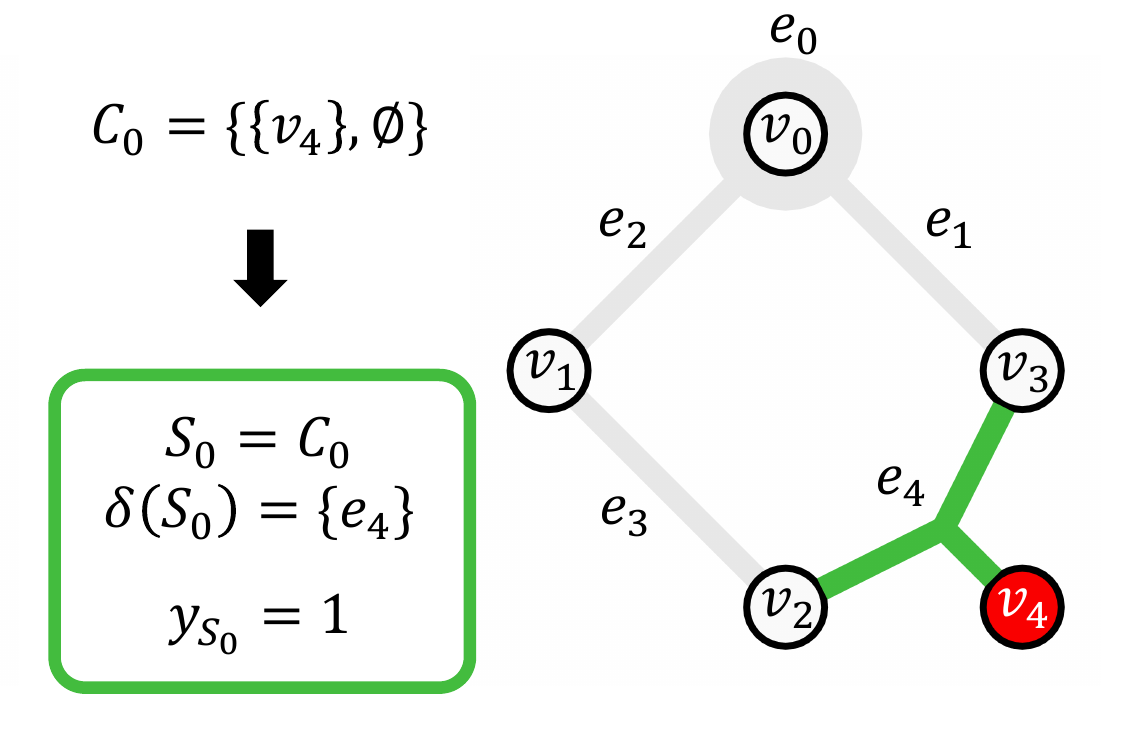}
    \caption{Grow \refdef{invalid} subgraph $S_0$}
    \label{fig:union-find-split-1}
  \end{subfigure}
  \begin{subfigure}[t]{0.268\linewidth}
    \centering
    \includegraphics[width=\textwidth]{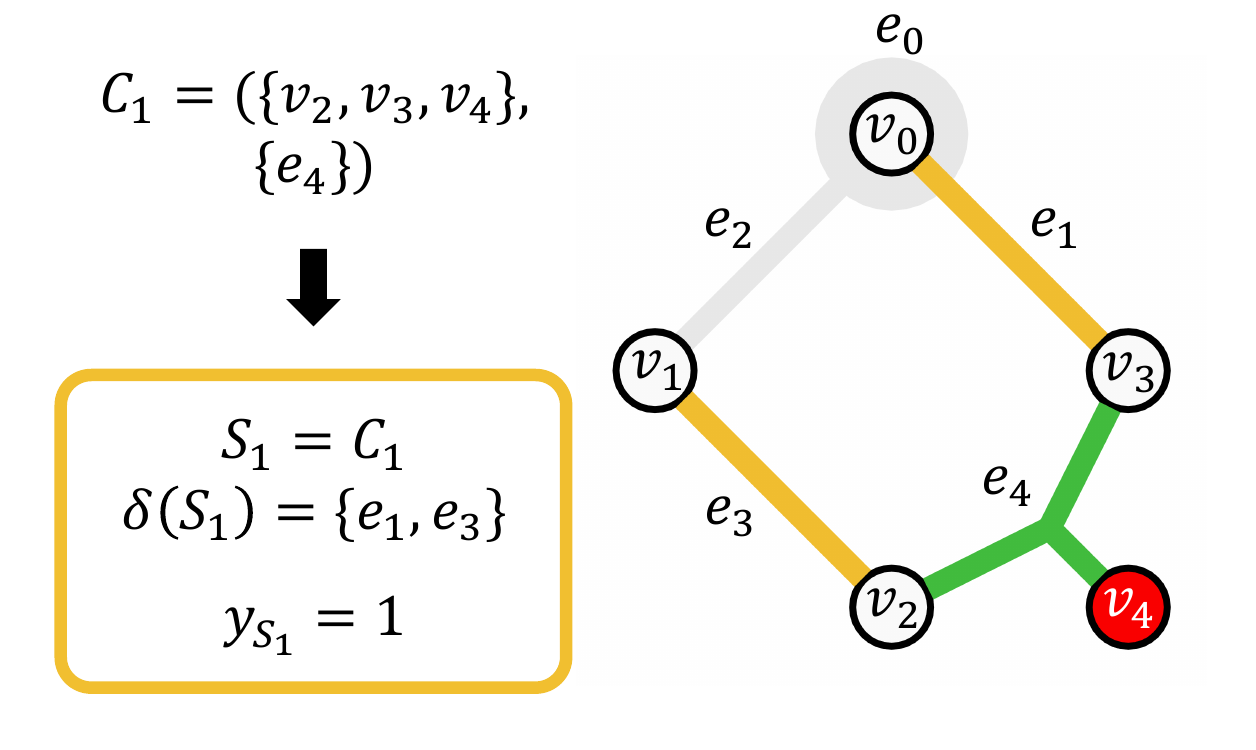}
    \caption{Grow \refdef{invalid} subgraph $S_1$}
    \label{fig:union-find-split-2}
  \end{subfigure}
  \begin{subfigure}[t]{0.303\linewidth}
    \centering
    \includegraphics[width=\textwidth]{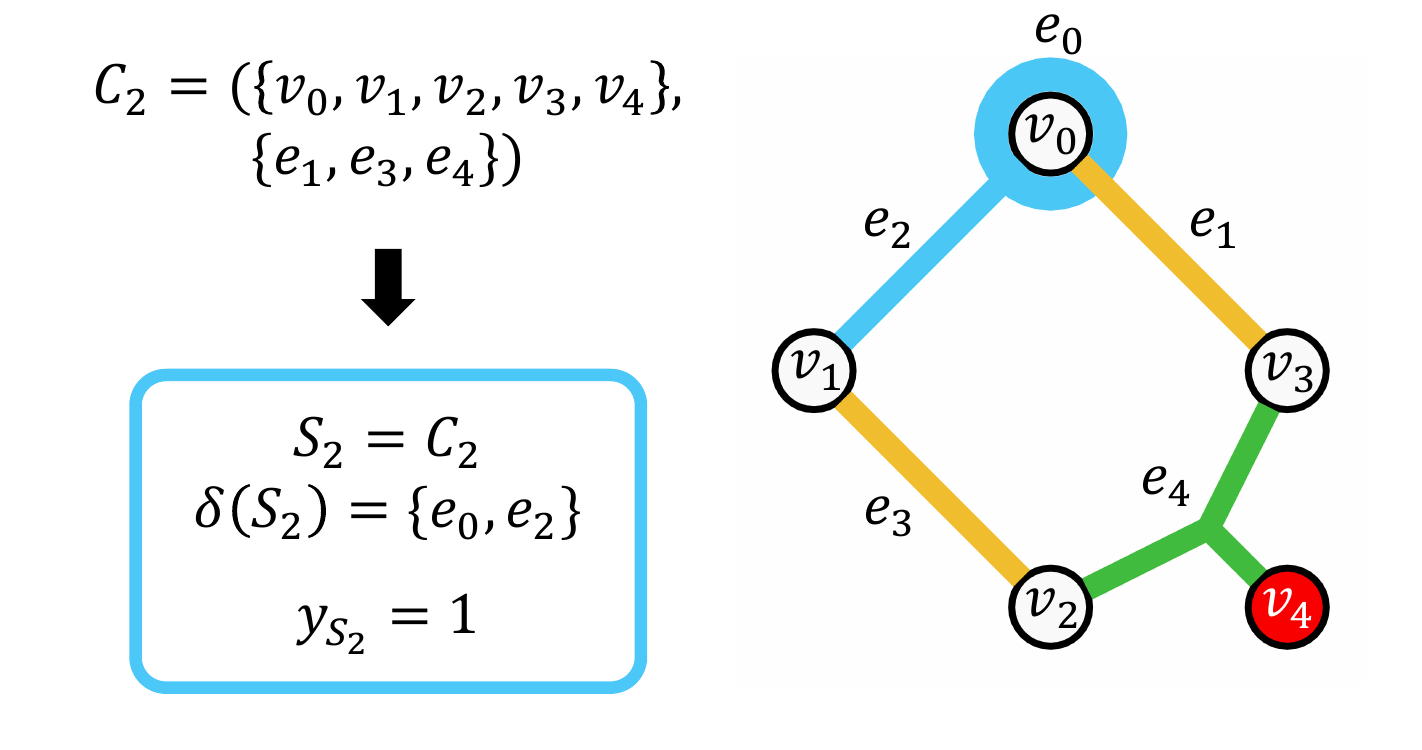}
    \caption{Grow $S_2$ into a \refdef{valid} \refdef{cluster}}
    \label{fig:union-find-split-3}
  \end{subfigure}
  \vspace{2ex}

  \begin{subfigure}[t]{0.179\linewidth}
    \centering
    \includegraphics[width=\textwidth]{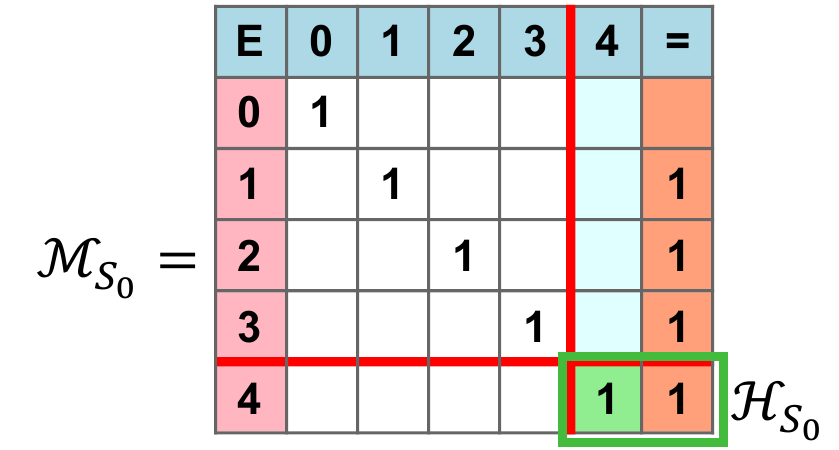}
    \caption{$\mathcal{H}_{S_0}$: an Odd row}
    \label{fig:hyperblossom-matrices-0}
  \end{subfigure}
  \begin{subfigure}[t]{0.461\linewidth}
    \centering
    \includegraphics[width=\textwidth]{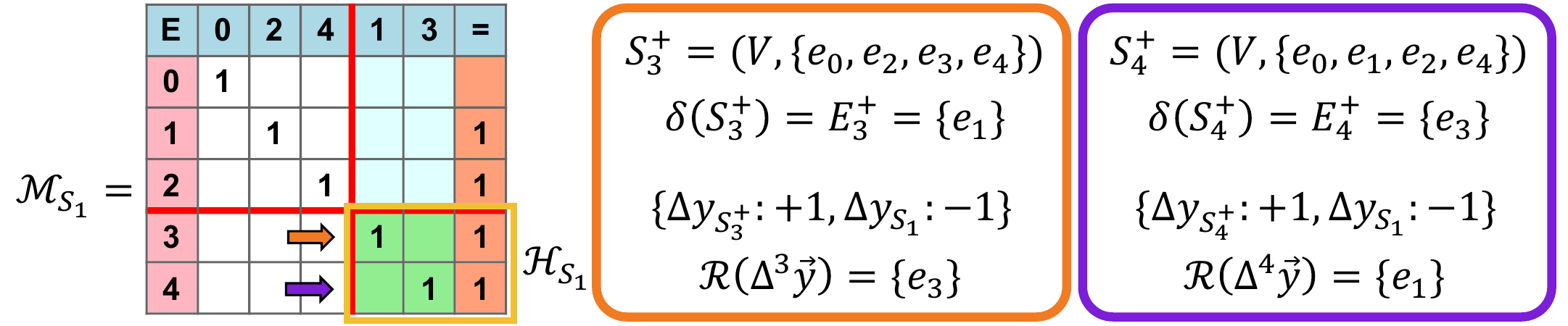}
    \caption{$\mathcal{H}_{S_1}$ finds two \refdef{relaxers} $\Delta^3\vec{y}$,  $\Delta^4\vec{y}$ for each Odd row}
    \label{fig:hyperblossom-matrices-1}
  \end{subfigure}
  \begin{subfigure}[t]{0.320\linewidth}
    \centering
    \includegraphics[width=\textwidth]{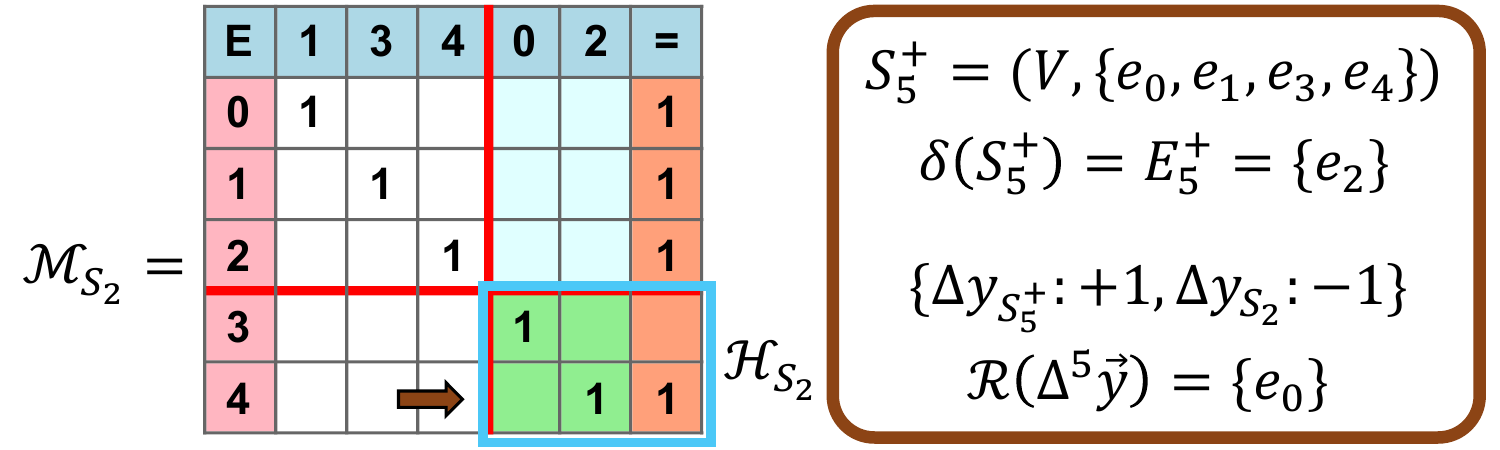}
    \caption{$\mathcal{H}_{S_2}$ finds one \refdef{relaxer} $\Delta^5\vec{y}$}
    \label{fig:hyperblossom-matrices-2}
  \end{subfigure}
  \vspace{2ex}

  \begin{subfigure}[t]{0.160\linewidth}
    \centering
    \includegraphics[width=\textwidth,page=1]{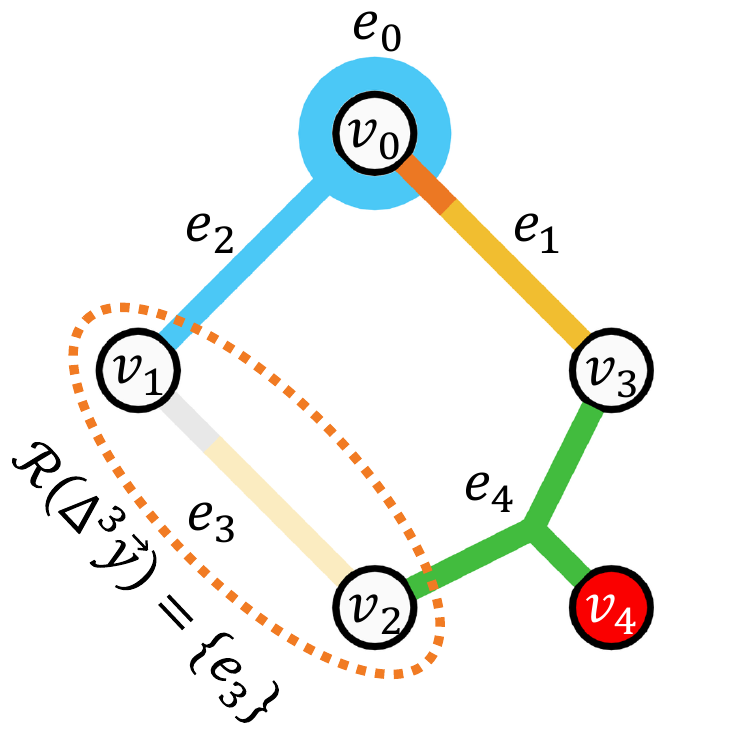}
    \caption{\refdef{relaxer} $\Delta^3\vec{y}$}
    \label{fig:relaxers-optimal-solutions-split-0}
  \end{subfigure}
  \begin{subfigure}[t]{0.160\linewidth}
    \centering
    \includegraphics[width=\textwidth,page=2]{figures/example-single-hair/relaxers-optimal-solutions.pdf}
    \caption{\refdef{relaxer} $\Delta^4\vec{y}$}
    \label{fig:relaxers-optimal-solutions-split-1}
  \end{subfigure}
  \begin{subfigure}[t]{0.160\linewidth}
    \centering
    \includegraphics[width=\textwidth,page=3]{figures/example-single-hair/relaxers-optimal-solutions.pdf}
    \caption{\refdef{relaxer} $\Delta^5\vec{y}$}
    \label{fig:relaxers-optimal-solutions-split-2}
  \end{subfigure}
  \begin{subfigure}[t]{0.160\linewidth}
    \centering
    \includegraphics[width=\textwidth,page=4]{figures/example-single-hair/relaxers-optimal-solutions.pdf}
    \caption{\refeqs{DLP} solution $\vec{y}$}
    \label{fig:relaxers-optimal-solutions-split-3}
  \end{subfigure}
  \begin{subfigure}[t]{0.160\linewidth}
    \centering
    \includegraphics[width=\textwidth,page=6]{figures/example-single-hair/relaxers-optimal-solutions.pdf}
    \caption{parity factor $\mathcal{E}$}
    \label{fig:relaxers-optimal-solutions-split-5}
  \end{subfigure}
  \begin{subfigure}[t]{0.160\linewidth}
    \centering
    \includegraphics[width=\textwidth,page=5]{figures/example-single-hair/relaxers-optimal-solutions.pdf}
    \caption{MWPF Proof}
    \label{fig:relaxers-optimal-solutions-split-4}
  \end{subfigure}
  \vspace{1ex}

  \caption{
    Example of the \hyperblossom algorithm with the \emph{SingleHair} \refdef{relaxer}-finding algorithm.
    The decoding hypergraph has a uniform weight $w_e = 1$ for all edges $e \in E$.
    (a-d) When the \refdef{cluster} $C$ is \refdef{invalid}, we keep growing the dual variable $y_C$.
    The \refdef{cluster} merges more vertices and \refdef{tight-edges} until it becomes \refdef{valid}.
    In this example, the \refdef{valid} \refdef{cluster} is $C = (V, E)$ and its \refdef{hyperblossoms} are $\mathcal{B}_C = \{ S_0, S_1, S_2 \}$.
    (e-g) For each \refdef{hyperblossom} $S \in \mathcal{B}$, the \emph{SingleHair} \refdef{relaxer}-finding algorithm constructs a \refdef{hyperblossom-matrix} $\mathcal{M}_S$ and a \refdef{hair-matrix} $\mathcal{H}_S$.
    Each Odd row in the \refdef{hair-matrix} corresponds to a \refdef{relaxer} $\Delta^i\vec{y}, \forall i \in \{ 3, 4, 5 \}$.
    Each \refdef{relaxer} grows a new dual variable $y_{S^+_i}$ and shrinks an existing dual variable $y_{S_j}$.
    (h-j) Growing along the \refdef{direction} $\Delta^i\vec{y}$ for a small length of $1 / 4$ relaxes the edges $\mathcal{R}(\Delta^i\vec{y})$ in the dotted circle.
    (k) We find a better \refeqs{DLP} solution $\vec{y}$ by solving the linear programming problem on the new \refdef{hyperblossoms} $\mathcal{B}_C^H = \{ S_0, S_1, S_2, S^+_3, S^+_4, S^+_5 \}$.
    (l) We find a parity factor (blue edges) among the \refdef{tight-edges}.
    (m) We prove the optimality of both the parity factor and $\vec{y}$ by \ref{theorem:provable-optimum}: the primal sum $\sum_{e \in E} w_e x_e$ is equal to the dual sum $\sum_{S \in \mathcal{O}} y_S$.
  }
  \label{fig:single-hair-example}
\end{figure*}

There are two transformations of the parity matrix that preserve the solution space of $\vec{x}$.
That is, $M_G (\vec{x}, 1)^T = 0$ if and only if $M'_G (\vec{x}, 1)^T = 0$ where $M'_G$ is derived from $M_G$ based on these transformations.
In other words, a parity factor for $G$ is also one for $G'$ and vice versa.
First, we can re-order the rows and columns, which correspond to renaming the vertices and edges, respectively.

Second, we can add one row to another row in $\mathbb{F}_2$, as demonstrated in \autoref{fig:example-matrix}.
Given the linearity of the  $\mathbb{F}_2$ constraints, adding one constraint to another will not change the solution space~\cite{leon1998linear}.

With the \emph{Row Addition}, we can use Gauss-Jordan elimination to obtain the \emph{reduced row echelon form}~\cite{leon1998linear} of the parity matrix, as shown in \autoref{fig:example-matrix}.
We call it \emph{Echelon Matrix}.
The columns that correspond to the row-leading 1s in the \emph{Echelon Matrix} are called \textit{pivot variables}, while the rest are \textit{free variables}.
The row containing the leading 1 of a pivot variable is called the \textit{pivot row} of the variable.
In the example of $M''_G$ in \autoref{fig:example-matrix}, all the variables are pivot and the solution space is singleton.
However, general QEC codes usually have a large solution space with a lot of free variables due to quantum degeneracy~\cite{demarti2024decoding}.
That is, given a parity factor $\mathcal{E}$ and any logical operator $\mathcal{E}_L \subseteq E, \mathcal{D}(\mathcal{E}_L)=\varnothing$, $\mathcal{E'} = \mathcal{E} \oplus \mathcal{E}_L$ is also a parity factor because $\mathcal{D}(\mathcal{E}') = \mathcal{D}(\mathcal{E}) \oplus \mathcal{D}(\mathcal{E}_L) = D$.
The time complexity of Gauss-Jordan elimination is $O(|V||E|^2)$, or $O(|V|^3)$ for sparse decoding hypergraphs from qLDPC codes.
We denote equivalent parity matrices as $M_G \cong M'_G$ and and their decoding hypergraphs as $G \cong G'$.

\subsubsection{Important Concepts}

\definitionlabelmatrix{hyperblossom}{Hyperblossom} Given a \refdef{hyperblossom} $S \in \mathcal{B}$, the \refdef{hyperblossom-matrix} $\mathcal{M}_{S}$ is an \emph{Echelon Matrix} (\S\ref{ssec:parity-matrix}) of the subgraph $(V, T)$ that places all the columns corresponding to tight \refdef{hairs} of $S$, ($\delta(S) \cap T$), on the rightmost.

\vspace{1ex}
As an example, we show \refdef{hyperblossom-matrices} in \autoref{fig:union-find-split-3} with three \refdef{hyperblossoms} $\mathcal{B} = \{ S_0, S_1, S_2 \}$ (\autoref{fig:hyperblossom-matrices-0} to \autoref{fig:hyperblossom-matrices-2}).
By defining the \refdef{hyperblossom-matrix} only on the \refdef{tight-edges} $T$, we implicitly assume that non-tight edges are never selected as a heuristics (\cref{eq:cs-must-tight}).

\vspace{1ex}
\definitionlabelmatrix{hair}{Hair} Given a \refdef{hyperblossom} $S \in \mathcal{B}$, the \refdef{hair-matrix} $\mathcal{H}_{S}$ is a submatrix of $\mathcal{M}_{S}$, consisting of the rightmost $|\delta(S) \cap T| + 1$ columns and all the rows below the last pivot row of the variables $T \setminus \delta(S)$.

\vspace{1ex}
The \refdef{hair-matrix} represents the solution space for the primal variables corresponding to the tight \refdef{hairs} of $S$, i.e., $x_e, \forall e \in \delta(S) \cap T$.
For example, \autoref{fig:hyperblossom-matrices-0} shows a \refdef{hair-matrix} $\mathcal{H}_{S_0}$ that constrains the parity factor solution space to have $x_{e_4} = 1$.

\vspace{1ex}
\lemmaHairMatrixOddRowExistence{hair-matrix-odd-row-existence}

\vspace{1ex}
For example, \autoref{fig:hyperblossom-matrices-2} shows an Odd row in the \refdef{hair-matrix} $\mathcal{H}_{S_2}$.
We prove the above lemma in \S\ref{ssec:hair-matrix-odd-row-existence}.

\subsubsection{Relaxer Finding Algorithm}
The algorithm enumerates through all \refdef{hyperblossoms}.
For each \refdef{hyperblossom} $S \in \mathcal{B}$, the algorithm constructs its \refdef{hair-matrix}.
According to the Lemma above, the algorithm finds all the Odd rows.
An Odd row poses an odd parity constraint on a subset of primal variables $E^+ \subseteq \delta(S) \cap T$.
That is, $E^+$ consists of the edges whose corresponding columns in the Odd row have value 1.
For example, $E^+ = \{ e_2 \}$ for the Odd row in $\mathcal{H}_{S_2}$ in \autoref{fig:hyperblossom-matrices-2}.

The algorithm then constructs $S^+ = (V, T \setminus E^+)$.
Because edges in $E^+$ are essential to satisfy the odd parity constraint of the Odd row, $S^+$ must be an \refdef{invalid} subgraph, i.e., $S^+ = (V, T \setminus E^+) \in \mathcal{O}$.

We show that $\Delta\vec{y} = \{ \Delta y_{S}: -1, \Delta y_{S^+}: +1 \}$ is a \refdef{feasible-direction}.
As $S$ shrinks, it will relax the DLP constraint \cref{eq:dual-constraint-2} for edges in $\delta(S)$ and therefore those in $E^+\subseteq \delta(S)$.
As a result, when $S^+$ grows and tightens the constraint for $\delta(S^+) \cap T = E^+$,  there will not be any violation of \cref{eq:dual-constraint-2}.

Finally, because $\Delta\vec{y}$ relaxes edges $E^- = \delta(S) \cap T \setminus E^+$, it is a \refdef{relaxer} if $E^- \neq \varnothing$. And the algorithm will continue to construct $S^+$ from remaining Odd rows and check if $\Delta\vec{y} = \{ \Delta y_{S}: -1, \Delta y_{S^+}: +1 \}$ is a \refdef{relaxer}.
If $E^-=\varnothing$, there will be no more Odd rows according the following lemma proved in \S\ref{ssec:hair-matrix-unique-row}.

\vspace{1ex}
\lemmaUniqueRowIfNotARelaxer{unique-row-if-not-a-relaxer}

\subsubsection{Time Complexity and Limitations}

When the algorithm has enumerated all \refdef{hyperblossoms} and could find no more \refdef{relaxers}, the \refdef{hair-matrices} of all \refdef{hyperblossoms} must have become a row of all 1s, as shown in \autoref{fig:hyperblossom-matrices-0}.
When this happens, given a tight \refdef{hair} $e$ of $S$ ($e\in \delta(S) \cap T$), there exists a feasible parity factor that includes $e$ but not other tight \refdef{hairs} of $S$, i.e., $\exists \mathcal{E} \subseteq T, \mathcal{D}(\mathcal{E}) = D, \mathcal{E} \cap \delta(S) = \{ e \}$, according to the property of an Echelon Matrix (\S\ref{ssec:parity-matrix}).
We say the \refeqs{DLP} solution is in a \emph{Single-Hair} state.
That is, the \emph{SingleHair} \refdef{relaxer} finder will not be able to find more \refdef{relaxers} if the \refeqs{DLP} solution is in a Single-Hair state.

\emph{SingleHair} runs in polynomial time of $O(\text{poly}(|V| + |T| + |\mathcal{B}|))$.
However, when using \emph{SingleHair}, the number of iterations of the \hyperblossom algorithm (\cref{algo:hyperblossom}, line \ref{line:dual-loop-hyperblossom})  may not be polynomially bounded, given the NP-hardness of the general \refeqs{MWPF} problem~\cite{berlekamp1978inherent}.
Also, the \hyperblossom algorithm equipped with \emph{SingleHair} alone may not eventually find an optimal \refeqs{DLP} solution.

Being a very simple \refdef{relaxer} finder algorithm, \emph{SingleHair} has some limitations, as explained in \S\ref{sec:analytic}.
Nonetheless, our empirical evaluation shows that it achieves the same accuracy as the MWPM decoder on simple graphs (\autoref{fig:eva-rsc-bit-flip}), where MWPM decoder is an optimal MLE decoder.
Moreover, it achieves a higher accuracy than existing UF~\cite{delfosse2021almost}, MWPM~\cite{dennis2002topological} and HUF~\cite{delfosse2022toward} decoders on hypergraphs (\S\ref{sec:evaluation}).
\emph{SingleHair} does not assume or exploit any special property of the decoding hypergraph.
We will discuss two \refdef{relaxer} finders that exploit the special properties of the decoding hypergraph in \S\ref{ssec:blossom-subroutine} and \S\ref{ssec:biased-single-dof}.

\begin{figure}[t]
  \begin{algorithm}[H]
    \caption{\emph{SingleHair} \refdef{relaxer} finder}\label{algo:single-hair}
    \begin{algorithmic}[1]
      \Require{$G$ (hypergraph), $\mathcal{B}$ (\refdef{hyperblossoms}), $T$ (\refdef{tight-edges})}
      \Ensure{$R[G,\mathcal{B},T]$ (\refdef{relaxer}) or $\textsc{Nil}$ (if not found)}
      \Procedure{SingleHairRelaxerFinder}{$G, \mathcal{B}, T$}
      \If{$(V, T)$ is an \refdef{invalid} subgraph}
      \State \Return $\textsc{Nil}$
      \EndIf
      \State $M_G \gets \Call{ParityMatrix}{V, T}$ \Comment{parity matrix \S\ref{ssec:parity-matrix}}
      \For{$S \in \mathcal{B}$}
      \State $\mathcal{M}_S \gets \Call{HyperBlossomMatrix}{M_S, G}$
      \State $\mathcal{H}_S \gets \Call{HairMatrix}{\mathcal{M}_G, S}$ \Comment{\refdef{hair-matrix}}
      \If{$\mathcal{H}_S$ is a single row of all 1s}
      \State \textbf{continue}
      \EndIf
      \State $r \gets \text{any Odd row of $\mathcal{H}_S$}$
      \State $E^+ \gets \text{the columns of $r$ that are 1}$
      \State $S^+ \gets (V, T \setminus E^+)$
      \State \Return $\{ \Delta y_S: -1, \Delta y_{S^+}: +1 \}$ \Comment{\refdef{relaxer}}
      \EndFor
      \State \Return $\textsc{Nil}$
      \EndProcedure
      \\
      \Procedure{HyperBlossomMatrix}{$M_G, S$}
      \State $\mathcal{M}_S \! \gets \! \Bigg(
        \begin{gathered}
          \forall e_j \notin \delta(S),\\ M_G(:, j)
        \end{gathered} \Bigg|
        \begin{gathered}
          \forall e_j \in \delta(S),\\ M_G(:, j)
      \end{gathered} \Bigg| M_G(:, 1+|T\cap \delta(S)|) \Bigg)$ \Comment{reorder the columns according to \refdef{hyperblossom-matrix}}
      \State \Return $\Call{GaussJordanElimination}{\mathcal{M}_S}$ \Comment{\S\ref{ssec:parity-matrix}}
      \EndProcedure
      \\
      \Procedure{HairMatrix}{$\mathcal{M}_G, S$}
      \State $i_0 \gets 1 + (\text{last pivot row among columns $T \setminus \delta(S)$})$
      \State $j_0 \gets 1 + |T \setminus \delta(S)|$
      \State $\mathcal{H}_S \gets \mathcal{M}_S(i_0\!:\ ,j_0\!:)$ \Comment{by definition of \refdef{hair-matrix}}
      \EndProcedure
    \end{algorithmic}
  \end{algorithm}
\end{figure}

\subsection{\emph{UnionFind} Relaxer Finder}\label{ssec:union-find-subroutine}

The \hyperblossom framework can implement the Union-Find decoders~\cite{delfosse2021almost,delfosse2022toward} with a trivial \refdef{relaxer} finder that does nothing but returns \textsc{Nil}, as shown in \cref{algo:union-find}.
The \refdef{clusters} in the \hyperblossom framework corresponds to the clusters in the UF decoders.
This reduces the Primal phase (\cref{algo:hyperblossom-primal-phase}) to a simple logic.
It first calls the \refdef{relaxer} finder via \cref{algo:relaxing} (line \ref{line:find-relaxers}), which returns $Rs \gets \varnothing$ (\cref{algo:hyperblossom-primal-phase}, line \ref{line:call-relaxing-hyperblossom}).
It then checks if the \refdef{cluster} $C$ is \refdef{invalid} ($C \in \mathcal{O}$, line \ref{line:trivial-direction-hyperblossom}).
If so, it grows the \refdef{invalid} \refdef{cluster} uniformly with a \refdef{direction} $\{ \Delta y_C: +1\}$ that expands on all incident non-tight edges $\delta(C) = E[V_C] \setminus E_C$.
This behavior is identical to that of a weighted UF decoder for simple graphs~\cite{huang2020fault,wu2022interpretation}.
More importantly, the \hyperblossom framework with \emph{UnionFind} generalizes the Hypergraph Union-Finder (HUF) decoder~\cite{delfosse2022toward} for weighted decoding hypergraphs.

\begin{figure}[t]
  \begin{algorithm}[H]
    \caption{\emph{UnionFind} \refdef{relaxer} finder}\label{algo:union-find}
    \begin{algorithmic}[1]
      \Require{$G$ (hypergraph), $\mathcal{B}$ (\refdef{hyperblossoms}), $T$ (\refdef{tight-edges})}
      \Ensure{$R[G,\mathcal{B},T]$ (\refdef{relaxer}) or $\textsc{Nil}$ (if not found)}
      \Procedure{UnionFindRelaxerFinder}{$G, \mathcal{B}, T$}
      \State \Return $\textsc{Nil}$
      \EndProcedure
    \end{algorithmic}
  \end{algorithm}
\end{figure}

\subsection{\emph{Blossom} Relaxer Finder}\label{ssec:blossom-subroutine}

For simple graphs, there exists a \refdef{relaxer} finder such that the \hyperblossom algorithm equipped with it terminates with optimal \refeqs{MWPF} and \refeqs{DLP} solutions, according to \reftheorem{simple-graph-optimality} and \reftheorem{hyperblossom-algorithm-optimality}.
We describe such a polynomial-time and optimal \refdef{relaxer} finder called \emph{Blossom} using ideas from the blossom algorithm.
In practice, the \hyperblossom algorithm with \emph{Blossom} will be less efficient than the highly optimized blossom algorithm.
Our goal here is to show that the \hyperblossom framework can implement an MWPM decoder for simple decoding graphs.

To show that the \hyperblossom algorithm with \emph{Blossom} indeed implements the MWPM decoder, we look at the progression of their \refeqs{DLP} solutions.
We show that the two sequences of \refeqs{DLP} solutions are step-by-step equivalent.
That is, there exists a bijective function $f$ such that the \refdef{blossom-dlp} $\vec{y^*}$ and the the \refeqs{DLP} solution $\vec{y}$ satisfy $\vec{y} = f(\vec{y^*})$ and $\sum\vec{y} = \sum\vec{y^*}$ throughout the whole algorithm.

\subsubsection{Important Concepts}\label{sssec:blossom-concepts}

We first provide some background on how the blossom algorithm works.
As mentioned in~\S\ref{ssec:algo-overview}, the blossom algorithm also works with two interactive phases as illustrated by \autoref{fig:primal-dual-interface}: the Primal phase solves the MWPM problem while the Dual phase solves a related DLP problem. Like in the \hyperblossom algorithm, the Primal phase finds \refdef{directions} for the Dual phase to adjust the dual variables and moves toward a better solution.

Unlike the \hyperblossom algorithm, the blossom algorithm uses a data structure called \emph{alternating tree} to find \refdef{directions}.
Before introducing the alternating tree, we must first introduce \emph{blossoms} and \emph{nodes}, which are its vertices.
A \emph{blossom} represents a set of an odd number of defect vertices $S^* \subseteq D, |S^*| = 1\mod 2$~\cite{wu2023qce}.
The set of all possible blossoms is denoted as $\mathcal{O}^*$.
Each $S^* \in \mathcal{O}^*$ corresponds to a dual variable $y^*_{S^*}$ in the blossom algorithm formulation, similar to the dual variables $y_S$ in the \refeqs{DLP}.
Those $S^* \in \mathcal{O}^*$ with positive dual variables $y^*_{S^*} > 0$ are called \emph{blossoms}, similar to the \refdef{hyperblossoms}.
We use $S^*$ with a superscript $*$ to denote the blossoms $S^* \subseteq D$ to avoid confusion with the \refdef{invalid} subgraphs $S = (V_S, E_S)$.
In general, $*$ superscript indicates a notion in the blossom algorithm.
The blossoms have a hierarchical structure: a parent blossom is constructed from a cycle of children blossoms.
A blossom without any parent blossom is called a \emph{node}.

We define the blossom DLP solutions that are intermediate states of the blossom algorithm as below.

\vspace{1ex}
\definitionlabel{blossom-dlp}{Blossom DLP solution} A feasible intermediate DLP solution $\vec{y^*}$ of the blossom algorithm.

\vspace{1ex}
\definitionlabel{good-direction}{Good Direction} Growing a \refdef{blossom-dlp} $\vec{y^*}$ along a \emph{good direction} $\Delta\vec{y^*}$ results in a \refdef{blossom-dlp} solution. That is, there exists a positive length $l > 0$ such that any small growth $0\le l' \le l$ results in a \refdef{blossom-dlp} $\vec{y^*} + l' \Delta\vec{y^*}$.

\vspace{1ex}
\definitionlabel{blossom-useful-direction}{Blossom Useful Direction} A \refdef{good-direction} $\Delta\vec{y^*}$ is \emph{Useful} if
$\sum \Delta\vec{y^*} = \sum_{S^* \in \mathcal{O}^*} \Delta y^*_{S^*} > 0$.

\begin{figure}[h]
  \centering
  \begin{minipage}{\linewidth}
    \begin{subfigure}[t]{0.48\linewidth}
      \centering
      \includegraphics[width=\textwidth,page=1]{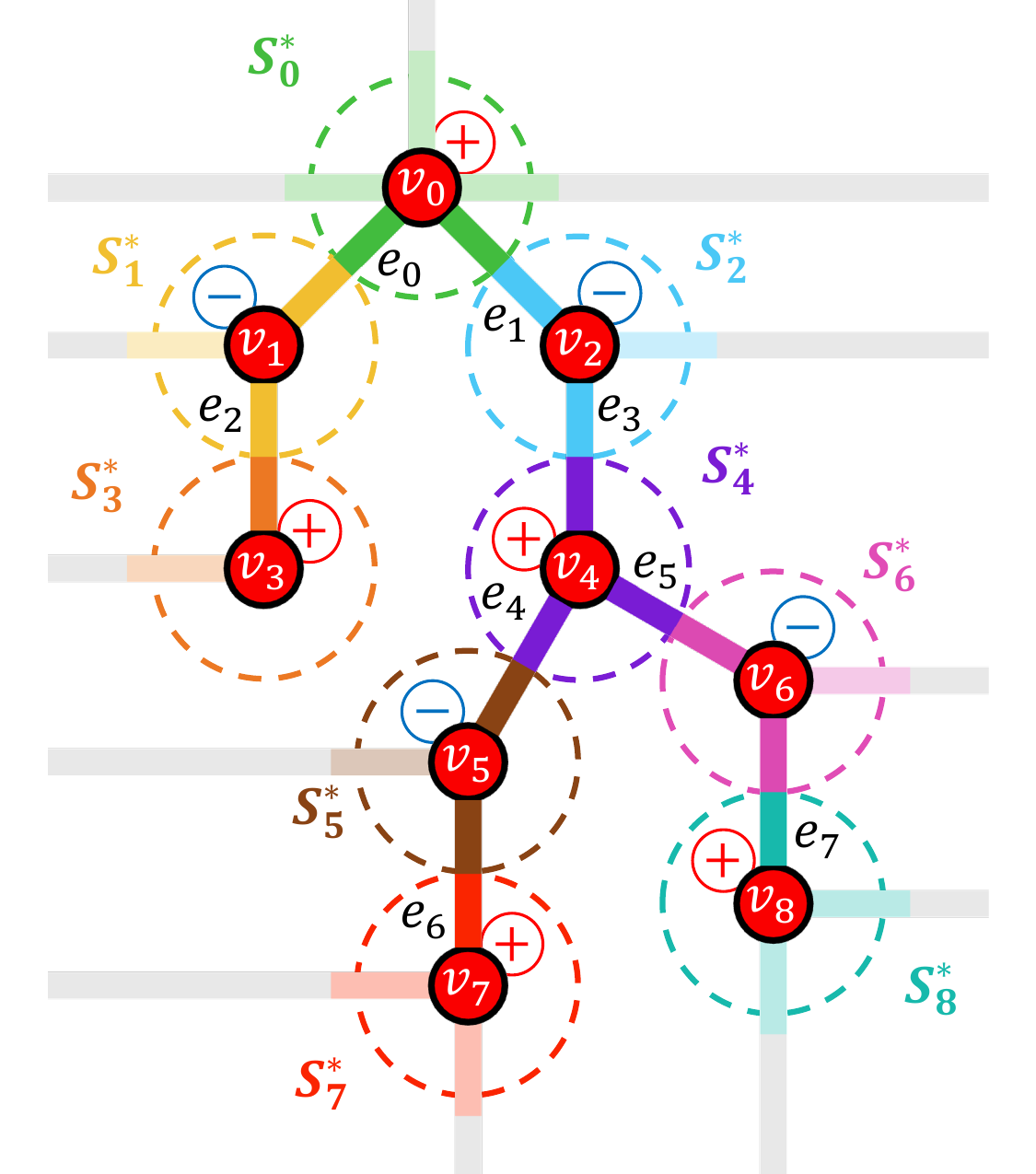}
      \caption{Alternating Tree}
      \label{fig:alternating-tree-direction-1}
    \end{subfigure}
    \begin{subfigure}[t]{0.48\linewidth}
      \centering
      \includegraphics[width=\textwidth,page=2]{figures/alternating-tree-direction/alternating-tree-direction.pdf}
      \caption{\refdef{direction} $\Delta\vec{y^*}$}
      \label{fig:alternating-tree-direction-2}
    \end{subfigure}
    \caption{Alternating tree in the blossom algorithm finds a \refdef{blossom-useful-direction} of alternating $\Delta y^*_{S^*} = +1$ and $\Delta y^*_{S^*} = -1$. Note that an alternating tree represents the relationship between nodes, which is visualized in the above example on top of the decoding graph, similar to that in~\cite{higgott2025sparse}.}
    \label{fig:alternating-tree-example-blossom-algorithm}
  \end{minipage}
\end{figure}

\vspace{1ex}
An \textit{alternating tree} consists of an odd number of nodes, where each node $S^*$ is marked by either ``$+$'' or ``$-$''~\cite{wu2023qce}, as shown in \autoref{fig:alternating-tree-direction-1}.
A ``$+$'' node can have an arbitrary number of children and a ``$-$'' node has exactly one child.
A parent and child nodes are connected by \refdef{tight-edges} in the syndrome graph~\cite{wu2023qce}.
The root node and leaf nodes are ``$+$'' nodes.
Any path from the root to a leaf has an alternating ``$+$'' and ``$-$'' pattern.
Each alternating tree represents a \refdef{blossom-useful-direction} $\Delta\vec{y^*}$ defined by $\Delta y^*_{S^*} = +1$ for ``$+$'' nodes and $\Delta y^*_{S^*} = -1$ for ``$-$'' nodes.
Since there are more ``$+$'' nodes than ``$-$'' nodes in an alternating tree, $\Delta\vec{y^*}$ is \emph{Useful} with $\sum_{S^* \in \mathcal{O}^*} \Delta y^*_{S^*} \ge 1$.
We show an example of an alternating tree growing along $\Delta\vec{y^*}$ for a small length in \autoref{fig:alternating-tree-direction-2}.

The blossom algorithm grows along a sequence of such \refdef{blossom-useful-directions} $\Delta \vec{y^*}$ from the alternating trees, until $\sum \vec{y^*} = W(\mathcal{E})$ where $\mathcal{E}$ is the parity factor that the MWPM decoder finds.

\vspace{1ex}
A \refdef{blossom-dlp} $\vec{y^*} \in \mathbb{R}^{|\mathcal{O}^*|}$ is related to a feasible \refeqs{DLP} solution $\vec{y} \in \mathbb{R}^{|\mathcal{O}|}$ given the following theorem:

\theoremBlossomMapping{blossom-mapping}

\vspace{1ex}
Note that the domain of the function $f(\cdot)$ is all \refdef{blossom-dlps}, but not $\mathbb{R}^{|\mathcal{O}^*|}$ or all feasible dual solutions to the DLP problem of the blossom algorithm.
Similarly, the codomain of $f(\cdot)$ (domain of $f^{-1}(\cdot)$) is not $\mathbb{R}^{|\mathcal{O}|}$ but only those $\vec{y} = f(\vec{y^*})$.

In \S\ref{ssec:dual-mapping-definition}, we provide such a function $f$ (\refdef{dual-mapping}) and use it as a constructive proof for the above theorem in \S\ref{ssec:dual-mapping-properties}.
We then define the direction-specific gradient of $f$, which can map a \refdef{good-direction} $\Delta\vec{y^*}$ to a \refdef{feasible-direction} $\Delta\vec{y}$.

\begin{align*}
  \Delta\vec{y} = (\Delta\vec{y^*} \circ \nabla) f(\vec{y^*}) \coloneq \lim_{l \to 0^+} \frac{f(\vec{y^*} + l \Delta\vec{y^*}) - f(\vec{y^*})}{l}
\end{align*}

In practice, we can use a sufficiently small $l$ to compute the mapped \refdef{feasible-direction} efficiently.
We present the algorithm in \cref{algo:blossom} line \ref{line:direction-mapping-implementation}.

\subsubsection{Relaxer Finding Algorithm}

We now describe the \emph{Blossom} \refdef{relaxer}-finding algorithm based on the \reftheorem{blossom-mapping}, as shown in \cref{algo:blossom}.
The \hyperblossom algorithm equipped with \emph{Blossom} always terminates with an optimal \refeqs{DLP} solution.
\emph{Blossom} first maps the current \refeqs{DLP} solution $\vec{y}$ to a \refdef{blossom-dlp} $\vec{y^*}$, reconstructs an alternating tree from $\vec{y^*}$, and then uses the alternating tree to find a \refdef{good-direction} that maps to a \refdef{relaxer} in certain cases.

\emph{Blossom} starts with the decoding graph $G$ and a \refeqs{DLP} solution $\vec{y}$. It first computes the corresponding DLP solution for the blossom algorithm $\vec{y^*} = f^{-1}(\vec{y})$ (\cref{algo:blossom}, line \ref{line:blossom-inverse-mapping}).

From $\vec{y^*}$, the algorithm then constructs the alternating trees (\cref{algo:blossom}, line \ref{line:alternating-tree-reconstruction}), using an algorithm included in a constructive proof of \reftheorem{alternating-tree-reconstruction-with-proof} in \S\ref{ssec:alternating-tree-reconstruction}.

If all the alternating trees consist of only one node, the algorithm returns \textsc{Nil}. Otherwise, it computes a \refdef{good-direction} as below.
As noted in \S\ref{sssec:blossom-concepts}, alternating trees represent a \refdef{blossom-useful-direction} $\Delta\vec{y^*}$ for the blossom algorithm (line \ref{line:blossom-tree-direction}).
Suppose the root node of this alternating tree is $S^*$ and one of its children is $S^*_c$ .
The algorithm takes $\Delta\vec{y^*}$ and changes $\Delta y^*_{S^*} = +1$ to $\Delta y^*_{S^*} = 0$ (line \ref{line:blossom-tree-direction-change}).
This results in a \refdef{good-direction} $\Delta\vec{y'^*}$ with $\sum_{S^* \in \mathcal{O}^*} \Delta y'^*_{S^*} = \sum_{S^* \in \mathcal{O}^*} \Delta y^*_{S^*} - 1 \ge 0$.

The algorithm then computes a \refdef{feasible-direction} $\Delta \vec{y'} = (\Delta\vec{y'^*} \circ \nabla) f(\vec{y^*})$ from the \refdef{good-direction} $\Delta\vec{y'^*}$.

Finally, we show that $\Delta \vec{y'}$ is a \refdef{relaxer} by showing that it relaxes some \refdef{tight-edges} between vertices in $S^*$ and vertices in $S^*_c$.
According to \reftheorem{blossom-mapping}, there is a minimum-weight path consisting of all \refdef{tight-edges} in $\vec{y} = f(\vec{y^*})$ between a pair of vertices in $S^*$ and $S^*_c$, given that there exists a tight syndrome-graph edge between the nodes $S^*$ and $S^*_c$ in the alternating tree by definition.
As $\Delta y'^*_{S^*} = 0$ and $\Delta y'^*_{S^*_c} = -1$, an edge on the minimum-weight path is relaxed.
Thus, $\Delta \vec{y'}$ is a \refdef{relaxer}.

\begin{figure}
  \centering
  \scalebox{1}{
    \begin{tikzpicture}%
\node[color=black,align=center,font=\small] at (2.0,4.0) {\refdef{blossom-dlps}};%
\node[color=black,align=center,font=\small] at (6.2,4.0) {\hyperblossom \refeqs{DLP} solutions};%
\path[dotted,draw] (1.2,3.0) to[bend left=10] (5.8,3.0);%
\path[dotted,draw] (1.2,1.0) to[bend left=10] (5.8,1.0);%
\path[dotted,draw] (2.2,2.0) to[bend left=10] (6.8,2.0);%
\path[fill=black,start angle=0,end angle=360,radius=0.1,draw] (1.3,3.0) arc;%
\node[color=black,align=center,font=\small] at (1.2,3.5) {$\vec{y^*}$};%
\path[fill=black,start angle=0,end angle=360,radius=0.1,draw] (1.3,1.0) arc;%
\node[color=black,align=center,font=\small] at (1.2,0.5) {$\vec{y^*} + l \Delta \vec{y^*}$};%
\path[fill=black,start angle=0,end angle=360,radius=0.1,draw] (2.3000000000000003,2.0) arc;%
\path[fill=black,start angle=0,end angle=360,radius=0.1,draw] (5.8999999999999995,3.0) arc;%
\node[color=black,align=center,font=\small] at (5.8,3.5) {$\vec{y} = f(\vec{y^*})$};%
\path[fill=black,start angle=0,end angle=360,radius=0.1,draw] (5.8999999999999995,1.0) arc;%
\node[color=black,align=center,font=\small] at (5.8,0.5) {$\vec{y} + l \Delta \vec{y} = f(\vec{y^*} + l \Delta \vec{y^*})$};%
\path[fill=black,start angle=0,end angle=360,radius=0.1,draw] (6.8999999999999995,2.0) arc;%
\path[thick,-{Stealth[scale=1.2]},solid,draw] (1.2,3.0) -- (1.2000000000000002,1.8399999999999999);%
\path[thick,solid,draw] (1.2,1.0) -- (1.2000000000000002,1.8399999999999999);%
\node[color=black,align=center,font=\small] at (0.7,2.0) {$\Delta\vec{y^*}$};%
\path[thick,-{Stealth[scale=1.2]},dashed,draw] (1.2,3.0) -- (1.7800000000000002,2.42);%
\path[thick,dashed,draw] (2.2,2.0) -- (1.7800000000000002,2.42);%
\node[color=black,align=center,font=\small] at (2.2,2.6) {$\Delta\vec{y'^*}$};%
\path[thick,-{Stealth[scale=1.2]},dashed,draw] (2.2,2.0) -- (1.62,1.42);%
\path[thick,dashed,draw] (1.2,1.0) -- (1.62,1.42);%
\node[color=black,align=center,font=\small] at (2.8,1.4) {$\{ \Delta y^*_{S^*}: +1 \}$};%
\path[thick,-{Stealth[scale=1.2]},solid,draw] (5.8,3.0) -- (5.800000000000001,1.8399999999999999);%
\path[thick,solid,draw] (5.8,1.0) -- (5.800000000000001,1.8399999999999999);%
\node[color=black,align=center,font=\small] at (5.3,2.0) {$\Delta\vec{y}$};%
\node[color=black,align=center,font=\small] at (4.8999999999999995,1.5) {(composed \\ \refdef{direction})};%
\path[thick,-{Stealth[scale=1.2]},dashed,draw] (5.8,3.0) -- (6.380000000000001,2.42);%
\path[thick,dashed,draw] (6.8,2.0) -- (6.380000000000001,2.42);%
\node[color=black,align=center,font=\small] at (7.4,2.6) {$\Delta\vec{y'}$ (\refdef{relaxer})};%
\path[thick,-{Stealth[scale=1.2]},dashed,draw] (6.8,2.0) -- (6.220000000000001,1.42);%
\path[thick,dashed,draw] (5.8,1.0) -- (6.220000000000001,1.42);%
\node[color=black,align=center,font=\small] at (6.8,1.5) {$\Delta\vec{y^t}$};%
\node[color=black,align=center,font=\small] at (7.4,1.0499999999999998) {(\refdef{trivial-direction})};%
\path[thick,dashed,draw] (3.9,3.7) -- (3.9,0.0);%
\path[thick,-{Stealth[scale=1.2]},draw] (3.05,3.3000000000000003) -- (3.75,3.3000000000000003);%
\path[thick,-{Stealth[scale=1.2]},draw] (3.75,3.1) -- (3.05,3.1);%
\node[color=black,align=center,font=\small] at (3.4,3.6) {$f(\cdot)$};%
\node[color=black,align=center,font=\small] at (3.4,2.8) {$f^{-1}(\cdot)$};%
\node[color=blue,align=center,font=\small] at (7.1,3.5) {\tikz[baseline=(char.base)]{\node[shape=circle,draw,inner sep=1pt] (char) {1};}$\,\text{\footnotesize line \ref{line:blossom-relaxer-finder}}$};%
\node[color=blue,align=center,font=\small] at (2.0,3.5) {\tikz[baseline=(char.base)]{\node[shape=circle,draw,inner sep=1pt] (char) {2};}$\,\text{\footnotesize line \ref{line:blossom-inverse-mapping}}$};%
\node[color=blue,align=center,font=\small] at (0.85,1.65) {\tikz[baseline=(char.base)]{\node[shape=circle,draw,inner sep=1pt] (char) {3};}$\,\text{\footnotesize line \ref{line:blossom-tree-direction}}$};%
\node[color=blue,align=center,font=\small] at (2.9000000000000004,2.35) {\tikz[baseline=(char.base)]{\node[shape=circle,draw,inner sep=1pt] (char) {4};}$\,\text{\footnotesize line \ref{line:blossom-tree-direction-change}}$};%
\node[color=blue,align=center,font=\small] at (7.6000000000000005,2.2) {\tikz[baseline=(char.base)]{\node[shape=circle,draw,inner sep=1pt] (char) {5};}$\,\text{\footnotesize line \ref{line:blossom-relaxer-mapping}}$};%
\end{tikzpicture}
  }
  \newcommand{\circledstep}[1]{\circledtext[height=1.5ex,charshrink=0.65]{#1}}
  \caption{The \refdef{blossom-dlps} (dots on the left side) correspond to the \refeqs{DLP} solutions (dots on the right side) with a bijective function $f$. The \emph{Blossom} \refdef{relaxer} finder finds a \refdef{relaxer} $\Delta \vec{y'}$ following the path of $\vec{y} \rightarrow \vec{y^*} \rightarrow \Delta\vec{y^*} \rightarrow \Delta\vec{y'^*} \rightarrow \Delta\vec{y'}$. The \hyperblossom algorithm with \emph{Blossom} progresses from $\vec{y} = f(\vec{y^*})$ to $\vec{y} + l\Delta \vec{y} = f(\vec{y^*} + l\Delta \vec{y^*})$, thus preserving the correspondence throughout the algorithm.}
  \label{fig:dual-mapping}
\end{figure}

We next show that the \emph{Blossom} \refdef{relaxer} finder is optimal.
When there exists an alternating tree with more than one node, the \hyperblossom algorithm finds a \refdef{trivial-direction} $\Delta \vec{y^t} = (\{ \Delta y^*_{S^*} = +1 \} \circ \nabla) f(\vec{y^*} + l \Delta \vec{y'^*})$ (line \ref{line:trivial-direction-hyperblossom} of \cref{algo:hyperblossom-primal-phase}) and composes a \refdef{useful-direction} $\Delta \vec{y} = \Delta \vec{y^t} + \Delta \vec{y'}$ (\cref{algo:compose}) that is identical to the \refdef{useful-direction} mapped from the blossom algorithm $(\Delta \vec{y^*} \circ \nabla) f(\vec{y^*})$, as shown in \autoref{fig:dual-mapping}.
Otherwise, the \hyperblossom algorithm falls back to the Union-Find decoder and it still behaves the same as an MWPM decoder: odd \refdef{clusters} (isolated nodes) grow uniformly while even \refdef{clusters} (matched nodes) stay.
In both cases, the \hyperblossom algorithm progresses from a \refeqs{DLP} solution $\vec{y} = f(\vec{y^*})$ to a new \refeqs{DLP} solution $\vec{y} + l\Delta \vec{y} = f(\vec{y^*} + l\Delta \vec{y^*})$, preserving the correspondence between the \refeqs{DLP} solutions and the \refdef{blossom-dlps} as shown in \autoref{fig:dual-mapping}.
Also given that the blossom algorithm terminates at a \refdef{blossom-dlp} with $\sum \vec{y^*} = W(\mathcal{E})$, the final \refeqs{DLP} solution is $\vec{y} = f(\vec{y^*})$ with $\sum \vec{y} = \sum \vec{y^*} = W(\mathcal{E})$. Given \reftheorem{provable-optimum}, $\vec{y}$ is an optimal \refeqs{DLP} solution and thus the \emph{Blossom} \refdef{relaxer} finder is optimal.

\begin{figure}[t]
  \begin{algorithm}[H]
    \caption{\emph{Blossom} \refdef{relaxer} finder}\label{algo:blossom}
    \begin{algorithmic}[1]
      \Require{$G$ (hypergraph), $\mathcal{B}$ (\refdef{hyperblossoms}), $T$ (\refdef{tight-edges})}
      \Ensure{$R[G,\mathcal{B},T]$ (\refdef{relaxer}) or $\textsc{Nil}$ (if not found)}
      \Procedure{BlossomRelaxerFinder}{$G, \mathcal{B}, T$} \label{line:blossom-relaxer-finder}
      \State $\vec{y^*} \gets f^{-1}(\vec{y})$ \label{line:blossom-inverse-mapping}
      \State reconstruct alternating trees from $\vec{y^*}$ \label{line:alternating-tree-reconstruction}
      \If{$\exists$ an alternating tree with multiple nodes} \label{line:blossom-exists-alternating-tree}
      \State $S^* \gets \text{the root node of the alternating tree}$
      \State $\Delta \vec{y^*} \gets \text{\refdef{useful-direction} of the alternating trees\dns}$ \label{line:blossom-tree-direction}
      \State $\Delta \vec{y'^*} \gets \Delta \vec{y^*} - \{ \Delta y^*_{S^*} = +1 \}$ \label{line:blossom-tree-direction-change}
      \State \Return $\Call{MapFeasibleDirection}{G, \vec{y}, \vec{y^*}, \Delta \vec{y'^*}}$ \label{line:blossom-relaxer-mapping}
      \EndIf
      \State \Return $\textsc{Nil}$
      \EndProcedure
      \\
      \Procedure{MapFeasibleDirection}{$G, \vec{y}, \vec{y^*}, \Delta \vec{y^*}$} \label{line:direction-mapping-implementation}
      \State $L \gets \{ w_e - \sum_{S \in \mathcal{O} | e \in E} y_S | e \in E \} \cup \{ y_S | S \in \mathcal{O} \}$
      \State $l \gets \min(\{ l \in L | l > 0 \}) / 2$ \Comment{sufficiently small length} \label{line:sufficiently-small-length}
      \State \Return $[f(\vec{y^*} + l \Delta \vec{y^*}) - f(\vec{y^*})]/l$
      \EndProcedure
    \end{algorithmic}
  \end{algorithm}
\end{figure}

\subsection{\emph{Nullity$_{\le 1}$} Relaxer Finder}\label{ssec:biased-single-dof}

\begin{figure}[t]
  \centering
  \begin{subfigure}[t]{0.32\linewidth}
    \centering
    \includegraphics[width=\textwidth]{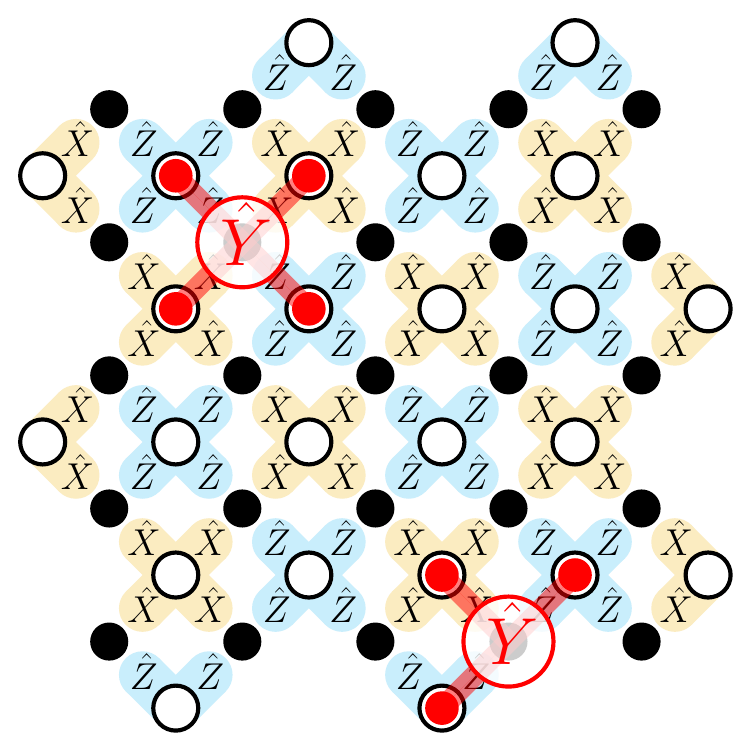}
    \caption{Surface Code.}
    \label{fig:single-dof-surface-code}
  \end{subfigure}
  \begin{subfigure}[t]{0.32\linewidth}
    \centering
    \includegraphics[width=\textwidth]{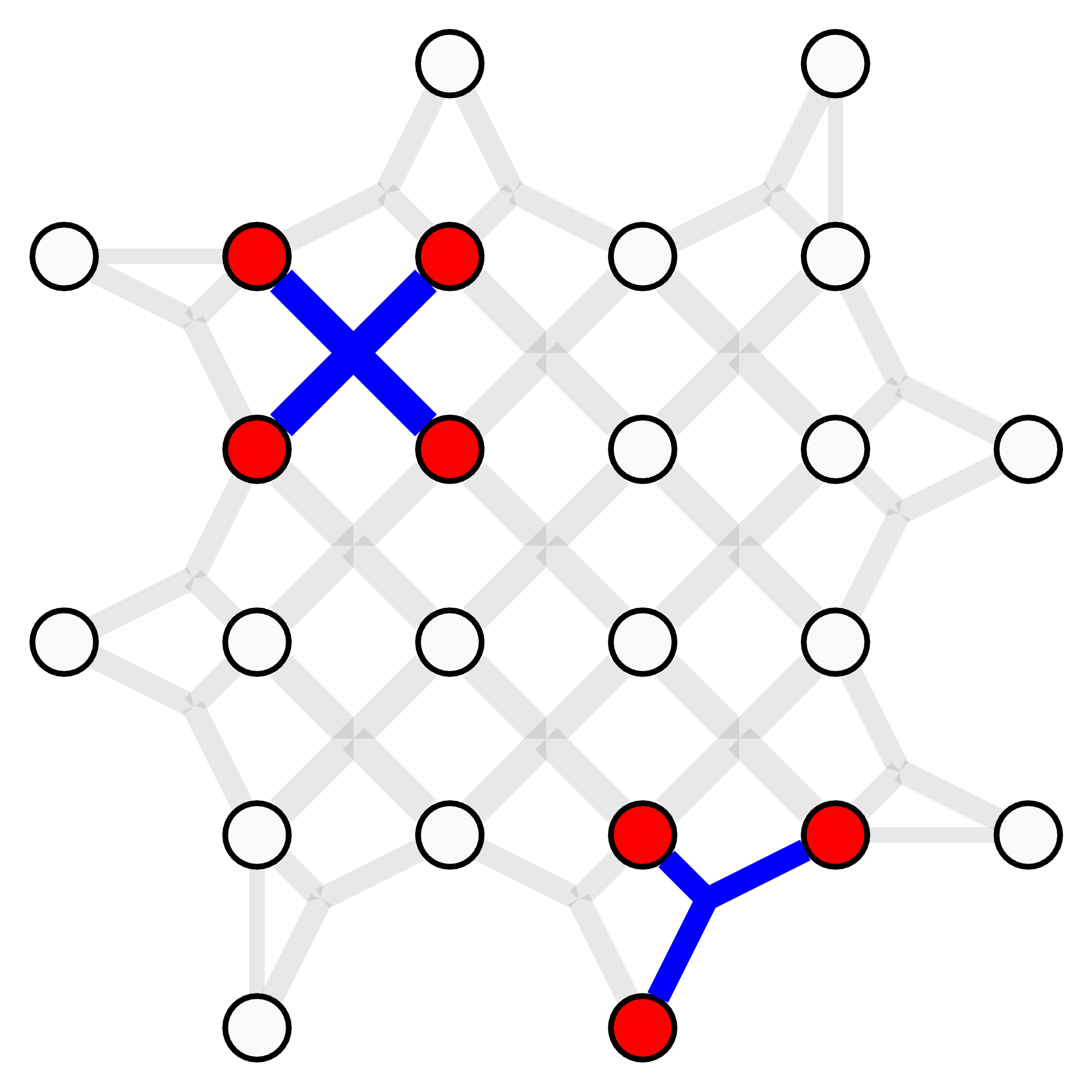}
    \caption{\refeqs{MWPF} $\mathcal{E}_1$.}
    \label{fig:single-dof-mwpf}
  \end{subfigure}
  \begin{subfigure}[t]{0.32\linewidth}
    \centering
    \includegraphics[width=\textwidth]{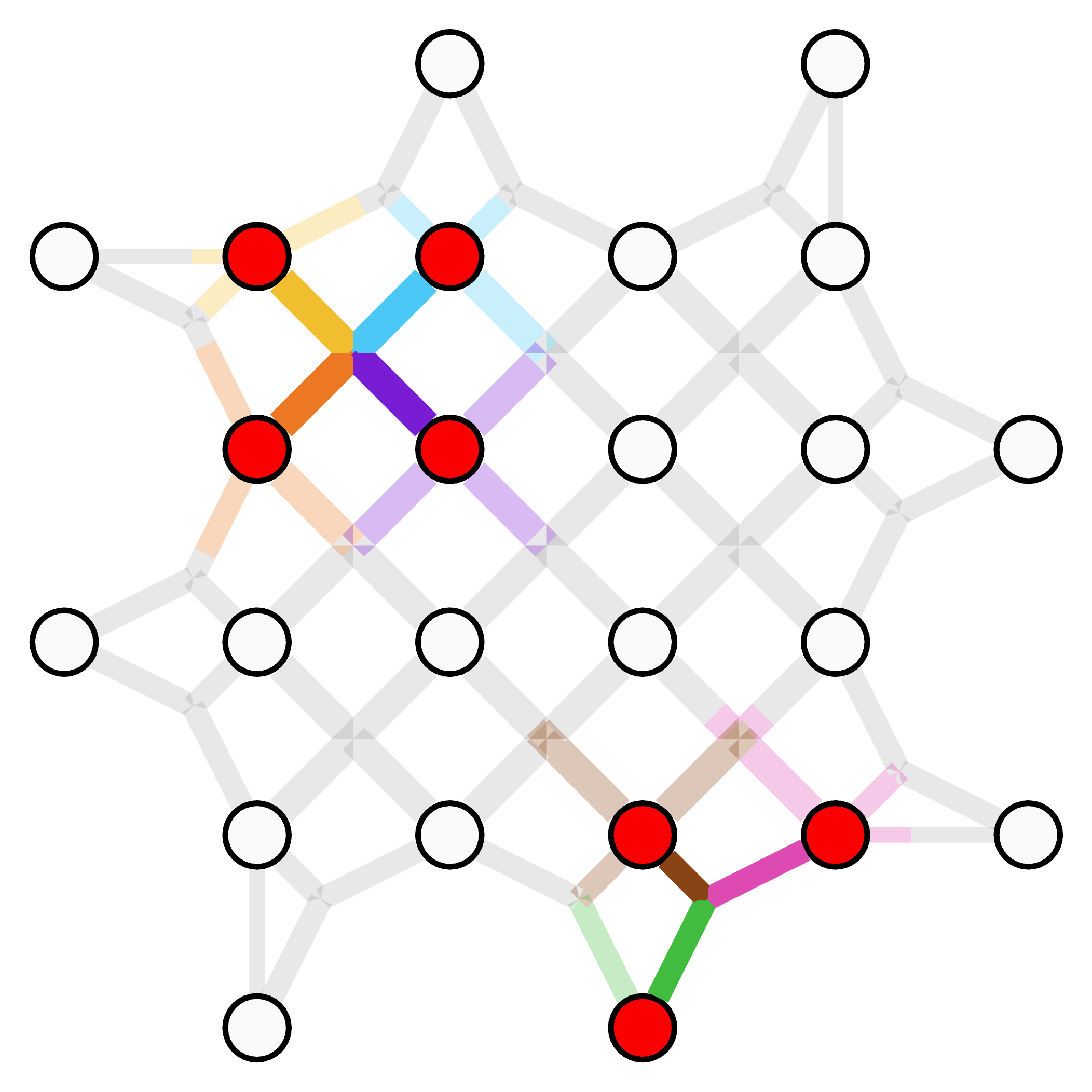}
    \caption{\refeqs{DLP} $\vec{y}$.}
    \label{fig:single-dof-dual}
  \end{subfigure}
  \caption{(a) The decoding hypergraph of a rotated surface code with biased $Y$ noise is a nullity$_{\le 1}$ hypergraph. (b,c) The optimal \refeqs{MWPF} and \refeqs{DLP} solutions.}
  \label{fig:single-dof}

  \vspace{1ex}
  \begin{subfigure}[t]{0.32\linewidth}
    \centering
    \includegraphics[width=\textwidth]{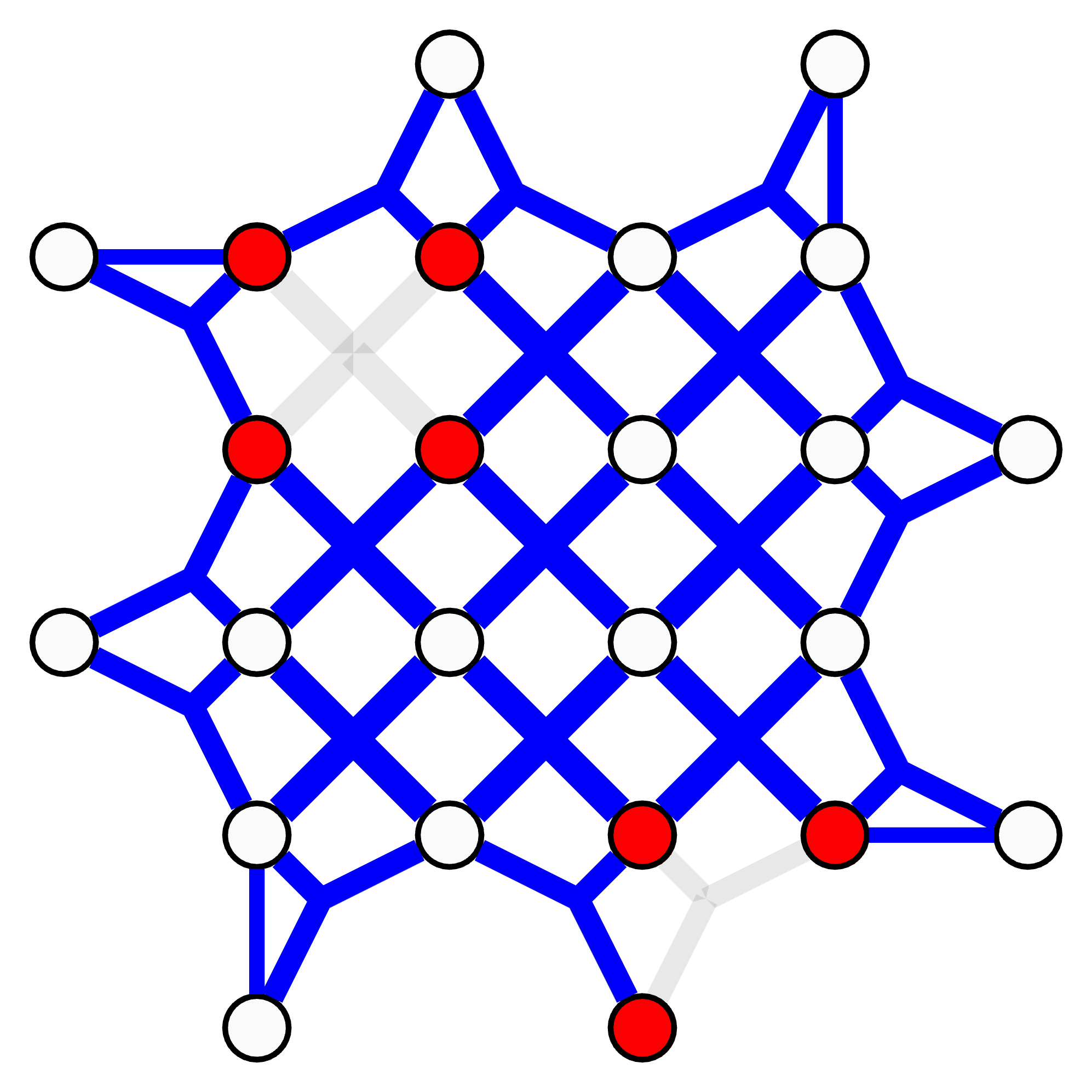}
    \caption{parity factor $\mathcal{E}_2$.\dns}
    \label{fig:single-dof-primal2}
  \end{subfigure}
  \begin{subfigure}[t]{0.64\linewidth}
    \centering
    \includegraphics[width=\textwidth]{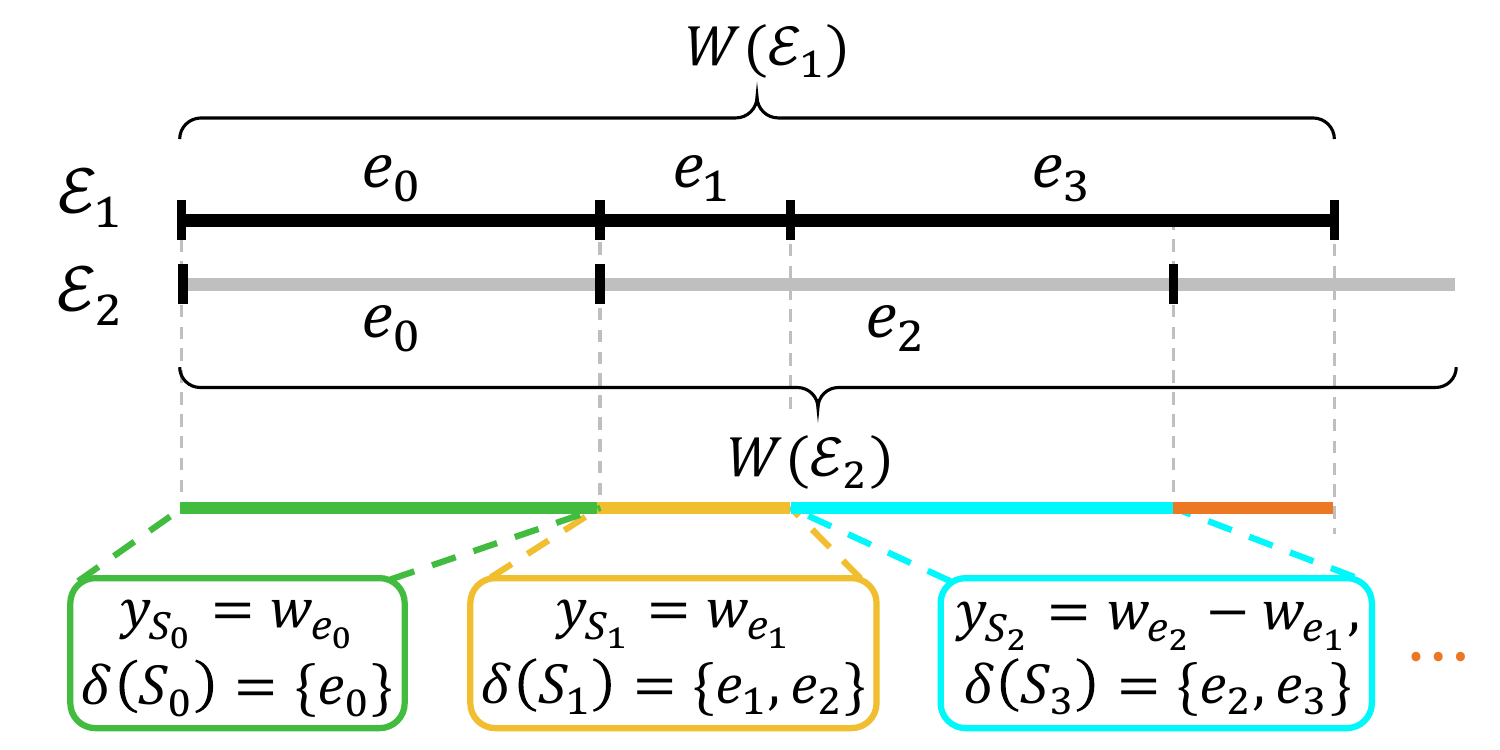}
    \caption{optimal \refeqs{DLP} $\vec{y}$ with $|\mathcal{B}| \le |E|$.}
    \label{fig:single-dof-dual-variables}
  \end{subfigure}
  \caption{
    \emph{Nullity$_{\le 1}$} is efficient and optimal.
    (a) Given the other parity factor $\mathcal{E}_2$, (b) it finds an optimal \refeqs{DLP} solution $\sum \vec{y} = W(\mathcal{E}_1)$ efficiently using no more than $|E|$ \refdef{hyperblossoms}.
  }
  \label{fig:single-dof-subroutine}
\end{figure}

The nullity$_{\le 1}$ hypergraphs are an important class of hypergraphs that satisfy \ref{condition:mwpf}, as proved with explicit constructions of optimal \refeqs{DLP} solutions in \S\ref{sec:singledof-algo}.
These hypergraphs have at most two parity factor solutions and, as a result, the \refeqs{MWPF} problem is trivial for them.
Nevertheless, the \emph{Nullity$_{\le 1}$} \refdef{relaxer} finder demonstrates the possibility of designing more efficient \refdef{relaxer} finders exploiting special hypergraph properties.
It is also practically important because some QEC codes with high noise bias tend to produce \refdef{clusters} that predominantly has a nullity of no more than 1.
An example of this type of decoding hypergraph comes from the rotated surface code with a biased $Y$ noise model, as shown in \autoref{fig:single-dof-surface-code}.
The number of independent parity checks is $d^2 - 1$ while there are $d^2$ different $Y$ error locations, leading to a nullity of 1.
For example, given a valid syndrome in \autoref{fig:single-dof-surface-code}, there are only two possible parity factors \autoref{fig:single-dof-mwpf} and \autoref{fig:single-dof-primal2}.
More generally, even if the decoding hypergraph is not a nullity$_{\le 1}$ hypergraph, its small \refdef{valid} \refdef{clusters} tends to be nullity$_{\le 1}$ hypergraphs because the number of \refdef{tight-edges} is usually small.
In such cases, \emph{Nullity$_{\le 1}$} can complement other \refdef{relaxer} finders to refine these \refdef{clusters}.

\subsubsection{Important Concepts}\label{ssec:nullity-le-1-concepts}

\vspace{1ex}
In the following section, we will construct an optimal \refeqs{DLP} solution $\vec{y}$ for every nullity$_{\le 1}$ hypergraph that has $\sum \vec{y} = W(\mathcal{E})$, which proves \reftheorem{single-dof-graph-optimality}.

Together with \reflemma{nullity-is-hereditary-with-proof} (\S\ref{ssec:optimal-1-dof}) and \reftheorem{hyperblossom-algorithm-optimality}, there exists an optimal \refdef{relaxer} finder such that the \hyperblossom algorithm finds the optimal \refeqs{MWPF} and \refeqs{DLP} solutions.
We will construct such an efficient and optimal \refdef{relaxer} finder in the following section.

\subsubsection{Relaxer Finding Algorithm}
\label{sec:singledof-algo}

We present an efficient \refdef{relaxer} finder for nullity$_{\le 1}$ hypergraphs.
The algorithm assumes that $(V, T)$ is a \refdef{valid} subgraph and that the hypergraph has a nullity 0 or 1.
If not, it returns \textsc{Nil}.
The basic idea of the algorithm is to find an optimal \refeqs{DLP} solution $\vec{y^o}$ first and then construct a \refdef{relaxer} for any suboptimal \refeqs{DLP} solution $\vec{y}$.
We will first explain how to find optimal \refeqs{DLP} solutions for nullity$_{=0}$ and nullity$_{=1}$ hypergraphs, respectively.

\nosection{Nullity$_{=0}$ Hypergraphs.}
When the nullity is 0, there exists only one parity factor $\mathcal{E}$.
In this case, the optimal \refeqs{DLP} solution has $\sum \vec{y^o} = W(\mathcal{E})$.
We construct a trivial optimal \refeqs{DLP} solution: $y^o_{S_e} = w_e, \forall e \in \mathcal{E}$ where $S_e = (V, E \setminus \{ e \}) \in \mathcal{O}$.
All the other dual variables are zero in $\vec{y^o}$.

\nosection{Nullity$_{=1}$ Hypergraphs.}
When the nullity is 1, there exist two parity factors $\mathcal{E}_1$ and $\mathcal{E}_2$.
Without losing generality, we assume $W(\mathcal{E}_1) \le W(\mathcal{E}_2)$.
In this case, the optimal \refeqs{DLP} solution has $\sum \vec{y} = W(\mathcal{E}_1)$.
Similar to the nullity$_{=0}$ case, we explicitly construct an optimal \refeqs{DLP} solution $\vec{y^o}$.
As shown in \autoref{fig:single-dof-dual-variables}, we visualize each edge in a parity factor with a length of its weight and arrange all edges in a straight line for $\mathcal{E}_1$ and $\mathcal{E}_2$, respectively.
We place the shared edges $\mathcal{E}_1 \cap \mathcal{E}_2$ in the front of each line and align the fronts of the two lines.
In each line, where two edges meet defines a point of joint. The points of joint in the two lines collectively define a series of \emph{intervals}: an interval is defined by two nearest joints regardless which line they are from.
Each interval with a length of $w_i$ and a pair of edges $e_{1,i} \in \mathcal{E}_1, e_{2,i} \in \mathcal{E}_2$ corresponds to an \refdef{invalid} subgraph $S_i = (V, E \setminus \{ e_{1,i}, e_{2,i} \})$ with $\delta(S_i) = \{ e_{1,i}, e_{2,i} \}$.
$e_{1,i}$ and $e_{2,i}$ can be the same as for $S_0$ in \autoref{fig:single-dof-dual-variables}.
By letting $y^o_{S_i} = w_i$ for every such overlapping interval and $y^o_{S} = 0$ for others, we have $\vec{y^o}$ as an optimal \refeqs{DLP} solution because $\sum \vec{y^o} = \sum_{i} w_i = W(\mathcal{E}_1)$.

Once we have obtained the optimal \refeqs{DLP} solution $\vec{y^o}$, we can then construct a \refdef{relaxer} for every suboptimal \refeqs{DLP} solution $\vec{y}$.
By definition of suboptimality, $\sum_{S \in \mathcal{O}} y_S < \sum_{S \in \mathcal{O}} y^o_S$.
In another word, $\Delta \vec{y} = \vec{y^o} - \vec{y}$ is a \refdef{useful-direction}.
There must exist at least one positive-weighted \refdef{tight-edge} $e \in T, w_e > 0$ in the suboptimal solution.
This is because $(V, T)$ is a \refdef{valid} subgraph: when no such edge exists, $(V, \{e \in E | w_e = 0\})$ is a \refdef{valid} subgraph and the optimal \refeqs{MWPF} $\mathcal{E}$ has $W(\mathcal{E}) = 0$.
This contradicts with the assumption that $\vec{y}$ is suboptimal given that the initial $\vec{y} = \textbf{0}$ is already optimal.
There exists at least one \refdef{hyperblossom} $S \in \mathcal{B}, y_S > 0$ that contributes to $e$ to make it tight, i.e., $e \in \delta(S)$.
Thus, reducing the value of $y_S$ will relax $e$.
We can then easily construct a \refdef{relaxer} $\Delta \vec{y'} = \Delta \vec{y} - \{ \Delta y_S : \sum_{S \in \mathcal{O}} \Delta y_S \}$.

\begin{figure}[t]
  \begin{algorithm}[H]
    \caption{\emph{Nullity$_{\le 1}$} \refdef{relaxer} finder}\label{algo:nullity-le-1}
    \begin{algorithmic}[1]
      \Require{$G$ (hypergraph), $\mathcal{B}$ (\refdef{hyperblossoms}), $T$ (\refdef{tight-edges})}
      \Ensure{$R[G,\mathcal{B},T]$ (\refdef{relaxer}) or $\textsc{Nil}$ (if not found)}
      \Procedure{Nullity$_{\le 1}$RelaxerFinder}{$G, \mathcal{B}, T$}
      \If{$(V, T)$ is an \refdef{invalid} subgraph}
      \State \Return $\textsc{Nil}$
      \EndIf
      \If{$G$ is a nullity$_{=0}$ hypergraph}
      \State $\mathcal{E} \gets \text{the parity factor}$
      \If{$\sum \vec{y} < W(\mathcal{E})$}
      \State $\vec{y^o} \gets \{ y_{(V, E \setminus \{ e \})}: w_e, \forall e \in \mathcal{E} \}$
      \State \Return $\Call{RelaxerFromOptimalDLP}{\vec{y}, \vec{y^o}}$
      \EndIf
      \EndIf
      \If{$G$ is a nullity$_{=1}$ hypergraph}
      \State $\mathcal{E}_1, \mathcal{E}_2 \gets \text{the parity factors s.t. $W(\mathcal{E}_1) \le W(\mathcal{E}_2)$}$
      \If{$\sum \vec{y} < W(\mathcal{E}_1)$}
      \State $\vec{y^o} \gets \textbf{0}$
      \State arrange the edges of $\mathcal{E}_1$ and $\mathcal{E}_2$ like \cref{fig:single-dof-dual-variables}
      \For{$i$-th interval with $e_{1,i} \in \mathcal{E}_1, e_{2,i} \in \mathcal{E}_2$}
      \State  $y^o_{(V, E \setminus \{ e_{1,i}, e_{2,i} \})} \gets w_i$
      \EndFor
      \State \Return $\Call{RelaxerFromOptimalDLP}{\vec{y}, \vec{y^o}}$
      \EndIf
      \EndIf
      \State \Return $\textsc{Nil}$
      \EndProcedure
      \\
      \Procedure{RelaxerFromOptimalDLP}{$\vec{y} (\mathcal{B}, T), \vec{y^o}$}
      \State $\Delta \vec{y} \gets \vec{y^o} - \vec{y}$
      \State $e \gets \text{pick one element from}\ \{ e \in T | w_e > 0 \}$
      \State $S \gets \text{pick one element from}\ \{ S \in \mathcal{B} | e \in \delta(S) \}$
      \State \Return $\Delta \vec{y} - \{ \Delta y_S: \sum_{S \in \mathcal{O}} \Delta y_S \}$
      \EndProcedure
    \end{algorithmic}
  \end{algorithm}

\end{figure}

\section{Implementation}\label{sec:implementation}

We present \hyperion, a software implementation of the \hyperblossom framework (\S\ref{sec:math}) and two \refdef{relaxer}-finding algorithms: \emph{SingleHair} (\S\ref{ssec:single-hair-subroutine}) and \emph{UnionFind} (\S\ref{ssec:union-find-subroutine}).
We implement \hyperion in Rust programming language (\S\ref{ssec:rust-impl}) and expose a Python library (\S\ref{ssec:python-interface}) to support easy integration into other tools like stim~\cite{gidney2021stim}, a widely used quantum error correction simulator.

\subsection{System Implementation in Rust}\label{ssec:rust-impl}

We implement \hyperion in Rust for high performance and flexible generics at no runtime cost.

\hyperion features two data types of the edge weights and dual variables: rational and floating-point.
The former provides infinite accuracy when solving the \refeqs{DLP} problem, but usually runs slower than the latter.
The two data types can be switched with a compile-time flag.
The implementation are the same for the two data types, except for the linear-programming solver invoked by line \ref{line:partial-dlp-hyperblossom} of \cref{algo:hyperblossom-dual-phase}.
We employ the HiGHS~\cite{huangfu2018parallelizing} solver for the floating-point data type and the SLP~\cite{kumar2020slp} solver for the rational data type.

To improve runtime efficiency, we split the decoding algorithm into two stages, the \emph{search} stage and the \emph{refine} stage.
The major difference is that the \emph{search} stage works on \refdef{invalid} \refdef{clusters} while the \emph{refine} stage works on \refdef{valid} \refdef{clusters}.
They use different data structures to update the \refeqs{DLP} variables.
In the \emph{search} stage, we ensure that every \refdef{cluster} finds a pair of feasible \refeqs{MWPF} and \refeqs{DLP} solutions.
In the \emph{refine} stage, we refine each \refdef{cluster} individually.

In the \emph{search} stage, all \refdef{invalid} \refdef{clusters} $C \in \mathcal{C} \cap \mathcal{O}$ grow simultaneously.
This has the benefit of minimizing their average size ($|V_C| + |E_C|$), as explained in~\cite{higgott2025sparse}.
\hyperion use a priority queue data structure to update the \refeqs{DLP} solution, inspired by~\cite{higgott2025sparse}.
The \emph{search} stage terminates when all \refdef{clusters} are \refdef{valid}, i.e., we find a feasible parity factor $\mathcal{E} = \cup_{C \in \mathcal{C}} \mathcal{E}_C$ at the end of the \emph{search} stage.

The subsequent \emph{refine} stage improves the local \refeqs{MWPF} $\mathcal{E}_C$ and the \refeqs{DLP} solution for each \refdef{cluster} individually.
\hyperion iteratively call the \refdef{relaxer} finders (\S\ref{sec:subroutine}) which introduces more \refdef{hyperblossoms} $\mathcal{B}_C$ to each \refdef{cluster} $C \in \mathcal{C}$.
The more \refdef{hyperblossoms} in the \refdef{cluster}, the slower the LP solver runs (line \ref{line:partial-dlp-hyperblossom} of \cref{algo:hyperblossom-dual-phase}).
Thus, to improve decoding efficiency, we assign each \refdef{cluster} $C \in \mathcal{C}$ a priority score $\text{Priority}(C)$ empirically:
\begin{align*}
  \text{Priority}(C) = \left(\sum_{S \in \mathcal{B}_C} y_S - \sum_{e \in E_C} w_e x_e \right) / \left( |E_C| + |\mathcal{B}_C| \right)^3
\end{align*}
The score favors small \refdef{clusters} with large primal-dual gaps between the \refeqs{MWPF} and \refeqs{DLP} objectives, enabling more efficient convergence per unit time.

Given the NP-hardness of the MWPF problem~\cite{berlekamp1978inherent}, the algorithm may not terminate in a polynomial time.
To mitigate this problem, we introduce a per-\refdef{cluster} limit on the number of \refdef{hyperblossoms} $c = \max|\mathcal{B}_C|$.
Once a \refdef{valid} \refdef{cluster} hits the limit ($|\mathcal{B}_C| \ge c$), \hyperion stop invoking \refdef{relaxer} finders for this \refdef{cluster}.
We denote an MWPF decoder with the per-\refdef{cluster} limit $c$ as ``MWPF(c)''.
We note that ``MWPF(c=0)'' is equivalent to the Hypergraph Union-Find (HUF) decoder~\cite{delfosse2022toward}.
The per-\refdef{cluster} limit prevents spending excessive amount of time on intractable \refdef{clusters}.
It improves the latency for real-time decoding at the cost of lower accuracy.

\subsection{Python Library and Stim Integration}\label{ssec:python-interface}

Through the Python library, users can use MWPF solvers with either rational or floating-point data type.
Each solver includes an interactive visualization tool and an adaptor for \texttt{stim}~\cite{gidney2021stim}.

The \texttt{mwpf} solver uses floating-point data type for fast decoding, while the alternative \texttt{mwpf-rational} solver uses rational data type for theoretical analysis.
The solvers share an identical interface and are interchangeable in most use cases.
They output a parity factor and an associated proximity bound as defined in \cref{eq:mwpf-chain}.

\hyperion also includes a 3D visualization tool that allows users to inspect each decoding step interactively.
It visualizes decoding hypergraphs in a 3D space and visualizes the dual variables as colored segments on the hyperedges.
In fact, we used it to generate the figures in this paper.
For portability, \hyperion supports both an interactive widget for Jupyter notebooks and a self-contained HTML file~\cite{wu_2025_15243982}.

\hyperion is compatible with the most popular QEC simulation library, \texttt{stim}~\cite{gidney2021stim}, and integrates seamlessly with its standard \texttt{sinter} simulation pipeline.
A \texttt{stim} user can incorporate \hyperion as the decoder with just a single line of code change.
Additionally, users can supply a Clifford circuit to \hyperion so that it exploits circuit-level information beyond the Detector Error Model (DEM) of \texttt{stim}, such as heralded erasure error channels.

\section{Evaluation}\label{sec:evaluation}

\begin{figure*}
  \begin{subfigure}[t]{\linewidth}
    \centering
    \textbf{Surface Code}
  \end{subfigure}
  \vspace{-2ex}

  \centering
  \begin{subfigure}[t]{.023\linewidth}
    \centering
    \includegraphics[width=1\textwidth]{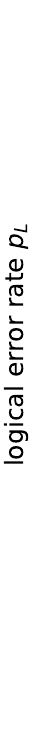}
  \end{subfigure}
  \hspace{-1.5ex}
  \begin{subfigure}[t]{.24\linewidth}
    \centering
    \includegraphics[width=1\textwidth]{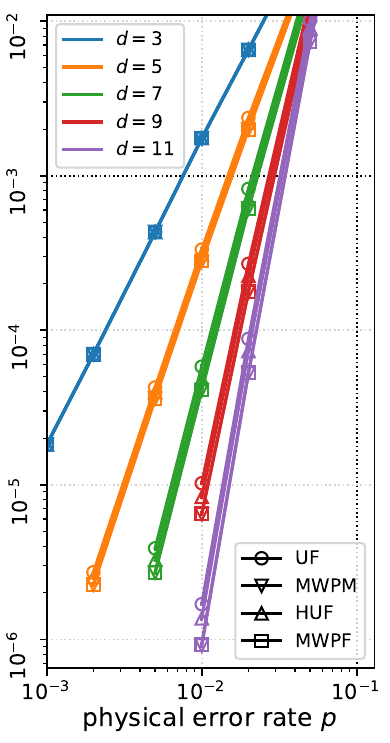}
    \caption{bit-flip noise}
    \label{fig:eva-rsc-bit-flip}
  \end{subfigure}
  \hspace{-2.5ex}
  \begin{subfigure}[t]{.24\linewidth}
    \centering
    \includegraphics[width=1\textwidth]{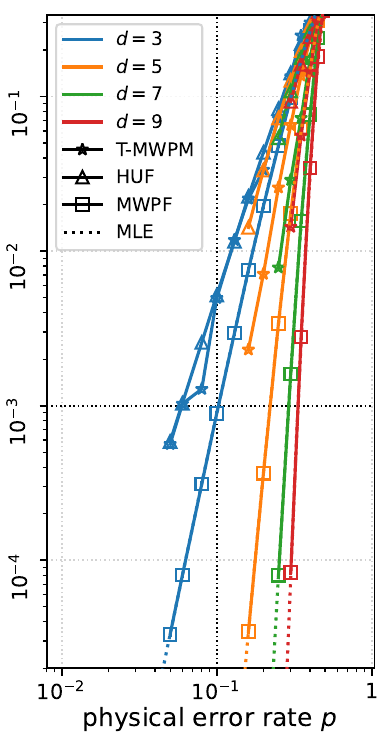}
    \caption{biased-$Y$ noise}
  \end{subfigure}
  \hspace{-2.5ex}
  \begin{subfigure}[t]{.24\linewidth}
    \centering
    \includegraphics[width=1\textwidth]{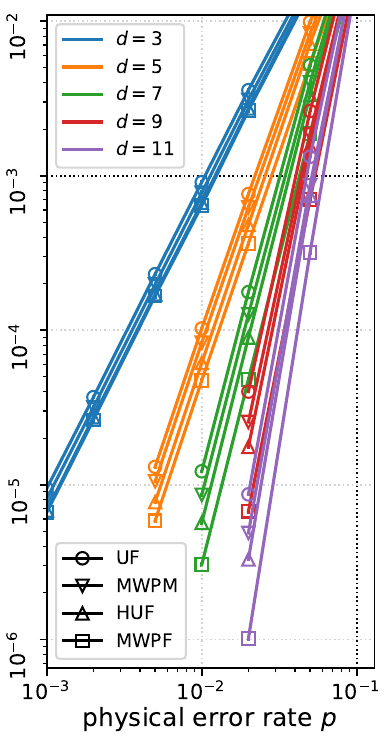}
    \caption{depolarizing noise}
  \end{subfigure}
  \hspace{-2.5ex}
  \begin{subfigure}[t]{.24\linewidth}
    \centering
    \includegraphics[width=1\textwidth]{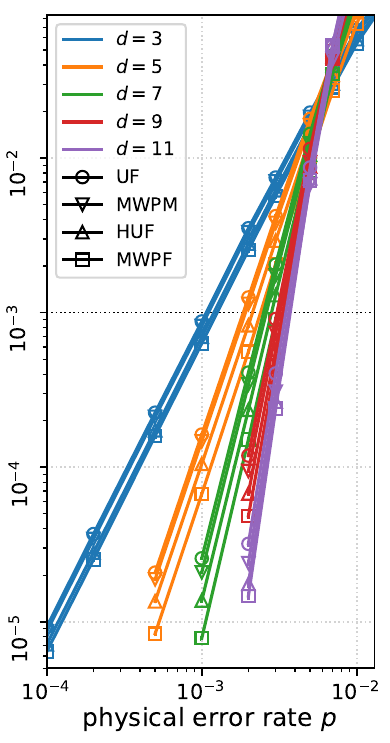}
    \caption{circuit-level noise}
  \end{subfigure}
  \caption{\hyperion (MWPF) achieves the highest accuracy on surface code under a variety of noise models. Only (a) involves a simple graph decoding, while (b)-(d) involve decoding hypergraphs. (a) For the bit-flip noise, it achieves the same accuracy as the MWPM decoder, which is an optimal MLE decoder. (b) For the biased-$Y$ noise, which results in a nullity$_{\le 1}$ decoding hypergraph, it achieves orders of magnitude higher accuracy than the state-of-the-art T-MWPM decoder~\cite{tuckett2019tailoring}, achieving the same accuracy as the optimal MLE decoder (dotted line) that we implement for nullity$_{\le 1}$ hypergraphs. (c) For the depolarizing noise and (d) circuit-level noise, it achieves higher accuracy than all other decoders.}
  \label{fig:eva-rsc}

\end{figure*}

We evaluate \hyperion with the floating-point data type and the \emph{SingleHair} \refdef{relaxer} finder (\S\ref{ssec:single-hair-subroutine}), denoted as ``MWPF(c)'' given a tunable per-\refdef{cluster} limit $c$ (\S\ref{sec:implementation}).
Note that ``MWPF(c=0)'' is essentially the Hypergraph Union-Find (HUF) decoder.
We answer the following questions in our evaluation:

\begin{itemize}
  \item Accuracy: how does the logical error rate of \hyperion compare to existing decoders?
  \item Speed: how long does it take to decode a syndrome, and what is the distribution of decoding time?
  \item Accuracy-speed trade-off: how does the decoding time scale with the desired accuracy?
\end{itemize}

\subsection{Simulation Setup}\label{ssec:eva-setup}

We use \texttt{stim}~\cite{gidney2021stim} to generate the syndrome given a circuit and noise model.
We generate enough syndromes to ensure the error bar corresponding to 95\% confidence interval is sufficiently small that the relative accuracies of different decoders are obvious.
For each setting, we use the same sequence of syndromes across all decoders when applicable.
The number of syndromes depends on the logical error rate, from $1000$ for the highest to $4.6\times10^9$ for the lowest.
We consider three types of qLDPC codes: surface code~\cite{dennis2002topological}, color code~\cite{bombin2006topological} and Bivariate Bicycle (BB) code~\cite{bravyi2024high}.
We employ code-capacity noise models including bit-flip noise ($p_X = p$, $p_Y = p_Z = 0$), depolarizing noise ($p_X = p_Y = p_Z = p/3$) and biased-$Y$ noise ($p_Y = p$, $p_X = p_Z = 0$).
We also consider circuit-level noise models defined in~\cite{gidney2021stim},~\cite{gidney2023new} and~\cite{gong2024toward} for the surface code, the color code and the BB code, respectively.
Among the combination of the above codes and noise models, only the surface code with the bit-flip noise model has a simple decoding graph, while the rest have decoding hypergraphs.
We compare \hyperion with existing MWPM decoder~\cite{wu2023qce,higgott2025sparse} and BPOSD decoder~\cite{roffe2020decoding,Roffe_LDPC_Python_tools_2022}.
We also consider variants of the MWPM decoder optimized for the color code (denoted as ``C-MWPM'') as implemented by the authors of~\cite{gidney2023new,chromobius} and for the tailored surface code with biased-$Y$ noise~\cite{tuckett2019tailoring} (denoted as ``T-MWPM''), which we implement in~\cite{qec-playground-t-mwpm}.
These variants are optimized for the hypergraphs from their intended codes and therefore, not general.
They cleverly, through approximation, convert the MWPF problem on the decoding hypergraph into an MWPM problem on a simple graph, which loses accuracy.

For a fair comparison, we ensure that all decoders for the same setting are decoding the same sequence of syndromes, except for T-MWPM, which does not support \texttt{stim} but has its own syndrome generator.

We choose Parity Blossom~\cite{wu2023qce} for the baseline MWPM decoder for two reasons.
First, it allows tuning between UF and MWPM decoders using a parameter similar to the per-\refdef{cluster} limit $c$ in \hyperion (\S\ref{ssec:rust-impl}), denoted as ``MWPM(c)''.
We note that MWPM(c=0) is equivalent to the UF decoder and MWPM(c=$\infty$) is equivalent to the MWPM decoder.
Second, like \hyperion, Parity Blossom is also implemented in Rust and has the same overhead adapting to the Python interface.
For speed benchmarks, all decoders use a single thread on an Apple M4 Pro CPU with 24 GB of memory.
Because Parity Blossom only works for simple graphs, it ignores the hyperedges when working on decoding hypergraphs, which inevitably leads to accuracy loss.

For \hyperion, we only use the \emph{SingleHair} \refdef{relaxer} finder (\S\ref{ssec:single-hair-subroutine}), which does not exploit any special property of the decoding hypergraph.
We use MWPF(c=200) as the \hyperion default and MWPF(c=0) as the HUF decoder.

We omit data near or above the threshold because (1) \hyperblossom relies on \emph{clustering}, which becomes prohibitively slow at high error rates, and (2) sub-threshold scaling is more relevant for QEC, as QEC is not profitable when physical error rates are too high.

\subsection{Decoding Accuracy}\label{ssec:eva-accuracy}

\hyperion achieves higher decoding accuracy compared to the MWPM decoder and its variants when decoding topological codes like surface code and color code.
However, for non-topological codes like the BB code, its accuracy is lower than the BPOSD decoder at high error rates or large code distances.

For the \textbf{surface code}, we consider three types of code-capacity noise models (\autoref{fig:eva-rsc} (a-c)) and a circuit-level noise model (\autoref{fig:eva-rsc} (d)).
Different noise models lead to different types of decoding hypergraphs, which impact the performance of decoders.
This highlights the importance of evaluating decoders with the right noise model.
For the bit-flip noise, the decoding hypergraph is a simple graph and as a result, the MWPM decoder is an optimal MLE decoder.
In this case, \hyperion achieves the same accuracy as the MWPM decoder, despite that theoretically the \emph{SingleHair} \refdef{relaxer} finder may not find the optimal solution (\S\ref{ssec:single-hair-subroutine}).
For the biased-$Y$ noise, the decoding hypergraph is a nullity$_{\le 1}$ hypergraph.
Although the T-MWPM decoder~\cite{tuckett2019tailoring} has a near-optimal threshold, it has been shown to be suboptimal in the logical error rate scaling~\cite{tsai2024mitigating}.
On the other hand, MWPF(c=$\infty$) is optimal for nullity$_{\le 1}$ decoding hypergraphs  (\S\ref{ssec:optimal-1-dof}) and as shown in \autoref{fig:eva-rsc} (b), MWPF(c=200) already achieves an accuracy orders of magnitude higher than that of T-MWPM.
For depolarizing noise and circuit-level noise, the decoding hypergraph is a general hypergraph.
We note that the (weighted) HUF decoder achieves a higher accuracy than the MWPM decoder, because it considers the hyperedges introduced by the $\hat{Y}$ errors.
\hyperion further achieves 3.3x lower logical error rate than the HUF decoder at $d=11$ with the code-capacity depolarizing noise, a total of 4.8x lower logical error rate compared to the MWPM decoder.

\begin{figure}[t]
  \begin{subfigure}[t]{\linewidth}
    \centering
    \textbf{Color Code}
  \end{subfigure}
  \vspace{-2ex}

  \centering
  \begin{subfigure}[t]{.046\linewidth}
    \centering
    \includegraphics[width=1\textwidth]{figures/mwpf-evaluation/logical-error-rates/y-label.pdf}
  \end{subfigure}
  \hspace{-1.5ex}
  \begin{subfigure}[t]{.48\linewidth}
    \centering
    \includegraphics[width=1\textwidth]{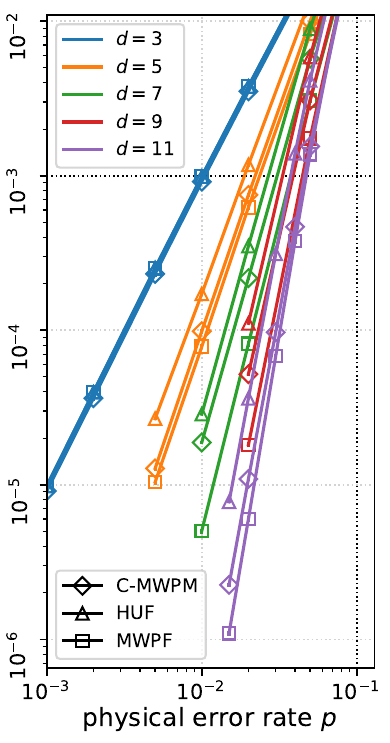}
    \caption{depolarizing noise}
  \end{subfigure}
  \hspace{-2.5ex}
  \begin{subfigure}[t]{.48\linewidth}
    \centering
    \includegraphics[width=1\textwidth]{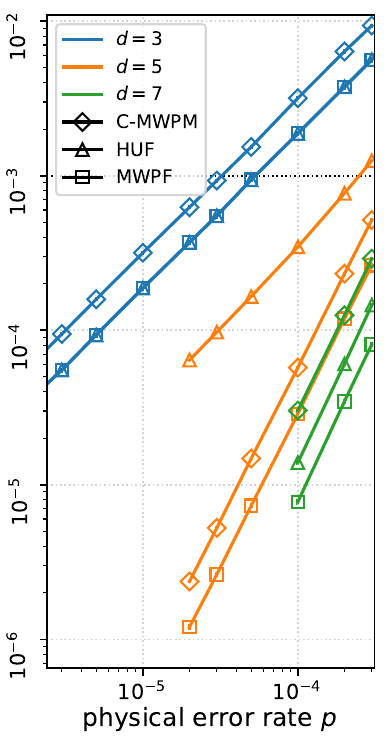}
    \caption{circuit-level noise}
    \label{fig:eva-color-circuit}
  \end{subfigure}
  \caption{\hyperion (MWPF) achieves a higher accuracy than that of the C-MWPM decoder on color code under both code-capacity depolarizing noise and circuit-level noise.}
  \label{fig:eva-color}
\end{figure}

For \textbf{color code}, we consider the depolarizing noise and circuit-level noise, which results in general hypergraphs.
While the HUF decoder is generally not as accurate as the C-MWPM decoder~\cite{gidney2023new}, \hyperion achieves a higher accuracy than that of the C-MWPM decoder, as shown in~\autoref{fig:eva-color}.
The circuit-level noise of the color code is suboptimal in terms of the effective code distance, as shown by the slope of the logical error rate curve in \autoref{fig:eva-color-circuit}.
Nonetheless, \hyperion achieves the same effective code distance as C-MWPM.
The HUF decoder, on the other hand, fails to increase the effective code distance when $d$ increases from $3$ to $5$.

\begin{figure}[t]
  \begin{subfigure}[t]{\linewidth}
    \centering
    \textbf{BB Code}
  \end{subfigure}
  \vspace{-2ex}

  \centering
  \begin{subfigure}[t]{.48\linewidth}
    \centering
    \includegraphics[width=1\textwidth]{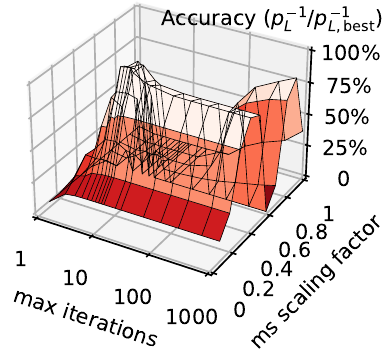}
  \end{subfigure}
  \begin{subfigure}[t]{.48\linewidth}
    \centering
    \includegraphics[width=1\textwidth]{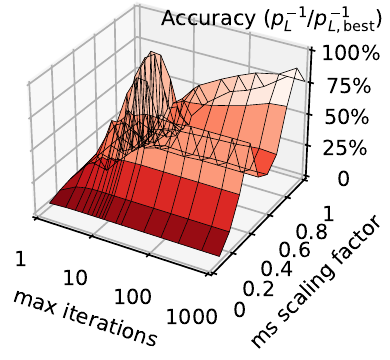}
  \end{subfigure}
  \begin{subfigure}[t]{.046\linewidth}
    \centering
    \includegraphics[width=1\textwidth]{figures/mwpf-evaluation/logical-error-rates/y-label.pdf}
  \end{subfigure}
  \hspace{-1.5ex}
  \begin{subfigure}[t]{.48\linewidth}
    \centering
    \includegraphics[width=1\textwidth]{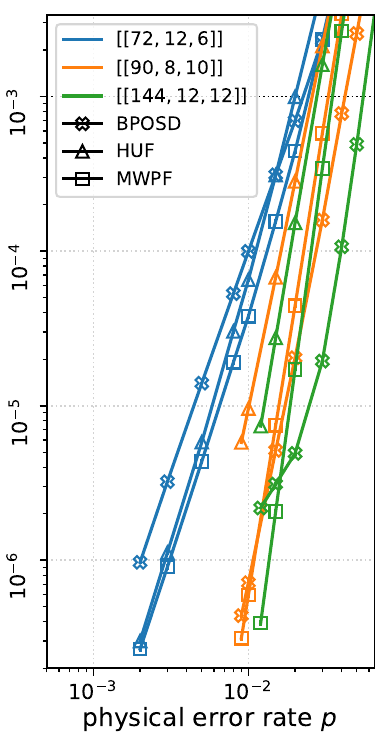}
    \caption{depolarizing noise}
  \end{subfigure}
  \hspace{-2.5ex}
  \begin{subfigure}[t]{.48\linewidth}
    \centering
    \includegraphics[width=1\textwidth]{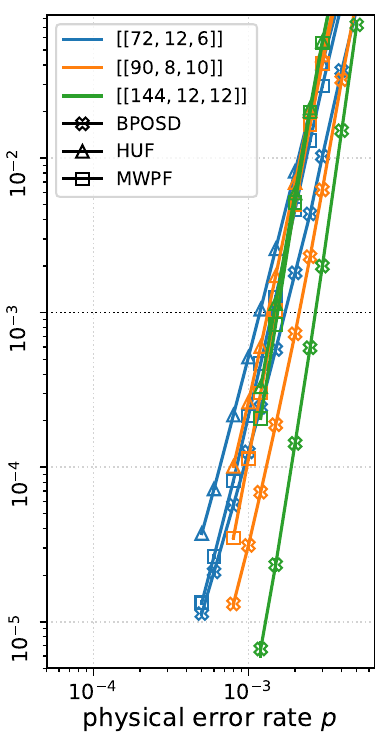}
    \caption{circuit-level noise}
  \end{subfigure}
  \caption{Comparison of \hyperion (MWPF), HUF and BPOSD decoders on BB code under code-capacity and circuit-level noise. (top) exhaustive search for optimal BP parameters on the $[[90,8,10]]$ BB code at $p=2.5\%$ code-capacity depolarizing noise and $p=0.1\%$ circuit-level noise.
  (bottom) using these optimal BP parameters, \hyperion is more accurate than the BPOSD decoder on small BB codes and at low physical error rates. However, at larger code sizes or higher error rates, BPOSD yields better accuracy.}
  \label{fig:eva-bb}
\end{figure}

For the \textbf{BB code}, a non-topological code, we consider the code-capacity depolarizing and circuit-level noise models.
As the baseline, we use the BPOSD decoder from ~\cite{roffe2020decoding,Roffe_LDPC_Python_tools_2022}, which is the state-of-the-art decoder for non-topological codes.
The BPOSD decoder first uses Belief Propagation (BP) to estimate a posterior distribution of the hyperedges.
It then finds a parity factor by enumerating hyperedges by their weights, which indicates the error probabilities, henceforth Ordered Statistics Decoding (OSD).
Since the BP process is suboptimal in decoding quantum codes,
tuning the BP parameters is essential to achieve a high accuracy.
As shown at the top of \autoref{fig:eva-bb}, we exhaustively search for the optimal BP parameters for the $[[90, 8, 10]]$ code.
We use the physical error rate of $p=2\%$ and $p=0.1\%$ for the bit-flip noise model and the circuit-level noise, respectively.
We find that a min-sum scaling factor of $0.9$ and a large number of iterations ($\ge 1000$) achieves the highest accuracy.
Note that this fine-tuned parameter is not necessarily optimal for other code distances and physical error rates.
We compare \hyperion with this fine-tuned BPOSD decoder for BB codes of three configurations including $[[90, 8, 10]]$.
As shown at the bottom of \autoref{fig:eva-bb}, with the bit-flip noise model, \hyperion achieves a higher accuracy than that of BPOSD for small code distances ($d=6$ and $d=10$), while the BPOSD decoder achieves higher accuracy than \hyperion for $d=12$.
With the more realistic circuit-level noise model,
BPOSD is more accurate than \hyperion by orders of magnitude.
It remains future work to explore better \refdef{relaxer} finders for BB codes.

\subsection{Decoding Speed}\label{ssec:eva-speed}

We evaluate the decoding speed of \hyperion on the surface code and the color code with both depolarizing noise and circuit-level noise.
We do not include the BB code in our evaluation, as it lacks a single tunable parameter (such as code distance) that can generate a continuous family of codes.

\begin{figure}
  \begin{subfigure}[t]{\linewidth}
    \centering
    \textbf{Surface Code}
  \end{subfigure}
  \vspace{-1.5ex}

  \centering
  \begin{subfigure}[t]{.046\linewidth}
    \centering
    \includegraphics[width=1\textwidth]{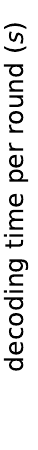}
  \end{subfigure}
  \hspace{-3.5ex}
  \begin{subfigure}[t]{.54\linewidth}
    \centering
    \includegraphics[width=0.8889\textwidth]{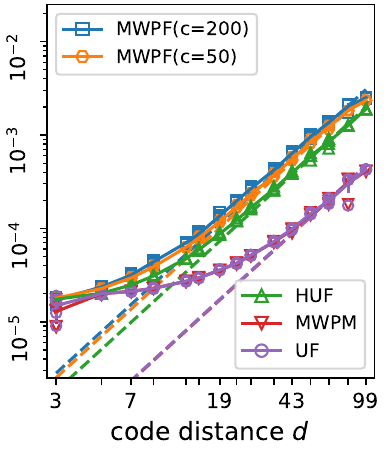}
    \caption{depolarizing $p=1\%$}
  \end{subfigure}
  \hspace{-3.5ex}
  \begin{subfigure}[t]{.48\linewidth}
    \centering
    \includegraphics[width=1\textwidth]{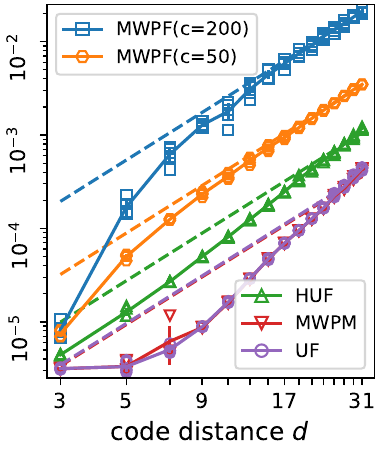}
    \caption{circuit-level $p=0.1\%$}
  \end{subfigure}
  \caption{\hyperion (MWPF) achieves an almost-linear time scaling on surface code. Each dashed line fits the linear decoding time scaling with the number of physical qubits.}
  \label{fig:eva-speed-surface}

  \begin{subfigure}[t]{\linewidth}
    \centering
    \textbf{Color Code}
  \end{subfigure}
  \vspace{-1.5ex}

  \begin{subfigure}[t]{.046\linewidth}
    \centering
    \includegraphics[width=1\textwidth]{figures/mwpf-evaluation/speed/y-label.pdf}
  \end{subfigure}
  \hspace{-3.5ex}
  \begin{subfigure}[t]{.54\linewidth}
    \centering
    \includegraphics[width=0.8889\textwidth]{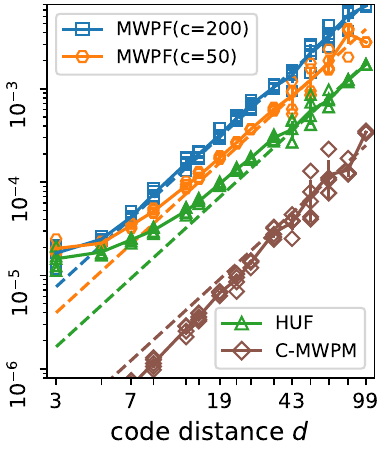}
    \caption{depolarizing $p=1\%$}
  \end{subfigure}
  \hspace{-3.5ex}
  \begin{subfigure}[t]{.48\linewidth}
    \centering
    \includegraphics[width=1\textwidth]{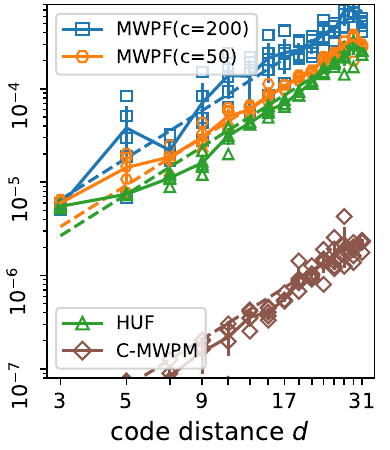}
    \caption{circuit-level $p=10^{-4}$}
  \end{subfigure}
  \caption{\hyperion (MWPF) achieves an almost-linear time scaling on color code.}
  \label{fig:eva-speed-color}
\end{figure}

As shown in \autoref{fig:eva-speed-surface} and \autoref{fig:eva-speed-color}, \hyperion with a fixed per-\refdef{cluster} limit of $c=50$ and $c=200$ achieves an almost-linear average decoding time as the code distance increases in all four code-noise combinations.
This can be explained by the fact that larger \refdef{clusters} are exponentially rare when the physical error rate is sufficiently low.
For each $d$ and $p$ combination, we repeat the measurement for 5 times to show the variance of the average decoding time.
For each data point, we sample at least 300k and 10k shots for the code-capacity noise and circuit-level noise, respectively.

\begin{figure}
  \centering
  \begin{subfigure}[t]{.046\linewidth}
    \centering
    \includegraphics[width=1\textwidth]{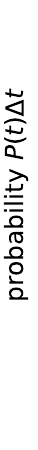}
  \end{subfigure}
  \hspace{-1.5ex}
  \begin{subfigure}[t]{.48\linewidth}
    \centering
    \includegraphics[width=1\textwidth]{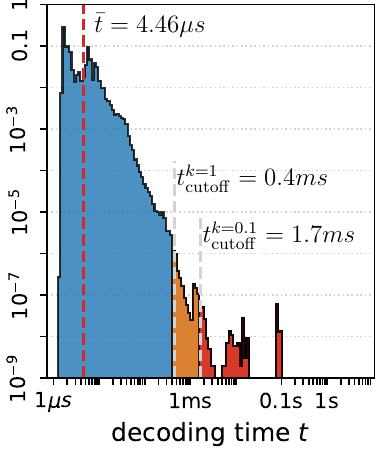}
    \caption{HUF or MWPF(c=0)}
  \end{subfigure}
  \hspace{-2.5ex}
  \begin{subfigure}[t]{.48\linewidth}
    \centering
    \includegraphics[width=1\textwidth]{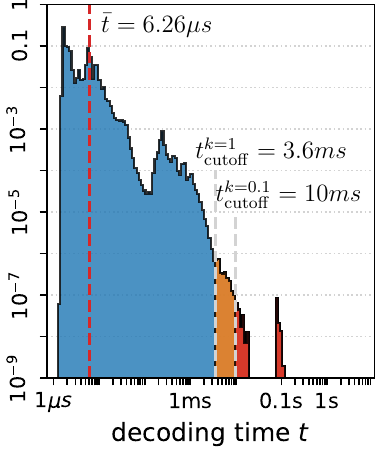}
    \caption{MWPF(c=15)}
  \end{subfigure}
  \\\vspace{1ex}
  \begin{subfigure}[t]{.046\linewidth}
    \centering
    \includegraphics[width=1\textwidth]{figures/mwpf-evaluation/distribution/y-label.pdf}
  \end{subfigure}
  \hspace{-1.5ex}
  \begin{subfigure}[t]{.48\linewidth}
    \centering
    \includegraphics[width=1\textwidth]{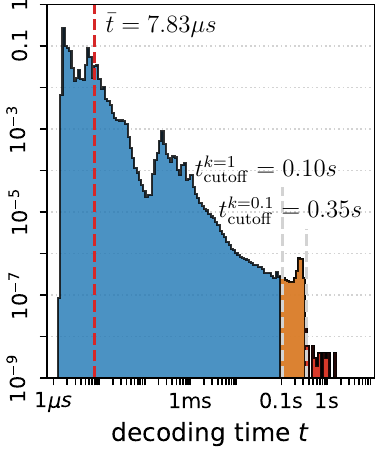}
    \caption{MWPF(c=200)}
  \end{subfigure}
  \hspace{-2.5ex}
  \begin{subfigure}[t]{.48\linewidth}
    \centering
    \includegraphics[width=1\textwidth]{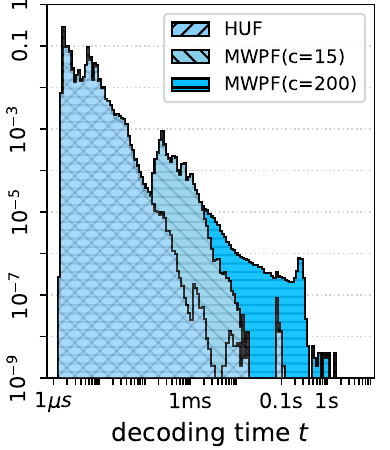}
    \caption{Comparison}
  \end{subfigure}
  \caption{Decoding runtime distribution ($d=7$ surface code with $p=1\%$ depolarizing noise). The MWPF decoder family (\hyperion) exhibits an exponential distribution of decoding time, with a noticeable tail lift at higher runtimes when the per-\refdef{cluster} limit $c$ is large. Note that the figure is plotted in log-log scale, so the area under the curve does not correspond to cumulative probability.}
  \label{fig:eva-distribution}
\end{figure}

We further look into the distribution of the decoding time.
We plot the histogram of the decoding time in \autoref{fig:eva-distribution} for various $c$ choices, each with $10^9$ samples.
We adopt the concept of $k$-cutoff decoding time in Micro Blossom~\cite{wu2025asplos} which defines the maximum decoding time when at most $(1+k)p_L$ logical error rate can be tolerated.
The cutoff decoding time is orders of magnitude larger than the average decoding time, resulting in milliseconds to seconds worst-case latency to achieve the desired accuracy.
This is not favorable for real-time applications but we hope hardware acceleration like that in~\cite{wu2025asplos} can speed up MWPF decoders in future work.
One interesting observation is that the larger $c$ increases the probability of higher decoding time while keeping low decoding time distribution largely unchanged in the log-scale plotting.
This is because the \hyperblossom algorithm skips working on the \refdef{locally-optimal-clusters} and focuses on those suboptimal \refdef{clusters} that require more refinement, i.e., a large primal-dual gap.

\subsection{Accuracy-Speed Trade-off}\label{ssec:eva-tradeoff}

\begin{figure}
  \centering
  \begin{subfigure}[t]{.046\linewidth}
    \centering
    \includegraphics[width=1\textwidth]{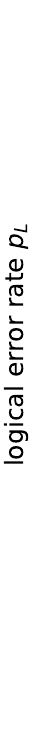}
  \end{subfigure}
  \hspace{-2.5ex}
  \begin{subfigure}[t]{.5\linewidth}
    \centering
    \includegraphics[width=0.96\textwidth]{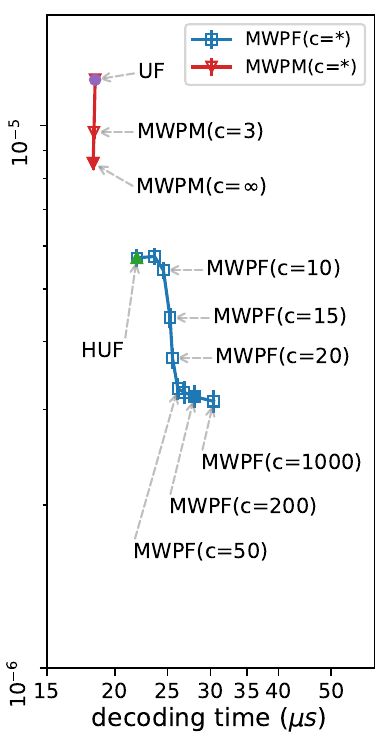}
    \caption{$d=7$ Surface Code}
    \label{fig:eva-trade-off-surface}
  \end{subfigure}
  \hspace{-2.5ex}
  \begin{subfigure}[t]{.48\linewidth}
    \centering
    \includegraphics[width=1\textwidth]{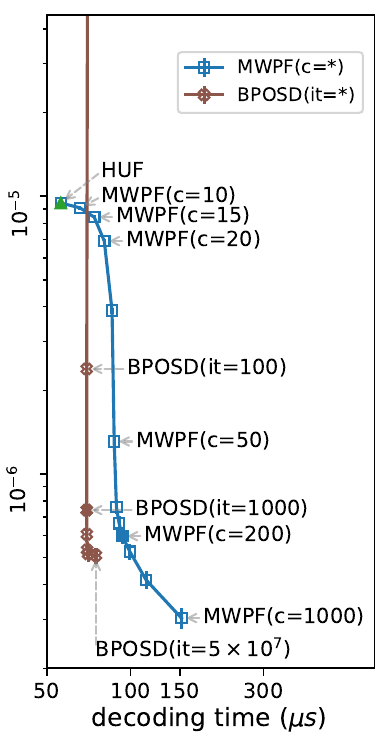}
    \caption{$[[90, 8, 10]]$ BB Code}
    \label{fig:eva-trade-off-bb}
  \end{subfigure}
  \caption{Trade-off between speed and decoding accuracy under code-capacity depolarizing noise of $p=1\%$. (a) for surface code, MWPF decoders are more accurate but slower compared to MWPM decoders. (b) for BB code, MWPF decoders achieve lower logical error rate than BPOSD decoders at the cost of longer decoding time.}
\end{figure}

With the tunable parameter of the per-\refdef{cluster} limit $c$, \hyperion provides an easy trade-off between decoding accuracy and decoding speed.
We evaluate the trade-off for the surface code and the BB code, both with the code-capacity noise model.

For the surface code, we look at the MWPF(c=*) decoder family (\hyperion) and the MWPM(c=*) decoder family (Parity Blossom) (\S\ref{ssec:eva-setup}).
We choose an intermediate code distance of $d=7$ and a depolarizing error rate of $p=1\%$.
As shown in \autoref{fig:eva-trade-off-surface}, within the MWPF decoder family, the MWPF(c=200) decoder achieves 1.8x lower logical error rate at the cost of 1.3x longer decoding time compared to the MWPF(c=0) decoder.
This demonstrates the usefulness of the \emph{SingleHair} \refdef{relaxer} finder beyond the trivial \emph{UnionFind} \refdef{relaxer} finder.
Furthermore, we see an S-shaped curve for the MWPF decoder family.
This is because for small $c$, \hyperion may not invoke the \emph{SingleHair} \refdef{relaxer} finder before a \refdef{cluster} hits the limit.
On the other hand, the logical error rate saturates at a large $c \ge 1000$ due to the limitation of the \emph{SingleHair} \refdef{relaxer} finder (\S\ref{ssec:failure-hypergraph}).
Comparing between the two decoder families, we see that the MWPF decoder family is more accurate but slower compared to the MWPM decoder family.

For the BB code, we look at the MWPF(c=*) decoder family and the BPOSD(max\_iter=*) decoder family.
We use a min-sum scaling factor of 0.9 for the BPOSD decoders, which is optimal under depolarizing noise of $p=2.5\%$ in an exhaustive search as shown in \autoref{fig:eva-bb}.
We choose an intermediate code size of $[[90,8,10]]$ and a depolarizing error rate of $p=1\%$, as shown in \autoref{fig:eva-trade-off-bb}.
Within the MWPF decoder family, the MWPF(c=1000) decoder achieves 31x lower logical error rate at the cost of 2.7x longer decoding time compared to the HUF decoder.
We observe that the curve is not saturating at $c=1000$, which suggests that larger $c$ may further improve the accuracy.
On the other hand, the BPOSD(max\_iter) decoder family saturates at a maximum iteration of $5\times 10^7$ yet still falls behind MWPF(c$\ge$500) decoders in terms of the decoding accuracy.
For this specific code and noise model, the MWPF decoder is more accurate than the BPOSD decoder.
We note that larger BB codes beyond $[[90,8,10]]$ may observe that the BPOSD decoder is more accurate than the MWPF decoder, as shown in \autoref{fig:eva-bb}.

\section{Related Work}\label{sec:related}

\nosection{Graph-based Decoders} model the decoding problem as a graph problem and use graph algorithms to solve it.
MWPM (\S\ref{ssec:mwpm}) and Union-Find~\cite{delfosse2021almost} are two popular graph-based decoders on simple graphs.
For simple graphs, fast software~\cite{higgott2025sparse,wu2023qce,higgott2023improved} and hardware~\cite{liyanage2023qce,ziad2024local,liyanage2024fpga,barber2025real,wu2025asplos,liyanage2025qce,das2022lilliput,overwater2022neural,riste2020real} implementations have reduced the decoding time to sub-microseconds for real-time decoding on superconducting qubits with $\mu s$-level measurement round~\cite{google2021exponential}.
HyperBlossom (\S\ref{sec:math}) and Hypergraph Union-Find (HUF)~\cite{delfosse2022toward} generalize these decoders for hypergraphs. Although they are generally orders of magnitude slower, they achieve higher decoding accuracy. They are particularly suitable for qubits with $ms$-level measurement round such as trapped-ion~\cite{egan2021fault,webber2022impact,ryan2021realization} and neutral-atom~\cite{bluvstein2024logical,henriet2020quantum}.
Importantly, \hyperblossom provides a unified framework for all graph-based decoders.

\nosection{Linear-Programming Decoders} achieve a decoding accuracy comparable to that of the belief propagation (BP) decoder while providing stronger theoretical guarantees on classical LDPC codes~\cite{feldman2005using}.
The authors of~\cite{barman2013decomposition} accelerate the decoding process by decomposing the decoding problem into smaller ones.
We use similar ideas to provide a strong proximity bound (\S\ref{ssec:problem-definitions}) and use \emph{clustering} (\S\ref{ssec:algo-cluster}) to decompose the decoding problem adaptively.

Generalizing LP decoders to quantum codes is non-trivial.
Recent works~\cite{fawzi2021linear,li2018lp,gu2025power} build on LP formulations that have a non-zero integrality gap ($\min\text{LP} < \min\text{ILP}$).
\hyperblossom advances these works by designing and solving a more complicated LP problem that closes the integrality gap for certain classes of hypergraphs, although it is still unclear whether it closes the gap for all hypergraphs.

\nosection{ILP and MaxSAT Decoders} use integer constraints to solve the MWPF problem and as a result, are generally slower.
For example, the authors of~\cite{landahl2011fault,lacroix2025scaling,cain2024correlated} introduce extra integer variables to describe the MWPF problem as an integer linear programming problem.
The authors of~\cite{berent2024decoding} employ MaxSAT solvers that natively support boolean variables.
These decoders rely on third-party solvers, which provide little control over the decoding process.
In contrast, \hyperion provides a user with fine control over the decoding process through \emph{clustering} (\S\ref{ssec:algo-cluster}) and prioritized \refdef{cluster} refinement (\S\ref{sec:implementation}): the user can make trade-offs between decoding accuracy and speed using a single parameter $c$.

\nosection{Heuristic Decoders} provide fewer theoretical guarantees but are usually faster and can be tuned to achieve a high accuracy.
Belief Propagation (BP) is near-optimal on some classical LDPC codes but is not suitable for QEC decoding alone, due to quantum degeneracy~\cite{poulin2008iterative,panteleev2021degenerate,raveendran2021trapping,morris2023absorbing}.
Thus, QEC decoders usually employ it as a pre-processing step to generate a posterior distribution and then apply another heuristic post-processing decoder to determine a correction~\cite{grospellier2021combining,panteleev2021degenerate,wolanski2024ambiguity,hillmann2024localized,gong2024toward,iolius2024almost,roffe2020decoding,fossorier1998reliability,fossorier2002soft,maan2025machine}.
Among these post-processing algorithms, \cite{wolanski2024ambiguity,hillmann2024localized} also use the idea of clustering to speed up the decoding process.
Neural network decoders~\cite{varsamopoulos2017decoding,gong2024graph,bausch2023learning,lange2025data,varbanov2025neural} are another promising, data-driven approach but they require prohibitively expensive training for large code distances.
Search-based decoders~\cite{ott2025decision,beni2025tesseract} also use heuristics to accelerate exploration of the parity factor solution space.
These decoders rely on approximations to achieve practical decoding speeds, and consequently provide no analytical guarantees on the accuracy of their outputs.
In contrast, \hyperblossom achieves a high speed while also producing an analytical bound of proximity to optimality, which can serve as logical-level soft information~\cite{pattison2021improved} for decoding concatenated codes~\cite{gidney2025yoked}.

\section{Conclusion and Future Work}

\hyperblossom is a new framework of certifying MLE decoding for quantum LDPC codes.
It is a generalization of the blossom algorithm and existing graph-based decoders of UF, MWPM and HUF.
We propose two important algorithmic designs: \emph{relaxing} for simplifying the problem and \emph{clustering} for faster decoding.
By producing a certified proximity bound for each syndrome it decodes, \hyperblossom provides additional information that is helpful for theoretical analysis of QEC codes.

Several directions remain open for future work.
First, developing specialized \refdef{relaxer}-finding algorithms tailored to specific code families may further improve both accuracy and speed.
Second, exploring parallelization strategies, including coarse-grained parallelization with fusion~\cite{wu2023qce,yang2024parallel} and fine-grained parallelization with hardware accelerator~\cite{liyanage2023qce,liyanage2025qce,wu2025asplos}, may future boost the decoding speed.
Finally, it is of theoretical interest to investigate whether \ref{condition:mwpf} holds for all or some types of hypergraphs.
This question is important because a proof that it holds  would justify pursuing optimal relaxer-finding algorithms for general hypergraphs, whereas counterexamples would reveal a fundamental limitation of the \hyperblossom framework: it would not be able to certify certain MLE decoding problems.

\section{Acknowledgement}

This work was supported in part by Yale University and NSF MRI Award \#2216030.
Yue Wu is also grateful for the support from Unitary Foundation.
The authors thank Shilin Huang, Arshpreet Singh Maan and Anqi Gong for insightful discussions on how to tune the BP decoder parameters.


\bibliographystyle{IEEEtran}
\providecommand{\noopsort}[1]{}\providecommand{\singleletter}[1]{#1}%

\clearpage
\appendix

\section{HyperBlossom Framework Proofs}

In this section, we prove the theorems presented in the \hyperblossom framework (\S\ref{sec:math}).
We first prove \reftheorem{ilp-equal-mpwf} (\S\ref{ssec:problem-definitions}) in \S\ref{ssec:minilp-equal-mwpf}.
We then prove the optimality of the \emph{relaxing} algorithm (\S\ref{ssec:cascaded-relaxing}) in \S\ref{ssec:optimality-of-relaxing} and the optimality of the \emph{clustering} algorithm (\S\ref{ssec:algo-cluster}) in \S\ref{ssec:cluster-optimality-proof}.
Putting together, we prove the optimality of the \hyperblossom algorithm (\S\ref{ssec:hyperblossom-algorithm}) in \S\ref{ssec:hyperblossom-optimality}.
Finally, we show the necessity of $E_S$ in the definition of \refdef{invalid} subgraph $S = (V_S, E_S)$ in \S\ref{ssec:why-need-ES}.

\subsection{minILP = minMWPF}\label{ssec:minilp-equal-mwpf}

\theoremilpequalmwpf{ilp-equal-mwpf-with-proof}

\begin{proof}
  To prove the equality of the optimal values of \refeqs{ILP} and \refeqs{MWPF}, we prove that $\min\text{\refeqs{ILP}} \le \min\text{\refeqs{MWPF}}$ and $\min\text{\refeqs{MWPF}} \le \min\text{\refeqs{ILP}}$.

  Given \reflemma{ilp-relax-mwpf-with-proof}, \refeqs{ILP} is a relaxation of \refeqs{MWPF}, i.e., $\min\text{\refeqs{ILP}} \le \min\text{\refeqs{MWPF}}$.

  We then prove \reflemma{smaller-weight-parity-factor-with-proof}.
  In this way, the minimum value of \refeqs{MWPF} is no larger than the minimum value of \refeqs{ILP}, i.e., $\min\text{\refeqs{MWPF}} \le \min\text{\refeqs{ILP}}$.

  Together, we prove $\min\text{\refeqs{ILP}} = \min\text{\refeqs{MWPF}}$.
\end{proof}

\vspace{1ex}
\definelemma{ilp-relax-mwpf-with-proof}{Any parity factor is a feasible \refeqs{ILP} solution}

\begin{proof}
  Given any parity factor $\vec{x} \in \mathbb{F}_2^{|E|}$, we prove the lemma by contradiction.
  Suppose $\vec{x}$ is not a feasible \refeqs{ILP} solution, then there exists an \refdef{invalid} subgraph $S \in \mathcal{O}$ where the solution $\vec{x}$ violates the corresponding \refeqs{ILP} constraint \cref{eq:ilp-constraint-2}, i.e.,
  \begin{align*}
    \exists S = (V_S, E_S) \in \mathcal{O}, \sum_{e \in \delta(S)} x_e < 1
  \end{align*}
  Since $x_e \in \mathbb{F}_2 = \{ 0, 1 \}$ in \refeqs{ILP}, we have $$x_e = 0, \forall e \in \delta(S)$$

  Given $\vec{x}$ is a parity factor, it must satisfy the parity constraints of $V_S$.
  That is, let $\mathcal{E}_S = \{ e \in E(V_S) | x_e = 1 \}$, we have $$\mathcal{D}(\mathcal{E}_S) \cap V_S = D \cap V_S$$

  We then transform the left hand side into a simpler form.
  By definition of \refdef{hair} and \refdef{invalid} subgraph, we have $\delta(S) = E(V_S) \setminus E_S$ and $E_S \subseteq E[V_S] \subseteq E(V_S)$.
  Together, we have $E(V_S) = E_S \cup \delta(S)$.
  Given $x_e = 0, \forall e \in \delta (S)$, we have $\mathcal{E}_S = \{ e \in E(V_S) | x_e = 1 \} = \{ e \in E_S | x_e = 1 \}$.
  Since $\mathcal{E}_S \subseteq E_S \subseteq E[V_S]$, it only generates defects within $V_S$, i.e., $\mathcal{D}(\mathcal{E}_S) \cap V_S = \mathcal{D}(\mathcal{E}_S)$.
  Thus, we have $$\mathcal{D}(\mathcal{E}_S) = D \cap V_S$$

  However, this violates the definition of \refdef{invalid} subgraph: $\forall \mathcal{E} \subseteq E_S, \mathcal{D}(\mathcal{E}) \neq D \cap V_S$.
  Thus, the assumption that $\vec{x}$ violates an \refeqs{ILP} constraint is wrong, i.e., $\vec{x}$ satisfies all \refeqs{ILP} constraints and thus it is a feasible \refeqs{ILP} solution.
\end{proof}

\vspace{1ex}
\definelemma{smaller-weight-parity-factor-with-proof}{For every feasible \refeqs{ILP} solution $\vec{x}$, there exists a parity factor $\mathcal{E}$ with no greater weight, i.e., $W(\mathcal{E}) \le \sum_{e \in E} w_e x_e$}

\begin{proof}

  We first prove that for every feasible \refeqs{ILP} solution $\vec{x}$, there exists $\mathcal{E} \subseteq E_S$ where $E_S = \{ e \in E | x_e = 1 \}$, such that $\mathcal{D}(\mathcal{E}) = D$.
  We prove this by contradiction.
  Suppose $\forall \mathcal{E} \subseteq E_S, \mathcal{D}(\mathcal{E}) \neq D$.
  Given $E_S \subseteq E = E[V]$, we have $S = (V, E_S)$ is an \refdef{invalid} subgraph according to the definition, i.e., $S \in \mathcal{O}$.
  There is an \refeqs{ILP} constraint \cref{eq:ilp-constraint-2} for every \refdef{invalid} subgraph $S \in \mathcal{O}$ that says $\sum_{e \in \delta(S)} x_e \ge 1$.
  However, according to the definition of \refdef{hair}, $\delta(S) = E(V_S) \setminus E_S$, we have $\sum_{e \in \delta(S)} x_e = 0$.
  This contradicts the assumption that $\vec{x}$ satisfies all the \refeqs{ILP} constraints \cref{eq:ilp-constraint-2}, including the one corresponding to $S$.
  Thus, there must exist $\mathcal{E} \subseteq E_S$ such that $\mathcal{D}(\mathcal{E}) = D$, i.e., $\mathcal{E}$ is a feasible \refeqs{MWPF} solution.

  We then prove that $W(\mathcal{E}) \le \sum_{e \in E} w_e x_e$.
  This is trivial because the weights of the hypergraph are non-negative $w_e \ge 0, \forall e \in E$.
  Thus, given $\mathcal{E} \subseteq E_S$, we have $W(\mathcal{E}) = \sum_{e \in \mathcal{E}} w_e \le \sum_{e \in E_S} w_e = \sum_{e \in E} w_e x_e$.

  We shall note that not all feasible \refeqs{ILP} solutions are parity factors, in contrast to \reflemma{ilp-relax-mwpf-with-proof}.
  For example, $x_e = 1, \forall e \in \mathcal{E}$ is clearly a feasible \refeqs{ILP} solution, but $E = \{ e \in E | x_e = 1 \}$ is usually not a parity factor.
  One might instead expect that optimal \refeqs{ILP} solutions are parity factors, but this is not true either.
  For example, given an optimal \refeqs{ILP} solution $\vec{x}$, changing $x_e = 0$ to $x_e = 1$ for any zero-weighted hyperedge $e_0 \in E, w_{e_0} = 0$ leads to another optimal \refeqs{ILP} solution $\vec{x'}$.
  This is because $\vec{x'}$ is clearly a feasible \refeqs{ILP} solution, and also $\sum_{e \in E} w_e x'_e = \sum_{e \in E} w_e x_e + w_{e_0} = \sum_{e \in E} w_e x_e$ thus $\vec{x'}$ is optimal.
  However, $\mathcal{E}' = \{ e \in E | x'_e = 1 \}$ is not a parity factor, because $\mathcal{D}(\mathcal{E}') = \mathcal{D}(\{ e \in E | x_e = 1 \}) \oplus \{e_0\} = D \oplus \{ e_0 \} \neq D$.
\end{proof}

\subsection{Optimality of Relaxing}\label{ssec:optimality-of-relaxing}

\theoremRelaxerExistenceOrTrivialDirection{relaxing-with-proof}
\begin{proof}
  Since $\vec{y}$ is suboptimal, there exists an optimal \refeqs{DLP} solution $\vec{y^o}$ with $\sum \vec{y^o} > \sum \vec{y}$.
  We use $\vec{y^o}$ to construct either a \refdef{relaxer} or a \refdef{trivial-direction}.

  Given the linearity of \refeqs{DLP}, $\Delta\vec{y} \coloneqq \vec{y^o} - \vec{y}$ is a \refdef{feasible-direction}, i.e., we can grow along $\Delta\vec{y}$ for a positive length $l > 0$.
  Since $\sum \Delta\vec{y} = \sum \vec{y^o} - \sum \vec{y} > 0$, there exists at least one \refdef{invalid} subgraph $S^+ \in \mathcal{O}$ with $\Delta y_{S^+} > 0$.
  We can then decompose the \refdef{direction} $\Delta\vec{y}$ into two: $\Delta^a\vec{y}$ and $\Delta^b\vec{y}$.
  The \refdef{direction} $\Delta^a\vec{y}$ consists of a single \refdef{invalid} subgraph $S^+$ growing at a rate of $\Delta^a y_{S^+} \coloneqq \min(\Delta y_{S^+}, \sum \Delta\vec{y}) > 0$.
  The other \refdef{direction} is defined as $\Delta^b\vec{y} \coloneqq \Delta\vec{y} - \Delta^a\vec{y}$.
  Depending on whether $\Delta^b\vec{y}$ relaxes any \refdef{tight-edges}, we prove that either $\Delta^b\vec{y}$ is a \refdef{relaxer}, or $\Delta^a\vec{y}$ is a \refdef{trivial-direction} up to a scaling factor.

  First, if $\Delta^b\vec{y}$ relaxes at least one \refdef{tight-edge}, i.e., $\exists e \in T$, $\sum_{S \in \mathcal{O} | e \in \delta(S)} \Delta^b y_S < 0$ (\cref{eq:relaxer-a}), then we prove that $\Delta^b\vec{y}$ is a \refdef{relaxer} by satisfying the other conditions \cref{eq:relaxer-b}, (\ref{eq:feasible-a}) and (\ref{eq:feasible-b}).
  The first condition \cref{eq:relaxer-b} requires $\sum_{S \in \mathcal{O}} \Delta^b y_S \ge 0$.
  \begin{align*}
    \Delta^b\vec{y} &= \Delta\vec{y} - \Delta^a\vec{y} \\
    \sum_{S \in \mathcal{O}} \Delta^b y_S &= \left( \sum_{S \in \mathcal{O}} \Delta\vec{y} \right) - \Delta^a y_S \\
    &= \left( \sum_{S \in \mathcal{O}} \Delta y_S \right) - \min\Big(\Delta y_S, \sum_{S \in \mathcal{O}} \Delta y_S\Big) \\
    &\ge 0
  \end{align*}

  \noindent We then prove \cref{eq:feasible-a}: $\forall S \in \mathcal{O} \setminus \mathcal{B}, \Delta^b y_S \ge 0$.
  We consider two cases when iterating every $S \in \mathcal{O} \setminus \mathcal{B}$.
  When $S = S^+$, we have $\Delta^b y_{S^+} = \Delta y_{S^+} - \Delta^a y_{S^+} = \Delta y_{S^+} - \min(\Delta y_{S^+}, \sum \Delta\vec{y}) \ge 0$.
  Otherwise, $S \in \mathcal{O} \setminus \mathcal{B} \setminus \{ S^+ \}$ thus $\Delta^a y_S = 0$ and $y_S = 0$.
  We have $\Delta^b y_S = \Delta y_S - \Delta^a y_S = \Delta y_S = y^o_S - y_S = y^o_S \ge 0$.
  Thus, in both cases, we have
  \begin{align*}
    \forall S \in \mathcal{O} \setminus \mathcal{B}, \Delta^b y_S \ge 0
  \end{align*}

  \noindent Lastly \cref{eq:feasible-b}: $\forall e \in T, \sum_{S \in \mathcal{O} | e \in \delta(S)} \Delta^b y_S \le 0$.
  Since $\Delta^b\vec{y} = \Delta\vec{y} - \Delta^a\vec{y}$ and $\forall S \in \mathcal{O}, \Delta^a y_S \ge 0$, we have $\forall S \in \mathcal{O}, \Delta^b y_S = \Delta y_S - \Delta^a y_S \le \Delta y_S$.
  \begin{align*}
    \forall e \in T, \sum_{S \in \mathcal{O} | e \in \delta(S)} \Delta^b y_S &= \sum_{S \in \mathcal{O} | e \in \delta(S)} \left( \Delta y_S - \Delta^a y_S \right) \\
    &\le \sum_{S \in \mathcal{O} | e \in \delta(S)} \Delta y_S \\
    &\le 0 \quad\text{(given $\Delta\vec{y}$ is \emph{Feasible})}
  \end{align*}

  On the other hand, if $\Delta^b\vec{y}$ does not relax any \refdef{tight-edge}, i.e., $\forall e \in T, \sum_{S \in \mathcal{O} | e \in \delta(S)} \Delta^b y_S \ge 0$, we prove that $\Delta^a\vec{y}$ is a \refdef{trivial-direction} up to a scaling factor.
  Clearly $\Delta^a \vec{y} = \{ \Delta y_S: \min(\Delta y_S, \sum \Delta\vec{y}) \}$ trivially contains a single growing \refdef{invalid} subgraph $S$.
  We then only need to prove that $\Delta^a \vec{y}$ is a \refdef{feasible-direction}, i.e., it satisfies \cref{eq:feasible-a,eq:feasible-b}.

  $\Delta^a \vec{y}$ clearly satisfies \cref{eq:feasible-a} because all the elements are non-negative: $\forall S \in \mathcal{O}, \Delta^a y_S \ge 0$.

  We then prove $\Delta^a \vec{y}$ satisfies \cref{eq:feasible-b}: it must not grow on any \refdef{tight-edge}, i.e., $\forall e \in T, \sum_{S \in \mathcal{O} | e \in \delta(S)} \Delta^a y_S \le 0$.
  Given $\Delta\vec{y}$ is a \refdef{feasible-direction}, it must satisfy \cref{eq:feasible-b}.
  Also, since $\Delta^b\vec{y}$ does not relax any \refdef{tight-edge}, nor can it grow on any \refdef{tight-edge}, we have $\forall e \in T, \sum_{S \in \mathcal{O} | e \in \delta(S)} \Delta^b y_S = 0$.
  Thus, we have
  \begin{align*}
    \forall e \in T, \qns\sum_{S \in \mathcal{O} | e \in \delta(S)}\qns \Delta^a y_S &= \sum_{S \in \mathcal{O} | e \in \delta(S)} \left( \Delta y_S - \Delta^b y_S \right) \nonumber \\
    &\le \sum_{S \in \mathcal{O} | e \in \delta(S)} - \Delta^b y_S \nonumber \\
    &= 0
  \end{align*}

  Overall, we have proved that either $\Delta^b\vec{y}$ is a \refdef{relaxer}, or $\Delta^a\vec{y} / \Delta^a y_{S^+}$ is a \refdef{trivial-direction}.

\end{proof}

\theoremBatchedRelaxing{cascaded-relaxing-with-proof}
\begin{proof}

  We explicitly construct the composed \refdef{direction} $\Delta'\vec{y}$ and prove $\sum \Delta'\vec{y} \ge \sum \Delta\vec{y}$.

  By definition of \refdef{relaxer}, each edge $e \in \mathcal{R}'_i$ is relaxed when growing along $\Delta^i\vec{y} = R'_i$ for a positive length:
  \begin{align}
    \sum_{S \in \mathcal{O} | e \in \delta(S)} \Delta^i y_S < 0, \qquad &\forall e \in \mathcal{R}(\Delta^i\vec{y}) \label{eq:cascaded-relaxing-i-relax}
  \end{align}

  For every edge $e \in \cup_i \mathcal{R}'_i$, there exists a relaxer $\Delta^e\vec{y}$ such that $e \in \mathcal{R}(\Delta^e\vec{y})$.
  Given \cref{eq:cascaded-relaxing-i-relax}, we have $k_e = \sum_{S \in \mathcal{O}| e \in \delta(S)} \Delta^e y_S < 0$.
  That is, growing along the \refdef{direction} $\Delta^e\vec{y}$ for a small positive length $l > 0$ will relax the edge $e$ by $-k_e l > 0$.

  For the \refdef{feasible-direction} $\Delta\vec{y}[T \setminus \cup_i \mathcal{R}'_i]$, we have
  \begin{align*}
    \sum_{S \in \mathcal{O} | e \in \delta(S)} \Delta y_S \le 0, \qquad &\forall e \in T \setminus \cup_i \mathcal{R}'_i
  \end{align*}

  However, $\Delta\vec{y}$ may violate the above inequality for \refdef{tight-edges} $\cup_i \mathcal{R}'_i \cap T$.
  We define the amount of violation as $\alpha_e \coloneqq \sum_{S \in \mathcal{O} | e \in \delta(S)} \Delta y_S$, and use $\Delta^e\vec{y}$ to fix the violation of the edge $e$ if the violation is positive $\alpha_e > 0$.
  We compose the new \refdef{direction} $\Delta'\vec{y}$ as follows:
  \begin{align}
    \Delta'\vec{y} = \Delta\vec{y} + \sum_{e \in \cup_i \mathcal{R}'_i \cap T | \alpha_e > 0} \frac{\alpha_e}{-k_e} \Delta^e\vec{y} \label{eq:composed-direction}
  \end{align}

  Before proving that $\Delta'\vec{y}$ satisfies the definition of a \refdef{feasible-direction}, we first prove \cref{eq:cascaded-relaxing-sum-sum} below.
  \begin{align}
    &\forall e \in T, \dns\sum_{S \in \mathcal{O} | e \in \delta(S)}\qns \Delta^i y_S \le 0 \quad\text{(given $\{ \Delta^i\vec{y} \}$ are \refdef{relaxers})} \nonumber \\
    &\;\ \downarrow\text{(given $\forall e' \subseteq \cup_i \mathcal{R}'_i, k_{e'} < 0$)} \nonumber \\
    &\Longrightarrow \forall e \in T, \dns\sum_{e' \in \cup_i \mathcal{R}'_i \cap T | \alpha_{e'} > 0} \frac{\alpha_{e'}}{-k_{e'}} \sum_{S \in \mathcal{O} | e \in \delta(S)}\qns \Delta^{e'} y_S \le 0 \nonumber \\
    &\Longrightarrow \forall e \in T, \sum_{S \in \mathcal{O} | e \in \delta(S)} \ \sum_{e' \in \cup_i \mathcal{R}'_i \cap T | \alpha_{e'} > 0} \frac{\alpha_{e'}}{-k_{e'}} \Delta^{e'} y_S \le 0 \label{eq:cascaded-relaxing-sum-sum}
  \end{align}

  We now prove that $\Delta'\vec{y}$ is a \refdef{feasible-direction}, i.e., it satisfies \cref{eq:feasible-a,eq:feasible-b}.
  We first prove \cref{eq:feasible-a}: $\forall S \in \mathcal{O} \setminus \mathcal{B}, \Delta' y_S \ge 0$.
  Since $\Delta\vec{y}$ and $\{ \Delta^e\vec{y} \}$ are both \refdef{feasible-directions}, we have $\forall S \in \mathcal{O} \setminus \mathcal{B}, \Delta y_S \ge 0, \Delta^e y_S \ge 0$ given \cref{eq:feasible-a}.
  Thus, we have
  \begin{align*}
    \forall S \in \mathcal{O} \setminus \mathcal{B}, \Delta' y_S &= \Delta y_S + \sum_{e \in \cup_i \mathcal{R}'_i \cap T | \alpha_e > 0} \frac{\alpha_e}{-k_e} \Delta^{e} y_S \\
    &\ge 0 \quad\text{(every term is non-negative)}
  \end{align*}

  \noindent We then prove \cref{eq:feasible-b}: $\forall e \in T, \sum_{S \in \mathcal{O} | e \in \delta(S)} \Delta' y_S \le 0$.
  \begin{align*}
    \forall e \in T, &\sum_{S \in \mathcal{O} | e \in \delta(S)} \Delta' y_S = \sum_{S \in \mathcal{O} | e \in \delta(S)} \Delta y_S \ + \\
    &\quad\quad \sum_{S \in \mathcal{O} | e \in \delta(S)}\ \sum_{e' \in \cup_i \mathcal{R}'_i \cap T | \alpha_{e'} > 0} \frac{\alpha_{e'}}{-k_{e'}} \Delta^{e'} y_S \\
    &\,\downarrow\text{(given \cref{eq:cascaded-relaxing-sum-sum})} \\
    &\le \sum_{S \in \mathcal{O} | e \in \delta(S)} \Delta y_S\\
    &\,\downarrow\text{(given $\Delta\vec{y}$ is a \refdef{feasible-direction})} \\
    &\le 0
  \end{align*}

  Now we have proved that $\Delta'\vec{y}$ is a \refdef{feasible-direction} with $\sum \Delta'\vec{y} \ge \sum \Delta\vec{y}$.
  The theorem also states that when $\Delta\vec{y}$ is a \refdef{relaxer}, $\Delta'\vec{y}$ is also a \refdef{relaxer} that relaxes a superset of \refdef{tight-edges} $\mathcal{R}(\Delta'\vec{y}) \supseteq \mathcal{R}(\Delta\vec{y})$.
  We now prove $\Delta'\vec{y}$ is a \refdef{relaxer}, i.e., it satisfies \cref{eq:relaxer-a,eq:relaxer-b}.
  First, we prove \cref{eq:relaxer-a}.
  Since $\mathcal{R}(\Delta\vec{y}) \neq \varnothing$ given $\Delta\vec{y}$ is a \refdef{relaxer}, we then only need to prove $\forall e \in \mathcal{R}(\Delta\vec{y}), \sum_{S \in \mathcal{O} | e \in \delta(S)} \Delta' y_S < 0$.
  \begin{align*}
    \forall e \in \mathcal{R}(\Delta\vec{y}), &\sum_{S \in \mathcal{O} | e \in \delta(S)}\dqns \Delta' y_S = \sum_{S \in \mathcal{O} | e \in \delta(S)} \Delta y_S \ +  \\
    &\quad\quad \sum_{S \in \mathcal{O} | e \in \delta(S)}\ \sum_{e' \in \cup_i \mathcal{R}'_i \cap T | \alpha_{e'} > 0} \frac{\alpha_{e'}}{-k_{e'}} \Delta^{e'} y_S \\
    &\,\downarrow\text{(given \cref{eq:cascaded-relaxing-sum-sum})} \\
    &\le \sum_{S \in \mathcal{O} | e \in \delta(S)} \Delta y_S \\
    &\,\downarrow\text{(given $\Delta\vec{y}$ is a \refdef{relaxer} and $e \in \mathcal{R}(\Delta\vec{y})$)} \\
    &< 0
  \end{align*}

  \noindent We next prove \cref{eq:relaxer-b}: $\sum_{S \in \mathcal{O}} \Delta' y_S \ge 0$.
  Given that $\Delta\vec{y}$ and $\{ \Delta^e\vec{y} \}$ are \refdef{relaxers}, we have $\sum_{S \in \mathcal{O}} \Delta y_S \ge 0$ and $\sum_{S \in \mathcal{O}} \Delta^e y_S \ge 0$.
  \begin{align*}
    \sum_{S \in \mathcal{O}} \Delta' y_S &= \sum_{S \in \mathcal{O}} \Delta y_S + \sum_{e \in \cup_i \mathcal{R}'_i \cap T | \alpha_e > 0} \frac{\alpha_e}{-k_e} \sum_{S \in \mathcal{O}} \Delta^{e} y_S \\
    &\,\downarrow\text{(given $\forall e \subseteq \cup_i \mathcal{R}'_i, k_{e} < 0$)} \nonumber \\
    & \ge \sum_{S \in \mathcal{O}} \Delta y_S \\
    &\,\downarrow\text{(given $\Delta\vec{y}$ is a \refdef{relaxer})} \\
    & \ge 0
  \end{align*}

  Together, we prove that $\Delta'\vec{y}$ is a \refdef{relaxer} $R'[T]$, relaxing a superset of the \refdef{tight-edges} $\mathcal{R}(\Delta'\vec{y}) \supseteq \mathcal{R}(\Delta\vec{y})$.
\end{proof}

\subsection{Optimality Criteria of Clusters}\label{ssec:cluster-optimality-proof}

In this section, we prove the \reftheorem{cluster-optimality-criteria} (\S\ref{ssec:algo-cluster}) and introduce some lemmas about the properties of \refdef{clusters} that we use to prove the optimality of the \hyperblossom algorithm in \S\ref{ssec:hyperblossom-optimality}.

\theoremClusterOptimalityCriteria{cluster-optimality-criteria-with-proof}

\begin{proof}
  To prove the theorem, we first show that the union of all local parity factors $\mathcal{E} = \bigcup_{C \in \mathcal{C}} \mathcal{E}_C$ is a global parity factor according to \reflemma{cluster-parity-factor}.
  Also, the union of all local \refeqs{DLP} solutions $\vec{y} = \bigcup_{C \in \mathcal{C}} \{ S: y_S | \forall S \in \mathcal{B}_C \}$ is a feasible \refeqs{DLP} solution according to \reflemma{cluster-dlp-solution}.
  Thus, we only need to prove that $W(\mathcal{E}) = \sum_{S \in \mathcal{O}} y_S$ and according to \reftheorem{provable-optimum}, $\mathcal{E}$ and $\vec{y}$ are optimal \refeqs{MWPF} and \refeqs{DLP} solutions, respectively.

  According to the definition of \refdef{locally-optimal-cluster}, we have $W(\mathcal{E}_C) = \sum_{S \in \mathcal{B}_C} y_S$ for every \refdef{cluster} $C \in \mathcal{C}$.
  Thus, we have
  \begin{align*}
    W(\mathcal{E}) &= W(\bigcup_{C \in \mathcal{C}} \mathcal{E}_C) \\
    &\,\downarrow\text{(by \reflemma{cluster-non-overlapping})} \\
    &= \sum_{C \in \mathcal{C}} W(\mathcal{E}_C) \\
    &= \sum_{C \in \mathcal{C}} \sum_{S \in \mathcal{B}_C} y_S \\
    &\,\downarrow\text{(by \reflemma{cluster-non-overlapping})} \\
    &= \sum_{S \in \mathcal{B}} y_S \\
    &= \sum_{S \in \mathcal{O}} y_S
  \end{align*}
  Together, we prove that $\mathcal{E}$ and $\vec{y}$ are optimal \refeqs{MWPF} and \refeqs{DLP} solutions, respectively.

\end{proof}

\vspace{1ex}
\lemmaClusterNonOverlapping{cluster-non-overlapping}

\begin{proof}
  By definition of the \refdef{cluster}, every \refdef{tight-edge} or \refdef{hyperblossom} merges two \refdef{clusters}, thus $E_{C_1} \cap E_{C_2} = \varnothing$ and $\mathcal{B}_{C_1} \cap \mathcal{B}_{C_2} = \varnothing$ when $C_1 \neq C_2$.
  A vertex is merged into a \refdef{cluster} only when a \refdef{tight-edge} or \refdef{hyperblossom} touches it.
  Therefore, we have $V_{C_1} \cap V_{C_2} = \varnothing$.
\end{proof}

\vspace{1ex}
\lemmaClusterOfDefect{cluster-of-defect}

\begin{proof}
  In the beginning of the algorithm, a defect vertex $v \in D$ belongs to a \refdef{cluster} according to the definition of \refdef{cluster}.
  Given \reflemma{cluster-non-overlapping}, we have $V_{C_1} \cap V_{C_2} = \varnothing$ when $C_1 \neq C_2$.
  Thus, $v$ belongs to a unique \refdef{cluster}, denoted as $C = \mathcal{C}(v)$ where $v \in V_C$.
\end{proof}

\vspace{1ex}
\lemmaClusterofHyperblossom{cluster-of-hyperblossom}

\begin{proof}
  According to the definition of \refdef{cluster}, each \refdef{hyperblossom} $S \in \mathcal{B}$ merges all the \refdef{clusters} that it touches, i.e., $V_S \subseteq V_C$.
  It touches at least one \refdef{cluster} because $S$ is an \refdef{invalid} subgraph with at least one defect vertex, i.e., $D \cap V_S \neq \varnothing$.
  Given \reflemma{cluster-of-defect}, such a defect vertex $v \in D \cap V_S$ belongs to a unique \refdef{cluster} $C = \mathcal{C}(v)$.
  Thus, the \refdef{hyperblossom} $S$ belongs to a unique \refdef{cluster}, denoted as $C = \mathcal{C}(S)$.
  Also, $S \in \mathcal{B}_C$ according to the definition of the \refdef{hyperblossoms} of a \refdef{cluster}.
\end{proof}

\vspace{1ex}
\lemmaSubgraphOSubset{subgraph-O-subset}
\begin{proof}
  We prove $\mathcal{O}' \subseteq \mathcal{O}$ by proving $\forall S' \in \mathcal{O}'$, $S' \in \mathcal{O}$.

  According to the definition of \refdef{invalid} subgraphs, every $S' = (V_{S'}, E_{S'}) \in \mathcal{O}'$ satisfies the following condition:
  \begin{gather*}
    \forall \mathcal{E} \subseteq E_{S'}, \mathcal{D}(\mathcal{E}) \neq D \cap V_{S'}
  \end{gather*}

  Given $V_{S'} \subseteq V' \subseteq V$ and $E_{S'} \subseteq E'[V_{S'}] \subseteq E[V_{S'}]$, $S'$ satisfies the definition of \refdef{invalid} subgraphs for the original decoding hypergraph $G = (V, E)$, i.e., $S' \in \mathcal{O}$.

\end{proof}

\definelemma{cluster-parity-factor}{Cluster Parity Factor}
The union of all the local parity factors $\mathcal{E} = \bigcup_{C \in \mathcal{C}} \mathcal{E}_C$ is a global parity factor, i.e., $\mathcal{D}(\mathcal{E}) = D$.

\begin{proof}
  According to the definition of \refdef{locally-optimal-cluster}, $\mathcal{D}(\mathcal{E}_C) = D \cap V_C$ where $\mathcal{E}_C \subseteq E_C$.
  Also, given \reflemma{cluster-non-overlapping}, $\mathcal{E}_C \subseteq E_C$ are disjoint, so we have $\bigcup_{C \in \mathcal{C}} \mathcal{E}_C = \bigoplus_{C \in \mathcal{C}} \mathcal{E}_C$ where $\bigoplus$ is the symmetric difference operator.
  Putting together, we have
  \begin{align*}
    \mathcal{D}(\mathcal{E}) &= \mathcal{D}(\bigcup_{C \in \mathcal{C}} \mathcal{E}_C) \\
    &\,\downarrow\text{(by \reflemma{cluster-non-overlapping})} \\
    &= \mathcal{D}(\bigoplus_{C \in \mathcal{C}} \mathcal{E}_C) \\
    &= \bigoplus_{C \in \mathcal{C}} \mathcal{D}(\mathcal{E}_C) \\
    &\,\downarrow\text{(by definition of \refdef{locally-optimal-cluster})} \\
    &= \bigoplus_{C \in \mathcal{C}} (D \cap V_C) \\
    &\,\downarrow\text{(by \reflemma{cluster-non-overlapping})} \\
    &= \bigcup_{C \in \mathcal{C}} (D \cap V_C) \\
    &= D \cap \Big( \bigcup_{C \in \mathcal{C}} V_C \Big)\\
    &\,\downarrow\text{(by \reflemma{cluster-of-defect})} \\
    &= D
  \end{align*}

\end{proof}

\definelemma{cluster-dlp-solution}{Cluster DLP Solution}
The union of all the local \refeqs{DLP} solutions $\vec{y} = \bigcup_{C \in \mathcal{C}} \{ S: y_S | \forall S \in \mathcal{B}_C \}$ is a feasible \refeqs{DLP} solution, i.e., $\vec{y}$ satisfies all the \refeqs{DLP} constraints.

\begin{proof}
  We prove that $\vec{y}$ satisfies \cref{eq:dual-constraint-1} and \cref{eq:dual-constraint-2}.

  For \cref{eq:dual-constraint-1}, we need to prove that $y_S \ge 0, \forall S \in \mathcal{O}$.
  This is trivial because $y_S \ge 0, \forall S \in \mathcal{B}_C$ for every \refdef{cluster} $C \in \mathcal{C}$, and the rest of the dual variables are $y_S = 0, \forall S \in \mathcal{O} \setminus \cup_{C \in \mathcal{C}} \mathcal{B}_C$.

  For \cref{eq:dual-constraint-2}, we need to prove that $\sum_{S \in \mathcal{O}} y_S \le w_e$ for every hyperedge $e \in E$.
  For each hyperedge $e \in E$, we consider two cases: whether it belongs to $E_C$ of some \refdef{cluster} or not.

  If $\exists C \in \mathcal{C}, e \in E_C$, then according to \reflemma{cluster-non-overlapping}, none of the \refdef{hyperblossoms} from other \refdef{clusters} contribute to $e$.
  This is because $\forall C' \in \mathcal{C} \setminus \{ C \}, \forall S \in \mathcal{B}_{C'}, V_S \cap e = \varnothing$ and thus $e \notin E(V_S) \supseteq E(V_S) \setminus E_S = \delta(S)$.
  Given that the local \refeqs{DLP} solution is feasible and no other dual variables contribute to $e$, we have $\sum_{S \in \mathcal{O}} y_S \le w_e$.

  Otherwise, if $e$ does not belong to any \refdef{cluster}, then it must not be a \refdef{tight-edge}, according to the definition of \refdef{cluster}.
  In other words, $\sum_{S \in \mathcal{O}} y_S < w_e$.

  In both cases, $\vec{y}$ satisfies \cref{eq:dual-constraint-2} for every hyperedge $e \in E$.

\end{proof}

\subsection{HyperBlossom Algorithm Optimality}\label{ssec:hyperblossom-optimality}

\theoremRelaxerExistenceOrInvalidCluster{relaxer-existence-invalid-cluster-with-proof}

\begin{proof}
  This theorem is a minor extension of \reftheorem{relaxing-with-proof}.
  The only additional step required in the proof is to show that if there exists a \refdef{trivial-direction} $\Delta\vec{y} = \{ \Delta y_S: +1 \}$ for some \refdef{invalid} subgraph $S \in \mathcal{O}$, then there must also exist an \refdef{invalid} \refdef{cluster} $\mathcal{C} \cap \mathcal{O} \neq \varnothing$.
  To do so, we first construct an \refdef{invalid} subgraph $S_3$ that is the union of some \refdef{clusters} $\mathcal{C}' \subseteq \mathcal{C}$, and then prove that at least one of the \refdef{clusters} in $\mathcal{C}'$ must be \refdef{invalid}.

  Given that the \refdef{trivial-direction} is a \refdef{feasible-direction}, $S$ must not grow on any \refdef{tight-edge}, i.e., $\delta(S) \cap T = \varnothing$.
  We can construct another \refdef{invalid} subgraph $S_1 = (V_S, E_S \cap T)$ where $\Delta\vec{y'} = \{ \Delta y_{S_1}: +1 \}$ is also a \refdef{trivial-direction}.
  To see why $S_1$ is an \refdef{invalid} subgraph, we use the definition of \refdef{invalid} subgraph:
  \begin{align*}
    &\forall \mathcal{E} \subseteq E_S, \mathcal{D}(\mathcal{E}) \neq V_S \cap D \\
    \Longrightarrow &\forall \mathcal{E} \subseteq E_S \cap T \subseteq E_S, \mathcal{D}(\mathcal{E}) \neq V_S \cap D
  \end{align*}
  To see why $\Delta\vec{y'}$ is a \refdef{feasible-direction}, we show that it does not grow on any \refdef{tight-edge}:
  \begin{align*}
    \delta(S_1) \cap T &= \Big( E(V_{S_1}) \setminus E_{S_1} \Big) \cap T \\
    &= \Big( E(V_S) \setminus (E_S \cap T) \Big) \cap T \\
    &\,\downarrow\text{(given $A \setminus (B \cap C) = (A \setminus B) \cup (A \setminus C)$)} \\
    &=\Big( (E(V_S) \setminus E_S) \cup (E(V_S) \setminus T) \Big) \cap T \\
    &=\Big( (E(V_S) \setminus E_S) \cap T \Big) \cup \Big( (E(V_S) \setminus T) \cap T \Big) \\
    &\,\downarrow\text{(given $\delta(S) \cap T = (E(V_S) \setminus E_S) \cap T = \varnothing$)} \\
    &=(E(V_S) \setminus T) \cap T \\
    &= \varnothing
  \end{align*}

  We can then remove the isolated non-defect vertices from $V_{S_1}$ to get another \refdef{invalid} subgraph $$S_2 = (\{ v \in V_S | v \in D \lor v \notin \cup_{e \in E_{S_1}} e \}, E_{S_1})$$
  To see why $S_2$ is an \refdef{invalid} subgraph, we use the definition of \refdef{invalid} subgraph:
  \begin{align*}
    &\forall \mathcal{E} \subseteq E_{S_1}, \mathcal{D}(\mathcal{E}) \neq V_{S} \cap D \\
    &\dqns\downarrow\text{(given $V_S \setminus V_{S_2} \subseteq \overline{D}$ is not incident to any $e \in E_{S_1}$)} \\
    \Longrightarrow &\forall \mathcal{E} \subseteq E_{S_1}, \mathcal{D}(\mathcal{E}) \neq V_{S_2} \cap D
  \end{align*}

  We further merge $S_2$ with any \refdef{hyperblossom} $S \in \mathcal{B}$ that touches $S_2$.
  In other words, when $V_S \cap V_{S_2} \neq \varnothing$, we merge $S_2$ and the \refdef{cluster} of the \refdef{hyperblossom} $\mathcal{C}(S)$ (\reflemma{cluster-of-hyperblossom}).
  This results in a new \refdef{invalid} subgraph $S_3$.

  Now we have constructed an \refdef{invalid} subgraph $S_3$ that satisfy the properties of the \refdef{cluster}: $E_{S_3}$ consists of \refdef{tight-edges} and $V_{S_3}$ includes defect vertices and non-defect vertices connected by an edge in $E_{S_3}$.
  Note that this doesn't mean $S_3$ is a \refdef{cluster}, rather, it can be the union of multiple \refdef{clusters} $\mathcal{C}' \subseteq \mathcal{C}$.
  That is,
  \begin{align*}
    V_{S_3} &= \cup_{C \in \mathcal{C}'} V_C \\
    E_{S_3} &= \cup_{C \in \mathcal{C}'} E_C
  \end{align*}

  Now we prove that there must exists an \refdef{invalid} \refdef{cluster} within $\mathcal{C}'$, i.e., $\mathcal{C}' \cap \mathcal{O} \neq \varnothing$.
  We prove it by contradiction.
  Suppose all the \refdef{clusters} in $\mathcal{C}'$ are \refdef{valid}, we have $\forall C \in \mathcal{C}', \exists \mathcal{E}_C \subseteq E_C, \mathcal{D}(\mathcal{E}_C) = D \cap V_C$.
  According to \reflemma{cluster-parity-factor}, the union of these local parity factors is a global parity factor, i.e., $\mathcal{D}(\bigcup_{C \in \mathcal{C}'} \mathcal{E}_C) = D \cap V_{S_3}$.
  This contradicts with the fact that $S_3$ is an \refdef{invalid} subgraph, because there exists a parity factor $\mathcal{E} = \bigcup_{C \in \mathcal{C}'} \mathcal{E}_C \subseteq E_{S_3}, \mathcal{D}(\mathcal{E}) = D \cap V_{S_3}$.
  Thus, there must exists an \refdef{invalid} \refdef{cluster} with $\mathcal{C}'$.

  Together, we prove that there exists an \refdef{invalid} \refdef{cluster} $C \in \mathcal{C}' \cap \mathcal{O}$ when there exists a \refdef{trivial-direction}.

\end{proof}

\theoremHyperBlossomOptimality{hyperblossom-algorithm-optimality-with-proof}

\begin{proof}

  According to the \reftheorem{cluster-optimality-criteria}, once all \refdef{clusters} are \refdef{locally-optimal-clusters}, the \hyperblossom algorithm finds the optimal \refeqs{MWPF} and \refeqs{DLP} solutions.
  Thus, we only need to prove that the \hyperblossom algorithm terminates only when all \refdef{clusters} are \refdef{locally-optimal-clusters}.

  We first prove that when the \refdef{cluster} is \refdef{invalid} or suboptimal, i.e., $\sum_{S \in \mathcal{B}_C} y_S < W(\mathcal{E}_C)$ where $\mathcal{E}_C$ is the \refeqs{MWPF} of $(V_C, E_C)$, the \hyperblossom algorithm finds a \refdef{useful-direction} such that the local dual objective $\sum_{S \in \mathcal{B}_C} y_S$ grows by a positive length $l > 0$.
  We then prove that the \hyperblossom algorithm terminates within a finite number of iterations.

  When the \refdef{cluster} is \refdef{invalid}, we prove that the \hyperblossom algorithm finds a \refdef{useful-direction}.
  At line \ref{line:trivial-direction-hyperblossom} of \cref{algo:hyperblossom-primal-phase}, it checks whether $(V_C, E'_C) \in \mathcal{O}$.
  Given that $E'_C \subseteq E_C$ and that $C$ is an \refdef{invalid} subgraph, we have
  \begin{align*}
    &\forall \mathcal{E} \subseteq E_C, \mathcal{D}(\mathcal{E}) \neq D \cap V_C \\
    \Longrightarrow &\forall \mathcal{E} \subseteq E'_C \subseteq E_C, \mathcal{D}(\mathcal{E}) \neq D \cap V_C
  \end{align*}
  Thus, the Primal phase finds a \refdef{trivial-direction} $\Delta\vec{y}[E'_C]$ under the assumption of \refdef{tight-edges} $E'_C$ (line \ref{line:trivial-direction-hyperblossom-value}).
  A \refdef{trivial-direction}, by definition, have $\sum \Delta\vec{y} > 0$.
  According to \reftheorem{cascaded-relaxing}, the composed \refdef{direction} at line \ref{line:compose-trivial-direction-hyperblossom} must be a \refdef{feasible-direction} with $\sum \Delta'\vec{y} \ge \sum \Delta\vec{y} > 0$.
  Thus, the \hyperblossom algorithm finds a \refdef{useful-direction} $\Delta'\vec{y}$.

  When the \refdef{cluster} is \refdef{valid} but suboptimal, we prove that the \hyperblossom algorithm also finds a \refdef{useful-direction}.
  Given the inequality chain \cref{eq:mwpf-chain} and \ref{condition:mwpf} for any subgraph of $G$, there exists an optimal \refeqs{DLP} solution $\vec{y^o}$ of the hypergraph $C = (V_C, E_C)$ that has $\sum_{S \in \mathcal{O}_C} y^o_S = W(\mathcal{E}_C)$ where $\mathcal{O}_C$ is the set of \refdef{invalid} subgraphs of the subgraph $C$.
  Note that the \refeqs{DLP} constraints are defined on $\mathcal{O}_C$ of the subgraph $C = (V_C, E_C)$ and thus $\vec{y^o}$ is not necessarily a feasible \refeqs{DLP} solution for the original graph $G$.
  Nonetheless, given \reflemma{subgraph-O-subset}, the dual variables still exist in the original problem $\mathcal{O}_C \subseteq \mathcal{O}$ and $\vec{y^o}$ is a (\emph{Feasible} or not) \refeqs{DLP} solution of the original graph $G$.
  To prove that $\Delta\vec{y} = \vec{y^o} - \vec{y}$ is a \refdef{useful-direction}, we only need to prove $\sum \Delta\vec{y} > 0$ and that it satisfies the conditions of \refdef{feasible-direction} \cref{eq:feasible-a,eq:feasible-b}.

  First, to prove $\sum \Delta\vec{y} > 0$, we have $$\sum \Delta\vec{y} = \sum \vec{y^o} - \sum \vec{y} = W(\mathcal{E}_C) - \sum \vec{y} > 0$$

  Second, we prove that $\Delta\vec{y}$ satisfies \cref{eq:feasible-a}: $\forall S \in \mathcal{O} \setminus \mathcal{B}, \Delta y_S \ge 0$.
  Both $\vec{y^o}$ and $\vec{y}$ only have non-zero values within $\mathcal{O}_C$.
  Thus, the \refdef{hyperblossoms} $\mathcal{B} \subseteq \mathcal{O}_C$.
  For those $S \in \mathcal{O} \setminus \mathcal{O}_C$, we have $\Delta y_S = y^o_S - y_S = 0$.
  For those $S \in \mathcal{O}_C \setminus \mathcal{B}$, we have $y_S = 0$ and thus $\Delta y_S = y^o_S - y_S = y^o_S \ge 0$ given $\vec{y^o}$ is a feasible \refeqs{DLP} solution of the subgraph $C = (V_C, E_C)$.
  Thus, $\Delta\vec{y}$ satisfies \cref{eq:feasible-a}.

  Third, we prove that $\Delta\vec{y}$ satisfies \cref{eq:feasible-b}: $\forall e \in T, \sum_{S \in \mathcal{O} | e \in \delta(S)} \Delta y_S \le 0$.
  For any \refdef{tight-edge} $e \in E_C$, by definition we have $\sum_{S \in \mathcal{O} | e \in \delta(S)} y_S = w_e$.
  Given that $\vec{y^o}$ is a feasible \refeqs{DLP} solution of the subgraph $C = (V_C, E_C)$, $\sum_{S \in \mathcal{O} | e \in \delta(S)} y_S \le w_e$.
  Together, we have $$\sum_{S \in \mathcal{O} | e \in \delta(S)} \Delta y_S = \sum_{S \in \mathcal{O} | e \in \delta(S)} y^o_S - w_e \le 0$$
  For those \refdef{tight-edges} $e \in T \setminus E_C$, neither $\vec{y^o}$ nor $\vec{y}$ contribute to the \refeqs{DLP} constraints of $e$ because $\forall S \in \mathcal{O}_C, \delta(S) \cap T \subseteq E(V_C) \cap T = E_C$.
  In this case, we have $\sum_{S \in \mathcal{O} | e \in \delta(S)} \Delta y_S = 0$.
  Thus, in both cases, we have $\Delta\vec{y}$ satisfies \cref{eq:feasible-b}.

  We now have finished the proof that when the \refdef{cluster} is \refdef{invalid} or suboptimal, the \hyperblossom algorithm finds a \refdef{useful-direction}.

  To prove the termination of the \hyperblossom algorithm, we prove that $\mathcal{B}^H_C$ is monotonically increasing for each iteration and $|\mathcal{B}^H_C| \le |\mathcal{O}|$.
  Each \refdef{cluster} maintains all the previously known \refdef{hyperblossoms} as $\mathcal{B}^H_C$ (line \ref{line:update-history-hyperblossom} of \cref{algo:hyperblossom-dual-phase}).
  Given that the \hyperblossom algorithm optimizes the partial \refeqs{DLP} solution (line \ref{line:partial-dlp-hyperblossom} of \cref{algo:hyperblossom-dual-phase}), any further progression of the dual objective must introduce new \refdef{hyperblossom} $S \in \mathcal{O} \setminus \mathcal{B}^H_C$.
  Thus, $|\mathcal{B}^H_C|$ is monotonically increasing for each iteration.
  Given that $|\mathcal{B}^H_C| \le |\mathcal{O}|$, the number of iterations is at most $|\mathcal{O}|$, finite but exponentially large.

  We have proved that the \hyperblossom algorithm terminates within a finite number of iterations, and it will not terminate until all the \refdef{clusters} become \refdef{locally-optimal-clusters}.
  Thus, the \hyperblossom algorithm is optimal under the condition that the decoding hypergraph and all its subgraphs satisfy \ref{condition:mwpf}.

\end{proof}

\subsection{Why $E_S$ is Necessary in the Definition of Invalid Subgraphs $S = (V_S, E_S)$}\label{ssec:why-need-ES}

In the \refdef{dual-mapping} for simple graphs (\S\ref{ssec:interoperability-mwpm}), we fix $E_S = E[V_S]$ for all the mapped \refdef{invalid} subgraphs, yet still achieves \ref{condition:mwpf}.
It raises the question of why not eliminating $E_S$ from the definition of \refdef{invalid} subgraphs and only use $V_S$ to define \refdef{invalid} subgraphs while fixing $E_S = E[V_S]$?
Although this works for simple graphs, we show that $E_S$ is necessary to ensure \ref{condition:mwpf} for even the simplest classes of nullity$_{=0}$ hypergraphs.
For general hypergraphs, adding $E_S$ helps reducing the gap between the minimum \refeqs{LP} and \refeqs{ILP} objective values.

\begin{figure}
  \centering
  \includegraphics[width=0.5\linewidth]{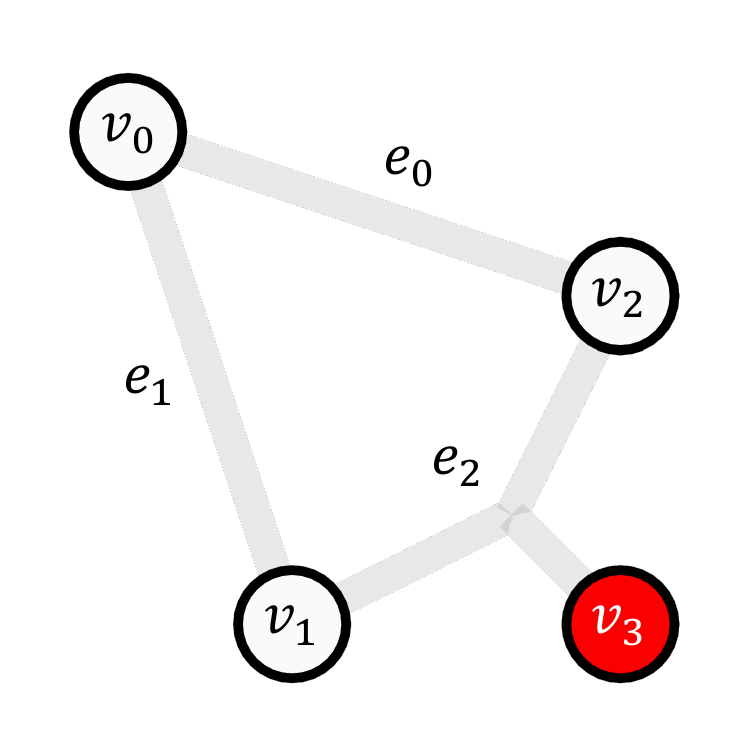}
  \caption{A nullity$_{=0}$ hypergraph. All the edge weights are 1. The syndrome is $D = \{v_3\}$.}
  \label{fig:why-need-ES}
\end{figure}

For the hypergraph in \cref{fig:why-need-ES}, which is a nullity$_{=0}$ hypergraph, we can prove that if we do not use $E_S$ when we define \refdef{invalid} subgraphs, i.e., fixing $E_S = E[V_S]$, then $\min\text{\refeqs{LP}} < \min\text{\refeqs{ILP}}$.

For this small hypergraph, we can manually list all the \refdef{invalid} subgraphs.
An \refdef{invalid} subgraph must contain at least one defect vertex, so every \refdef{invalid} subgraph must contain $v_3$.
Thus, we have 7 \refdef{invalid} subgraphs assuming $E_{S_i} = E[V_{S_i}]$:

\begin{itemize}
  \item $V_{S_1} = \{v_3\}, \delta(S_1) = \{e_2\}$
  \item $V_{S_2} = \{v_3, v_0\}, \delta(S_2) = \{ e_0, e_1, e_2 \}$
  \item $V_{S_3} = \{v_3, v_1\}, \delta(S_3) = \{ e_1, e_2 \}$
  \item $V_{S_4} = \{v_3, v_2\}, \delta(S_4) = \{ e_0, e_2 \}$
  \item $V_{S_5} = \{v_3, v_0, v_1\}, \delta(S_5) = \{ e_0, e_2 \}$
  \item $V_{S_6} = \{v_3, v_0, v_2\}, \delta(S_6) = \{ e_1, e_2 \}$
  \item $V_{S_7} = \{v_3, v_1, v_2\}, \delta(S_7) = \{ e_0, e_1 \}$
\end{itemize}

Since the only parity factor solution is $\mathcal{E} = \{ e_0, e_1, e_2 \}$ with $W(\mathcal{E}) = 3$, one can easily verify that there is no feasible modified-\refeqs{DLP} solution so that $\sum_{S \in \mathcal{O}} y_S = 3$.
Indeed, the optimal modified-\refeqs{DLP} solution is $y_{S_1} = y_{S_7} = 1$ with $\sum_{S \in \mathcal{O}} y_S = 2$.
Thus, removing $E_S$ from the definition of \refdef{invalid} subgraphs leads to a gap between the minimum \refeqs{LP} and \refeqs{ILP} objective values.

We can easily see that adding $E_S$ back will remove the gap between the minimum \refeqs{LP} and \refeqs{ILP} objective values, given an optimal \refeqs{DLP} solution:
\begin{itemize}
  \item $S_1 = (V, E \setminus \{ e_0 \}), \delta(S_1) = \{e_0\}, y_{S_1} = 1$
  \item $S_2 = (V, E \setminus \{ e_1 \}), \delta(S_2) = \{e_1\}, y_{S_2} = 1$
  \item $S_3 = (V, E \setminus \{ e_2 \}), \delta(S_3) = \{e_2\}, y_{S_3} = 1$
\end{itemize}

\section{Theoretical Analysis of \emph{SingleHair}}\label{sec:analytic}

We theoretically analyze the \emph{SingleHair} \refdef{relaxer}-finding algorithm (\S\ref{ssec:single-hair-subroutine}) with two parts.
The first part proves two lemmas about why the \emph{SingleHair} works (\S\ref{ssec:hair-matrix-odd-row-existence}) and when it stops finding \refdef{relaxers} (\S\ref{ssec:hair-matrix-unique-row}).
The second part analyzes the limitations of the \emph{SingleHair} on hypergraphs (\S\ref{ssec:failure-hypergraph}) and even simple graphs (\S\ref{ssec:failure-simple-graph}).
We also prove its optimality on nullity$_{\le 1}$ hypergraphs (\S\ref{ssec:optimal-1-dof}).

\subsection{Hair Matrix Odd Row Existence}\label{ssec:hair-matrix-odd-row-existence}

\lemmaHairMatrixOddRowExistence{hair-matrix-odd-row-existence-with-proof}
\begin{proof}
  We prove the lemma in two cases: the \refdef{hyperblossom-matrix} $\mathcal{M}_S$ has a feasible solution or not.

  First, if $\mathcal{M}_S$ does not have any feasible solution, the last row of the \refdef{hyperblossom-matrix} $\mathcal{M}_S$ must have $1$ on the last column and $0$ on all other columns because $\mathcal{M}_S$ is in the reduced row echelon form.
  Since this row does not have a pivot, it must be part of the \refdef{hair-matrix} $\mathcal{H}_S$.
  Thus, the last row of the \refdef{hair-matrix} must be an Odd row given that the last column of this row is $1$.

  Otherwise, $\mathcal{M}_S$ has at least one feasible parity factor solution.
  We prove the lemma by contradiction.
  That is, if all the constraints in the \refdef{hair-matrix} $\mathcal{H}_S$ are even parity constraints, then we prove that it contradicts with the fact that $S$ is an \refdef{invalid} subgraph.
  Since all the constraints in $\mathcal{H}_S$ are even parity constraints, $x_e = 0, \forall e \in E_C \cap \delta(S)$ corresponds to one feasible solution of $\mathcal{M}_S$.
  That is, there exists a feasible solution $\mathcal{E} \subseteq E_C \setminus \delta(S)$ that satisfies $\mathcal{D}(\mathcal{E}) = D \cap V_C$ but not using any of the edges in $\delta(S)$, i.e., $\mathcal{E} \cap \delta(S) = \varnothing$.
  We now prove that there exists $\mathcal{E}_S \subseteq E_S$ so that $\mathcal{D}(\mathcal{E}_S) = D \cap V_S$, i.e., it violates the assumption that $S$ is an \refdef{invalid} subgraph.

  We explicitly construct $\mathcal{E}_S = \mathcal{E} \cap E_S$.
  We then use the definition of \refdef{cluster} and \refdef{invalid} subgraph to prove that $\mathcal{D}(\mathcal{E}_S) = D \cap V_S$.
  \begin{align*}
    \mathcal{D}(\mathcal{E}_S) &= \mathcal{D}(\mathcal{E} \cap E_S) \\
    &\,\downarrow(\text{by definition of \refdef{defect-error-pattern}}) \\
    &= \Big\{ v \in V \Big|\ |\{ e \in \mathcal{E} \cap E_S | v \in e \}| = 1\ \text{mod}\ 2 \Big\} \\
    &\,\downarrow(\text{given $\forall e \in E_S \subseteq E[V_S], e \subseteq V_S$}) \\
    &= \Big\{ v \in V_S \Big|\ |\{ e \in \mathcal{E} \cap E_S | v \in e \}| = 1\ \text{mod}\ 2 \Big\} \\
    &\,\downarrow(\text{given $E(V_S) = E_S \cup \delta(S)$ and $\mathcal{E} \cap \delta(S) = \varnothing$}) \\
    &= \Big\{ v \in V_S \Big|\ |\{ e \in \mathcal{E} \cap E(V_S) | v \in e \}| = 1\ \text{mod}\ 2 \Big\} \\
    &\,\downarrow(\text{given $E(V_S)$ includes all edges incident to $V_S$}) \\
    &= \Big\{ v \in V_S \Big|\ |\{ e \in \mathcal{E} | v \in e \}| = 1\ \text{mod}\ 2 \Big\} \\
    &= \mathcal{D}(\mathcal{E}) \cap V_S \\
    &= (D \cap V_C) \cap V_S \\
    &\,\downarrow(\text{given $V_S \subseteq V_C$ according to \refdef{cluster}}) \\
    &= D \cap V_S
  \end{align*}
  Clearly, we have $\mathcal{E}_S = \mathcal{E} \cap E_S \subseteq E_S$.
  Thus, there exists $\mathcal{E}_S \subseteq E_S$ with $\mathcal{D}(\mathcal{E}_S) = D \cap V_S$, which contradicts with the fact that $S$ is an \refdef{invalid} subgraph.
  Therefore, $\mathcal{H}_S$ must contains at least one Odd row.

  In both cases, we have proved that the \refdef{hair-matrix} $\mathcal{H}_S$ must contains at least one Odd row.

\end{proof}

\subsection{Hair Matrix Stops at Unique Row}\label{ssec:hair-matrix-unique-row}

From the \emph{SingleHair} \refdef{relaxer}-finding algorithm (\S\ref{ssec:single-hair-subroutine}), we show that whenever $E^- \neq \varnothing$ for any \refdef{hyperblossom} $S \in \mathcal{B}$, \emph{SingleHair} will find a \refdef{relaxer}.
We prove the following lemma that shows the \emph{SingleHair} \refdef{relaxer} finder stops finding a \refdef{relaxer} only when all the \refdef{hair-matrices} consists of a single row.
Together with \reflemma{hair-matrix-odd-row-existence-with-proof}, the only row must be an Odd row.
Furthermore, the Odd row must consists of all 1s because an element of 0 means the corresponding hyperedge is totally unconstrained, which is impossible since a hyperedge connects to at least one vertex.

\vspace{1ex}
\lemmaUniqueRowIfNotARelaxer{unique-row-if-not-a-relaxer-with-proof}

\begin{proof}
  We use the property of the reduced row echelon form to prove the lemma.

  First of all, \emph{SingleHair} only constructs \refdef{hair-matrices} when the \refdef{cluster} is \refdef{valid}.
  Thus, there is no Odd row that has a 1 on the last column but 0 on all other columns.
  Such a column causes the \refdef{cluster} to be \refdef{invalid} because no parity factor solution will ever satisfy the odd parity constraint.
  Since a row of 0s on all the columns is not considered a row in a reduced row echelon form, all the rows in the \refdef{hair-matrix} contains at least one 1 for the variable columns.

  We then prove that the \refdef{hair-matrix} must have a single row.
  We use contradiction to prove the lemma, i.e., assuming there are multiple rows and then prove $E^- \neq \varnothing$, which contradicts with the assumption that $E^- = \varnothing$.

  In the reduced row echelon form, each variable column must be either a pivot column or a free column.
  Each pivot column corresponds to a unique row and each row corresponds to a unique pivot column.
  If there exists more than one row in the \refdef{hair-matrix}, then there must be more than one pivot column.
  Suppose the pivot columns are $E_p \subseteq T \cap \delta(S)$ where $|E_p| \ge 2$.

  Given \reflemma{hair-matrix-odd-row-existence-with-proof}, there exists an Odd row in the \refdef{hair-matrix} $\mathcal{H}_S$.
  Suppose the pivot column of this Odd row is $e_p \in E_p$.

  We then prove that $E^- \neq \varnothing$.
  A pivot column must be 1 on its corresponding row and 0 on all other rows.
  Thus, $\forall e \in E_p \setminus \{ e_p \}$, the column corresponding to $e$ must be 0 on the Odd row.
  According to the definition of \refdef{hair-matrix}, the columns of $\mathcal{H}_S$ are $\delta(S) \cap T$.
  Given the definition of $E^- = \delta(S) \cap T \setminus E^+$ where $E^+$ are the columns whose value on the Odd row is 1, we can see that $E^-$ are the columns of the \refdef{hair-matrix} $\mathcal{H}_S$ whose value on the Odd row is 0.
  Thus, we have $E_p \setminus \{ e_p \} \subseteq E^-$.
  Given $|E_p| \ge 2$, we have $|E^-| \ge |E_p| - 1 \ge 1$, i.e.,$E^- \neq \varnothing$.

  This contradicts with the assumption that $E^- = \varnothing$.
  Therefore, a \refdef{hair-matrix} must consist of a single row when $E^- = \varnothing$.

\end{proof}

\subsection{\emph{SingleHair} is Suboptimal on Hypergraphs}\label{ssec:failure-hypergraph}

\begin{figure}
  \centering
  \begin{subfigure}[t]{0.32\linewidth}
    \centering
    \includegraphics[width=\textwidth]{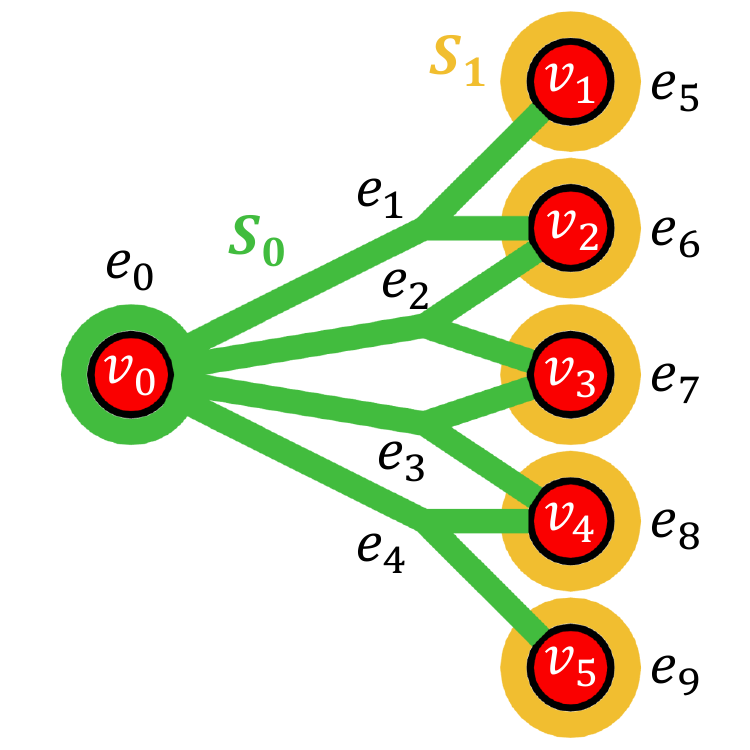}
    \caption{Suboptimal $\vec{y}$}
    \label{fig:hypergraph-failure-split-0}
  \end{subfigure}
  \begin{subfigure}[t]{0.64\linewidth}
    \centering
    \includegraphics[width=\textwidth]{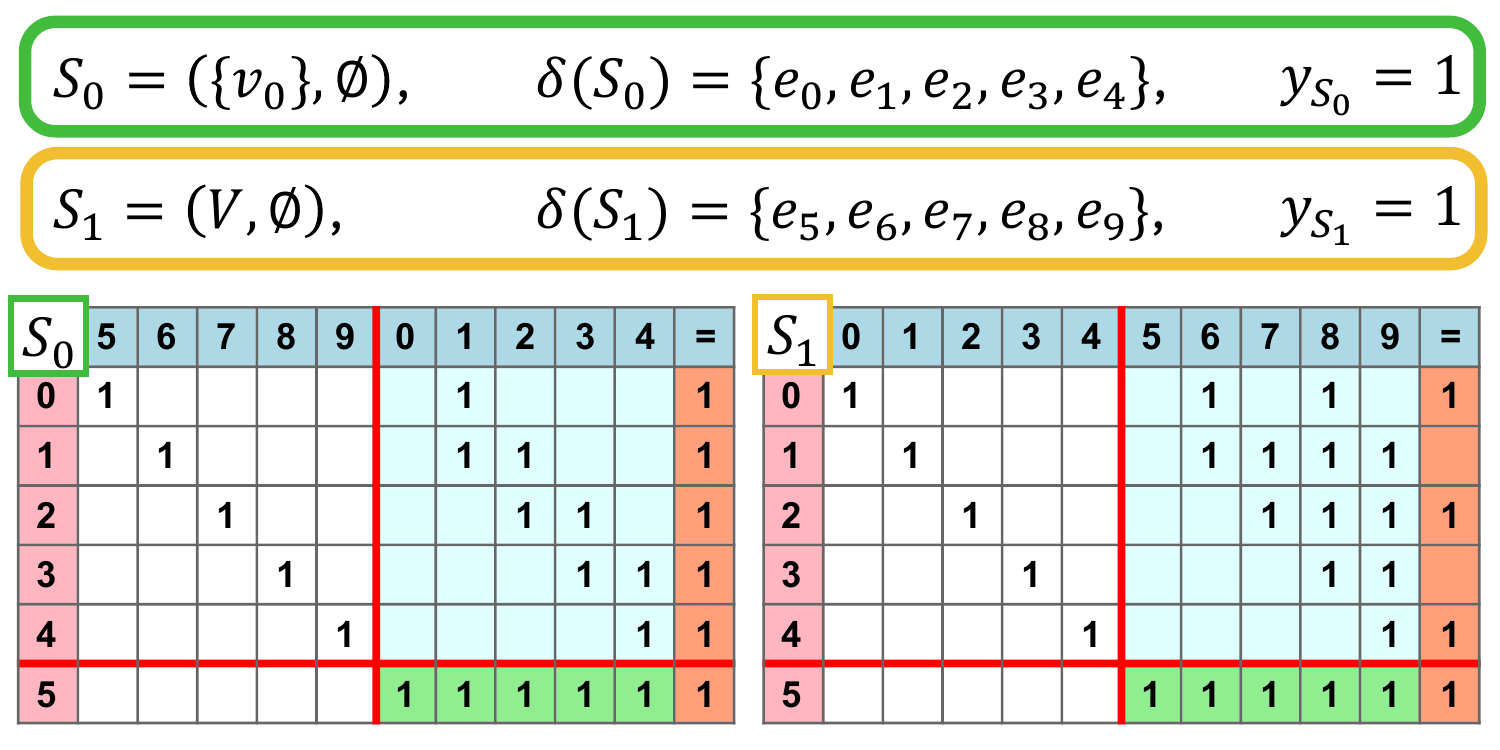}
    \caption{\refdef{hyperblossoms} and \refdef{hair-matrices}}
    \label{fig:hypergraph-failure-split-1}
  \end{subfigure}
  \begin{subfigure}[t]{0.32\linewidth}
    \centering
    \includegraphics[width=\textwidth]{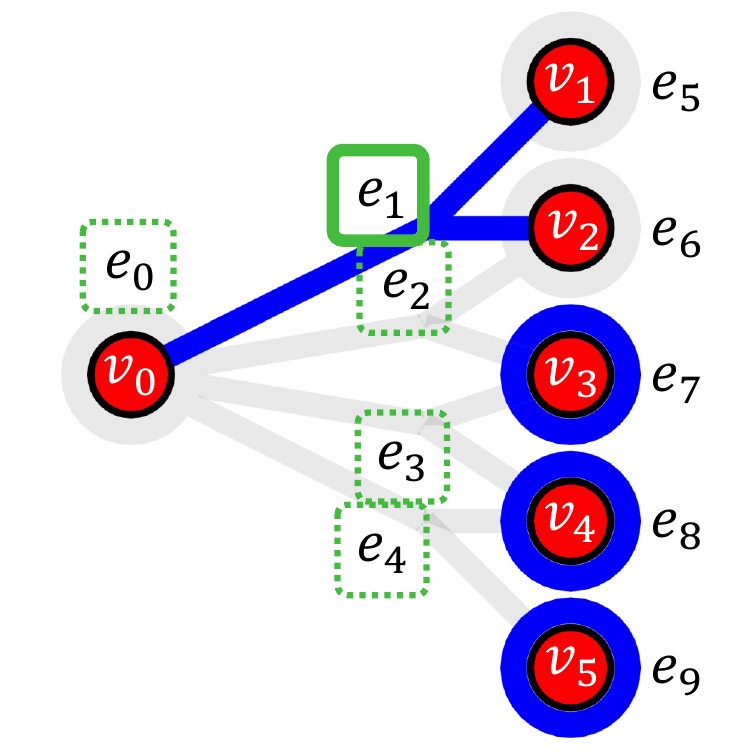}
    \caption{MWPF $\mathcal{E}_{S_0,e_1}$}
    \label{fig:hypergraph-failure-pf1}
  \end{subfigure}
  \begin{subfigure}[t]{0.32\linewidth}
    \centering
    \includegraphics[width=\textwidth]{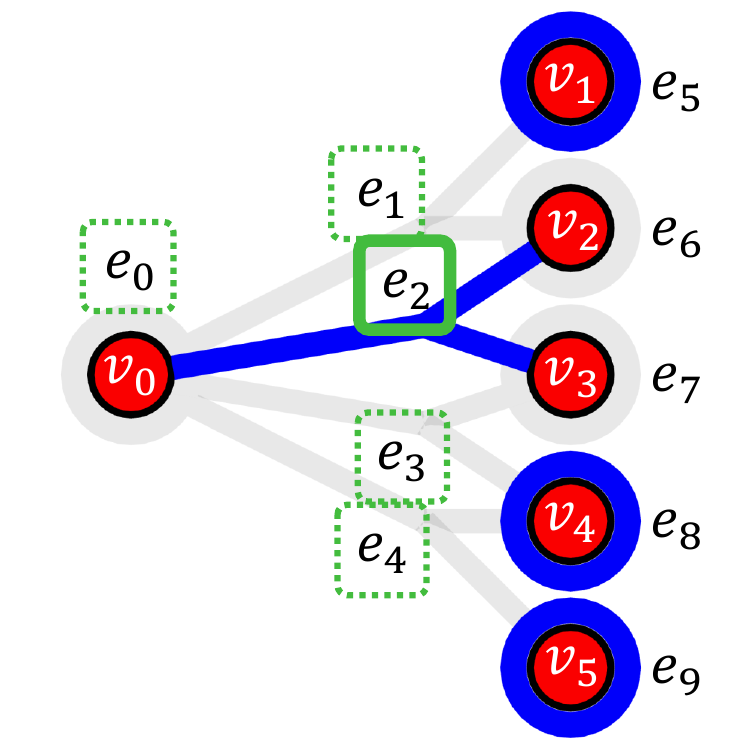}
    \caption{MWPF $\mathcal{E}_{S_0,e_2}$}
    \label{fig:hypergraph-failure-pf2}
  \end{subfigure}
  \begin{subfigure}[t]{0.32\linewidth}
    \centering
    \includegraphics[width=\textwidth]{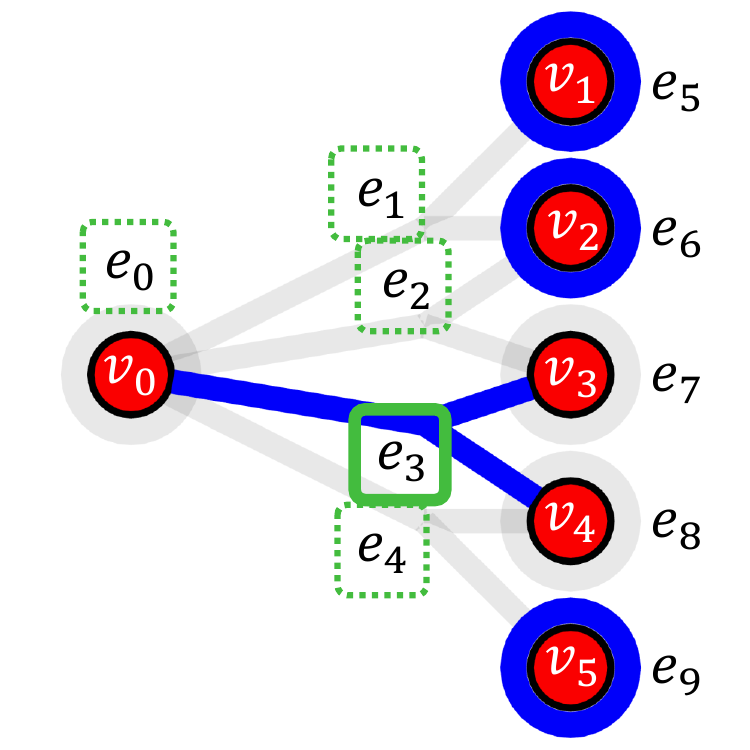}
    \caption{MWPF $\mathcal{E}_{S_0,e_3}$}
    \label{fig:hypergraph-failure-pf3}
  \end{subfigure}
  \begin{subfigure}[t]{0.32\linewidth}
    \centering
    \includegraphics[width=\textwidth]{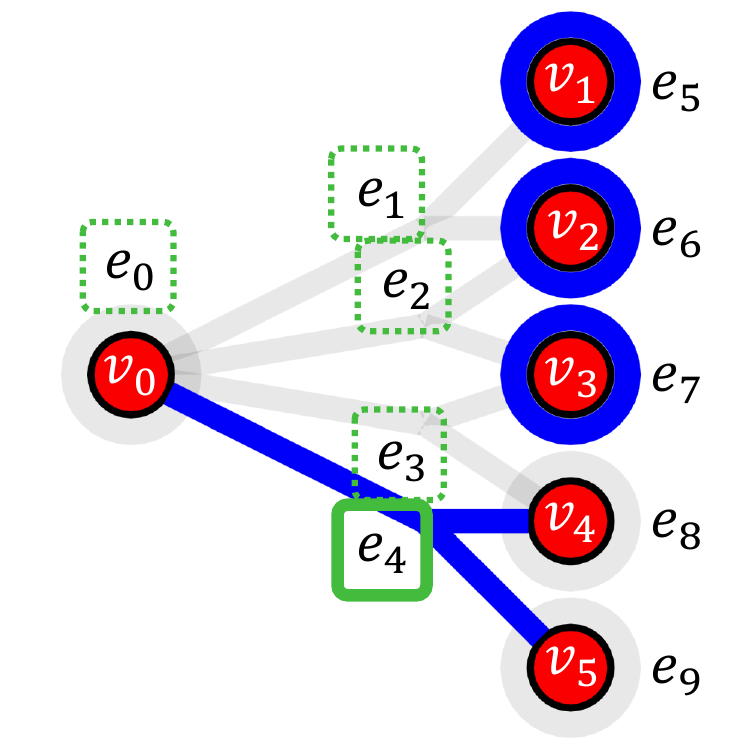}
    \caption{MWPF $\mathcal{E}_{S_0,e_4}$}
    \label{fig:hypergraph-failure-pf4}
  \end{subfigure}
  \begin{subfigure}[t]{0.32\linewidth}
    \centering
    \includegraphics[width=\textwidth]{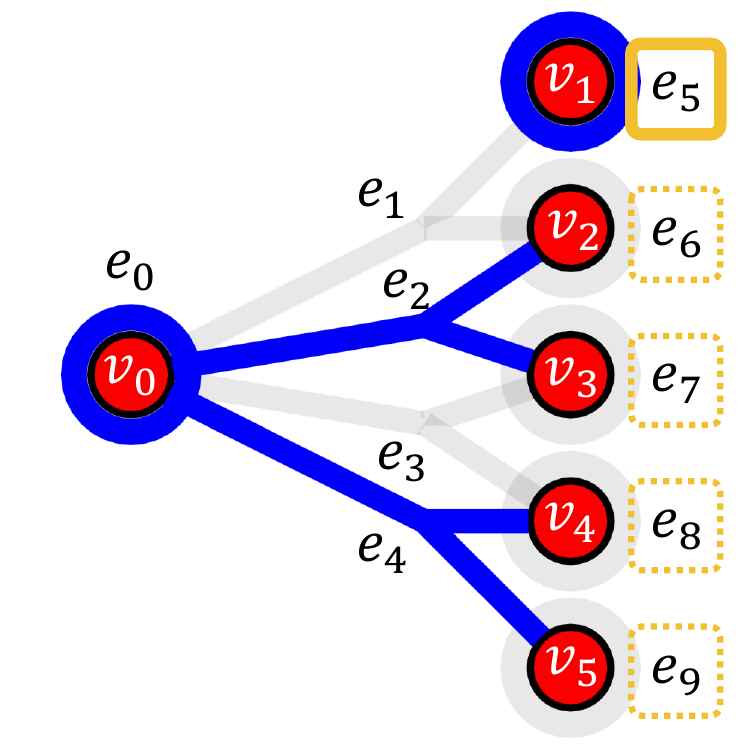}
    \caption{MWPF $\mathcal{E}_{S_1,e_5}$}
    \label{fig:hypergraph-failure-pf5}
  \end{subfigure}
  \begin{subfigure}[t]{0.32\linewidth}
    \centering
    \includegraphics[width=\textwidth]{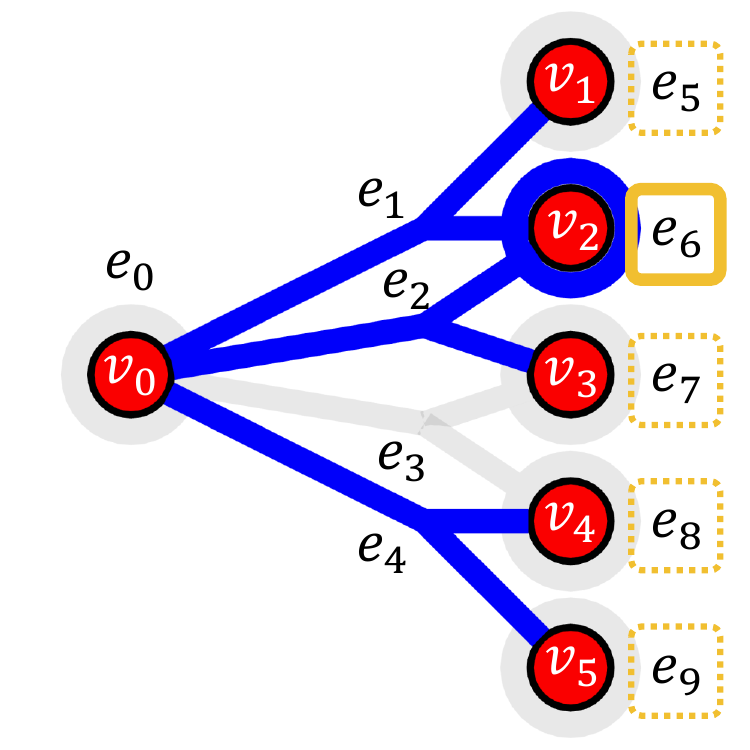}
    \caption{MWPF $\mathcal{E}_{S_1,e_6}$}
    \label{fig:hypergraph-failure-pf6}
  \end{subfigure}
  \begin{subfigure}[t]{0.32\linewidth}
    \centering
    \includegraphics[width=\textwidth]{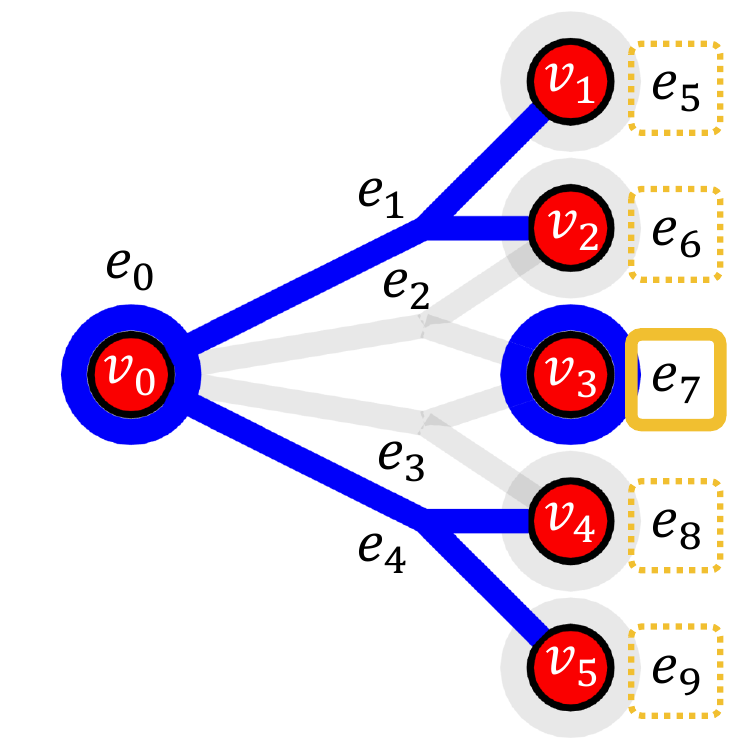}
    \caption{MWPF $\mathcal{E}_{S_1,e_7}$}
    \label{fig:hypergraph-failure-pf7}
  \end{subfigure}
  \begin{subfigure}[t]{0.32\linewidth}
    \centering
    \includegraphics[width=\textwidth]{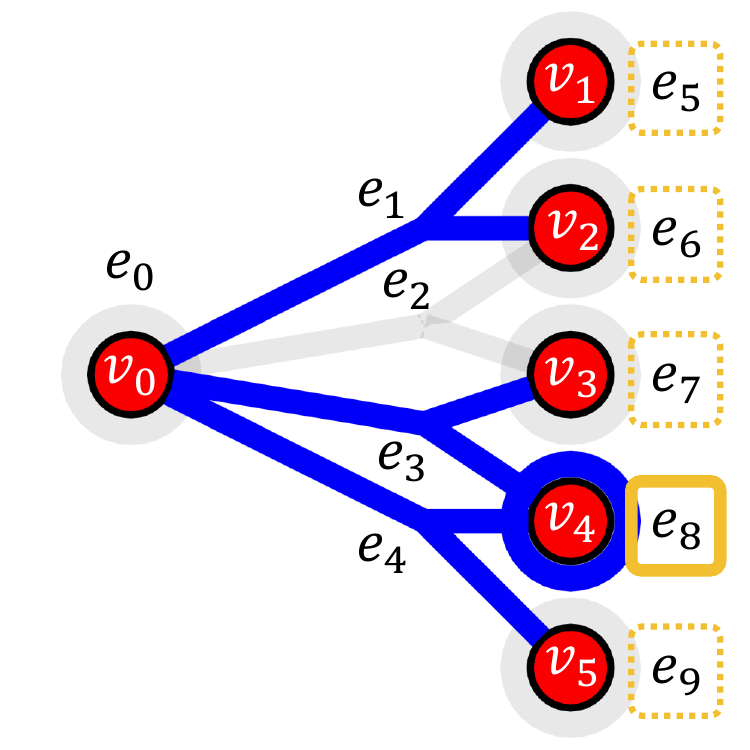}
    \caption{MWPF $\mathcal{E}_{S_1,e_8}$}
    \label{fig:hypergraph-failure-pf8}
  \end{subfigure}
  \begin{subfigure}[t]{0.32\linewidth}
    \centering
    \includegraphics[width=\textwidth]{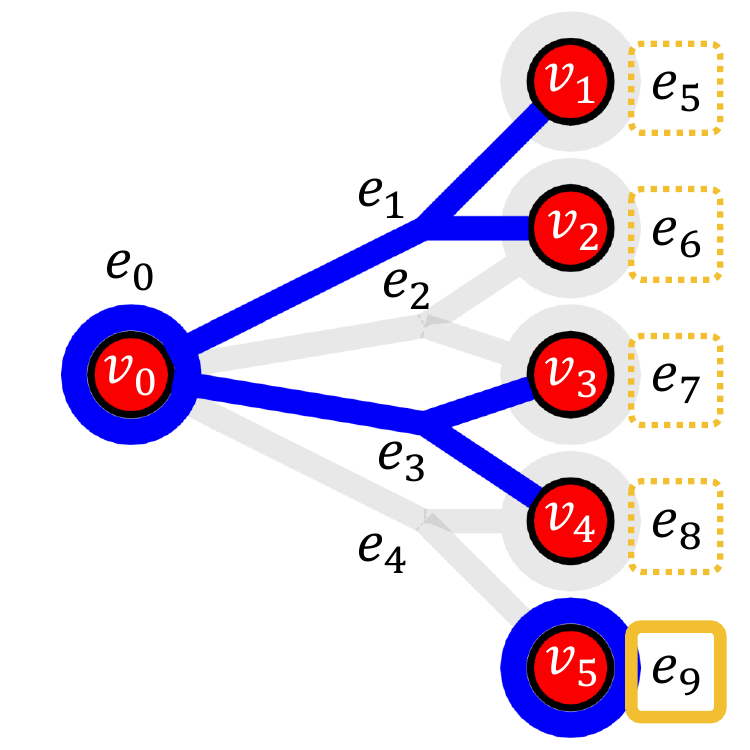}
    \caption{MWPF $\mathcal{E}_{S_1,e_9}$}
    \label{fig:hypergraph-failure-pf9}
  \end{subfigure}

  \vspace{2ex}
  \begin{subfigure}[t]{0.64\linewidth}
    \centering
    \includegraphics[width=\textwidth]{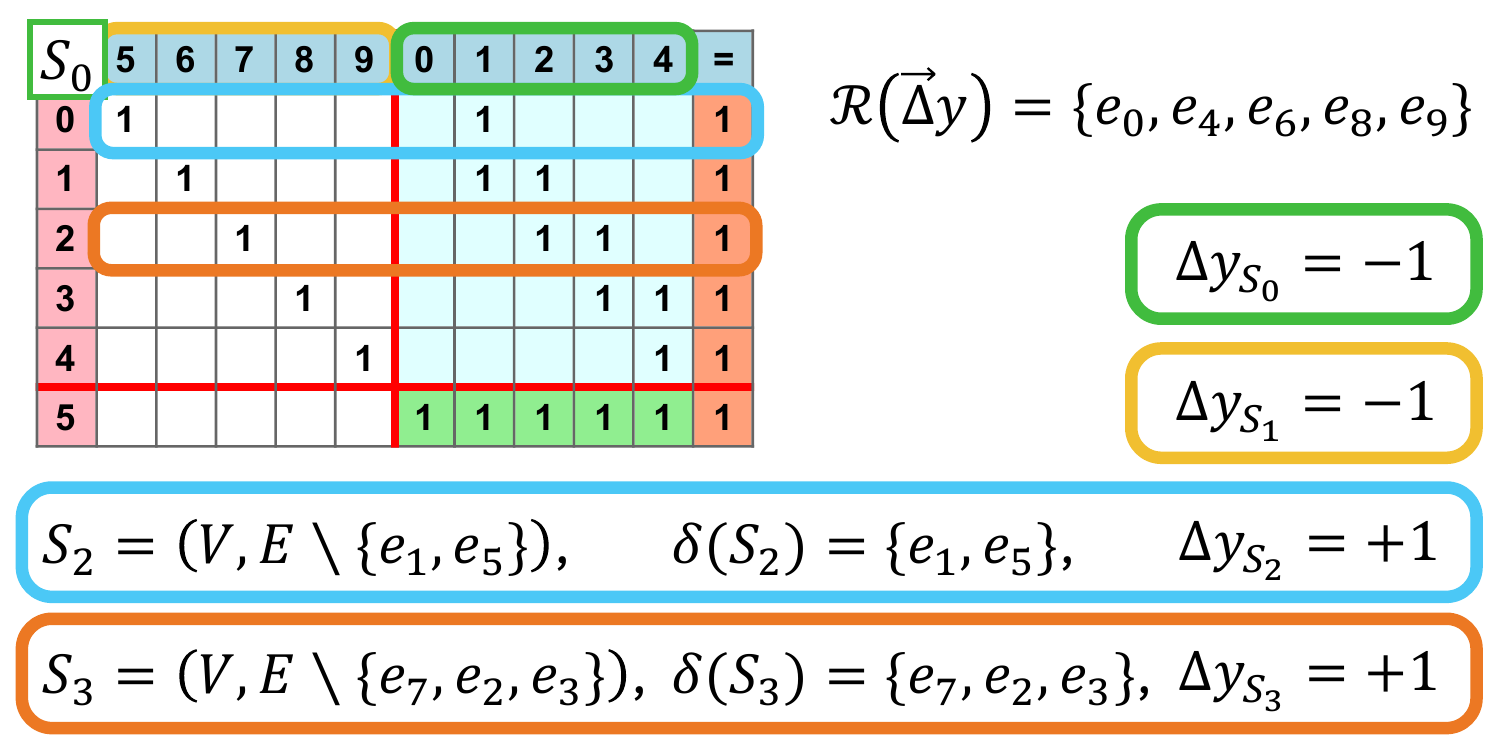}
    \caption{Finding \refdef{relaxer} $\Delta\vec{y}$ from $\mathcal{M}_{S_0}$}
    \label{fig:hypergraph-failure-fix-split-0}
  \end{subfigure}
  \begin{subfigure}[t]{0.32\linewidth}
    \centering
    \includegraphics[width=\textwidth]{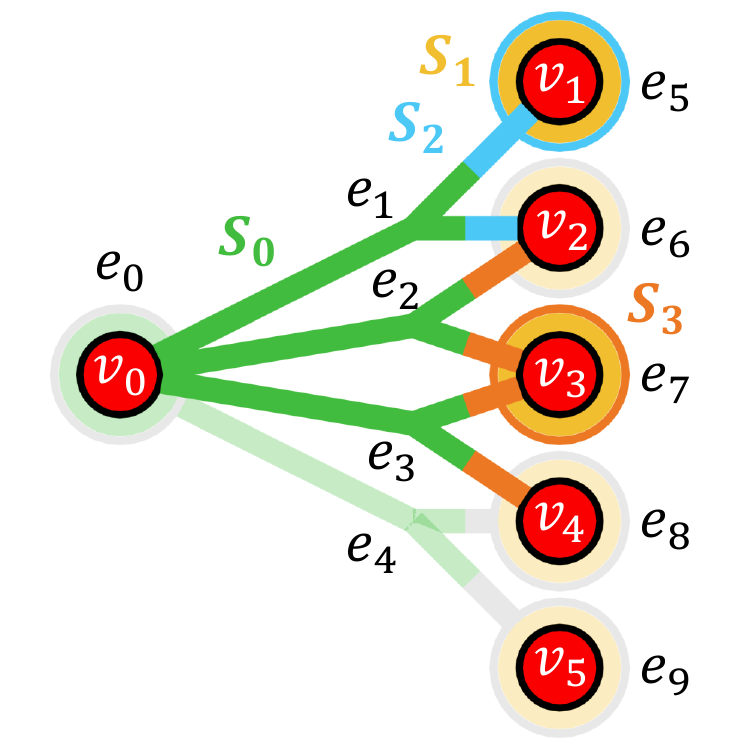}
    \caption{Relax $\mathcal{R}(\Delta\vec{y})$}
    \label{fig:hypergraph-failure-fix-split-1}
  \end{subfigure}
  \begin{subfigure}[t]{0.32\linewidth}
    \centering
    \includegraphics[width=\textwidth]{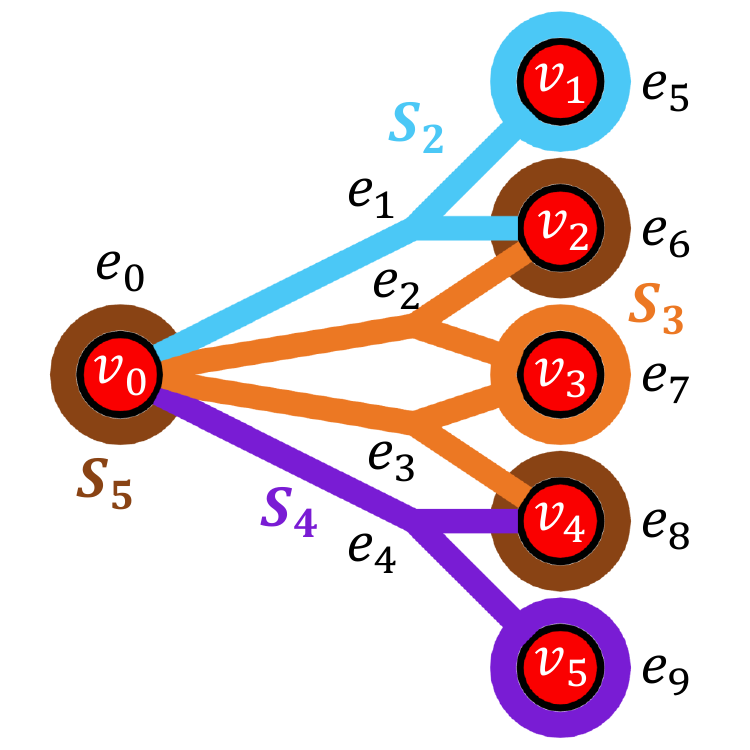}
    \caption{Optimal \refeqs{DLP} $\vec{y}$\dqns}
    \label{fig:hypergraph-failure-fix-split-2}
  \end{subfigure}
  \begin{subfigure}[t]{0.64\linewidth}
    \centering
    \includegraphics[width=\textwidth]{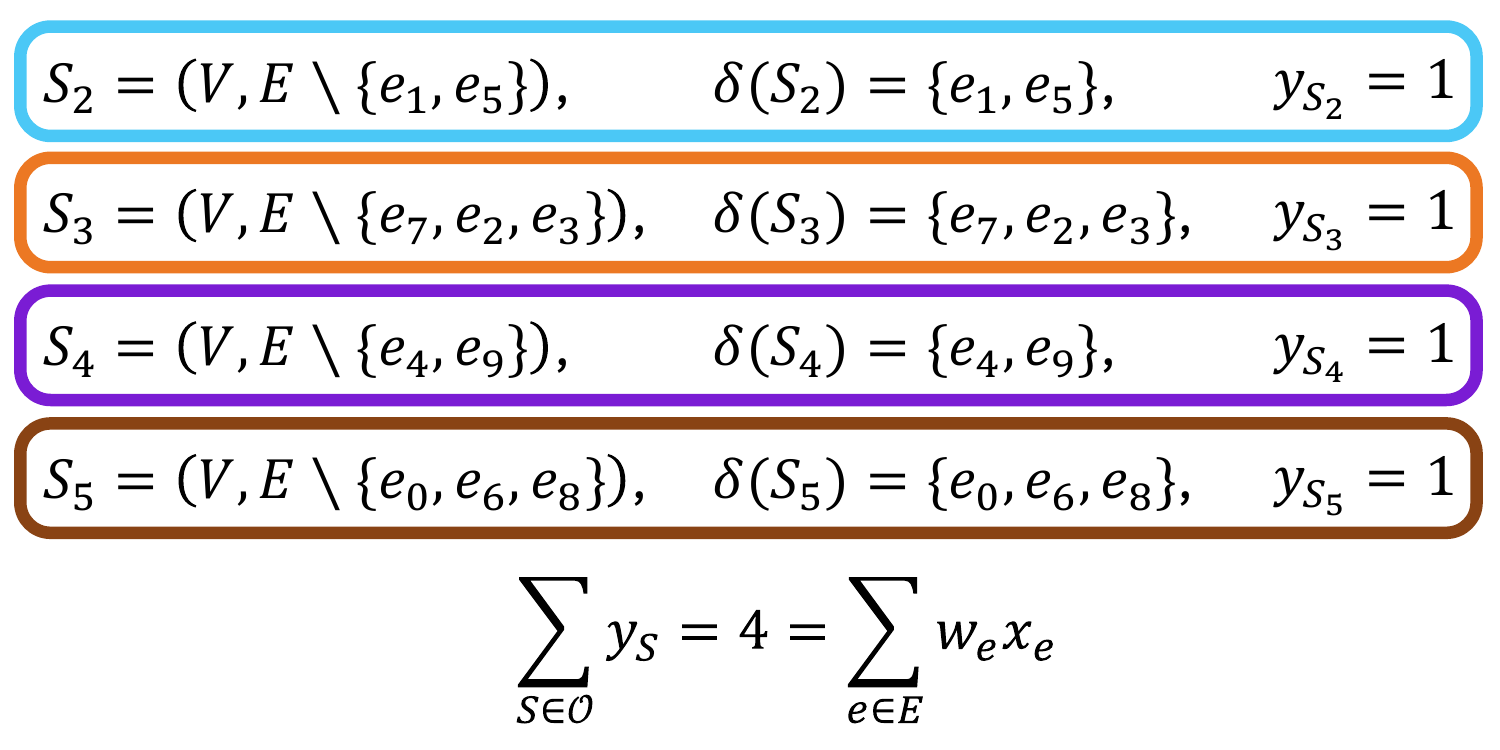}
    \caption{MWPF Optimality Certificate}
    \label{fig:hypergraph-failure-fix-split-3}
  \end{subfigure}

  \caption{
    An example of the \emph{SingleHair} \refdef{relaxer}-finding algorithm fails to find any \refdef{relaxer} given a suboptimal \refeqs{DLP} solution.
    The decoding hypergraph has a uniform weight of $w_e = 1$, with two \refdef{hyperblossoms} $S_0$ and $S_1$ of $y_{S_0} = y_{S_1} = 1$.
  }
  \label{fig:hypergraph-failure}
\end{figure}

We construct a failing case of the \emph{SingleHair} \refdef{relaxer}-finding algorithm by exploiting its limitation: the inability to consider dependency between \refdef{hyperblossoms}.
As shown in \cref{fig:hypergraph-failure}, the \emph{SingleHair} \refdef{relaxer} finder constructs the \refdef{hair-matrices} $\mathcal{H}_{S_0}$ and $\mathcal{H}_{S_1}$ in \cref{fig:hypergraph-failure-split-1}, both consisting of a single Odd row.
Thus, \emph{SingleHair} returns \textsc{Nil} with this suboptimal \refeqs{DLP} solution.

However, the MWPF solutions only satisfy the following complementary slackness theorem for one of the two \refdef{hyperblossoms} at a time, but never both.
\begin{align}
  y_S > 0 \ \Longrightarrow \ \sum_{e \in \delta(S)} x_e = 1, \qquad \forall S \in \mathcal{O} \tag{C2} \label{eq:cs-single-hair}
\end{align}
In particular, \cref{fig:hypergraph-failure-pf1,fig:hypergraph-failure-pf2,fig:hypergraph-failure-pf3,fig:hypergraph-failure-pf4} satisfy \cref{eq:cs-single-hair} for $S_0$ and \cref{fig:hypergraph-failure-pf5,fig:hypergraph-failure-pf6,fig:hypergraph-failure-pf7,fig:hypergraph-failure-pf8,fig:hypergraph-failure-pf9} satisfy \cref{eq:cs-single-hair} for $S_1$.

Although the \emph{SingleHair} \refdef{relaxer} finder cannot find more \refdef{relaxers}, we can manually find a \refdef{relaxer} to make progress in the \hyperblossom algorithm.
As shown in \cref{fig:hypergraph-failure-fix-split-0}, $\Delta\vec{y}$ is a \refdef{feasible-direction} that involves four \refdef{hyperblossoms} (two shrinking and two growing).
It is a \refdef{relaxer} because $\Delta\vec{y}$ relaxes edges $\mathcal{R}(\Delta\vec{y}) = \{e_0, e_4, e_6, e_8, e_9\}$, as shown in \cref{fig:hypergraph-failure-fix-split-1}.
After the relaxation, the \refdef{cluster} becomes \refdef{invalid} $C = (V, E\setminus\mathcal{R}(\Delta\vec{y})) \in \mathcal{O}$, so that the \hyperblossom algorithm can compose a \refdef{useful-direction} to grow.
The final \refeqs{DLP} solution $\vec{y}$ has a weight of 4, equal to the weight of all the \refeqs{MWPF} solutions \cref{fig:hypergraph-failure-pf1,fig:hypergraph-failure-pf2,fig:hypergraph-failure-pf3,fig:hypergraph-failure-pf4,fig:hypergraph-failure-pf5,fig:hypergraph-failure-pf6,fig:hypergraph-failure-pf7,fig:hypergraph-failure-pf8,fig:hypergraph-failure-pf9}.
In fact, all these parity factors satisfy \cref{eq:cs-single-hair} for the \refdef{hyperblossoms} $\mathcal{B} = \{ S_2, S_3, S_4, S_5 \}$ at the same time.
Thus, we prove the optimality of both the \refeqs{DLP} solution $\vec{y}$ and the \refeqs{MWPF} solutions.

This example shows that in order to achieve optimality, a \refdef{relaxer} finder may need to explore \refdef{relaxers} that involve more than a pair of \refdef{invalid} subgraphs.
Indeed, optimal \refdef{relaxer} finders like \emph{Blossom} (\S\ref{ssec:blossom-subroutine}) and \emph{Nullity$_{\le 1}$} (\S\ref{ssec:biased-single-dof}) find these more complicated \refdef{relaxers} to reach optimality.

\subsection{\emph{SingleHair} is Suboptimal on Simple Graphs}\label{ssec:failure-simple-graph}

\begin{figure*}
  \centering

  \begin{subfigure}[t]{0.24\linewidth}
    \centering
    \includegraphics[width=\textwidth,page=1]{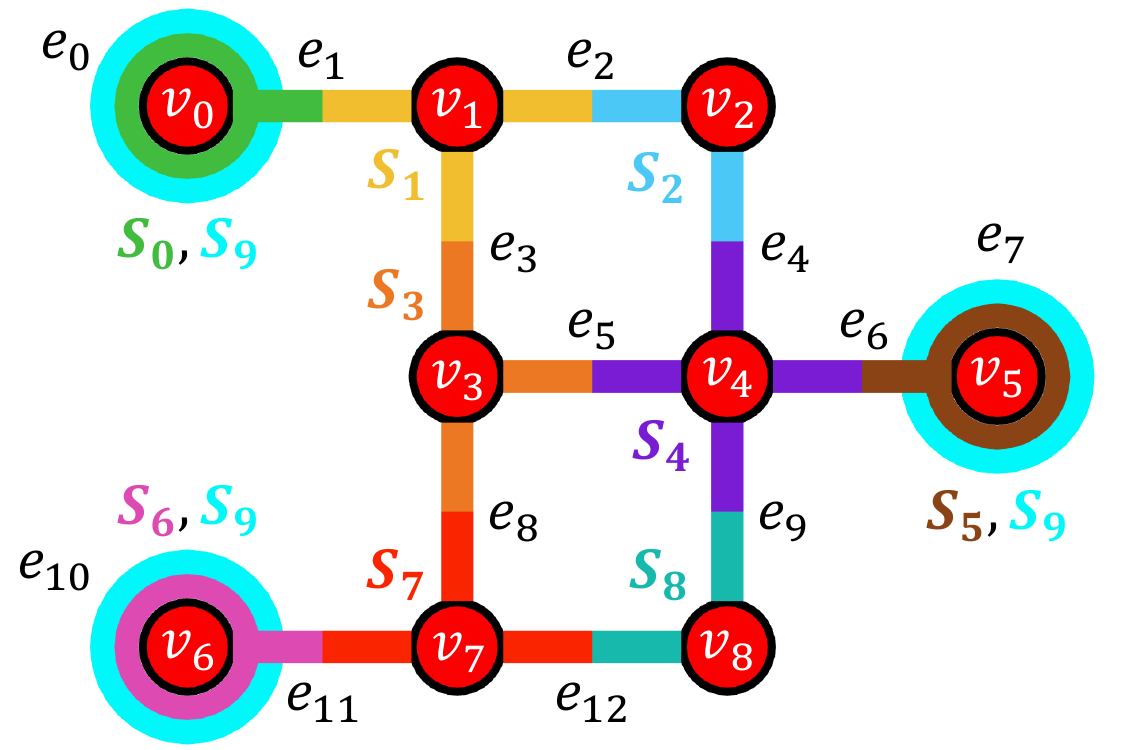}
    \caption{Suboptimal $\sum \vec{y} = 5$.}
    \label{fig:simple-graph-failure-1}
  \end{subfigure}
  \begin{subfigure}[t]{0.24\linewidth}
    \centering
    \includegraphics[width=\textwidth,page=2]{figures/simple-graph-failure/simple-graph-failure.pdf}
    \caption{MWPF $W(\mathcal{E}) = 6$.}
    \label{fig:simple-graph-failure-2}
  \end{subfigure}
  \begin{subfigure}[t]{0.24\linewidth}
    \centering
    \includegraphics[width=\textwidth,page=3]{figures/simple-graph-failure/simple-graph-failure.pdf}
    \caption{Another MWPF $\mathcal{E}'$.}
    \label{fig:simple-graph-failure-3}
  \end{subfigure}
  \begin{subfigure}[t]{0.24\linewidth}
    \centering
    \includegraphics[width=\textwidth,page=4]{figures/simple-graph-failure/simple-graph-failure.pdf}
    \caption{Optimal dual $\sum \vec{y'} = 6$.}
    \label{fig:simple-graph-failure-4}
  \end{subfigure}

  \vspace{1ex}
  \begin{subfigure}[t]{0.192\linewidth}
    \centering
    \includegraphics[width=\textwidth,page=1]{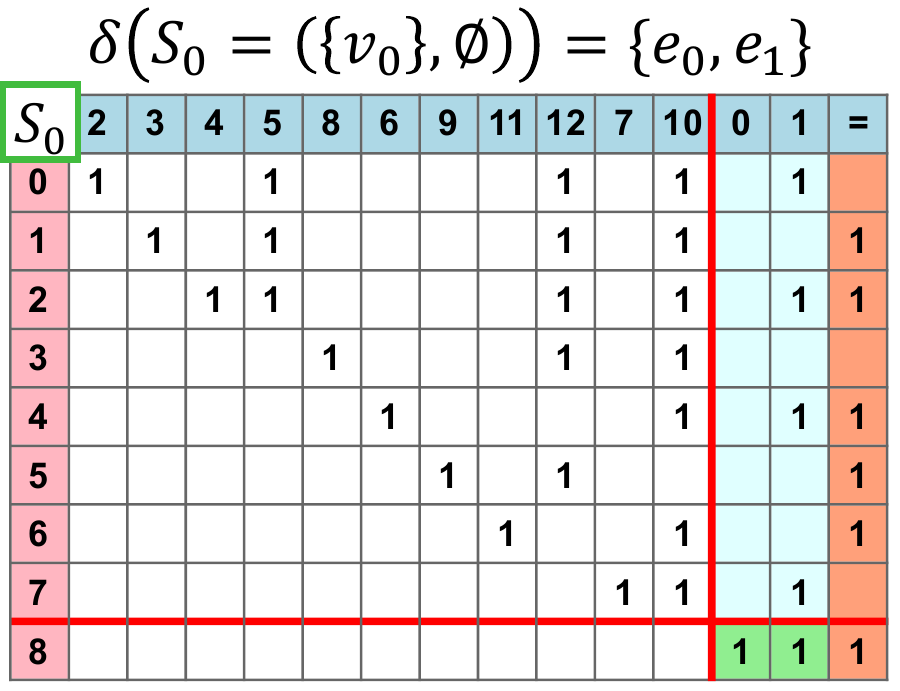}
    \caption{\refdef{hyperblossom} $S_0$.}
    \label{fig:simple-graph-failure-hair-1}
  \end{subfigure}
  \begin{subfigure}[t]{0.192\linewidth}
    \centering
    \includegraphics[width=\textwidth,page=2]{figures/simple-graph-failure/hair-matrices.pdf}
    \caption{\refdef{hyperblossom} $S_1$.}
    \label{fig:simple-graph-failure-hair-2}
  \end{subfigure}
  \begin{subfigure}[t]{0.192\linewidth}
    \centering
    \includegraphics[width=\textwidth,page=3]{figures/simple-graph-failure/hair-matrices.pdf}
    \caption{\refdef{hyperblossom} $S_2$.}
    \label{fig:simple-graph-failure-hair-3}
  \end{subfigure}
  \begin{subfigure}[t]{0.192\linewidth}
    \centering
    \includegraphics[width=\textwidth,page=4]{figures/simple-graph-failure/hair-matrices.pdf}
    \caption{\refdef{hyperblossom} $S_3$.}
    \label{fig:simple-graph-failure-hair-4}
  \end{subfigure}
  \begin{subfigure}[t]{0.192\linewidth}
    \centering
    \includegraphics[width=\textwidth,page=5]{figures/simple-graph-failure/hair-matrices.pdf}
    \caption{\refdef{hyperblossom} $S_4$.}
    \label{fig:simple-graph-failure-hair-5}
  \end{subfigure}

  \vspace{1ex}
  \begin{subfigure}[t]{0.192\linewidth}
    \centering
    \includegraphics[width=\textwidth,page=6]{figures/simple-graph-failure/hair-matrices.pdf}
    \caption{\refdef{hyperblossom} $S_5$.}
    \label{fig:simple-graph-failure-hair-6}
  \end{subfigure}
  \begin{subfigure}[t]{0.192\linewidth}
    \centering
    \includegraphics[width=\textwidth,page=7]{figures/simple-graph-failure/hair-matrices.pdf}
    \caption{\refdef{hyperblossom} $S_6$.}
    \label{fig:simple-graph-failure-hair-7}
  \end{subfigure}
  \begin{subfigure}[t]{0.192\linewidth}
    \centering
    \includegraphics[width=\textwidth,page=8]{figures/simple-graph-failure/hair-matrices.pdf}
    \caption{\refdef{hyperblossom} $S_7$.}
    \label{fig:simple-graph-failure-hair-8}
  \end{subfigure}
  \begin{subfigure}[t]{0.192\linewidth}
    \centering
    \includegraphics[width=\textwidth,page=9]{figures/simple-graph-failure/hair-matrices.pdf}
    \caption{\refdef{hyperblossom} $S_8$.}
    \label{fig:simple-graph-failure-hair-9}
  \end{subfigure}
  \begin{subfigure}[t]{0.192\linewidth}
    \centering
    \includegraphics[width=\textwidth,page=10]{figures/simple-graph-failure/hair-matrices.pdf}
    \caption{\refdef{hyperblossom} $S_9$.}
    \label{fig:simple-graph-failure-hair-10}
  \end{subfigure}

  \caption{
    The \emph{SingleHair} \refdef{relaxer}-finding algorithm fails to find an optimal \refeqs{DLP} solution.
    (a) The decoding graph has a uniform weight of $1$.
    The ten \refdef{hyperblossoms} $S_0, S_1, \ldots, S_9$ all have \refeqs{DLP} variable of $y_{S_i} = 1 / 2$.
    (b,c) Two \refeqs{MWPF} solutions $\mathcal{E}$ and $\mathcal{E}'$ have higher weights than the \refeqs{DLP} solution, indicating that the \refeqs{DLP} solution is suboptimal.
    (e-n) The \emph{SingleHair} \refdef{relaxer}-finding algorithm cannot find any \refdef{relaxers} because the \refdef{hair-matrices} $\mathcal{H}_{S_i}$ all consist of a single Odd row.
    Thus, the \hyperblossom algorithm with the \emph{SingleHair} \refdef{relaxer} finder terminates with this suboptimal \refeqs{DLP} solution.
    (d) The optimal \refeqs{DLP} solution $\vec{y'}$ has a weight of 6.
  }
  \label{fig:simple-graph-failure}
\end{figure*}

One might expect that the artificially constructed failing case in \S\ref{ssec:failure-hypergraph} only applies to hypergraphs, but we show in this section that it also applies to simple graphs.

We use numerical simulation to find a suboptimal case of the \emph{SingleHair} \refdef{relaxer}-finding algorithm on simple graphs, as shown in \cref{fig:simple-graph-failure}.
Given the initial \refeqs{DLP} solution with 10 \refdef{hyperblossoms} (\cref{fig:simple-graph-failure-1}), the \emph{SingleHair} \refdef{relaxer}-finding algorithm cannot find a \refdef{relaxer} any more because all \refdef{hair-matrices} consist of a single Odd row, as shown in \cref{fig:simple-graph-failure-hair-1,fig:simple-graph-failure-hair-2,fig:simple-graph-failure-hair-3,fig:simple-graph-failure-hair-4,fig:simple-graph-failure-hair-5,fig:simple-graph-failure-hair-6,fig:simple-graph-failure-hair-7,fig:simple-graph-failure-hair-8,fig:simple-graph-failure-hair-9,fig:simple-graph-failure-hair-10}.
However, any \refeqs{MWPF} solution violates the complementary slackness theorem \cref{eq:cs-single-hair} for at least one \refdef{hyperblossom}.
For example, $\mathcal{E}$ in \cref{fig:simple-graph-failure-2} violates \cref{eq:cs-single-hair} for $S_9$ while $\mathcal{E}'$ in \cref{fig:simple-graph-failure-3} violates \cref{eq:cs-single-hair} for $S_4$.
This is because the \refeqs{DLP} solution is suboptimal.
An optimal solution $\vec{y'}$ in \cref{fig:simple-graph-failure-4} satisfies \cref{eq:cs-single-hair} for all the \refdef{hyperblossoms}.

\subsection{\emph{SingleHair} is Optimal on Nullity$_{\le 1}$ Hypergraphs}\label{ssec:optimal-1-dof}

We prove that the \emph{SingleHair} \refdef{relaxer}-finding algorithm is optimal for the nullity$_{\le 1}$ hypergraphs, although it is not as efficient as the \emph{Nullity$_{\le 1}$} \refdef{relaxer} finder because the latter exploits the special property of nullity$_{\le 1}$.
Before looking into the \emph{SingleHair} \refdef{relaxer} finder, we first prove the following lemma.

\vspace{1ex}

\lemmaNullityIsHereditary{nullity-is-hereditary-with-proof}

\begin{proof}

Any subgraph is constructed by removing some edges and then removing some vertices that are not incident to any remaining edge.

Removing an edge corresponds to removing a column of the incidence matrix.
According to the rank-nullity theorem, the number of columns is equal to the sum of the rank and the nullity.
The rank of the incidence matrix decreases by at most 1 when removing a column.
Thus, the nullity does not increase.

When removing a vertex that is not incident to any remaining edge, the row corresponding to the vertex is all 0s.
Thus, the rank does not change by removing an all-0 row.
Also, given that the number of columns does not change, the nullity does not change.

Overall, the nullity of any subgraph is no larger than the nullity of the original hypergraph.
Therefore, nullity$_{\le 1}$ is a hereditary property of hypergraphs.

\end{proof}

\vspace{1ex}
\lemmaSingleHairOptimalityNullityLetwo{single-hair-optimality-single-dof-with-proof}
\begin{proof}
  We prove the lemma by explicitly proving the optimality of the final parity factor solution using \reftheorem{provable-optimum}.

  Given \reflemma{unique-row-if-not-a-relaxer-with-proof} (\S\ref{ssec:hair-matrix-unique-row}), when the \hyperblossom algorithm with the \emph{SingleHair} \refdef{relaxer} finder terminates, all the \refdef{hair-matrices} consist of a single Odd row of all 1s.

  Given that the decoding hypergraph is nullity$_{\le 1}$ and \reflemma{nullity-is-hereditary-with-proof}, any \refdef{hyperblossom-matrix} $\mathcal{M}_S$ has a nullity of at most 1, i.e., $\mathcal{M}_S$ has at most two parity factors.
  In the following, we consider two cases: nullity$_{=0}$ and nullity$_{=1}$.

  \nosection{Nullity$_{=0}$}.
  If the parity matrix contains a single parity factor solution $\mathcal{E}$, then we prove the parity factor and \refeqs{DLP} solution satisfies $\sum_{S \in \mathcal{O}} y_S = W(\mathcal{E})$.
  When the nullity is 0, there is no free variable.
  That is, all the \refdef{hair-matrices} consist of a single (pivot) column $e \in T$ with a fixed choice of $x_e = 1$ constrained by the only Odd row, i.e., $e \in \mathcal{E}$.
  Thus, every \refdef{hyperblossom} $S \in \mathcal{B}$ contributes to exactly one edge of $\mathcal{E}$ because $\delta(S) \cap T = \{ e \}$, the only column of the \refdef{hair-matrix}.
  We have
  \begin{align*}
    1.\ W(\mathcal{E}) &= \sum_{e \in \mathcal{E}} w_e \\
    &\,\downarrow\text{(given $\mathcal{E} \subseteq T$)} \\
    &= \sum_{e \in \mathcal{E}}\ \sum_{S \in \mathcal{B} | e \in \delta(S)} y_S \\
    &\,\downarrow\text{(a \refdef{hyperblossom} contributes at most once)} \\
    &\le \sum_{S \in \mathcal{B}} y_S = \sum_{S \in \mathcal{O}} y_S, \\
    2.\ W(\mathcal{E}) &\ge \sum_{S \in \mathcal{O}} y_S, \quad \text{(given \cref{eq:mwpf-chain})} \\
    \Longrightarrow W(\mathcal{E}) &= \sum_{S \in \mathcal{O}} y_S
  \end{align*}

  \nosection{Nullity$_{=1}$}.
  If the parity matrix contains two parity factor solutions $\mathcal{E}_1 \neq \mathcal{E}_2$, then we prove that they are both optimal, i.e., $\sum_{S \in \mathcal{O}} y_S = W(\mathcal{E}_1) = W(\mathcal{E}_2)$.

  We define \textit{common edges} as the edges that are shared by both parity factors, i.e., $E_{1,2}\coloneqq \mathcal{E}_1 \cap \mathcal{E}_2$.
  Similarly, the edges exclusive to $\mathcal{E}_1$ and $\mathcal{E}_2$ are $E_1 \coloneqq \mathcal{E}_1 \setminus \mathcal{E}_2$ and $E_2 \coloneqq \mathcal{E}_2 \setminus \mathcal{E}_1$, respectively.
  We prove that each \refdef{hair-matrix} $\mathcal{H}_S$ contains either a single variable of common edge $e \in E_{1,2}$, or two variables of exclusive edges $e_1 \in E_1, e_2 \in E_2$.

  When a \refdef{hair-matrix} $\mathcal{H}_S$ of a \refdef{hyperblossom} $S \in \mathcal{B}$ contains a single variable $e \in T$, it is a pivot variable of an Odd row with no free variables, i.e., $x_e = 1$.
  Thus, $e$ belongs to both parity factors $\mathcal{E}_1$ and $\mathcal{E}_2$, i.e., it is a common edge $e \in E_{1,2}$.

  When a \refdef{hair-matrix} $\mathcal{H}_S$ contains two variables $e_1, e_2 \in T$, none of them can be common edges, because otherwise there exists a feasible solution with $x_{e_i} = 0$ violating the definition of a common edge.
  Also, $e_1$ and $e_2$ cannot be exclusive edges of the same feasible parity factor solution, otherwise there exists a third feasible solution $\mathcal{E}_3$ that includes $e_1$ but not $e_2$.
  Thus, $e_1$ and $e_2$ must be exclusive edges of different parity factors $e_1 \in \mathcal{E}_1$ and $e_2 \in \mathcal{E}_2$.

  A \refdef{hair-matrix} cannot have more than two variables because it has only one single row and the nullity (the number of free variables) is no more than 1.

  Thus, each \refdef{hyperblossom} either contributes uniquely to a common edge, or contributes to two exclusive edges simultaneously.
  \begin{align*}
    1.\ W(\mathcal{E}_i) &= \sum_{e \in \mathcal{E}_i} w_e \\
    &\,\downarrow\text{(given $\mathcal{E}_i \subseteq T$)} \\
    &= \sum_{e \in \mathcal{E}_i}\ \sum_{S \in \mathcal{B} | e \in \delta(S)} y_S \\
    &\,\downarrow\text{(given $\mathcal{E}_i = E_{1,2} \cup E_i$)} \\
    &= \sum_{e \in E_{1,2} \cup E_i}\ \sum_{S \in \mathcal{B} | e \in \delta(S)} y_S \\
    &\,\downarrow\text{(a \refdef{hyperblossom} contributes at most once)} \\
    &\le \sum_{S \in \mathcal{B}} y_S = \sum_{S \in \mathcal{O}} y_S \quad \forall i \in \{1,2\}, \\
    2.\ W(\mathcal{E}_i) &\ge \sum_{S \in \mathcal{O}} y_S \quad \forall i \in \{1,2\}, \quad \text{(given \cref{eq:mwpf-chain})} \\
    \Longrightarrow W(\mathcal{E}_i) &= \sum_{S \in \mathcal{O}} y_S\quad \forall i \in \{1,2\}
  \end{align*}

\end{proof}

\section{Interoperability with MWPM Decoder}\label{ssec:interoperability-mwpm}

Given the similarity between the math behind MWPF (\hyperblossom framework in \S\ref{ssec:problem-definitions}) and MWPM (blossom algorithm in \S\ref{ssec:mwpm}), one would wonder whether they can interoperate with each other.
This is particularly useful for heterogeneous QEC architectures~\cite{stein2025hetec} where we could use a faster MWPM decoder on some portion of the decoding hypergraph and transfer the optimal \refdef{blossom-dlp} to the \hyperblossom algorithm.

In this section, we first extend the definitions of the blossom algorithm in~\cite{wu2023qce} so that we can use them to define the mapping (\S\ref{ssec:simple-graph-condition-blossom-definitions}).
We then define the bijective function $f$ that maps a \refdef{blossom-dlp} $\vec{y^*} \in \mathbb{R}^{|\mathcal{O}^*|}$ to a \refeqs{DLP} solution $\vec{y} \in \mathbb{R}^{|\mathcal{O}|}$ (\S\ref{ssec:dual-mapping-definition}).
We prove some important properties of the mapping function in \S\ref{ssec:dual-mapping-properties}.
Using the properties of $f$, we prove that all simple graphs satisfy \ref{condition:mwpf} in \S\ref{ssec:mwpf-condition-simple-graph}.
We then reconstruct the blossom algorithm data structures from a \refdef{blossom-dlp} $\vec{y^*}$ (\S\ref{ssec:alternating-tree-reconstruction}).
Finally, we explain how to decode heterogeneous QEC codes (\S\ref{ssec:heterogeneous-qec-decoding}) more efficiently, using a technique from Fusion Blossom~\cite{wu2023qce}.

\subsection{Blossom Algorithm Definitions}\label{ssec:simple-graph-condition-blossom-definitions}

We use the definitions and theorems in~\cite{wu2023qce} extensively to assist the proofs.
We carefully identify the conflicts of the notations in~\cite{wu2023qce} and this work, and minimize the confusion with the following conventions.
\begin{itemize}
  \item We use a superscript $*$ for the concepts in the blossom algorithm, e.g. $y^*$, $S^*$ and $\mathcal{O}^*$, to distinguish between similar concepts in~\cite{wu2023qce} and this work.
  \item ``$\mathcal{D}(S^*)$'' refers to the \emph{Progeny} of a blossom $S^* \in \mathcal{O}^*$ in~\cite{wu2023qce}. Since we use $\mathcal{D}(\mathcal{E})$ to refer to \refdef{defect-error-pattern} in this work, we will use ``$\mathcal{P}(S^*)$'' instead for the \emph{Progeny} of a blossom.
  \item $C(v)$ refers to the geometric concept of \emph{Circle} in~\cite{wu2023qce} while $C \in \mathcal{C}$ refers to the \emph{Cluster} in this paper. We will not use the \emph{Circle} $C(v)$ in this section to avoid confusions.
\end{itemize}

Fusion Blossom~\cite{wu2023qce} relates the variables of \refdef{blossom-dlp} $y^*_{S^*}, \forall S^* \in \mathcal{O^*}$ with geometric \refdef{covers} on the decoding graph.
The geometric concepts not only help to visualize the dual variables, but also help prove the correctness using their finite-overlapping properties, like \textbf{Theorem: Node Cover Finite Overlap}~\cite{wu2023qce}.
The theorem only describes a \emph{node}, which is a special kind of blossom who has no parent, i.e., not part of a larger blossom.
This is fine in the blossom algorithm because it only cares about \emph{nodes} in the alternating trees.
However, in this work, we need to consider the internal structure of a \refdef{cover}.
Namely, each blossom has its own \refdef{ring}, which is part of the \refdef{cover} that it solely contributes to.
The \refdef{rings} of different blossoms are finite-overlapping, and the union of all the \refdef{rings} of a node's \emph{Progeny} constitute its \refdef{cover}.
We extend the definitions in~\cite{wu2023qce} with the \refdef{descendants}, \refdef{rings} and the \refdef{descover-distance} that is useful to determine whether a point belongs to a \refdef{ring}.

\vspace{1ex}
\definitionlabel{descendants}{Descendants}
Given a blossom $S^* \in \mathcal{O}^*$, its \emph{Descendants} $\mathcal{J}(S^*)$ is the set of all blossoms with $S^*$ as their ancestor:
$\mathcal{J}(S^*) = \{ C^* \in \mathcal{O}^* | y^*_{C^*} > 0, C^* \subsetneq S^* \}$.

\vspace{1ex}
A related concept in~\cite{wu2023qce} is the \emph{Progeny} $\mathcal{P}(S^*)$ of a blossom $S^* \in \mathcal{O}^*$, which includes its \refdef{descendants} and itself, i.e., $\mathcal{P}(S^*) = \mathcal{J}(S^*) \cup \{ S^* \}$.
In this work, we need to clearly distinguish between the \emph{Progeny} $\mathcal{P}(S^*)$ and the \refdef{descendants} $\mathcal{J}(S^*)$ of a blossom.

Before defining the \refdef{descendants-cover} and \refdef{ring} for this work, we review the definition of \refdef{cover} in~\cite{wu2023qce}.

\vspace{1ex}
\definitionlabel{cover}{Cover}
The \emph{Cover} of a blossom $S^* \in \mathcal{O}^*$ includes the points $p \in G$ (vertices $V$ and all the points of the edges $E$ in the decoding graph $G$) within certain distance from the defect vertices $S^*$.
\begin{align*}
  \text{Cover}(S^*) = \bigcup_{v \in S^*} \Big\{ p \in G \Big| \text{Dist}(p, v) \leq \qns\sum_{D^* \in \mathcal{P}(S^*) | v \in D^*}\qns y^*_{D^*} \Big\}
\end{align*}

\vspace{1ex}
\definitionlabel{descendants-cover}{Descendants Cover}
For a blossom $S^* \in \mathcal{O}^*$, $\text{DesCover}(S^*)$ is the union of its \refdef{descendants}' \refdef{covers} and its defect vertices:
$$\text{DesCover}(S^*) = \bigcup_{C^* \in \mathcal{J}(S^*)}\dqns \text{Cover}(C^*) \cup S^*$$

\vspace{1ex}
\definitionlabel{ring}{Ring}
Given a blossom $S^* \in \mathcal{O}^*$, $\text{Ring}(S^*)$ is the exclusive part of its \refdef{cover} by excluding its \refdef{descendants-cover}:
$$\text{Ring}(S^*) = \text{Cover}(S^*) \setminus \text{DesCover}(S^*)$$

\definelemma{ring-finite-overlap}{Ring Finite Overlap}
Given blossoms $S^*_1, S^*_2 \in \mathcal{O}^*, S^*_1 \neq S^*_2$, $\text{Ring}(S^*_1) \cap \text{Ring}(S^*_2)$ is a finite set of points.

\begin{figure}[h]
  \centering
  \begin{subfigure}[t]{.49\linewidth}
    \centering
    \includegraphics[width=1\textwidth]{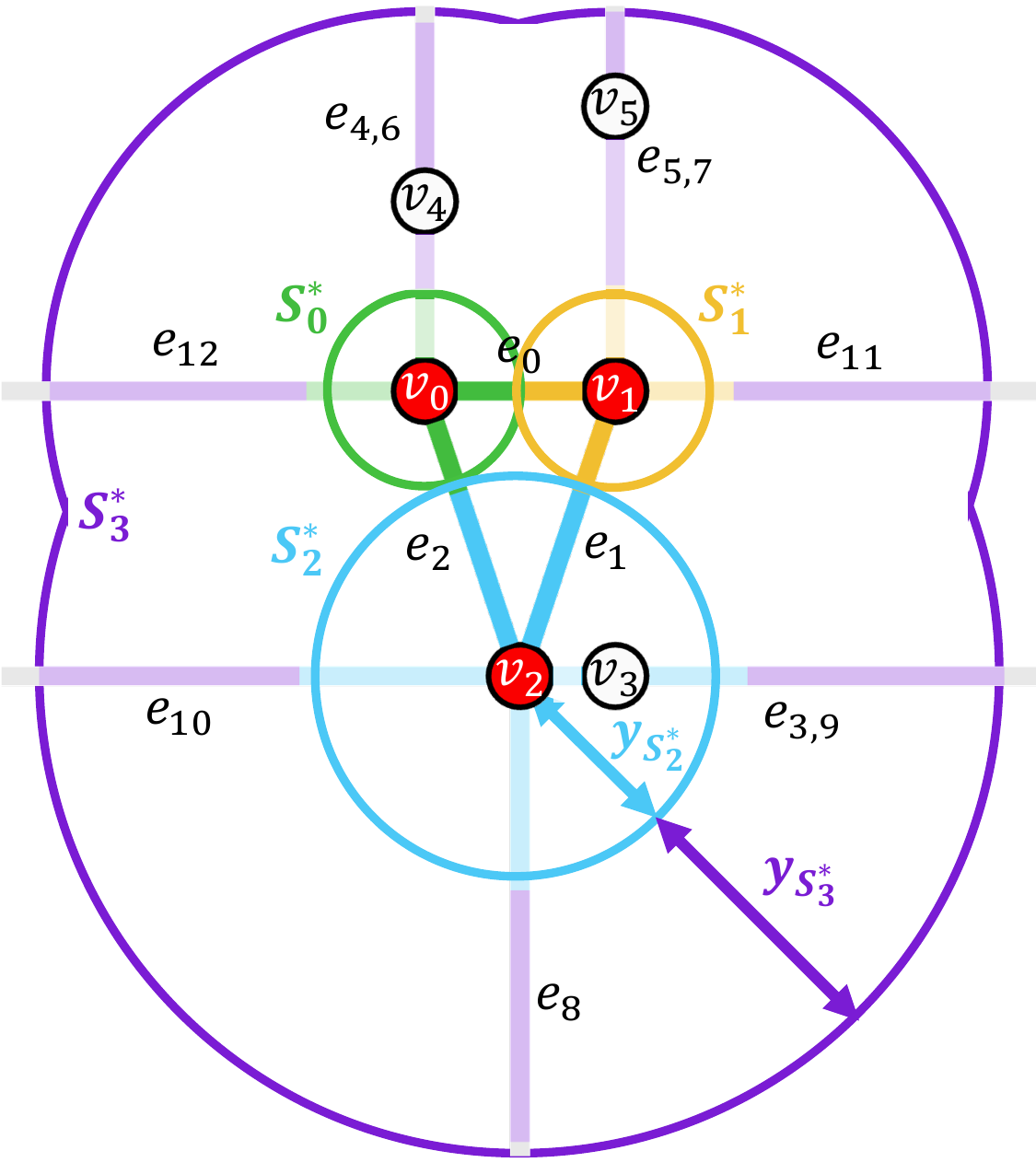}
    \caption{Blossom algorithm.}
    \label{fig:interoperability-mwpm}
  \end{subfigure}
  \begin{subfigure}[t]{.49\linewidth}
    \centering
    \includegraphics[width=1\textwidth]{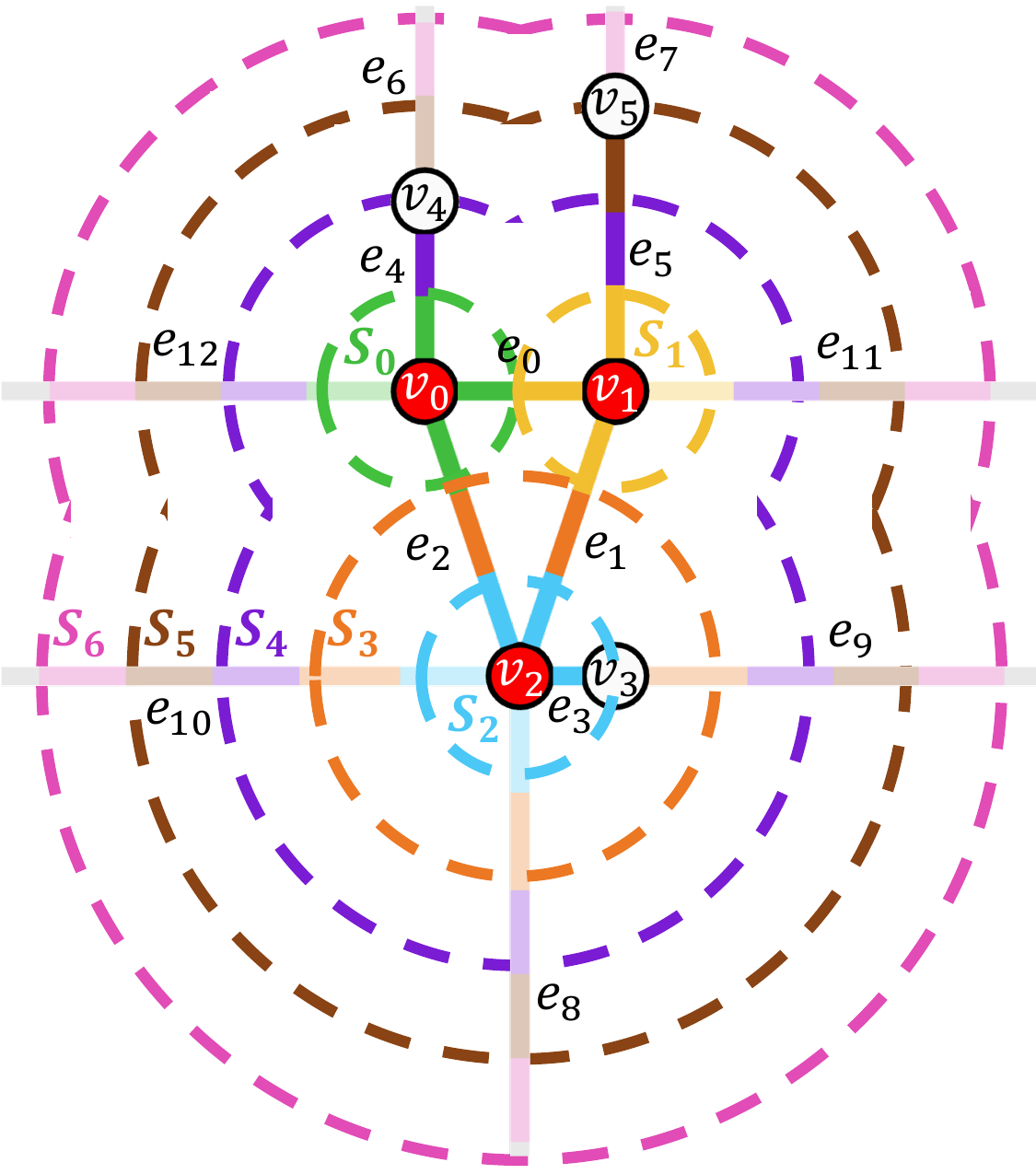}
    \caption{MWPF.}
    \label{fig:interoperability-mwpf}
  \end{subfigure}

  \vspace{1ex}
  \begin{subfigure}[t]{.433\linewidth}
    \centering
    \includegraphics[width=1\textwidth]{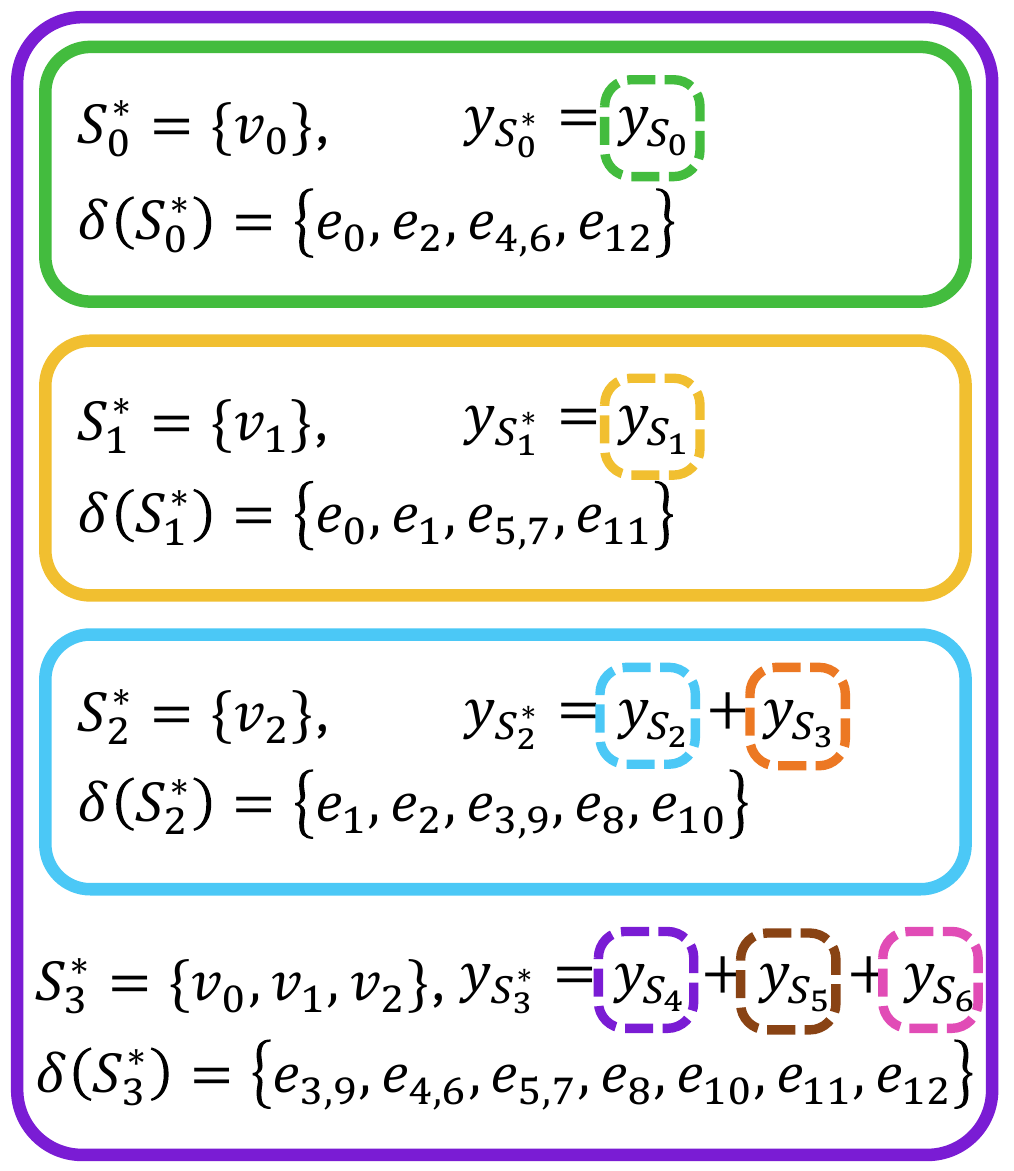}
    \caption{\refdef{blossom-dlp} $\vec{y^*}$.\qqns}
    \label{fig:interoperability-dual-variables-split-0}
  \end{subfigure}
  \begin{subfigure}[t]{.527\linewidth}
    \centering
    \includegraphics[width=1\textwidth]{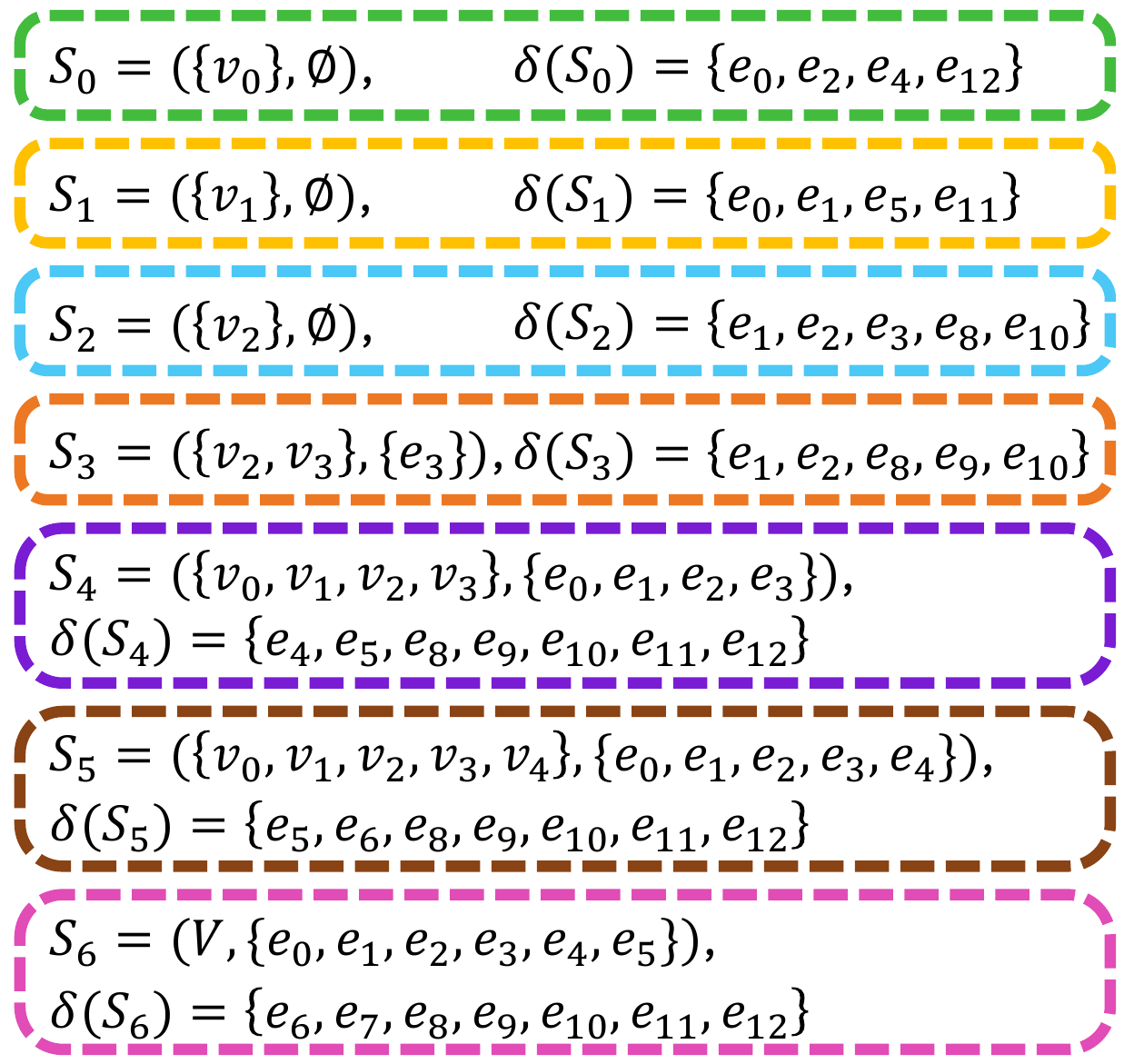}
    \caption{\refeqs{DLP} solution $\vec{y}$.}
    \label{fig:interoperability-dual-variables-split-1}
  \end{subfigure}
  \caption{
    An example of mapping a \refdef{blossom-dlp} $\vec{y^*}$ to a \refeqs{DLP} solution $\vec{y} = f(\vec{y^*})$.
    (a,b) Each dual variable $y^*_{S^*}$ ($y_S$) is visualized by a \refdef{ring} with a radius of $y^*_{S^*}$ ($y_S$).
    (c) The blossom algorithm works on the syndrome graph (\S\ref{ssec:mwpm}), which only includes the defect vertices $D = \{ v_0, v_1, v_2, \cdots \}$.
    For reference, we show the non-defect vertices $\{ v_3, v_4, v_5 \}$ that are on the syndrome graph edges incident to $v_i \in D$, but note that these non-defect vertices do not exist in the syndrome graph.
    Each blossom $S^* \in \mathcal{B}^*$ maps to multiple \refdef{hyperblossoms} $\mathcal{O}(S^*) \cap \mathcal{B}$.
    For example, $S^*_2$ maps to two \refdef{hyperblossoms} $\mathcal{O}(S^*) \cap \mathcal{B} = \{ S_2, S_3 \}$.
    (d) To understand the mapping, imagine whenever the \refdef{ring} of a blossom $S^* \in \mathcal{O}^*$ encounters a non-defect vertex $v \in \overline{D}$ when it grows, a new \refdef{hyperblossom} $S = (V_S, E[V_S]) \in \mathcal{O}$ is created to include $v$, i.e., $v \in V_S$.
  }
  \label{fig:interoperability}
\end{figure}

\begin{proof}
  The proof is intuitive from the visualization in \cref{fig:interoperability}.
  Here we prove it rigorously.

  If $S^*_1$ and $S^*_2$ have different root node, $\text{Root}(S^*_1) \neq \text{Root}(S^*_2)$, their \refdef{covers} have a finite overlap given \textbf{Theorem: Node Cover Finite Overlap} in~\cite{wu2023qce}.
  Given the definition of \refdef{rings}, $\text{Ring}(S^*) \subseteq \text{Cover}(S^*)$, we have
  \begin{gather}
    |\text{Cover}(S^*_1) \cap \text{Cover}(S^*_2)| \ \text{is finite}, \nonumber \\
    \text{Ring}(S^*_1) \subseteq \text{Cover}(S^*_1), \text{Ring}(S^*_2) \subseteq \text{Cover}(S^*_2) \nonumber \\
    \Longrightarrow |\text{Ring}(S^*_1) \cap \text{Ring}(S^*_2)| \ \text{is finite} \label{eq:ring-finite-overlap}
  \end{gather}

  If $S^*_1$ and $S^*_2$ belong to the same node, we only need to prove that their \refdef{rings} has a finite overlap at the moment they merge into the same node, since their \refdef{rings} do not change afterwards according to \reflemma{frozen-ring}.
  Before they merge into the same node, their \refdef{rings} have a finite overlap due to the same reason in \cref{eq:ring-finite-overlap}.

  In both cases, the overlap of the \refdef{rings} of $S^*_1$ and $S^*_2$ remains finite.

\end{proof}

\definelemma{frozen-ring}{Frozen Ring}
The \refdef{ring} of a blossom $S^*$ does not change once it merges into a parent blossom $S_p^*$.

\begin{proof}
  We use some properties of the blossom algorithm to prove this lemma.
  Once $S^*$ merges into another blossom $S_p^*$ where $S^* \subsetneq S_p^*$, according to the blossom algorithm procedure~\cite{wu2023qce}, the dual variables $y^*_{S^*}$ of these \refdef{descendants} $S^* \in \mathcal{J}(S^*_p)$ do not change, unless $S^*$ becomes a node again (under an event of blossom expansion).

  It then suffices to prove that the \refdef{ring} of $S^*$ does not change when it is among the \refdef{descendants} of $S_p^*$.
  By definition of \refdef{cover}, it only depends on the dual variables of its \emph{Progeny} $\mathcal{P}(S^*)$.
  The \emph{Progeny} of $S^*$ is among the \refdef{descendants} of $S^*_p$, i.e., $\mathcal{P}(S^*) = \mathcal{J}(S^*) \cup \{ S^*\}\subseteq \mathcal{J}(S^*_p)$.
  Thus, the \refdef{cover} of $S^*$ is frozen, i.e., $\text{Cover}(S^*)$ does not change.
  Similarly, the \refdef{descendants-cover} of $S^*$ is frozen, i.e., $\text{DesCover}(S^*)$ does not change.
  Thus, according to the definition, the \refdef{ring} of $S^*$ is frozen, i.e., $\text{Ring}(S^*) = \text{Cover}(S^*) \cap \text{DesCover}(S^*)$ does not change.

\end{proof}

\vspace{1ex}
Next, we define concepts that are useful to determine whether a vertex belongs to a \refdef{ring} or not.

\vspace{1ex}
\definitionlabel{descover-distance}{DesCover Distance}
Given a vertex $v \in V$ and a blossom $S^* \in \mathcal{B}^*$, the \emph{DesCover Distance}, denoted by $\text{DCDist}(S^*, v)$, is the minimum distance from $v$ to the points in the \refdef{descendants-cover} $\text{DesCover}(S^*)$.
$$\text{DCDist}(S^*, v) = \min_{p \in \text{DesCover}(S^*)} \text{Dist}(p, v)$$

\vspace{1ex}
\definelemma{DCDist-criteria}{DesCover Distance Criteria}
Given a blossom $S^* \in \mathcal{B}^*$ and a vertex $v \in V$, we have the criteria:
\begin{align}
  \text{DCDist}(S^*, v) = 0 &\Longleftrightarrow v \in \text{DesCover}(S^*) \label{eq:descover-criteria-1} \\
  0 < \text{DCDist}(S^*, v) \le y^*_{S^*} &\Longleftrightarrow v \in \text{Ring}(S^*) \label{eq:descover-criteria-2} \\
  \text{DCDist}(S^*, v) > y^*_{S^*} &\Longleftrightarrow v \notin \text{Cover}(S^*) \label{eq:descover-criteria-3}
\end{align}

\begin{proof}
  The proofs of these criteria are intuitive from the visualization in \cref{fig:interoperability}, where the \refdef{descendants-cover} $\text{DesCover}(S^*_3) = \text{Cover}(S^*_1) \cup \text{Cover}(S^*_2) \cup \text{Cover}(S^*_3)$ unions the region of the three circles of $S^*_0$, $S^*_1$ and $S^*_2$.
  Any point that has zero distance to this region falls in $\text{DesCover}(S^*_3)$ (\cref{eq:descover-criteria-1}).
  Any point that has a distance larger than $y_{S^*_3}$, marked by the radius of the \refdef{ring} in \cref{fig:interoperability-dual-variables-split-0}, is outside of the $\text{Cover}(S^*_3)$ (\cref{eq:descover-criteria-1}).
  The rest of the points are within $\text{Ring}(S^*_3)$.
  We prove them rigorously below.
  To prove \cref{eq:descover-criteria-1},
  \begin{align}
    &\text{DCDist}(S^*, v) = 0 \nonumber \\
    &\ \ \downarrow\text{(by definition of \refdef{descover-distance})} \nonumber \\
    &\Longleftrightarrow \exists p \in \text{DesCover}(S^*),  \text{Dist}(p, v) = 0 \nonumber \\
    &\Longleftrightarrow v \in \text{DesCover}(S^*) \label{eq:descover-criteria-1-proof}
  \end{align}
  To prove \cref{eq:descover-criteria-3}, we use the definition of \refdef{descendants-cover} and \refdef{cover}.
  We first prove \cref{eq:descover-criteria-3} for the case $\mathcal{J}(S^*) = \varnothing$ so that we only need to focus on $\mathcal{J}(S^*) \neq \varnothing$ later.
  When $S^*$ does not have any \refdef{descendants} $\mathcal{J}(S^*) = \varnothing$, it consists of a single defect vertex $S^* = \{ u \}$.
  Its \refdef{descendants-cover} is $\text{DesCover}(S^*) =  S^* = \{ u \}$ and thus $\text{DCDist}(S^*, v) = \text{Dist}(u, v)$.
  The \refdef{cover} of $S^*$ is $\text{Cover}(S^*) = \{ p \in G | \text{Dist}(p, u) \le y^*_{S^*} \}$ and thus $$v \notin \text{Cover}(S^*) \Longleftrightarrow \text{DCDist}(S^*, v) = \text{Dist}(u, v) > y^*_{S^*}$$
  We now prove \cref{eq:descover-criteria-3} for the case of $\mathcal{J}(S^*) \neq \varnothing$.
  In this case, every defect vertex $u \in S^*$ must belong to one of its \refdef{descendants}' \refdef{covers}, i.e., $\exists C^* \in \mathcal{J}(S^*), u \in \text{Cover}(C^*)$.
  we can then simplify the \refdef{descendants-cover} as $\text{DesCover}(S^*) = \bigcup_{C^* \in \mathcal{J}(S^*)} \text{Cover}(C^*)$.
  We have
  \begin{align}
    &\text{DCDist}(S^*, v) > y^*_{S^*} \nonumber \\
    &\Longleftrightarrow \forall p \in \text{DesCover}(S^*), \text{Dist}(p, v) > y^*_{S^*} \nonumber \\
    &\ \ \downarrow\text{(given the above simplification of $\text{DesCover}(S^*)$)} \nonumber \\
    &\Longleftrightarrow \forall p \in \bigcup_{C^* \in \mathcal{J}(S^*)}\dqns \text{Cover}(C^*) , \text{Dist}(p, v) > y^*_{S^*} \nonumber \\
    &\Longleftrightarrow \forall C^* \in \mathcal{J}(S^*), \forall p \in \text{Cover}(C^*) , \text{Dist}(p, v) > y^*_{S^*} \nonumber \\
    &\ \ \downarrow\text{(by definition of \refdef{cover})} \nonumber \\
    &\Longleftrightarrow \forall C^* \in \mathcal{J}(S^*), \forall u \in C^*, \forall p \in G, \nonumber \\
    &        \qquad \Big( \text{Dist}(p, u) \le \qns\sum_{D^* \in \mathcal{P}(C^*) | u \in D^*}\tqns y^*_{D^*} \Big) \rightarrow \Big(\text{Dist}(p, v) > y^*_{S^*} \Big) \nonumber \\
    &\Longleftrightarrow \forall C^* \in \mathcal{J}(S^*), \forall u \in C^*, \text{Dist}(u, v) > y^*_{S^*} + \qqns\sum_{D^* \in \mathcal{P}(C^*) | u \in D^*}\tqns y^*_{D^*} \nonumber \\
    &\ \ \downarrow\text{(given \reflemma{blossom-hierarchy-dual-sum})} \nonumber \\
    &\Longleftrightarrow \forall u \in S^*, \text{Dist}(u, v) > y^*_{S^*} + \sum_{D^* \in \mathcal{J}(S^*) | u \in D^*} y^*_{D^*} \nonumber \\
    &\ \ \downarrow\text{(given $\mathcal{P}(S^*) = \mathcal{J}(S^*) \cup \{ S^*\}$ and $S^* \notin \mathcal{J}(S^*)$)} \nonumber \\
    &\Longleftrightarrow \forall u \in S^*, \text{Dist}(u, v) > \sum_{D^* \in \mathcal{P}(S^*) | u \in D^*} y^*_{D^*} \nonumber \\
    &\ \ \downarrow\text{(by definition of \refdef{cover})} \nonumber \\
    &\Longleftrightarrow v \notin \text{Cover}(S^*) \label{eq:descover-criteria-3-proof}
  \end{align}

  To prove \cref{eq:descover-criteria-2}, we simply use \cref{eq:descover-criteria-1-proof,eq:descover-criteria-3-proof}:
  \begin{align*}
    &0 < \text{DCDist}(S^*, v) \le y^*_{S^*} \\
    &\ \ \downarrow\text{(given \cref{eq:descover-criteria-1-proof} and \cref{eq:descover-criteria-3-proof})} \\
    &\Longleftrightarrow v \notin \text{DesCover}(S^*), v \in \text{Cover}(S^*) \\
    &\ \ \downarrow\text{(by definition of \refdef{ring})} \\
    &\Longleftrightarrow v \in \text{Ring}(S^*)
  \end{align*}
\end{proof}

\vspace{1ex}
\definelemma{blossom-hierarchy-dual-sum}{Blossom Hierarchy Dual Sum}
Given a blossom $S^* \in \mathcal{B}^*$ with $\mathcal{J}(S^*) \neq \varnothing$ and any its vertex $u \in S^*$,
\begin{align*}
  \sum_{D^* \in \mathcal{J}(S^*) | u \in D^*} y^*_{D^*} = \max_{C^* \in \mathcal{J}(S^*)} \sum_{D^* \in \mathcal{P}(C^*) | u \in D^*} y^*_{D^*}
\end{align*}

\begin{proof}
  We prove the equation by first proving there exists $C^* \in \mathcal{J}(S^*)$ such that the dual sums above are equal, and then proving that for all $C^* \in \mathcal{J}(S^*)$, the right-hand side is no larger than the left-hand side.

  Given the hierarchy of blossoms, there exists an immediate child $C_u^* \in \mathcal{J}(S^*)$ of $S^*$ such that $u \in C_u^*$.
  All the blossoms that contains $u$ must be either \refdef{descendants} of $C_u^*$ or it has to be $S^*$ or $S^*$'s parent blossoms.
  Since $S^*$ and its parent blossoms are not in $\mathcal{J}(S^*)$, we have $\{ D^* \in \mathcal{J}(S^*) | u \in D^* \} = \{ D^* \in \mathcal{P}(C_u^*) | u \in D^* \}$.
  Thus, there exists $C^* = C_u^* \in \mathcal{J}(S^*)$ such that $$\sum_{D^* \in \mathcal{J}(S^*) | u \in D^*} y^*_{D^*} = \sum_{D^* \in \mathcal{P}(C^*) | u \in D^*} y^*_{D^*}$$

  We next prove that for all $C^* \in \mathcal{J}(S^*)$, the right-hand side is no larger than the left-hand side.
  We consider two cases: $u \in C^*$ and $u \notin C^*$.
  In the first case, $C^*$ must be one of the \refdef{descendants} of $C_u^*$, i.e., $C^* \in \mathcal{P}(C_u^*)$.
  Given the hierarchy, the \refdef{descendants} of $C^*$ must be a subset of the \refdef{descendants} of $C_u^*$, i.e., $\mathcal{P}(C^*) \subseteq \mathcal{P}(C_u^*)$.
  Thus, $$\sum_{D^* \in \mathcal{P}(C^*) | u \in D^*}\dqns y^*_{D^*} \le \sum_{D^* \in \mathcal{P}(C_u^*) | u \in D^*}\dqns y^*_{D^*} = \sum_{D^* \in \mathcal{J}(S^*) | u \in D^*}\dqns y^*_{D^*}$$
  In the second case when $u \notin C^*$, $u$ must not be in any of the \refdef{descendants} of $C^*$, i.e., $\forall D^* \in \mathcal{P}(C^*), u \notin D^*$.
  Thus $$\sum_{D^* \in \mathcal{P}(C^*) | u \in D^*} y^*_{D^*} = 0 \le \sum_{D^* \in \mathcal{J}(S^*) | u \in D^*} y^*_{D^*}$$
  In both cases, the right-hand side is no larger than the left-hand side.
  Thus, we have proved the lemma.

\end{proof}

\subsection{Dual Mapping Definition}\label{ssec:dual-mapping-definition}

\vspace{1ex}
\definitionlabel{dual-mapping}{Dual Mapping}
Given a \refdef{blossom-dlp} $\vec{y^*}$, we map each blossom dual variable $y^*_{S^*}, \forall S^* \in \mathcal{B}^*$ to a set of \refeqs{DLP} variables given the following rules.
Let the non-defect vertices in the \refdef{ring} be $\overline{D}(S^*) = \overline{D} \cap \text{Ring}(S^*)$, we sort these non-defect vertices based on their \refdef{descover-distance} to $S^*$.
That is,
\begin{align*}
  &\forall v_i, v_j \in \overline{D}(S^*), 1 \le i,j \le |\overline{D}(S^*)|, \\
  &i \le j \Longrightarrow \text{DCDist}(S^*, v_i) \le \text{DCDist}(S^*, v_j)
\end{align*}
Let all the vertices in the \refdef{descendants-cover} be $V(S^*) = V \cap \text{DesCover}(S^*)$.
Then $S^*$ maps to a set of \refdef{invalid} subgraphs $\mathcal{O}(S^*) \subseteq \mathcal{O}$ defined below.
\begin{align*}
  \forall k,  0 \le k &\le |\overline{D}(S^*)|, \\
  V_{S_k} &\coloneqq V(S^*) \cup \{ v_i \in \overline{D}(S^*) | \forall i,  1 \le i \le k \} \\
  \mathcal{O}(S^*) &= \{ S_k = (V_{S_k}, E[V_{S_k}]) | \forall k, 0 \le k \le |\overline{D}(S^*)| \}
\end{align*}

Note that here we use the edge set induced by the subgraph $E[V'] = \{ e = (u, v) \in E | u \in V' \land v \in V' \}$ where both $u$ and $v$ must be vertices in the set $V' \subseteq V$.
This is different from the set of edges incident to any vertex $E(V') = \{ e = (u, v) \in E | u \in V' \lor v \in V' \}$.

We then construct the dual variables $y_{S_k}, \forall S_k \in \mathcal{O}(S^*)$.
When there is no non-defect vertex in the \refdef{ring}, we have $S_0 = (V(S^*), E[V(S^*)])$ and $y_{S_0} = y^*_{S^*}$.
In general, they are defined below:
\begin{align*}
  &\forall k, 0 \le k < |\overline{D}(S^*)|, \\
  &\qquad y_{S_k} \coloneqq \text{DCDist}(S^*, v_{k+1}) - \text{DCDist}(S^*, v_{k}) \\
  &\text{if}\ k = |\overline{D}(S^*)|, \\
  &\qquad y_{S_k} \coloneqq \left\{
    \begin{array}{ll}
      y^*_{S^*}, &\text{if } \overline{D}(S^*) = \varnothing\\
      y^*_{S^*} - \text{DCDist}(S^*, v_{k}), &\text{otherwise}
    \end{array}
    \right.
  \end{align*}

  We can simplify the above equations by defining $v_0$ as a point on the boundary of $\text{DesCover}(S^*)$ and $v_{|\overline{D}(S^*)| + 1}$ as a point on the boundary of $\text{Cover}(S^*)$.
  By definition, we have $ \text{DCDist}(S^*, v_0) = 0$ and  $\text{DCDist}(S^*, v_{|\overline{D}(S^*)| + 1}) = y^*_{S^*}$.
  In this way, the equations are simplified to:
  \begin{align*}
    &\forall k, 0 \le k \le |\overline{D}(S^*)|, \\
    &\qquad y_{S_k} \coloneqq \text{DCDist}(S^*, v_{k+1}) - \text{DCDist}(S^*, v_{k})
  \end{align*}

\vspace{1ex}
\definitionlabel{inverse-dual-mapping}{Inverse Dual Mapping}
Given any \refeqs{DLP} solution $\vec{y}$, we define the inverse of \refdef{dual-mapping} $f^{-1}$:
\begin{align*}
  \vec{y^*} = f^{-1}(\vec{y}): \quad y^*_{S^*} = \sum_{S \in \mathcal{O} | V_S \cap D = S^*} y_S, \quad\forall S^* \in \mathcal{O}^*
\end{align*}

\subsection{Dual Mapping Properties}\label{ssec:dual-mapping-properties}

We prove the following theorem about the \refdef{dual-mapping}, with a couple lemmas and their proofs following.

\theoremBlossomMapping{blossom-mapping-with-proof}

\begin{proof}
  We prove that the \refdef{dual-mapping} defined in \S\ref{ssec:dual-mapping-definition} satisfies all the properties of $f$ above, using several lemmas proved later in this section.
  \begin{itemize}
    \item $f$ is bijective because we explicitly construct the \refdef{inverse-dual-mapping} $f^{-1}$ and prove that $f^{-1}(f(\vec{y^*})) = \vec{y^*}$ for any \refdef{blossom-dlp} $\vec{y^*}$ in \reflemma{blossom-inverse-mapping-proof}.
    \item any feasible \refdef{blossom-dlp} $\vec{y^*}$ maps to a feasible \refeqs{DLP} solution $\vec{y} = f(\vec{y^*})$ given \reflemma{dual-mapping-feasibility}.
    \item the dual objectives are the same, i.e., $\sum_{S \in \mathcal{O}} y_S = \sum_{S^* \in \mathcal{O}^*} y^*_{S^*}$, given \reflemma{dual-mapping-sum}.
    \item the existence of tight edge (syndrome graph) implies the existence of minimum-weight tight path (decoding graph) and vice versa, given \reflemma{dual-mapping-tight-edge-eq-tight-path}. Note that the lemma requires the two vertices belong to different nodes. This requirement can be removed because according to \reflemma{frozen-dual-mapping}, once they are merged into the same node, the \refdef{tight-edges} remain tight.
  \end{itemize}
\end{proof}

\definelemma{blossom-inverse-mapping-proof}{Blossom Inverse Mapping}
The \refdef{inverse-dual-mapping} $f^{-1}$ satisfies $f^{-1}(f(\vec{y^*})) = \vec{y^*}$ for any \refdef{blossom-dlp} $\vec{y^*}$.

\begin{proof}
  According to \reflemma{hyperblossom-unique-mapping}, every \refdef{hyperblossom} $S = (V_S, E_S)$ is mapped uniquely from a blossom $S^* = V_S \cap D$.

  We prove the that dual variables are equal for every $S^* \in \mathcal{O}^*$.
  If $S^*$ is a blossom, i.e., $y^*_{S^*} > 0$, then the inverse mapped $y'_{S^*}$ is equal to the original $y^*_{S^*}$:
  \begin{align*}
    y'_{S^*} &= \sum_{S \in \mathcal{O} | V_S \cap D = S^*} y_S \\
    &= \sum_{S \in \mathcal{O}(S^*)} y_S \\
    &\,\downarrow\text{(given \reflemma{dual-mapping-matches-dual-sum})} \\
    &= y^*_{S^*}
  \end{align*}
  On the other hand, if $S^*$ is not a blossom, i.e., $y^*_{S^*} = 0$, then it does not map to any \refdef{invalid} subgraphs and thus any \refdef{invalid} subgraph must have $y_S = 0$ if $V_S \cap D = S^*$.
  The inverse mapped $y'_{S^*} = 0 = y^*_{S^*}$ which is also equal to the original dual variable.
  Since the inverse mapped dual variable is equal to the original dual variable for every $S^* \in \mathcal{O}^*$, we have $f^{-1}(f(\vec{y^*})) = \vec{y^*}$

\end{proof}

\definelemma{hyperblossom-unique-mapping}{Hyperblossom Unique Mapping}
Given a \refdef{blossom-dlp} $\vec{y^*}$ and its corresponding \refdef{dual-mapping} $\vec{y} = f(\vec{y^*})$, a \refdef{hyperblossom} $S = (V_S, E_S)$ of $\vec{y}$ is mapped from a unique blossom $S^* = V_S \cap D$.

\begin{proof}
  Given $S$ is a \refdef{hyperblossom} in $\vec{y} = f(\vec{y^*})$, it must be constructed as the $k$-th \refdef{invalid} subgraph $S = S_k = (V_{S_k}, E[V_{S_k}])$ of some blossom $S^* \in \mathcal{O}^*$ according to the definition of \refdef{dual-mapping}.
  We then prove that $V_{S_k} \cap D = S^*$ and thus $S$ is mapped uniquely from $S^*$:
  \begin{align*}
    V_{S_k} \cap D &= \Big( V(S^*) \cup \{ v_i \in \overline{D}(S^*) | \forall k, 1 \le i \le k \} \Big) \cap D \\
    &= \Big(V(S^*) \cap D \Big) \cup \\
    &\qquad\qquad \Big( \{ v_i \in \overline{D}(S^*) | \forall k, 1 \le i \le k \} \cap D \Big) \\
    &\,\downarrow\text{(given $\overline{D}(S^*) \cap D = \overline{D} \cap D \cap \text{Ring}(S^*) = \varnothing$)} \\
    &= V(S^*) \cap D \\
    &= V \cap \text{DesCover}(S^*) \cap D \\
    &\,\downarrow\text{(given $D \subseteq V$)} \\
    &= \text{DesCover}(S^*) \cap D \\
    &\,\downarrow\text{(by definition of \refdef{descendants-cover})} \\
    &= \Big( \cup_{C^* \in \mathcal{J}(S^*)} \text{Cover}(C^*) \cup S^* \Big) \cap D \\
    &\,\downarrow\text{(given $S^* \subseteq D$)} \\
    &= \bigg( D \cap \Big( \cup_{C^* \in \mathcal{J}(S^*)} \text{Cover}(C^*) \Big) \bigg) \cup S^* \\
  \end{align*}

  We then prove that the term in the big parenthesis of the last step is a subset of $S^*$.
  It suffices to prove that the \refdef{covers} of the \refdef{descendants} of $S^*$ do not contain any other defect vertices apart from $S^*$, i.e., $\forall C^* \in \mathcal{J}(S^*), \text{Cover}(C^*) \cap D \subseteq S^*$.

  We use contradiction to prove the above.
  Suppose there exists $C^* \in \mathcal{J}(S^*)$ and a defect vertex $v \in D \setminus S^*$ so that $v \in \text{Cover}(C^*)$.
  By definition of \refdef{descover-distance}, we have $\text{DCDist}(S^*, v) = 0$.
  On the other hand, given \textbf{Theorem: Node Cover Finite Overlap} in~\cite{wu2023qce}, such an external defect vertex $v \notin S^*$ must not be an internal point of $\text{Cover}(S^*)$.
  According to \reflemma{DCDist-criteria}, we have $\text{DCDist}(S^*, v) \ge y^*_{S^*}$.
  To conclude a contradiction, we simply need to prove $y^*_{S^*} > 0$ because the \refdef{descover-distance} $\text{DCDist}(S^*, v)$ cannot be both $0$ and no less than some positive value $y^*_{S^*}$.
  According to \reflemma{dual-mapping-matches-dual-sum}, we have $y^*_{S^*} = \sum_{S_i \in \mathcal{O}(S^*)} y_{S_i} \ge y_{S_k}$.
  Given $S_k$ is a \refdef{hyperblossom}, we have $y^*_{S^*} \ge y_{S_k} > 0$.
  Thus, we have found a contradiction.

  Applying the conclusion $\forall C^* \in \mathcal{J}(S^*), \text{Cover}(C^*) \cap D \subseteq S^*$ to the previous equation, we have
  \begin{align*}
    V_{S_k} \cap D &= \bigg( D \cap \Big( \cup_{C^* \in \mathcal{J}(S^*)} \text{Cover}(C^*) \Big) \bigg) \cup S^* \\
    &= S^*
  \end{align*}

\end{proof}

\vspace{1ex}
\definelemma{dual-mapping-disjoint-sets}{Dual Mapping Disjoint Sets}
Given two blossoms $S^*_1, S^*_2 \in \mathcal{O}^*, S^*_1 \neq S^*_2$ of a \refdef{blossom-dlp}, their \refdef{dual-mappings} involve two disjoint sets of \refdef{invalid} subgraphs, i.e., $\mathcal{O}(S^*_1) \cap \mathcal{O}(S^*_2) = \varnothing$.

\begin{proof}
  According to \reflemma{hyperblossom-unique-mapping}, all the mapped \refdef{invalid} subgraphs $\mathcal{O}(S^*_i)$ must contain exactly the defect vertices $S^*_i$, i.e., $\forall S \in \mathcal{O}(S^*_i), V_S \cap D = S^*_i$.
  Thus,
  \begin{align*}
    \mathcal{O}(S^*_1) \cap \mathcal{O}(S^*_2) &\subseteq \{ S \in \mathcal{O} | V_S \cap D = S^*_1 \land V_S \cap D = S^*_2 \} \\
    &\,\downarrow\text{(given $S^*_1 \neq S^*_2$)} \\
    &= \varnothing
  \end{align*}

\end{proof}

\vspace{1ex}
\definelemma{DCDist-order}{DesCover Distance Order}
Given a blossom $S^* \in \mathcal{O}^*$, a vertex $v \in V$ and the $k$-th non-defect vertex of the \refdef{dual-mapping} $v_k \in \text{Ring}(S^*)$,
\begin{align*}
  &v \in V_{S_k} \Longrightarrow \text{DCDist}(S^*, v) \le \text{DCDist}(S^*, v_k)\\
  &v \notin V_{S_k} \Longrightarrow \text{DCDist}(S^*, v) \ge \text{DCDist}(S^*, v_{k+1})
\end{align*}

\begin{proof}
  According to \refdef{dual-mapping}, if $v \in V_{S_k} = V(S^*) \cup \{ v_i \in \overline{D}(S^*) | \forall i, 1 \le i \le k \}$, then $v$ is either in $V(S^*)$ or it belongs to $\overline{D}(S^*)$ no later than the position of $v_k$.
  In the first case, $\text{DCDist}(S^*, v) = 0 \le \text{DCDist}(S^*, v_k)$ given \reflemma{DCDist-criteria} (\cref{eq:descover-criteria-1}).
  In the second case, $\text{DCDist}(S^*, v) \le \text{DCDist}(S^*, v_k)$ according to the definition of the sorting in \refdef{dual-mapping}.
  Thus, $$v \in V_{S_k} \Longrightarrow \text{DCDist}(S^*, v) \le \text{DCDist}(S^*, v_k)$$

  Similarly, if $v \notin V_{S_k}$, then $v$ must not be in $V(S^*)$ and must be behind $v_k$ in the sorted list of $\overline{D}(S^*)$ if $v \in \overline{D}(S^*)$.
  We consider two cases: $v \notin \text{Cover}(S^*)$ and $v \in \text{Cover}(S^*)$.
  If $v \notin \text{Cover}(S^*)$, then $\text{DCDist}(S^*, v) > y^*_{S^*} \ge \text{DCDist}(S^*, v_{k+1})$ given \reflemma{DCDist-criteria} (\cref{eq:descover-criteria-2} and \cref{eq:descover-criteria-3}).
  If $v \in \text{Cover}(S^*)$, we further split into two cases.
  First, if $v \in D$, then this external defect vertex $v \notin S^*$ must appear on the boundary of $\text{Cover}(S)$ because otherwise it will violate the \textbf{Theorem: Node Cover Finite Overlap} in~\cite{wu2023qce}.
  In this case, $\text{DCDist}(S^*, v) = y^*_{S^*} \ge \text{DCDist}(S^*, v_{k+1})$.
  Second, if $v \notin D$, then $v$ must be behind $v_k$ in the sorted $\overline{D}(S^*)$, i.e., $\text{DCDist}(S^*, v) \ge \text{DCDist}(S^*, v_{k+1})$.
  Together, we have $$v \notin V_{S_k} \Longrightarrow \text{DCDist}(S^*, v) \ge \text{DCDist}(S^*, v_{k+1})$$

\end{proof}

\vspace{1ex}
\definelemma{dual-mapping-matches-dual-sum}{Blossom Dual Mapping Sum}
Given a blossom $S^* \in \mathcal{O}^*$ from the blossom algorithm, its \refdef{dual-mapping} has the same dual sum.
$$\forall S^* \in \mathcal{O}^*, \sum_{S \in \mathcal{O}(S^*)} y_S = y^*_{S^*}$$

\begin{proof}
  According to the definition of \refdef{dual-mapping},
  \begin{align*}
    \sum_{S \in \mathcal{O}(S^*)} y_S &= \sum_{0 \le k \le |\overline{D}(S^*)|} y_{S_k} \\
    &= \text{DCDist}(S^*, v_{|\overline{D}(S^*)| + 1}) - \text{DCDist}(S^*, v_0) \\
    &= y^*_{S^*} - 0 = y^*_{S^*}
  \end{align*}
\end{proof}

\vspace{1ex}
\definelemma{dual-mapping-sum}{Dual Mapping Sum}
Given a feasible \refdef{blossom-dlp} $\vec{y^*}$, its \refdef{dual-mapping} $\vec{y} = f(\vec{y^*})$ has the same dual sum.
$$\sum_{S \in \mathcal{O}} y_S = \sum_{S^* \in \mathcal{O}^*} y^*_{S^*}$$

\begin{proof}
  According to \reflemma{dual-mapping-disjoint-sets}, each blossom $S^* \in \mathcal{O}^*$ maps to a disjoint subset of dual variables $\mathcal{O}(S^*) \in \mathcal{O}$.
  The other dual variables default to 0, so we have
  \begin{align*}
    \sum_{S \in \mathcal{O}} y_S &= \sum_{S^* \in \mathcal{O}^*} \sum_{S \in \mathcal{O}(S^*)} y_S \\
    &\,\downarrow\text{(\reflemma{dual-mapping-matches-dual-sum})} \\
    &= \sum_{S^* \in \mathcal{O}^*} y^*_{S^*} \\
  \end{align*}
\end{proof}

\vspace{1ex}
\definelemma{hair-dcdist-order}{Hair DesCover Distance Order}
Given a blossom $S^*$ and a \refdef{hyperblossom} $S_k \in \mathcal{O}(S^*) \cap \mathcal{B}$ from its \refdef{dual-mapping}, a \refdef{hair} edge $e = (u, v) \in \delta(S_k)$ satisfies:
\begin{align*}
  \text{DCDist}(S^*, u) &\le \text{DCDist}(S^*, v_k) \\
  &< \text{DCDist}(S^*, v_{k+1}) \le \text{DCDist}(S^*, v)
\end{align*}

\begin{proof}
  We first prove that $u \in V_{S_k}$ and $v \notin V_{S_k}$, and then use \reflemma{DCDist-order} to prove the inequality.

  According to the definition of \refdef{dual-mapping}, $S_k = (V_{S_k}, E[V_{S_k}])$.
  Its edges $E_{S_k} = E[V_{S_k}]$ includes all the edges between vertices $V_{S_k}$.
  Given $\delta(S_k) = E(V_{S_k}) \setminus E[V_{S_k}]$, one of the incident vertex of a \refdef{hair} $e = (u, v) \in \delta(S_k)$ must be in $V_{S_k}$ and the other must not be.
  Without losing generality, we assume $u \in V_{S_k}$, $v \notin V_{S_k}$.

  According to the definition of \refdef{dual-mapping}, the \refdef{invalid} subgraph $S_k$ corresponds to two points $v_k$ and $v_{k+1}$:
  \begin{align*}
    &y_{S_k } = \text{DCDist}(S^*, v_{k+1}) - \text{DCDist}(S^*, v_k) > 0 \\
    &\Longrightarrow \text{DCDist}(S^*, v_k) < \text{DCDist}(S^*, v_{k+1})
  \end{align*}

  According to \reflemma{DCDist-order} and $u \in V_{S_k}$, we have $\text{DCDist}(S^*, u) \le \text{DCDist}(S^*, v_k)$.
  Similarly, given $v \notin V_{S_k}$, we have $\text{DCDist}(S^*, v) \ge \text{DCDist}(S^*, v_{k+1})$.
  Putting together, we have
  \begin{align*}
    \text{DCDist}(S^*, u) &\le \text{DCDist}(S^*, v_k) \\
    &< \text{DCDist}(S^*, v_{k+1}) \le \text{DCDist}(S^*, v)
  \end{align*}

\end{proof}

\vspace{1ex}
\definelemma{ring-contribution-to-edge}{Ring Contributes to Edge}
If a blossom $S^* \in \mathcal{B}^*$ contributes to the edge constraint \cref{eq:dual-constraint-2} of $e \in E$, then its \refdef{ring} overlaps with $e$ for a non-zero length:
$$\sum_{S \in \mathcal{O}(S^*) | e \in \delta(S)} y_S > 0 \Longrightarrow \text{Length}\Big(e \cap \text{Ring}(S^*)\Big) > 0$$

\begin{proof}
  We prove that there exists a non-zero overlapping $e \cap \text{Ring}(S^*)$ when $\sum_{S \in \mathcal{O}(S^*) | e \in \delta(S)} y_S > 0$.
  It suffices to prove that there exists an infinite number of points of $e$ that belongs to the \refdef{ring} of $S^*$.

  Given $\sum_{S \in \mathcal{O}(S^*) | e \in \delta(S)} y_S > 0$, there exists at least one $S_k \in \mathcal{O}(S^*)$ such that $e = (u, v) \in \delta(S_k)$ and $y_{S_k} > 0$.
  According to \reflemma{hair-dcdist-order},
  \begin{align*}
    \text{DCDist}(S^*, u) &\le \text{DCDist}(S^*, v_k) \\
    &< \text{DCDist}(S^*, v_{k+1}) \le \text{DCDist}(S^*, v)
  \end{align*}

  Since the function $\text{DCDict}(S^*, p)$ is a continuous function given the points on the edge $e = (u, v)$, there exists an infinite number of points $p \in e$ such that
  \begin{align*}
    \text{DCDict}(S^*, v_k) < \text{DCDict}(S^*, p) < \text{DCDist}(S^*, v_{k+1})
  \end{align*}

  According to \reflemma{DCDist-criteria} and $\text{DCDict}(S^*, v_k) \ge 0$, $\text{DCDist}(S^*, v_{k+1}) \le y^*_{S^*}$, all such points belong to the \refdef{ring} $p \in \text{Ring}(S^*)$.
  Thus, we have $\text{Length}\Big(e \cap \text{Ring}(S^*)\Big) > 0$.

\end{proof}

\vspace{1ex}
\definelemma{max-ring-contribution}{Max Ring Contribution}
The contribution of a blossom $S^* \in \mathcal{B}^*$ on the constraint \cref{eq:dual-constraint-2} of an edge $e \in E$ is upper bounded by the length of the overlapping segment of $e$ with the \refdef{ring} of $S^*$.
$$\sum_{S_k \in \mathcal{O}(S^*) | e \in \delta(S_k)} y_{S_k} \le \text{Length}(e \cap \text{Ring}(S^*))$$

\begin{proof}
  We consider two cases: $\text{Length}(e \cap \text{Ring}(S^*)) = 0$ or $\text{Length}(e \cap \text{Ring}(S^*)) > 0$.

  If $\text{Length}(e \cap \text{Ring}(S^*)) = 0$, then according to \reflemma{ring-contribution-to-edge}, none of its mapped dual variables $y_{S_k}$ contributes to the edge constraint \cref{eq:dual-constraint-2} of $e$, i.e., $\sum_{S_k \in \mathcal{O}(S^*) | e \in \delta(S_k)} y_{S_k} = 0$.
  In this case, both sides of the equation are zero and they satisfy the inequality.

  If $\text{Length}(e \cap \text{Ring}(S^*)) > 0$, there might be multiple $S_k \in \mathcal{O}(S^*)$ that contribute to the edge, i.e., $e \in \delta(S_k)$ and $y_{S_k} > 0$.
  We first prove that each contributing $S_k$ corresponds to a disjoint segment of $e \cap \text{Ring}(S_k)$ whose length is at least $y_{S_k}$ if $e \in \delta(S_k)$.
  We then prove that the union of all such segments is $e \cap \text{Ring}(S^*)$ up to a finite number of points.
  With these, we can conclude the inequality.

  We define the corresponding segment of $S_k$ as all the points $p \in e$ whose \refdef{descover-distance} is in between $\text{DCDict}(S^*, v_k)$ and $\text{DCDict}(S^*, v_{k+1})$. That is,
  \begin{align*}
    E_k = \{ p \in e | \text{DCDict}(S^*, v_k) &< \text{DCDict}(S^*, p) \\ &< \text{DCDist}(S^*, v_{k+1}) \}
  \end{align*}

  We prove that the segments are disjoint, i.e., $E_i \cap E_j = \varnothing$ for all $i \neq j$.
  Without loss of generality, we assume $i < j$.
  According to the definition of \refdef{dual-mapping}, we have $\text{DCDict}(S^*, v_{i+1}) \le \text{DCDict}(S^*, v_j)$ given $i+1 \le j$.
  That is, the open upper bound of \refdef{descover-distance} in $E_i$ is no larger than the open lower bound of that in $E_j$.
  Thus, the segments are disjoint.

  We then prove that the union of the segments is $e \cap \text{Ring}(S^*)$ up to a finite number of points.
  Given \reflemma{DCDist-criteria} (\cref{eq:descover-criteria-2}), $e \cap \text{Ring}(S^*) = \{ p \in e | 0 < \text{DCDict}(S^*, p) \le y^*_{S^*} \}$.
  According to the definition of \refdef{dual-mapping}, we have $\text{DCDict}(S^*, v_0) = 0$ and $\text{DCDict}(S^*, v_{|\overline{D}(S^*)| + 1}) = y^*_{S^*}$.
  Thus, we have
  \begin{align*}
    \cup_{0 \le k \le |\overline{D}(S^*)|} E_k &= \bigcup_{0 \le k \le |\overline{D}(S^*)|} \Big\{ p \in e \Big| \text{DCDict}(S^*, v_k) <  \\
    &\quad \text{DCDict}(S^*, p) < \text{DCDist}(S^*, v_{k+1}) \Big\} \\
    &\dqns\qqns\qqns\text{($\approx$ up to a finite number of points)} \\
    &\dqns\approx \Big\{ p \in e \Big| \text{DCDict}(S^*, v_0) < \text{DCDict}(S^*, p) \\
    &\dqns\qquad  \le \text{DCDist}(S^*, v_{|\overline{D}(S^*)|+1}) \Big\} \\
    &\dqns= \{ p \in e | 0 < \text{DCDict}(S^*, p) \le y^*_{S^*} \} \\
    &\dqns= e \cap \text{Ring}(S^*)
  \end{align*}

  Finally, we prove that when $e = (u, v) \in \delta(S_k)$, the length of the corresponding segment $E_k$ is at least $y_{S_k}$, i.e., $\text{Length}(E_k) \ge y_{S_k}$.
  According to \reflemma{hair-dcdist-order},
  \begin{align*}
    \text{DCDist}(S^*, u) &\le \text{DCDist}(S^*, v_k) \\
    &< \text{DCDist}(S^*, v_{k+1}) \le \text{DCDist}(S^*, v)
  \end{align*}
  The function $\text{DCDict}(S^*, p)$ is a continuous function for the points on the edge $e = (u, v)$.
  Thus, for every value $l$ between $\text{DCDist}(S^*, v_k)$ and $\text{DCDist}(S^*, v_{k+1})$, there exists a point $p \in e$ whose \refdef{descover-distance} is $l$:
  \begin{align*}
    &\forall l, \text{DCDict}(S^*, v_k) \le l \le \text{DCDict}(S^*, v_{k+1}),\\
    &\quad \exists p \in e, \text{DCDict}(S^*, p) = l
  \end{align*}
  Thus, the length of $E_k$ satisfies
  \begin{align*}
    \text{Length}(E_k) &= \text{Length}\Big(\Big\{ p \in e \Big| \text{DCDict}(S^*, v_k) \\
    &\ \ \quad< \text{DCDict}(S^*, p) < \text{DCDist}(S^*, v_{k+1}) \Big\} \Big) \\
    &\ge \text{DCDist}(S^*, v_{k+1}) - \text{DCDist}(S^*, v_k) \\
    &= y_{S_k}
  \end{align*}

\end{proof}

\vspace{1ex}
\definelemma{dual-mapping-feasibility}{Dual Mapping Feasibility}
Given a \refdef{blossom-dlp} $\vec{y^*}$, its \refdef{dual-mapping} $\vec{y} = f(\vec{y^*})$ is a feasible \refeqs{DLP} solution.

\begin{proof}
  We prove the feasibility of the dual variables $y_S, \forall S \in \mathcal{O}$ by proving that they satisfy both constraints \cref{eq:dual-constraint-1} and \cref{eq:dual-constraint-2} in \refeqs{DLP}.

  First, the dual variables default to $y_S = 0$ satisfying \cref{eq:dual-constraint-1}.
  If a variable $y_{S_k}$ is mapped from a blossom $S^* \in \mathcal{O}^*$, then $y_{S_k} = \text{DCDict}(S^*, v_{k+1}) - \text{DCDict}(S^*, v_{k}) \ge 0$ also satisfies \cref{eq:dual-constraint-1}.

  Second, we prove that for each edge $e \in E$, the constraint \cref{eq:dual-constraint-2} is also satisfied:
  \begin{align*}
    &\sum_{S \in \mathcal{O} | e \in \delta(S)} y_S \\
    &\,\downarrow\text{(given \reflemma{dual-mapping-disjoint-sets})}\\
    &= \sum_{S^* \in \mathcal{O}^*}\ \ \sum_{S_k \in \mathcal{O}(S^*) | e \in \delta(S_k)} y_{S_k} \\
    &\,\downarrow\text{(given \reflemma{max-ring-contribution})} \\
    &\le \sum_{S^* \in \mathcal{O}^*} \text{Length}(e \cap \text{Ring}(S^*)) \\
    &\,\downarrow\text{(given \reflemma{ring-finite-overlap})} \\
    &\le\ w_e
  \end{align*}

\end{proof}

\definelemma{dual-mapping-tight-edge-eq-tight-path}{Dual Mapping Tight Edge and Path}
Given a \refdef{blossom-dlp} $\vec{y^*}$ and \refdef{dual-mapping} $f$, if a syndrome-graph edge $(u, v)$ is tight in $\vec{y^*}$ between two nodes $S^*_1, S^*_2 \in \mathcal{O}^*$ with $u \in S^*_1, v \in S^*_2$, then the minimum-weight path between $u$ and $v$ consists of all \refdef{tight-edges} in the \refeqs{DLP} solution $\vec{y} = f(\vec{y^*})$.

\begin{proof}
  We first prove that all the edges on the minimum-weight path must be fully covered, and then use \reflemma{length-cover-equals-contrib} to show the equality.

  We first prove that $e \in \text{minPath}(u, v)$ must be fully covered.
  This is because every point $p \in \text{minPath}(u, v)$ has $\text{Dist}(p, u) + \text{Dist}(p, v) = w_{(u, v)}$ by definition of the minimum-weight path.
  By definition of a tight edge in the syndrome graph, we have
  \begin{align*}
    \sum_{S^* \in \mathcal{O}^* | (u, v) \in \delta(S^*)} y^*_{S^*} &= \sum_{S^* \in \mathcal{O}^* | u \in S^*} y^*_{S^*} + \sum_{S^* \in \mathcal{O}^* | v \in S^*} y^*_{S^*} \\
    &\,\downarrow\text{($\text{Root}(u) = S^*_1, \text{Root}(v) = S^*_2$)} \\
    &= \qns\sum_{D^* \in \mathcal{P}(S^*_1) | u \in D^*}\qns y^*_{D^*} + \qns\sum_{D^* \in \mathcal{P}(S^*_2) | v \in D^*}\qns y^*_{D^*} \\
    &= w_{(u, v)} \\
    &= \text{Dist}(p, u) + \text{Dist}(p, v)
  \end{align*}
  Thus, we have either $\text{Dist}(p, u) \le \sum_{D^* \in \mathcal{P}(S^*_1) | u \in D^*} y^*_{D^*}$ or $\text{Dist}(p, v) \le \sum_{D^* \in \mathcal{P}(S^*_2) | v \in D^*} y^*_{D^*}$, or both.
  According to the definition of \refdef{cover}, $p$ either belongs to $\text{Cover}(S^*_1)$ or $\text{Cover}(S^*_2)$, or both.
  Thus, we have $\text{Length}((\text{Cover}(S^*_1) \cup \text{Cover}(S^*_2)) \cap e) = w_e$.

  We then prove $\text{Contrib}(S_1^*) + \text{Contrib}(S_2^*) = w_e$:
  \begin{align*}
    &\text{Contrib}(S_1^*) + \text{Contrib}(S_2^*) \\
    &\,\downarrow\text{(by \reflemma{length-cover-equals-contrib})} \\
    &= \text{Length}(\text{Cover}(S^*_1) \cap e) + \text{Length}(\text{Cover}(S^*_2) \cap e) \\
    &\,\downarrow\text{(by \textbf{Theorem: Node Cover Finite Overlap}~\cite{wu2023qce})} \\
    &= \text{Length}((\text{Cover}(S^*_1) \cup \text{Cover}(S^*_2)) \cap e) \\
    &= w_e
  \end{align*}

  Given \reflemma{dual-mapping-disjoint-sets}, we have $\mathcal{O}(D_1^*) \cap \mathcal{O}(D_2^*) = \varnothing$ for any two blossoms $D^*_1, D^*_2 \in \mathcal{B}^*, D^*_1 \neq D^*_2$, and thus
  \begin{align*}
    \sum_{S \in \mathcal{O} | e \in \delta(S)} y_S &= \sum_{S^* \in \mathcal{B}^*}\ \sum_{S \in \mathcal{O}(S^*) | e \in \delta(S)} y_S \\
    &\ge \sum_{S^* \in \mathcal{P}(S^*_1) \cup \mathcal{P}(S^*_2)}\ \sum_{S \in \mathcal{O}(S^*) | e \in \delta(S)} y_S \\
    &\,\downarrow\text{($\mathcal{P}(S^*_1) \cap \mathcal{P}(S^*_2) = \varnothing$ for nodes $S^*_1 \neq S^*_2$)} \\
    &= \sum_{S^* \in \mathcal{P}(S^*_1)}\ \sum_{S \in \mathcal{O}(S^*) | e \in \delta(S)} y_S + \\
    &\quad\quad\quad \sum_{S^* \in \mathcal{P}(S^*_2)}\ \sum_{S \in \mathcal{O}(S^*) | e \in \delta(S)} y_S \\
    &= \text{Contrib}(S^*_1) + \text{Contrib}(S^*_2) \\
    &= w_e
  \end{align*}
  Also, we have $\sum_{S \in \mathcal{O} | e \in \delta(S)} y_S \le w_e$ because $\vec{y} = f(\vec{y^*})$ is a feasible \refeqs{DLP} solution satisfying \cref{eq:dual-constraint-2}, according to \reflemma{dual-mapping-feasibility}.
  Together, we have $\sum_{S \in \mathcal{O} | e \in \delta(S)} y_S = w_e$, i.e., $e$ is a \refdef{tight-edge}.
  This concludes the proof that every edge on the minimum-weight path between $u$ and $v$ is a \refdef{tight-edge}.
\end{proof}

\definelemma{length-cover-equals-contrib}{Length Cover Equals Contrib}
Given a node $S^* \in \mathcal{B}^*$ and any edge $e \in \text{minPath}(u, v)$ between an internal defect $u \in S^*$ and an external defect $v \in D \setminus S^*$,
\begin{align*}
  &\text{Length}(\text{Cover}(S^*) \cap e) = \text{Contrib}(S^*), \\
  &\quad\text{where}\ \text{Contrib}(S^*) = \sum_{D^* \in \mathcal{P}(S^*)}\ \ \sum_{S \in \mathcal{O}(D^*) | e \in \delta(S)} y_S
\end{align*}

\begin{proof}
  We prove it using mathematical induction.
  At the beginning of the algorithm, they are equal because $\text{Length}(\text{Cover}(S^*) \cap e) = \text{Contrib}(S^*) = 0$.
  When the covered segment is starting to increase because of the growth of the node $S^*$, we prove that by growing for a small length $l$, both $\text{Length}(\text{Cover}(S^*) \cap e)$ and $\text{Contrib}(S^*)$ increase by the same amount $l$.

  First, given that the edge $e$ is on the minimum-weight path between $u$ and $v$, the distance $\text{Dist}(u, p)$ is monotonically increasing for the points on the edge $e$.
  Thus, according to the definition of \refdef{cover}, the covered segment $\text{Cover}(S^*) \cap e$ is also increasing for the same length $l$.

  Second, according to the blossom algorithm, the only changing dual variable among $\mathcal{P}(S^*)$ is the node $S^*$ itself.
  Thus, $y^*_{S^*}$ increases by $l$.
  According to the definition of \refdef{dual-mapping}, only the outmost $S_k \in \mathcal{O}(S^*)$ changes for the same amount of $l$.
  Together, we have $$\Delta \text{Contrib}(S^*) = \Delta \dqns \sum_{S \in \mathcal{O}(S^*) | e \in \delta(S)} \dqns y_S =
  \begin{cases}
    \Delta y_{S_k} &\text{if } e \in \delta(S_k)\\
    0 &\text{otherwise},
  \end{cases}$$
  Thus, we only need to prove $e \in \delta(S_k)$ in order to prove that $\text{Contrib}(S^*)$ also increases by $l$.

  Without losing generality, we assume $e = (a, b)$ where $a \in \text{Cover}(S^*)$ and $b \notin \text{Cover}(S^*)$.
  Given $a$ cannot be on the boundary of $\text{Cover}(S^*)$ after it grows for a small length $l$, we have $a \in \overline{D}$ according to \textbf{Theorem: Node Cover Finite Overlap}~\cite{wu2023qce}.
  The vertices of $S_k$ includes all the vertices in the \refdef{descendants-cover} of $S^*$ as well as those non-defect vertices within a \refdef{descover-distance} of $y^*_{S^*}$.
  Thus, $a \in V_{S_k}$ and as a result, $e \in E(V_{S_k})$.
  Similarly, $b \notin V_{S_k}$ and as a result, $e \notin E[V_{S_k}]$.
  Given $E_{S_k} \subseteq E[V_{S_k}]$, we have $e \notin E_{S_k}$.
  Thus, by the definition of \refdef{hair}, we have $e \in E(V_{S_k}) \setminus E_{S_k} = \delta(S_k)$.

  Together, $\text{Length}(\text{Cover}(S^*) \cap e) = \text{Contrib}(S^*)$ for any edge $e \in \text{minPath}(u, v)$.
\end{proof}

\definelemma{frozen-dual-mapping}{Frozen Dual Mapping}
The \refdef{dual-mapping} of a blossom $S^*$, including $\mathcal{O}(S^*)$ and their corresponding dual variables $\{ y_{S_k}, S_k \in \mathcal{O}(S^*) \}$, does not change once $S^*$ merges into a parent blossom $S_p^*$.

\begin{proof}
  We use a similar proof to that of \reflemma{frozen-ring}, with the properties of the blossom algorithm.
  Once $S^*$ merges into another blossom $S_p^*$ where $S^* \subsetneq S_p^*$, according to the blossom algorithm~\cite{wu2023qce}, the dual variables $y^*_{S^*}$ of these \refdef{descendants} $S^* \in \mathcal{J}(S^*_p)$ do not change, unless $S^*$ becomes a node again (under an event of blossom expansion).

  It then suffices to prove that the $\mathcal{O}(S^*)$ and their dual variables do not change when it is among the \refdef{descendants} of $S_p^*$.
  According to the definition of \refdef{dual-mapping}, $\mathcal{O}(S^*)$ and their dual variables only depends on the \refdef{descendants-cover} of $S^*$ and the value of $y^*_{S^*}$.
  The value of $y^*_{S^*}$ does not change because $S^* \in \mathcal{J}(S^*_p)$.
  Thus, we only need to prove that the \refdef{descendants-cover} of $S^*$ does not change.

  By definition of \refdef{cover}, it only depends on the dual variables of its \emph{Progeny} $\mathcal{P}(S^*)$.
  The \emph{Progeny} of $S^*$ is among the \refdef{descendants} of $S^*_p$, i.e., $\mathcal{P}(S^*) = \mathcal{J}(S^*) \cup \{ S^*\}\subseteq \mathcal{J}(S^*_p)$.
  Thus, the \refdef{covers} of all \refdef{descendants} of $S^*$ does not change.
  According to the definition of \refdef{descendants-cover}, $\text{DesCover}(S^*)$ does not change.

  Together, the \refdef{dual-mapping} of $S^*$ is frozen, i.e., $\mathcal{O}(S^*)$ and their corresponding dual variables $\{ y_{S_k}, S_k \in \mathcal{O}(S^*) \}$ do not change.

\end{proof}

\subsection{minLP = minILP for Simple Graphs}\label{ssec:mwpf-condition-simple-graph}

\theoremSimpleGraphOptimality{simple-graph-optimality-with-proof}

\begin{proof}
  We prove \ref{condition:mwpf} using the inequality chain \cref{eq:mwpf-chain} by finding a pair of feasible parity factor $\vec{x}$ and feasible \refeqs{DLP} solution $\vec{y}$ with the same weight $\sum_{e \in E} w_e x_e = \sum_{S \in \mathcal{O}} y_S$ for any simple graph $G = (V, E)$.

  An MWPM decoder finds a pair of feasible parity factor $\vec{x}$ and \refdef{blossom-dlp} $\vec{y^*}$ with the same weight $\sum_{e \in E} w_e x_e = \sum_{S^* \in \mathcal{O}^*} y^*_{S^*}$~\cite{dennis2002topological,wu2022interpretation}.
  We then use \refdef{dual-mapping} to map $\vec{y^*}$ to a \refeqs{DLP} solution $\vec{y} = f(\vec{y^*})$.
  Given \reflemma{dual-mapping-sum},
  $$\sum_{S \in \mathcal{O}} y_S = \sum_{S^* \in \mathcal{O}^*} y^*_{S^*} = \sum_{e \in E} w_e x_e$$
  Also, given \reflemma{dual-mapping-feasibility}, $\vec{y}$ is a feasible \refeqs{DLP} solution.
\end{proof}

\subsection{Alternating Tree Reconstruction}\label{ssec:alternating-tree-reconstruction}

We prove that an alternating tree data structure can be reconstructed from a \refdef{blossom-dlp} $\vec{y^*}$ alone, without any other information.
In this way, we have access to the blossom algorithm data structures even though the \refdef{relaxer} finder only has $\vec{y} = f(\vec{y^*})$ as input.

\theoremAlternatingTreeReconstruction{alternating-tree-reconstruction-with-proof}

\begin{proof}

  We use a constructive proof by explicitly describing such an algorithm.
  We first reconstruct the hierarchy of blossoms from the \refdef{blossom-dlp} $\vec{y^*}$, and then use the tight syndrome-graph edges between the reconstructed nodes to rebuild alternating trees and matched pairs, as shown in \cref{algo:alternating-tree-reconstruction}.

  Given all the blossoms $\mathcal{B}^* = \{ y^*_{S^*} > 0 | S^* \in \mathcal{O}^* \}$ of $\vec{y^*}$, we can reconstruct the hierarchy of blossoms.
  As shown in \cref{algo:alternating-tree-reconstruction} line \ref{line:iterate-blossoms}, we do a bottom-up reconstruction by iterating through $\mathcal{B^*}$ from small blossoms to large blossoms.
  In this way, a child blossom is always visited before its parent blossom.
  Initially, each defect vertex corresponds to a node $\{v\}, \forall v \in D$ (line \ref{line:initial-nodes}).
  Once a blossom $S^*$ is visited, its children blossoms are removed from $\text{Nodes}$ and their parent blossom $S^*$ is added to $\text{Nodes}$ (line \ref{line:update-nodes}).

  \begin{figure}[t]
    \begin{algorithm}[H]
      \caption{Alternating Tree Reconstruction}\label{algo:alternating-tree-reconstruction}
      \begin{algorithmic}[1]
        \Require{$G$, $D$ (defects), $\vec{y^*}$ (\refdef{blossom-dlp})}
        \Ensure{nodes in alternating trees and matched pairs}
        \Procedure{AlternatingTreeReconstruct}{$G, D, \vec{y^*}$}
        \State $\mathcal{B}^* \gets \{ y^*_{S^*} > 0 | S^* \in \mathcal{O}^* \}$ \label{line:remaining-blossoms}
        \State $\text{Nodes} \gets \{ \{v\} | v \in D \}$ \label{line:initial-nodes}
        \While{$\mathcal{B}^* \neq \varnothing$}
        \State $S^* \gets \arg\min_{S^* \in \mathcal{B}^*} |S^*|$ \Comment{the smallest blossom} \label{line:iterate-blossoms}
        \State $\text{Des} \gets \{ S^*_c \in \text{Nodes} | S_c^* \subseteq S^* \}$ \Comment{children}
        \State $\text{assert}\ S^* \equiv \cup_{S^*_c \in \text{Des}} S_c^*$ \Comment{form a blossom}
        \State $\text{Nodes} \gets (\text{Nodes} \setminus \text{Des}) \cup \{ S^* \}$ \label{line:update-nodes}
        \State $\mathcal{B}^* \gets \mathcal{B}^* \setminus \{ S^* \}$
        \EndWhile
        \State $T^* \gets \{ e \in E^* | \sum_{S^* \in \mathcal{O}^* | e \in \delta(S^*)} y^*_{S^*} = w_e \}$ \Comment{compute tight syndrome-graph edges}
        \State $\Delta y_{S^*} \gets +1, \forall S^* \in \text{Nodes}$ \Comment{nodes grow by default} \label{line:grow-nodes}
        \For{$e \in T^*$} \Comment{resolve \emph{Conflicts}~\cite{wu2025asplos} between nodes}
        \State $\Call{PrimalPhaseResolve}{\text{Nodes}, \text{Conflict}(e)}$ \label{line:resolve-conflicts}
        \EndFor
        \State \Return $\text{Nodes}$
        \EndProcedure
      \end{algorithmic}
    \end{algorithm}
  \end{figure}

  With the nodes, we reconstruct the alternating trees and matched pairs by iterating over the tight syndrome-graph edges.
  To do that, we first let all the nodes grow $\Delta y^*_{S^*} = +1, \forall S^* \in \text{Nodes}$ (line \ref{line:grow-nodes}).
  We then call the Primal Phase of the Parity Blossom algorithm~\cite{wu2023qce} to address the \emph{Conflicts}~\cite{wu2025asplos} between the nodes (line \ref{line:resolve-conflicts}), each corresponding to a tight syndrome-graph edge.
  During this process, the nodes are organized into alternating trees and matched pairs.

\end{proof}

\subsection{More Efficient Heterogeneous QEC Decoding}\label{ssec:heterogeneous-qec-decoding}

Although the \hyperblossom algorithm with the \emph{Blossom} \refdef{relaxer} finder (\S\ref{ssec:blossom-subroutine}) is step-by-step equivalent to the MWPM decoder, it is not as efficient as running the MWPM decoder directly.
Here we describe a more efficient way of decoding a heterogeneous QEC code~\cite{stein2025hetec}, where the decoding graph is partially simple.
The idea is to directly run the MWPM decoder for the matchable subgraph, and then map the optimal \refdef{blossom-dlp} solution to an initial feasible \refeqs{DLP} solution, so that the \hyperblossom algorithm can make progress on top of the initial solution.

A \emph{matchable subgraph} $G' = (V', E[V'])$ is an induced subgraph of the decoding hypergraph $G = (V, E)$ where $|e| \le 2, \forall e \in E[V']$.
We treat the vertices $V_b = \{ v \in V' | E(v) \not\subseteq E[V'] \}$ as temporary boundaries~\cite{wu2023qce}, since they could be incident to some external hyperedges.
There must exists a feasible \refdef{blossom-dlp} for $G'$ because we only relax the parity constraints on the boundary vertices $V_b$ and that $G$ is a \refdef{valid} graph.
In this case, we can use an MWPM decoder to find an optimal \refdef{blossom-dlp} for $G'$.

Once the MWPM decoder finishes, we then map the optimal \refdef{blossom-dlp} $\vec{y^*}$ to a feasible \refeqs{DLP} solution of $G'$ using \refdef{dual-mapping}.
In this case, a feasible \refeqs{DLP} solution $\vec{y} = f(\vec{y^*})$ of the subgraph $G'$ is also a \refeqs{DLP} solution of $G$, according to \reflemma{subgraph-O-subset}.

To see why $\vec{y}$ is a feasible \refeqs{DLP} solution on $G$, we use the property of the temporary boundaries $V_b$: any boundary vertex $v \in V_b$ is not strictly inside any \refdef{cover}, i.e., either $v$ does not belong to any \refdef{cover} or $v$ is on the boundary of a \refdef{cover}.
Since the \refdef{covers} never grow beyond any boundary vertex $v \in V_b$, all the external hyperedges $E \setminus E[V']$ are not propagated by any blossom, i.e., $\forall e \in E \setminus E[V'], \forall S^* \in \mathcal{B}^*, \text{Length}(e \cap \text{Ring}(S^*)) = 0$.
According to \reflemma{max-ring-contribution}, we have $\sum_{S_k \in \mathcal{O}(S^*) | e \in \delta(S_k)} y_{S_k} = 0$.
That is, the \refeqs{DLP} solution $\vec{y}$ satisfies \cref{eq:dual-constraint-2} for every hyperedge, i.e., $\forall e \in E \setminus E[V'], \sum_{S \in \mathcal{O} | e \in \delta(S)} y_S = 0$.
The \refeqs{DLP} solution $\vec{y}$ also satisfies \cref{eq:dual-constraint-2} for all edges in $E[V']$ because it is a feasible \refeqs{DLP} solution of $G'$.
Thus, $\vec{y}$ satisfies \cref{eq:dual-constraint-2} for all hyperedges $e \in E = (E \setminus E[V']) \cup E[V']$.
Also, $\vec{y}$ satisfies \cref{eq:dual-constraint-1} for all $S \in \mathcal{O}$ because $y_S = 0, \forall S \in \mathcal{O} \setminus \mathcal{O}'$.
Together, $\vec{y}$ is a feasible \refeqs{DLP} solution on $G$.

The \hyperblossom algorithm then takes the feasible \refeqs{DLP} solution $\vec{y} = f(\vec{y^*})$ as the initial solution, and makes progress on top of it.
This is more efficient than running the \emph{Blossom} \refdef{relaxer} finder, bypassing all the complication of the \hyperblossom algorithm for this matchable subgraph.

\end{document}